%% file: main-corr-withoutcomments.tex
\documentclass[titlepage,10pt,leqno]{article}

\usepackage{multirow}
\usepackage{eepic,epic}
\usepackage{epsfig}
\usepackage{graphicx}
\usepackage{color}

\usepackage{ifthen}
\usepackage{mathptmx}
\usepackage{amssymb}
\usepackage{amsmath}
\usepackage{pifont}



\newcommand{\OMIT}[1]{} %

\newcommand{\copelandalpha}{\mbox{\rm{}Copeland$^{\alpha}$}}
\newcommand{\copelandone}{\mbox{\rm{}Copeland$^{1}$}}
\newcommand{\copelandzero}{\mbox{\rm{}Copeland$^{0}$}}
\newcommand{\copelandalphairrational}{\mbox{\rm{}Copeland$^{\alpha}_{\rm{}Irrational}$}}
\newcommand{\copelandalphavarirrational}[1]{\mbox{\rm{}Copeland$^{{#1}}_{\rm{}Irrational}$}}
\newcommand{\copelandzeroscore}{\mathit{score}^0}
\newcommand{\copelandonescore}{\mathit{score}^1}
\newcommand{\copelandalphascore}{\mathit{score}^\alpha}
\newcommand{\copelandalphavarscore}[1]{\mathit{score}^{#1}}

\newenvironment{proofs}{\noindent{\bf Proof.}\hspace*{1em}}{\literalqed\bigskip}
\def\literalqed{{\ \nolinebreak\hfill\mbox{\qedblob\quad}}}

\newcommand{\sproof}{\noindent{\bf Proof.}\hspace*{1em}}
\newcommand{\sproofof}[1]{\noindent{\bf Proof of {#1}.}\hspace*{1em}}
\newcommand{\eproofof}[1]{\noindent{\hspace*{0.1in} \hfil \hfill \mbox{\literalqed{} {#1}}}\quad\bigskip}

\newcommand{\condition}{\,\mid \:}

\newcommand{\sbribery}[1]{{\probbf \mbox{\rm{}#1}\hbox{-}\allowbreak\mbox{bribery}}}
\newcommand{\sdestbribery}[1]{{\probbf \mbox{\rm{}#1}\hbox{-}\allowbreak\mbox{destructive}\hbox{-}\allowbreak\mbox{bribery}}}

\usepackage{ifthen}
\newcommand{\hugeDebug}{false}
\ifthenelse{\equal{\hugeDebug}{false}}{
\newcommand{\normalspacing}{\singlespacing}
}
{
\newcommand{\normalspacing}{\niceninespacing}
}
\ifthenelse{\equal{\hugeDebug}{false}}{
\setlength{\oddsidemargin}{0.25in}
\setlength{\evensidemargin}{\oddsidemargin}
\setlength{\textwidth}{6in}
\setlength{\textheight}{8in}
\setlength{\topmargin}{-0.0in}
}
{
\setlength{\oddsidemargin}{-0.4in}
\setlength{\evensidemargin}{\oddsidemargin}
\setlength{\textwidth}{5.3in}
}

\ifthenelse{\equal{\hugeDebug}{false}}{
\pagestyle{plain}
}
{
\pagestyle{headings}
}

\makeatletter
\newcommand{\singlespacing}{\let\CS=
\@currsize\renewcommand{\baselinestretch}{1}\tiny\CS}
\newcommand{\singlespacingplus}{\let\CS=
\@currsize\renewcommand{\baselinestretch}{1.25}\tiny\CS}
\newcommand{\doublespacing}{\let\CS=
\@currsize\renewcommand{\baselinestretch}{1.75}\tiny\CS}
\newcommand{\extradoublespacing}{\let\CS=
\@currsize\renewcommand{\baselinestretch}{1.9}\tiny\CS}
\newcommand{\draftspacing}{\let\CS=
\@currsize\renewcommand{\baselinestretch}{2.0}\tiny\CS}
\newcommand{\hugedraftspacing}{\let\CS=
\@currsize\renewcommand{\baselinestretch}{2.4}\tiny\CS}
\newcommand{\niceonespacing}{\let\CS=\@currsize\renewcommand{\baselinestretch}{1.1}\tiny\CS}
\newcommand{\nicetwospacing}{\let\CS=\@currsize\renewcommand{\baselinestretch}{1.2}\tiny\CS}
\newcommand{\nicethreespacing}{\let\CS=\@currsize\renewcommand{\baselinestretch}{1.3}\tiny\CS}
\newcommand{\singlespacingplusplus}{\let\CS=\@currsize\renewcommand{\baselinestretch}{1.35}\tiny\CS}
\newcommand{\nicefourspacing}{\let\CS=\@currsize\renewcommand{\baselinestretch}{1.4}\tiny\CS}
\newcommand{\nicefivespacing}{\let\CS=\@currsize\renewcommand{\baselinestretch}{1.5}\tiny\CS}
\newcommand{\nicesixspacing}{\let\CS=\@currsize\renewcommand{\baselinestretch}{1.6}\tiny\CS}
\newcommand{\nicesevenspacing}{\let\CS=\@currsize\renewcommand{\baselinestretch}{1.7}\tiny\CS}
\newcommand{\niceeightspacing}{\let\CS=\@currsize\renewcommand{\baselinestretch}{1.8}\tiny\CS}
\newcommand{\niceninespacing}{\let\CS=\@currsize\renewcommand{\baselinestretch}{1.9}\tiny\CS}
\makeatother
\normalspacing

\makeatletter \@beginparpenalty=10000 \makeatother

\mathcode`\0="0030      %
\mathcode`\1="0031
\mathcode`\2="0032
\mathcode`\3="0033
\mathcode`\4="0034
\mathcode`\5="0035
\mathcode`\6="0036
\mathcode`\7="0037
\mathcode`\8="0038
\mathcode`\9="0039

\newcount\hour  \newcount\minutes  \hour=\time  \divide\hour by 60
\minutes=\hour  \multiply\minutes by -60  \advance\minutes by \time
\def\mmmddyyyy{\ifcase\month\or Jan\or Feb\or Mar\or Apr\or May\or Jun\or Jul\or
  Aug\or Sep\or Oct\or Nov\or Dec\fi \space\number\day, \number\year}
\def\hhmm{\ifnum\hour<10 0\fi\number\hour :%
  \ifnum\minutes<10 0\fi\number\minutes}
\let\reversenotation\overleftarrow
\makeatletter
\def\@cite#1#2{[#1\if@tempswa , #2\fi]}
\makeatother

\makeatletter
\def\@citex[#1]#2{\if@filesw\immediate\write\@auxout{\string\citation{#2}}\fi
  \def\@citea{}\@cite{\@for\@citeb:=#2\do
    {\@citea\def\@citea{,\linebreak[0]}\@ifundefined
       {b@\@citeb}{{\bf ?}\@warning
       {Citation `\@citeb' on page \thepage \space undefined}}%
\hbox{\csname b@\@citeb\endcsname}}}{#1}}
\makeatother

\makeatletter
\def\@cite#1#2{[#1\if@tempswa , #2\fi]}
\def\@citex[#1]#2{\if@filesw\immediate\write\@auxout{\string\citation{#2}}\fi
  \def\@citea{}\@cite{\@for\@citeb:=#2\do
    {\@citea\def\@citea{,\kern1pt\linebreak[0]}\@ifundefined
       {b@\@citeb}{{\bf ?}\@warning
       {Citation `\@citeb' on page \thepage \space undefined}}%
\hbox{\csname b@\@citeb\endcsname}}}{#1}}
\makeatother

\newcommand\qedblob{\mbox{\ding{113}}}
\def\qedsymbol{{\ \nolinebreak\hfill\mbox{\qedblob\quad}}\smallskip}

\newcommand{\probbf}{\rm}

\newcommand{\spmicrobribery}[1]{{\probbf \mbox{\rm{}#1}\hbox{-}microbribery}}
\newcommand{\spdestmicrobribery}[1]{{\probbf \mbox{\rm{}#1}\hbox{-}destructive\hbox{-}microbribery}}

\newcommand{\scontrol}[2]{{\probbf \mbox{#1}\hbox{-}\mbox{\rm{}#2}}}

\newcommand{\naturals}{\mathbb{N}}
\newcommand{\integers}{\mathbb{Z}}

\newcommand{\cost}{\mathit{cost}}
\newcommand{\wincost}{\mathit{wincost}}
\newcommand{\tiecost}{\mathit{tiecost}}
\newcommand{\flowcost}{\mathit{flowcost}}
\newcommand{\flowvalue}{\mathit{flowvalue}}

\newcommand{\promote}{\mathit{promote}}
\newcommand{\demote}{\mathit{demote}}

\newcommand{\score}{\mathit{score}}
\newcommand{\llullscore}{\copelandonescore}

\newcommand{\versus}{\mathrm{vs}}

\newcommand{\electionsystem}{\ensuremath{\cal E}}
\newcommand{\calS}{{\cal S}}
\newcommand{\pad}{{\rm{}Pad}}

\newtheorem{theorem}{Theorem}[section]

\newtheorem{corollary}[theorem]{Corollary}
\newtheorem{definition}[theorem]{Definition}
\newtheorem{lemma}[theorem]{Lemma}
\newtheorem{observation}[theorem]{Observation}

\newtheorem{proposition}[theorem]{Proposition}
\newtheorem{example}[theorem]{Example}
\newtheorem{notation}[theorem]{Notation}

\newtheorem{claim}[theorem]{Claim}

\newcommand{\p}{\ensuremath{\mathrm{P}}}
\newcommand{\np}{\ensuremath{\mathrm{NP}}}

\title{%
Llull and Copeland Voting Computationally Resist Bribery and Control\thanks{%
Supported in part by DFG grants RO-1202/\{9-3, 11-1, 12-1\},
NSF grants CCR-0311021, CCF-0426761, and IIS-0713061,
the Alexander von Humboldt Foundation's TransCoop
program, 
the European Science Foundation's EUROCORES program LogICCC,
and two Friedrich Wilhelm Bessel Research Awards.
Work done in part during visits by the first three authors to
Heinrich-Heine-Universit\"at D\"usseldorf and during visits by the fourth
author to the University of Rochester.
This paper combines and extends University of Rochester Computer Science Department
Technical Reports TR-913 (February 2007) and TR-923 (October 2007).
Some results have been presented at the 22nd AAAI Conference on
Artificial Intelligence (AAAI-07)~\cite{fal-hem-hem-rot:c:llull},
at the October 2007 Dagstuhl Seminar on Computational Issues in Social Choice,
at the 4th International Conference on Algorithmic Aspects in Information
and Management (AAIM-08)~\cite{fal-hem-hem-rot:c:llull-aaim}, and
at the 2nd International Workshop on Computational Social Choice (COMSOC-08).} }

\author{Piotr Faliszewski \\ 
        Department of Computer Science \\
        University of Rochester \\
        Rochester, NY 14627, USA
\and
        Edith Hemaspaandra \\
        Department of Computer Science \\
        Rochester Institute of Technology \\
        Rochester, NY 14623, USA
\and
        Lane A. Hemaspaandra \\ 
        Department of Computer Science \\
        University of Rochester \\
        Rochester, NY 14627, USA
\and
        J{\"o}rg Rothe \\
        Institut f\"ur Informatik \\
        Heinrich-Heine-Universit{\"a}t D{\"u}sseldorf  \\
        40225 D\"usseldorf, Germany
       }

\date{September 28, 2008}

\begin{document}
\sloppy

\maketitle
\begin{abstract}  
Control and bribery are settings in which an external agent seeks to
influence the outcome of an election.
Constructive
control of elections refers to attempts by an agent to, via such
actions as addition/deletion/partition of candidates or voters, ensure
that a given candidate wins~\cite{bar-tov-tri:j:control}. 
Destructive control refers to attempts by an agent to, via the same
actions, preclude a given candidate's
victory~\cite{hem-hem-rot:j:destructive-control}.
An election system in which an agent 
can sometimes affect the result and it can be determined in polynomial time
on which inputs the agent can succeed is said to be 
vulnerable to the given type of control.
An election system in which an agent 
can sometimes affect the result, yet in which it is NP-hard 
to recognize the inputs on which the agent can succeed,
is said to be resistant
to the given type of control.

Aside from election systems with an $\np$-hard winner problem, 
the only systems previously known to be
resistant to all the standard control types were highly artificial
election systems created by hybridization~\cite{hem-hem-rot:c:hybrid}.
This paper studies a parameterized version
of Copeland voting, denoted by $\copelandalpha$, where the parameter
$\alpha$ is a rational number between $0$ and $1$ that specifies how
ties are
valued
in the pairwise comparisons of candidates. 
In every
previously studied
constructive or destructive
control scenario,
we determine which of 
resistance or vulnerability holds 
for $\copelandalpha$ for each
rational~$\alpha$, $0 \leq \alpha \leq 1$.  In particular, we prove that
Copeland$^{0.5}$, the system commonly referred to as ``Copeland
voting,'' provides full resistance to constructive control,
and we prove the same for $\copelandalpha$, for all rational $\alpha$,
$0 < \alpha < 1$.
Among systems with a polynomial-time winner problem, Copeland voting is the
first natural election system proven to have 
full resistance to constructive control.
In addition, we prove that both Copeland$^0$ and Copeland$^1$
(interestingly, $\copelandone$ is an election system developed by
the thirteenth-century mystic Ramon Llull)
are resistant to all standard types of
constructive control other than 
one variant of addition of candidates.  
Moreover, we show that
for each rational~$\alpha$, $0 \leq \alpha \leq 1$,
$\copelandalpha$ voting is
fully
resistant to bribery attacks,
and we establish fixed-parameter tractability of bounded-case control
for $\copelandalpha$.

We also study $\copelandalpha$ elections under more flexible
models such as microbribery and extended control,
we integrate the potential irrationality of
voter preferences into many of our results, and
we prove our results in both the unique-winner model and
the nonunique-winner model.  Our vulnerability results for
microbribery are proven via a novel technique involving min-cost
network flow.
\end{abstract}

\newpage
\tableofcontents
\newpage

\section{Introduction}
\label{sec:introduction}

\subsection{Some Historical Remarks: Llull's and Copeland's Election Systems}
\label{sec:introduction:llull-copeland}

Elections have played an important role in human 
societies for thousands of years.  
For example, elections were of central importance 
in the democracy of ancient Athens.
There citizens typically could only agree (vote \emph{yes\/})
or disagree (vote \emph{no\/}) with the speaker, and simple
majority-rule was in effect.
The mathematical study of elections, give or take a 
few discussions by the ancient Greeks and Romans, was 
until recently thought to have been
initiated only a few hundred years ago, namely in the breakthrough
work of Borda and Condorcet---later in part reinvented
by Dodgson
(see, e.g.,~\cite{mcl-urk:b:polsci:classics} for reprints of these
classic papers). 
One of the most interesting results of this early work is
Condorcet's observation~\cite{con:b:condorcet-paradox} 
that if one conducts elections with more than
two alternatives then even if all voters have rational
(i.e.,
transitive) preferences,
the society 
in aggregate can be irrational 
(indeed, can have 
cycles of 
strict preference).
Nonetheless, Condorcet
believed that if there exists a candidate $c$ such that $c$ 
defeats each other candidate in head-to-head contests then $c$ 
should win the election (see, e.g.,~\cite[p.~114]{mcl-urk:b:polsci:classics}).  Such a candidate is called a
Condorcet winner. Clearly, there can be at most one Condorcet
winner in any election, and there might be none. 

This understanding of history has been reconsidered during the 
past few decades, as it has been discovered that 
the study of
elections was considered deeply as
early as
the thirteenth
century 
(see
H\"agele and Pukelsheim~\cite{hae-puk:j:electoral-writings-ramon-llull} 
and the citations 
therein regarding Ramon Llull and the fifteenth-century
figure Cusanus, especially the citations 
that in~\cite{hae-puk:j:electoral-writings-ramon-llull} are numbered 3, 5, and 24--27). 
Ramon Llull (b.~1232, d.~1315), a Catalan 
mystic,
missionary, and
philosopher
developed an election system that (a) has an efficient 
winner-determination
procedure and (b) elects a Condorcet winner whenever
one exists and otherwise elects candidates that are, in a certain sense,
closest to being Condorcet winners.  

Llull's motivation for
developing an election system was to obtain a method of choosing 
abbesses, abbots, bishops, and perhaps even the pope. His
election ideas never gained public acceptance in medieval Europe and 
were long forgotten.

It is interesting to note that Llull allowed voters to have so-called
\emph{irrational\/} preferences. Given three candidates, $c$, $d$, and
$e$, it was perfectly acceptable for a voter to prefer $c$ to $d$, $d$
to~$e$, and $e$ to~$c$. 
On the other hand, in 
modern studies of voting and
election systems each voter's preferences are 
most typically 
modeled as a
linear order over all candidates.
(In this paper, as is common when discussing elections, ``linear
order'' implies strictness, i.e., no tie in the ordering; that is, by
``linear order'' we mean a strict, complete order, i.e., an
irreflexive, antisymmetric, complete, transitive relation.)
Yet allowing irrationality is very
tempting and natural.  Consider Bob, who likes to eat out but
is often in a hurry.  Bob prefers diners to fast food because he is
willing to wait a little longer to get better food. Also, given a choice
between a fancy restaurant and a diner he prefers the fancy restaurant,
again because he is willing to wait somewhat longer to get better
quality. However, given the choice between a fast-food place and
a fancy restaurant Bob might reason that he is not willing to wait 
so much longer to be served  at
the fancy restaurant and so will choose
fast food instead. Thus
regarding catering options, Bob's preferences are irrational in our
sense, i.e., intransitive. When voters make their
choices based on multiple criteria---a very common and natural 
occurrence
both among humans and software agents---such irrationalities can occur.  

Llull's election system 
is remarkably similar to 
what is now known as 
``Copeland elections''~\cite{cop:m:copeland},  
a more than half-century old voting procedure that is
based on pairwise comparisons of candidates: The winner (by a 
majority of votes---in this paper ``majority'' always, as is standard,
means strict majority) of each such head-to-head contest is 
awarded 
one
point and the loser is awarded zero points; 
in ties, both parties are (in the most common interpretation of
Copeland's meaning) awarded half a point;
whoever collects the most
points over all these contests (including tie-related points)
is the election's winner.  In fact, the
point value awarded 
for ties in such head-to-head majority-rule contests is treated
in two ways in the literature when speaking of Copeland elections:
half a point (most common) and zero points (less common). 
To provide a framework that
can capture both those notions, as well as capturing Llull's system and
the whole family of systems created by choices of how we value
ties, we
propose and introduce a parameterized
version of Copeland elections, denoted by $\copelandalpha$, where the
parameter $\alpha$ is a rational number, $0 \leq \alpha \leq 1$, and
in the case of a tie both candidates receive $\alpha$ points.  
So 
the system widely referred to in the literature as
``Copeland elections'' is Copeland$^{0.5}$, where tied candidates
receive half a point each (see, e.g., Merlin and
Saari~\cite{saa-mer:j:copeland1,mer-saa:j:copeland2}; the definition
used by Conitzer et al.~\cite{con-lan-san:j:when-hard-to-manipulate}
can be scaled to be equivalent to Copeland$^{0.5}$).
Copeland$^0$, where
tied candidates come away empty-handed, has sometimes 
also been referred to 
as
``Copeland elections''
(see, e.g., Procaccia, Rosenschein, and
Kaminka~\cite{pro-ros-kam:c:noisy} and an early version of this
paper~\cite{fal-hem-hem-rot:c:llull}).
The above-mentioned
election 
system proposed by 
Ramon Llull in the thirteenth century is in this notation
Copeland$^1$, where tied candidates are awarded one
point each, just like winners of head-to-head
contests.\footnote{%
\label{f:one}%
Page~23 of 
H\"agele and Pukelsheim~\cite{hae-puk:j:electoral-writings-ramon-llull} 
indicates in a way we 
find deeply convincing
(namely by a direct quote of Llull's 
in-this-case-very-clear words 
from 
his
\emph{Artifitium Electionis Personarum}---which was 
rediscovered by those authors in the year 2000) 
that 
at least one of 
Llull's election systems was 
Copeland$^1$, and
so in this paper we refer to the 
both-candidates-score-a-point-on-a-tie variant as Llull voting.  

In some settings Llull 
required the candidate and voter 
sets to be identical and had an elaborate two-stage tie-breaking
rule ending in randomization. We disregard these issues here
and cast his system into the modern idiom for election systems.
(However, we note in passing that there do exist some modern papers
in which the voter and candidate sets are taken to be identical, see for 
example the work of and references 
in~\cite{alt-ten:c:axiomatic-personalized-ranking}.)}
The group stage of the FIFA World Cup finals is in essence a collection
of Copeland$^{\frac{1}{3}}$ tournaments. %

At first glance, one might be tempted to think that the definitional
perturbation due to the parameter $\alpha$ in $\copelandalpha$
elections is negligible.  However, it in fact can make the dynamics of
Llull's system
quite different from those of,
for instance, Copeland$^{0.5}$ or Copeland$^0$.
Specific examples witnessing this claim, both regarding
complexity results and regarding their proofs, are given at the end
of Section~\ref{sec:introduction:results}.

Finally,
we mention that a probabilistic variant of
Copeland voting (known as the Jech method) was defined already in 1929 by
Zermelo~\cite{zer:j:jech} and later on was reintroduced by several
other researches (see, e.g., the paper of Levin and
Nalebuff~\cite{lev-nal:j:voting-rules} for further references and a
description of the Jech method). We note in passing that the
Jech method is applicable even when it is fed
incomplete information. In the present paper, however,
we do not consider incomplete-information or probabilistic scenarios,
although we commend such settings as interesting for future work.

\subsection{Computational Social Choice}
\label{sec:introduction:comsoc}

In general it is impossible to design a perfect
election system.  
In the 1950s 
Arrow~\cite{arr:b:polsci:social-choice} 
famously showed
that there is no social choice
system that satisfies a certain small set of arguably
reasonable
requirements, and later Gibbard~\cite{gib:j:polsci:manipulation},
Satterthwaite~\cite{sat:j:polsci:manipulation}, 
and 
Duggan and Schwartz~\cite{dug-sch:j:polsci:gibbard} 
showed that
any natural election system can sometimes be manipulated by strategic voting,
i.e., by a voter revealing different preferences than his or her true ones
in order to affect an election's result in his or her favor.  Also, no
natural election system with a polynomial-time winner-determination procedure
has yet been shown to be resistant to all
types of control via procedural changes.  
Control refers to attempts by an external agent (called ``the chair'')
to, via such actions as addition/deletion/partition of candidates or
voters, make a given candidate win the election (in the case of
constructive control~\cite{bar-tov-tri:j:control}) or preclude a
given candidate's victory (in the case of destructive
control~\cite{hem-hem-rot:j:destructive-control}).

These obstacles are very discouraging, but the field of computational
social choice theory grew in part from the realization that
computational complexity provides 
a potential shield against manipulation/control/etc.
In particular, 
around 1990, Bartholdi,
Tovey, and
Trick~\cite{bar-tov-tri:j:manipulating,bar-tov-tri:j:control}
and Bartholdi and Orlin~\cite{bar-oli:j:polsci:strategic-voting}
brilliantly observed that while we perhaps might not be able to make
manipulation (i.e., strategic voting) and control of elections
impossible, we could at least try to make such manipulation and control 
so computationally difficult that neither voters nor election
organizers will attempt it.  For example, if there is a way for
a committee's chair to set up an election within the committee in such a
way that his or her favorite option is guaranteed to win, but 
the chair's computational task would 
take a million years, then for all practical purposes
we may feel that the chair is prevented from finding such a set-up.

Since the seminal work of Bartholdi, Orlin, Tovey, and Trick, a large
body of research has
been dedicated to the study of computational
properties of election systems. Some topics that have received much
attention are the complexity of manipulating
elections \cite{con-san:c:voting-tweaks,con-san:c:nonexistence,con-lan-san:j:when-hard-to-manipulate,elk-lip:c:polsci:universal-tweaks-coalitions,hem-hem:j:dichotomy,pro-ros:j:juntas,pro-ros-zoh:c:multiwinner}
and of controlling elections via procedural
changes~\cite{hem-hem-rot:j:destructive-control,hem-hem-rot:c:hybrid,pro-ros-zoh:c:multiwinner,erd-now-rot:t-With-MFCS08-Ptr:sp-av}.
Recently, Faliszewski, Hemaspaandra, and
Hemaspaandra introduced the study of the complexity of bribery in elections
(\cite{fal-hem-hem:c:bribery}, see also~\cite{fal:c:nonuniform-bribery}). 
Bribery shares some features of manipulation and
some features of control.  In particular, the briber picks the voters he or she
wants to affect (as in voter control problems) and asks them to vote as he
or she wishes (as in manipulation). (For additional citations and pointers,
see the recent survey~\cite{fal-hem-hem-rot:btoappearWithTrPtr:richer}.)

In this paper we study
$\copelandalpha$ elections with respect to
the computational complexity of bribery and procedural control; 
see~\cite{fal-hem-sch:c:copeland-ties-matter}
for a study of manipulation within $\copelandalpha$.

The
study of election systems and their computational properties, such as
the complexity of their manipulation, control, and bribery problems,
is an important topic in  multiagent systems.
Agents/voters
may have different,
often conflicting, individual preferences over the given alternatives
(or candidates) and  voting rules (or, synonymously, election systems)
provide a useful method for agents to come to a ``reasonable'' 
decision
on which alternative to choose.
Thus elections can be employed in multiagent settings and also in
other contexts to solve many practical problems.
As just a few examples we mention the work of Ephrati
and Rosenschein~\cite{eph-ros:j:multiagent-planning} where
elections are used for planning, the work of
Ghosh et al.~\cite{gho-mun-her-sen:c:voting-for-movies} who developed
a recommender system for movies that is based on voting,
and the work of
Dwork et al.~\cite{dwo-kum-nao-siv:c:rank-aggregation} where
elections are used to aggregate results from multiple web-search
engines. In a multiagent setting we may have hundreds of elections
happening every minute and we cannot hope to carefully check in
each case whether the party that organized the election attempted some
procedural change to skew the results. However, if it is computationally
hard to find such procedural changes then we can hope it is
practically infeasible for the organizers to undertake them.

A standard technique for showing that a particular election-related
problem (for example, the problem of deciding whether the chair can make his
or her favorite candidate
a winner by influencing at most $k$
voters not to cast their votes) is computationally
intractable is to
show that it is $\np$-hard. This approach is taken in almost all
the papers on computational social choice
cited above, and it is the approach that we take in this paper.
One of the justifications for
using $\np$-hardness as a
barrier against manipulation and control of elections
is that in multiagent
settings any attempts to influence the election's outcome are made by
computationally bounded software agents that have neither human
intuition nor the computational ability to solve $\np$-hard problems.

Recently, such papers as 
\cite{con-san:c:nonexistence,pro-ros:j:juntas,hem-hom:jtoappearWithPtr:dodgson-greedy,mcc-pri-sli:j:dodgeson} 
have studied
the frequency (or sometimes, probability weight) of correctness
of heuristics for 
voting problems.  
Although this is a fascinating and important direction, it does not
at this point remove the need to study worst-case hardness. Indeed, we
view worst-case study as a natural prerequisite to a
frequency-of-hardness attack: After all, there is no point in seeking
frequency-of-hardness results if the problem at hand is in~$\p$ to
begin with.
And if one cannot even prove worst-case hardness for a problem, then
proving ``average-case'' hardness is even more beyond reach.
Also, 
current frequency results have 
debilitating 
limitations (for example, 
being locked into specific distributions;
depending on unproven assumptions; 
and adopting ``tractability'' notions that declare undecidable
problems tractable and that are not robust under 
even linear-time reductions).  
These models are 
arguably
not 
ready for prime time
and, contrary to some people's 
impression, 
fail to 
imply 
average-case polynomial runtime claims.  
\cite{erd-hem-rot-spa:c:lobbying,hem-hom:jtoappearWithPtr:dodgson-greedy} 
provide
discussions
of some of these issues.

\subsection{Outline of Our Results}
\label{sec:introduction:results}

The goal of this paper is to study
$\copelandalpha$ elections
from the point of view of computational social choice
theory, in the setting where voters are rational and in the setting 
where
voters are allowed to have irrational preferences.  (Note: 
When we henceforward say ``irrational 
voters,'' we mean that the voters may have irrational
preferences, not that they each must.) We study the issues of bribery
and control and we point the reader to the work of Faliszewski, Hemaspaandra,
and Schnoor~\cite{fal-hem-sch:c:copeland-ties-matter} for
work on manipulation.
(Very briefly summarized, the work of Faliszewski, Hemaspaandra, and
Schnoor~\cite{fal-hem-sch:c:copeland-ties-matter} on manipulation of
$\copelandalpha$ elections shows that for all rational $\alpha$, $0 <
\alpha < 1$, $\alpha \neq \frac{1}{2}$, the coalitional manipulation
problem in unweighted $\copelandalpha$ elections, even for
coalitions of just two manipulators, is $\np$-complete. Some of the
constructions of the present paper have been adopted or adapted in
that paper in order to prove results about manipulation.)

Bribery and control problems have some very natural real-life
interpretations.  For example, during presidential elections a
candidate might want to encourage as many of his or her supporters as
possible to vote (``get-out-the-vote'' efforts): control via addition
of voters; elections can be held at an inconvenient date for a group
of voters (e.g., a holiday) or at a hard-to-reach location (e.g.,
requiring one to own a car, or such that getting to the location
involves passing dangerous areas): control via deleting voters;
one can choose voting districts in a way favorable to a particular
candidate or party (gerrymandering): control via partitioning
voters;
one can introduce a new
candidate to the election in the hope that he or she will steal votes
away from the opponents of one's favorite candidate without affecting the favorite
candidate's performance: control via adding candidates.  All the other
control scenarios that we study also have natural interpretations.

Similarly, bribery is a natural and important issue in the context of
elections.  We stress, however, that bribery problems do not
necessarily need to correspond to cheating or any sort of illegal
action. One could view bribery problems as, for example,
problems of
finding the minimum
number of voters who can switch the result of the
election and, thus, as problems of finding coalitions, especially if
one assigns prices to voters to measure the difficulty of convincing a
particular voter to join the coalition (see, e.g., the paper of
Faliszewski~\cite{fal:c:nonuniform-bribery} for an example of a
bribery problem where such an interpretation is very natural).

It is quite natural to study control and bribery
both in constructive settings (where we want to make our favorite candidate
a winner) and in destructive settings (where we try to prevent a
candidate from winning). In the context of real-life elections, one
often hears voters speaking of which candidate they hope will win,
but one also often hears voters expressing the sentiment ``Anyone
but \emph{him}.''
The constructive and destructive settings correspond to
actions that agents belonging to these groups might be interested in.

One of the main achievements of this paper is to
classify which of resistance or vulnerability holds 
for $\copelandalpha$ in every
previously studied control scenario for each rational value 
of~$\alpha$, $0 \leq \alpha \leq 1$.
In doing so, we provide
an example of a control problem where the complexity of
Copeland$^{0.5}$ (which is the
system commonly referred to as ``Copeland'')
differs from that of both Copeland$^0$ and
Copeland$^1$: While the latter two problems are vulnerable to
constructive control by adding (an unlimited number of) candidates,
Copeland$^{0.5}$ is resistant to this control type (see
Section~\ref{sec:prelims} for definitions and Theorem~\ref{thm:ccacu}
for this result).

In fact, Copeland (i.e., Copeland$^{0.5}$) 
is the first natural election system (with a
polynomial-time winner problem)
proven to be resistant to every
type of constructive control that has been proposed in the literature
to date.  This result closes a 15-year quest for a natural election
system fully resistant to constructive control.
We also show that
$\copelandalpha$ is resistant to both constructive and destructive
bribery, for both the case of rational voters and the case of irrational voters. Our
hardness
proofs work for the case of unweighted voters without price tags
(see~\cite{fal-hem-hem:c:bribery}) and thus, naturally, apply as well to 
the more involved scenarios of weighted unpriced voters, unweighted priced
voters, and weighted priced voters. 

To prove our bribery results, we introduce a method of 
controlling the relative
performances of certain voters in such a way that, if one sets up
other restrictions appropriately, the legal possibilities for
bribery actions are sharply constrained.  We call our approach
``the UV technique,'' since it is based on dummy candidates $u$ and $v$\@.
The proofs of Theorems~\ref{thm:bribery:dest-copelandalpha} and~\ref{thm:bribery:cons-copelandalpha} are 
particular
applications of this method.
We feel that the UV
technique will be useful, even beyond the
scope of this paper, for the analysis of bribery in other
election systems based on head-to-head contests.

We also study $\copelandalpha$ elections under more flexible models
such as ``microbribery'' (see
Section~\ref{sec:microbribery:vulnerability}) and ``extended control''
(see Section~\ref {sec:control-fpt}).  
We show that $\copelandalpha$ (with irrational voters allowed) is
vulnerable to destructive microbribery and to destructive candidate
control via providing fairly simple greedy algorithms.
In contrast,
our polynomial-time algorithms for constructive microbribery are
proven via a technique involving min-cost network flows. To the best
of our knowledge, this is the first application of min-cost flows
to election problems. We believe that the range
of applicability of flow networks to election problems extends well
beyond microbribery for $\copelandalpha$ elections and we point the
reader to a recent, independent paper by Procaccia, Rosenschein, and
Zohar~\cite{pro-ros-zoh:j:proportional-representation}\footnote{Procaccia,
  Rosenschein, and
  Zohar~\cite{pro-ros-zoh:j:proportional-representation} independently
  of our work in~\cite{fal-hem-hem-rot:c:llull} used a similar
  technique in their work regarding the complexity of achieving
  proportional representation.}
and to a paper by Faliszewski~\cite{fal:c:nonuniform-bribery} for examples
of such applications.

We also mention that during our study of Copeland control
we noticed that the proof of an
important result of Bartholdi, Tovey, and 
Trick~\cite[Theorem~12]{bar-tov-tri:j:control} 
(namely, that
Condorcet voting is resistant to constructive control by
deleting voters) is invalid.  The invalidity is due to the proof
centrally using nonstrict voters, in violation of
Bartholdi, Tovey, and Trick's 
\cite{bar-tov-tri:j:control} (and our) model, and the invalidity
seems potentially daunting or impossible to fix with the proof approach taken there.  
We note also that Theorem~14 of the same paper has a similar flaw.
In Section~\ref{sec:control-condorcet} we
validly reprove their claimed results using our techniques.

As mentioned in Section~\ref{sec:introduction:llull-copeland},
$\copelandalpha$ elections may behave quite differently depending on
the value of the tie-rewarding parameter~$\alpha$. We now give
concrete examples to make this case.
Specifically, proofs of results for $\copelandalpha$ occasionally
differ considerably for distinct values of~$\alpha$,
and in some cases even the computational complexity of
various
control and manipulation problems (for the manipulation case 
see~\cite{fal-hem-sch:c:copeland-ties-matter})
may jump between $\p$ membership and $\np$-completeness depending
on~$\alpha$.
Regarding control, we have already noted that Theorem~\ref{thm:ccacu}
shows that some control problem
(namely, control by adding an unlimited number of candidates)
for $\copelandalpha$ is $\np$-complete
for each rational $\alpha$ with $0 < \alpha < 1$,
yet Theorem~\ref{thm:unlimited-adding-constructive}
shows that same control problem to be in $\p$ for $\alpha \in \{0,1\}$.
To give another example involving a different control problem,
namely control by partition of candidates with the ties-eliminate
tie-handling rule (see Section~\ref{sec:prelims}), we note that
the proofs of Theorem~\ref{thm:ccpc-te-1} (which applies to $\alpha = 1$
for this control problem within $\copelandalpha$)
and of Theorem~\ref{thm:ccpc-te-smaller-than-1} (which applies to all
rational $\alpha$ with $0 \leq \alpha < 1$ for the same problem)
differ substantially.
Regarding constructive microbribery, the vulnerability constructions for
$\alpha = 0$ (see Lemma~\ref{lem:bribery-prime-even-copeland-ties}) and
$\alpha = 1$ (see Lemma~\ref{lem:bribery-prime-even-llull-ties})
significantly
differ from each other, and neither of them works for tie-rewarding values
other than $0$ and $1$.
The above remarks notwithstanding, for most of our results we show that 
it is possible to obtain a unified---though due to this uniformity
sometimes rather involved---construction that works
for $\copelandalpha$ for every rational~$\alpha$, $0 \leq
\alpha \leq 1$.

\subsection{Organization}
\label{sec:introduction:organization}

This paper is organized as follows.  In Section~\ref{sec:prelims}, we
formalize the notion of elections and in particular of
$\copelandalpha$ elections, we introduce some useful notation, and we formally
define the control and bribery problems we are interested in.
In Section~\ref{sec:bribery},
we show that for each rational $\alpha$, $0 \leq \alpha \leq 1$,
$\copelandalpha$ elections are
fully resistant to bribery, both in the case 
of rational voters and in the case of irrational voters.
On the other hand, if one changes the 
bribery model 
to allow ``microbribes'' of 
voters 
(a fine-grained approach to bribery, in which the more one changes a voter's 
vote, the more one has to pay the voter), 
we prove vulnerability for each rational $\alpha$, $0 \leq \alpha \leq 1$, in the 
irrational-voters destructive case and for some specific values of $\alpha$ in the
irrational-voters constructive case.
In Sections~\ref{sec:control-candidate} and~\ref{sec:control-voter}, we present our results on
procedural control
for $\copelandalpha$ elections for each rational
$\alpha$ with $0 \leq \alpha \leq 1$. 
We will see that very broad resistance holds for the constructive-control cases.
Section~\ref{sec:control-fpt} presents
our results on fixed-parameter tractability of
bounded-case control for $\copelandalpha$.
Section~\ref{sec:control-condorcet} provides valid proofs for several
control problems for Condorcet elections (studied by Bartholdi, Tovey,
and Trick~\cite{bar-tov-tri:j:control}) whose original proofs were 
invalid due to being at odds with
the model of elections used in~\cite{bar-tov-tri:j:control}.  
We conclude the paper with a
brief summary in Section~\ref{sec:conclusions} and by stating some
open problems.

\section{Preliminaries}
\label{sec:prelims}

\subsection{Elections: The Systems of Llull and Copeland}

An \emph{election} $E = (C,V)$ consists of
a finite candidate set $C = \{c_1, \ldots, c_m\}$ and
a finite
collection
$V$ of 
voters, where each voter is represented
(individually, except later when we discuss succinct 
inputs) via his or her preferences over
the candidates.
An \emph{election system} (or an \emph{election rule}) 
is a rule that determines the winner(s) of each
given election, i.e., a mapping from pairs $(C,V)$ to
subsets of~$C$.

We consider two ways
in which voters can express their preferences.  In the \emph{rational} case (our
default case),
each voter's preferences are represented as a linear order
 over the set $C$,\footnote{In this paper, we take ``linear order'' to mean
a strict total order. This is a common convention within voting theory, see,
e.g., the book of Austen-Smith and Banks~\cite{aus-ban:b:positive-political-I}.
However, we mention that 
in the field of mathematics 
the term ``linear order'' is typically taken to allow nonstrictness, i.e.,
to allow ties.}
i.e.,
each voter 
$v_i$
has
a \emph{preference list} $c_{i_1} > c_{i_2} >
\cdots > c_{i_m}$, with
$\{i_1, i_2, \ldots , i_m\} = 
\{1, 2, \ldots , m\}$. 
In the \emph{irrational} case,
each voter's preferences are
represented as a \emph{preference table} that for every 
unordered
pair of 
distinct candidates
 $c_i$ and $c_j$ in~$C$
indicates whether the voter prefers $c_i$ to $c_j$ (i.e., $c_i >
c_j$) or prefers $c_j$ to $c_i$ (i.e., $c_j > c_i$).

Some well-known election rules for the case of rational voters are
plurality, Borda count, and Condorcet.  \emph{Plurality} elects the
candidate(s) that are ranked first by the largest number of
voters.  \emph{Borda count} elects the candidate(s) that receive 
the most points,
where each voter $v_i$ gives each candidate $c_j$ as many points as the 
number of 
candidates $c_j$ is preferred to with respect to $v_i$'s preferences.
A candidate $c_i$ is a \emph{Condorcet winner} if
for every other candidate $c_j$ 
it holds that
$c_i$ is preferred to $c_j$ by a majority of voters.
Note that each election instance will have at most one 
Condorcet winner. 

In this
paper, we introduce a parameterized version of Copeland's election
system~\cite{cop:m:copeland},
which we denote by $\copelandalpha$, where the
parameter $\alpha$ is a rational number between $0$ and $1$ that
specifies how ties are rewarded in the head-to-head majority-rule
contests between any two distinct candidates.

\begin{definition}
\label{def:copeland}
  Let $\alpha$, $0 \leq \alpha \leq 1$, be a rational number.
  In a $\copelandalpha$ election, for each head-to-head
  contest between
  two distinct candidates, if some candidate is preferred by a majority of
  voters then he or she obtains one point and the other candidate
  obtains zero points, and if a tie occurs then both candidates obtain
  $\alpha$ points.
  Let $E = (C,V)$ be an election.  For each $c \in C$,
  $\copelandalphascore_{E}(c)$ is (by definition) the sum of $c$'s $\copelandalpha$
  points in~$E$.  Every candidate $c$ with maximum
  $\copelandalphascore_{E}(c)$ 
  (i.e., every candidate $c$ satisfying  
  $(\forall d \in C)[\copelandalphascore_E(c) \geq \copelandalphascore_E(d)]$) wins.

  Let $\copelandalphairrational$ denote the same election system but
  with voters allowed to be irrational.
\end{definition}

As mentioned earlier, 
in the literature the term ``Copeland elections'' is most often used
for the system Copeland$^{0.5}$ (e.g., \cite{saa-mer:j:copeland1,mer-saa:j:copeland2} 
and a rescaled version of~\cite{con-lan-san:j:when-hard-to-manipulate}), 
but has occasionally been used for
Copeland$^0$ (e.g.,~\cite{pro-ros-kam:c:noisy} and an early version of this paper~\cite{fal-hem-hem-rot:c:llull}).  
As mentioned earlier, the system Copeland$^1$ was
proposed by Llull in the thirteenth century (see the
literature pointers given in the introduction) and so is called Llull voting.

We now define some notation to help in the discussion of
$\copelandalpha$ elections.  Informally put, if $E = (C,V)$ is an
election and if $c_i$ and $c_j$ are any two candidates in $C$ then by
$\versus_E(c_i,c_j)$ we mean the surplus of votes that candidate $c_i$
has over $c_j$. Formally, we define this notion as follows.

\begin{definition}
Let $E = (C,V)$ be an election and let $c_i$
and $c_j$ be two arbitrary candidates from~$C$.
Define the \emph{relative vote-score of $c_i$ with respect to $c_j$} by
\[
\versus_E(c_i,c_j) = \left\{
\begin{array}{ll}
0 & \mbox{if $c_i = c_j$} \\
\| \{v \in V \condition v \mbox{ prefers $c_i$ to $c_j$}\} \| - 
\| \{v \in V \condition v \mbox{ prefers $c_j$ to $c_i$}\} \|
  & \mbox{otherwise.}
\end{array}
\right.
\] 
\end{definition}
So, if $c_i$ defeats $c_j$ in a 
head-to-head contest in $E$ then
$\versus_E(c_i,c_j) > 0$, if they are tied then $\versus_E(c_i,c_j) =
0$, and if $c_j$ defeats $c_i$ 
then $\versus_E(c_i,c_j) < 0$. 
(Throughout this paper, ``defeats'' excludes the possibility of a tie,
i.e., ``defeats'' means ``(strictly) defeats.'' We will say
``ties-or-defeats''
when we wish to allow a tie to suffice.)
Clearly,
$\versus_E(c_i,c_j) = -\versus_E(c_j,c_i)$.
We often speak, in the plural, of relative vote-scores when we mean a
group of results of head-to-head contests between particular
candidates.

Let $\alpha$, $0 \leq \alpha \leq 1$, be a rational number.
Definition~\ref{def:copeland} introduced $\copelandalphascore_E(c)$, the
$\copelandalpha$ score of candidate $c$ in election~$E$. 
Note that for each candidate $c_i \in C$,
\begin{eqnarray*}
\copelandalphascore_E(c_i) & = &
\|\{ c_j \in C \condition c_i \neq c_j \mbox{ and } \versus_E(c_i,c_j) > 0\}\|
\\
& & {}+ \alpha
\|\{ c_j \in C \condition c_i \neq c_j \mbox{ and } \versus_E(c_i,c_j) = 0\}\|.
\end{eqnarray*}
In particular, we have $\copelandalphavarscore{0}_E(c_i) = \|\{ c_j
\in C \condition c_i \neq c_j \mbox{ and } \versus_E(c_i,c_j) > 0\}\|$, 
and $\copelandalphavarscore{1}_E(c_i) = \|\{ c_j \in
C \condition c_i \neq c_j \mbox{ and } \versus_E(c_i,c_j) \geq 0\}\|$.  
Note further that the highest
possible
$\copelandalpha$ score in any election $E = (C,V)$ is $\|C\|-1$.

Recall that a candidate $c_i \in C$ is a
$\copelandalpha$ winner of $E = (C,V)$ if for all
$c_j \in C$ it holds that
$\copelandalphascore_{E}(c_i) \geq \copelandalphascore_{E}(c_j)$.
(Clearly, some elections can have more than one $\copelandalpha$ winner.)
A candidate $c_i$ is
a Condorcet winner of $E$ if
$\copelandalphavarscore{0}_E(c_i) = \|C\|-1$, that is,
if $c_i$ defeats all other candidates
in head-to-head contests.

In many of our constructions to be presented in the upcoming proofs, 
we use the following notation for rational voters.

\begin{notation}
Within every election we fix some arbitrary order
over
the candidates.
Any occurrence of a subset $D$ of candidates in a preference
list means the candidates from $D$ are listed with respect to that
fixed order. Occurrences of $\reversenotation{D}$ mean the 
same except
that the candidates from $D$ are listed in the reverse order.
\end{notation}
For example, if $C = \{a,b,c,d,e\}$, with the alphabetical order being
used,
and $D = \{a,c,e\}$ then $b > D > d$
means $b > a > c > e > d$, and $b > \reversenotation{D} > d$ means $b >
e > c > a > d$.

\subsection{Bribery and Control Problems}
\label{ss:problems}

We now describe the computational problems that we study in this
paper.  Our problems come in two flavors: constructive and
destructive. In the constructive version the goal is to determine whether, 
via the
bribery or control action type under study, it is possible to make a
given candidate a
winner of the election.  In the destructive case
the goal is to determine whether it is possible to prevent a
given candidate from being a winner of the election.

Let $\electionsystem$ be an election system. In our case,
$\electionsystem$ will be either $\copelandalpha$ or
$\copelandalphairrational$, where $\alpha$, $0 \leq \alpha \leq 1$, is
a rational number.
The bribery problem for $\electionsystem$ with rational voters is
defined as follows~\cite{fal-hem-hem:c:bribery}.

\begin{description}
\item[Name:] $\sbribery{\electionsystem}$ and $\sdestbribery{\electionsystem}$.
\item[Given:] A set $C$ of candidates, a
collection
$V$ of voters 
specified via their preference lists over $C$, 
a distinguished candidate $p \in C$, and a nonnegative integer $k$.
\item[Question (constructive):] Is it possible to make $p$ a winner of the
  $\electionsystem$ election resulting from $(C,V)$ by
   modifying the preference lists of at most $k$ voters?
\item[Question (destructive):] Is it possible to ensure that $p$ is
  not a winner of the $\electionsystem$ election resulting from
  $(C,V)$ by modifying the preference lists of at most $k$ voters?
\end{description}

The version of this problem for elections with irrational voters allowed
is defined 
exactly like the rational one, with the only difference being that voters are
represented via preference tables rather than preference lists, and
the briber may completely change a voter's preference table at
unit cost.
At the end of the present section, Section~\ref{ss:problems}, we
will describe the variants based on seeking to make $p$ be
(or to preclude $p$ from being) a \emph{unique} winner.
Later in the paper we will study
another variant of bribery problems---a variant in which 
one is allowed to perform microbribes:
bribes for which the cost depends 
on each preference-table entry change,
and the briber pays separately for each such change.

Bribery problems seek to change the outcome of
elections via
modifying 
the reported preferences of some of the voters.
In contrast, control problems seek to change the outcome 
of an election
by modifying
the election's structure via adding/deleting/partitioning either candidates
or voters.  When formally defining these control types,
we use the following naming conventions for the corresponding control
problems.  The name of a control problem starts with the election
system used (when clear from context, it may be omitted), followed by
CC for ``constructive control'' or by DC for ``destructive control,''
followed by the acronym of the type of control: AC for ``adding (a
limited number of) candidates,'' AC$_{\rm u}$ for ``adding (an
unlimited number of) candidates,'' DC for ``deleting candidates,'' PC
for ``partition of candidates,'' RPC for ``run-off partition of
candidates,'' AV for ``adding voters,'' DV for ``deleting voters,''
and PV for ``partition of voters.''  
All the partitioning cases
(PC, RPC, and PV) are two-stage elections, and we here use both
tie-handling rules of Hemaspaandra, Hemaspaandra, and
Rothe~\cite{hem-hem-rot:j:destructive-control} for first-stage
subelections in these two-stage elections.  In particular, for all the
partitioning cases, the acronym PC, RPC, and PV, respectively, is
followed by the acronym of the tie-handling rule used in first-stage
subelections, namely TP for ``ties promote'' (i.e., all winners of 
first-stage subelections are promoted to the final round of the
election) and TE for ``ties eliminate'' (i.e., only unique winners
of first-stage subelections are promoted to the final round of
the election, so if there is more than one winner in a given
first-stage subelection or there is no winner in a given first-stage
subelection then that subelection does not move any of its candidates
forward).

We now formally define
our control problems. These definitions
are due to
Bartholdi, Tovey, and Trick~\cite{bar-tov-tri:j:control} for
constructive control and to
Hemaspaandra, Hemaspaandra, and
Rothe~\cite{hem-hem-rot:j:destructive-control} for destructive control.

Let $\electionsystem$ be an election system. Again, $\electionsystem$
will here be either $\copelandalpha$ or $\copelandalphairrational$,
where $\alpha$, $0 \leq \alpha \leq 1$, is a rational number.
We describe our control problems as if they were for the case 
of rational preferences, but the irrational cases are perfectly
analogous, except for replacing preference lists with preference tables.

\subsubsection*{Control via Adding Candidates}

We start with two versions of control via adding candidates. In the
unlimited version the goal of the election chair is to
introduce candidates from a pool of spoiler candidates so as to
make his or her favorite candidate a winner of the election
(in the constructive case) or prevent his or her despised candidate
from being a winner (in the destructive case).
As suggested by the name of the problem,
in the unlimited version the chair can introduce any subset
of the spoiler candidates (none, some, or all are all legal options)
into the election.

\begin{description}
\item[Name:] $\scontrol{\electionsystem}{CCAC$_{\rm u}$}$ and
  $\scontrol{\electionsystem}{DCAC$_{\rm u}$}$ (control via adding an
  unlimited number of candidates).
\item[Given:] Disjoint sets $C$ and $D$ of candidates, a 
collection
$V$
  of voters specified via their  preference lists over
  the candidates in the set $C \cup D$, and a distinguished candidate
  $p \in C$.
\item[Question ($\scontrol{\electionsystem}{CCAC$_{\rm u}$}$):] Is there
  a subset $E$ of $D$ such that $p$ is a winner of
  the $\electionsystem$ election with voters $V$ and candidates $C
  \cup E$?
\item[Question ($\scontrol{\electionsystem}{DCAC$_{\rm u}$}$):] Is there
  a subset $E$ of $D$ such that $p$ is not a winner
  of the $\electionsystem$ election with voters $V$ and candidates $C
  \cup E$?
\end{description}

The definition of
$\scontrol{\electionsystem}{CCAC$_{\rm u}$}$ was (using different notation)
introduced by Bartholdi, Tovey, and
Trick~\cite{bar-tov-tri:j:control}.
In contrast with
the other
control problems involving adding or deleting candidates or voters, in
the adding candidates problem Bartholdi, Tovey, and Trick did not
introduce a nonnegative integer $k$ that bounds the number of
candidates (from the set $D$) the chair is allowed to add.
We feel this
asymmetry 
in their definitions is not well justified,\footnote{Bartholdi, 
Tovey, and Trick~\cite{bar-tov-tri:j:control} are aware of this
asymmetry.  They write: ``To a certain extent the exact formalization
of a problem is a matter of taste.  [\ldots] we could equally well
have formalized [the problem of control via adding candidates] to be
whether there are $K$ or fewer candidates to be added [\ldots]
It does not much matter for the problems we discuss, since both
versions are of the same complexity.''  In contrast, the complexity
of the problems studied here crucially hinges on which formalization
is used, and we thus define both versions formally.}
and thus we
define a with-change-parameter
version of the control-by-adding-candidates problems, which 
we denote by $\rm AC_l$ (where the ``l'' stands for
the 
fact that part of the problem instance is 
a \emph{l}imit on the number 
of candidates that can be added, in contrast with the model of 
Bartholdi, Tovey, and Trick~\cite{bar-tov-tri:j:control},
which we denote by $\rm AC_u$ with the ``u'' standing for 
the fact that the number of added candidates is \emph{u}nlimited, 
at least in the sense of not being limited via a separately input 
integer).
The  with-parameter version is the 
long-studied case for AV, DV, and DC, and we in this paper
will use 
AC as being synonymous with $\rm AC_l$, and will thus use 
the notation AC for the rest of this paper when speaking
of $\rm AC_l$.
We suggest 
 this as a natural regularization of the definitions and 
 we hope this version 
 will 
 become the ``normal'' version of the adding-candidates problem for further
 study.  However,
 we caution the reader that in earlier papers AC is used to 
 mean $\rm AC_u$.  

In the present paper, we will
obtain results not just for $\rm AC_{\rm l}$ but also 
for the $\rm AC_u$ case, in order to allow comparisons
between the results of this paper and those of earlier works.

Turning now to what we mean by AC (equivalently, $\rm AC_{\rm l}$),
as per the above definition in $\scontrol{$\electionsystem$}{CCAC}$ 
(i.e., $\scontrol{$\electionsystem$}{CCAC$_{\rm l}$}$) 
we ask whether it is possible to
make the distinguished candidate $p$ a winner of
some $\electionsystem$ election obtained
by adding at most $k$ candidates from the spoiler
candidate set $D$.  (Note that $k$ is part of the input.)
We define the destructive version,
$\scontrol{$\electionsystem$}{DCAC}$ 
(i.e., $\scontrol{$\electionsystem$}{DCAC$_{\rm l}$}$), 
analogously.

\begin{description}
\item[Name:] $\scontrol{\electionsystem}{CCAC}$ and
  $\scontrol{\electionsystem}{DCAC}$ (control via adding a
  limited number of candidates).
\item[Given:] Disjoint sets $C$ and $D$ of candidates, a 
collection
$V$
  of voters specified via their  preference lists over
  the candidates in the set $C \cup D$, a distinguished candidate
  $p \in C$, and a nonnegative integer $k$.
\item[Question ($\scontrol{\electionsystem}{CCAC}$):] Is there
  a subset $E$ of $D$ such that $\|E\| \leq k$ and
  $p$ is a winner of the $\electionsystem$ election with voters $V$
  and candidates $C \cup E$?
\item[Question ($\scontrol{\electionsystem}{DCAC}$):] Is there
  a subset $E$ of $D$ such that $\|E\| \leq k$ and
  $p$ is not a winner of the $\electionsystem$ election with voters
  $V$ and candidates $C \cup E$?
\end{description}

\subsubsection*{Control via Deleting Candidates}

In constructive control via deleting candidates, the chair seeks
to ensure that
his or her favorite candidate $p$ is a winner of the election by
suppressing at most $k$ candidates.
In the destructive variant of this problem, the chair's goal is
to block $p$ from winning by suppressing at most $k$ candidates
other than~$p$.

\begin{description}
\item[Name:] $\scontrol{\electionsystem}{CCDC}$ and
$\scontrol{\electionsystem}{DCDC}$ (control via deleting candidates).
\item[Given:] A set $C$ of candidates, a 
collection
$V$ of voters represented via preference lists over $C$, a
  distinguished candidate $p \in C$, and a nonnegative integer $k$.
\item[Question ($\scontrol{\electionsystem}{CCDC}$):] Is it possible
to by deleting at most $k$ candidates ensure that $p$ is a winner of
the resulting $\electionsystem$ election?

\item[Question ($\scontrol{\electionsystem}{DCDC}$):] Is it possible
to by deleting at most $k$ candidates other than $p$ ensure that $p$
is not a winner of the resulting $\electionsystem$ election?
\end{description}

\subsubsection*{Control via Partition and Run-Off Partition of Candidates}

Bartholdi, Tovey, and Trick~\cite{bar-tov-tri:j:control} gave two
types of control of elections via partition of
candidates.  In both cases the candidate set $C$ is partitioned into two
groups, $C_1$ and $C_2$ (i.e., $C_1 \cup C_2 = C$ and 
$C_1 \cap C_2 = \emptyset$), and the 
election is conducted in two stages. 
For control via run-off partition of candidates,
the election's first stage is conducted
separately on each group of candidates, $C_1$ and $C_2$,
and the group winners that survive
the tie-handling rule compete against each other in the second stage.
In control via partition of candidates, the first-stage election is
performed on the candidate set $C_1$ and those of
that election's winners that survive the tie-handling rule compete
against all candidates from $C_2$ in the second stage.

In the ties-promote
(TP) model, all first-stage winners within a group are promoted to
the final round.  In the ties-eliminate (TE) model, a first-stage
winner within a group is promoted to the final round if and only if he
or she is the unique winner within that group.

\begin{description}
\item[Name:] $\scontrol{\electionsystem}{CCRPC}$ and
$\scontrol{\electionsystem}{DCRPC}$ (control via run-off partition of
candidates).
\item[Given:] A set $C$ of candidates, a 
collection
$V$ of voters
  represented via preference lists over $C$, and a distinguished candidate
  $p \in C$.
\item[Question ($\scontrol{\electionsystem}{CCRPC}$):] Is there
  a partition of $C$ into $C_1$ and $C_2$ such that $p$ is a winner of
  the two-stage election where the winners of subelection $(C_1,V)$
  that survive the tie-handling rule compete against the winners of
  subelection $(C_2,V)$ that survive the tie-handling rule? Each
  subelection (in both stages) is conducted using election
  system~$\electionsystem$.

\item[Question ($\scontrol{\electionsystem}{DCRPC}$):] Is there
  a partition of $C$ into $C_1$ and $C_2$ such that $p$ is not a winner
  of the two-stage election where the winners of subelection $(C_1,V)$
  that survive the tie-handling rule compete against the winners of
  subelection $(C_2,V)$ that survive the tie-handling rule? Each
  subelection (in both stages) is conducted using election
  system~$\electionsystem$.
\end{description}

The above description defines four computational problems
for a given election system~$\electionsystem$:
$\scontrol{\electionsystem}{CCRPC-TE}$,
$\scontrol{\electionsystem}{CCRPC-TP}$,
$\scontrol{\electionsystem}{DCRPC-TE}$, and
$\scontrol{\electionsystem}{DCRPC-TP}$.

\begin{description}
\item[Name:] $\scontrol{\electionsystem}{CCPC}$ and
$\scontrol{\electionsystem}{DCPC}$ (control via partition of
candidates).
\item[Given:] A set $C$ of candidates, a
collection
$V$ of voters
  represented via preference lists over $C$, and a distinguished candidate
  $p \in C$.
\item[Question ($\scontrol{\electionsystem}{CCPC}$):] Is there
  a partition of $C$ into $C_1$ and $C_2$ such that $p$ is a winner of
  the two-stage election where the winners of subelection $(C_1,V)$
  that survive the tie-handling rule compete against all candidates in
  $C_2$? Each subelection (in both stages) is conducted using election
  system~$\electionsystem$.

\item[Question ($\scontrol{\electionsystem}{DCPC}$):] Is there a
  partition of $C$ into $C_1$ and $C_2$ such that $p$ is not a winner
  of the two-stage election where the winners of subelection $(C_1,V)$
  that survive the tie-handling rule compete against all candidates in
  $C_2$? Each subelection (in both stages) is conducted using election
  system~$\electionsystem$.

\end{description}

This description defines four computational problems
for a given election system~$\electionsystem$:
$\scontrol{\electionsystem}{CCPC-TE}$,
$\scontrol{\electionsystem}{CCPC-TP}$,
$\scontrol{\electionsystem}{DCPC-TE}$, and
$\scontrol{\electionsystem}{DCPC-TP}$.

\subsubsection*{Control via Adding Voters}

In the scenario of control via adding voters, the chair's goal is to
either ensure that $p$ is a winner (in the constructive case) or ensure that
$p$ is not a winner (in the destructive case) via causing up to $k$
additional voters to participate in the election. The chair can draw the
voters to add to the election from a prespecified collection of voters
(with given preferences).

\begin{description}
\item[Name:] $\scontrol{\electionsystem}{CCAV}$ and
$\scontrol{\electionsystem}{DCAV}$ (control via adding voters).
\item[Given:] A set $C$ of candidates, two disjoint collections of
voters, $V$ and $W$,
  represented via preference lists over $C$, a distinguished candidate
  $p$, and a nonnegative integer~$k$.
\item[Question ($\scontrol{\electionsystem}{CCAV}$):] Is there
  a subset $Q$, $\|Q\| \leq k$, of voters in $W$ such that the
  voters in $V \cup Q$ jointly elect $p \in C$ as a
  winner according to system~$\electionsystem$?

\item[Question ($\scontrol{\electionsystem}{DCAV}$):] Is there
  a subset $Q$, $\|Q\| \leq k$, of voters in $W$ such that the
  voters in $V \cup Q$ do not elect $p$ as a winner according to 
  system~$\electionsystem$?

\end{description}

\subsubsection*{Control via Deleting Voters} 

In the control via deleting voters case the chair seeks to either ensure that
$p$ is a winner (in the constructive case) or prevent $p$ from being a winner
(in the destructive case) via blocking up to $k$ voters from participating 
in the election. (This loosely models vote suppression or disenfranchisement.)

\begin{description}
\item[Name:] $\scontrol{\electionsystem}{CCDV}$ and
$\scontrol{\electionsystem}{DCDV}$ (control via deleting voters).
\item[Given:] A set $C$ of candidates, a 
collection
$V$ of voters represented via preference lists over $C$, a
  distinguished candidate $p \in C$, and a nonnegative integer $k$.
\item[Question ($\scontrol{\electionsystem}{CCDV}$):] Is it possible
to by deleting at most $k$ voters ensure that $p$ is a winner of the
resulting $\electionsystem$ election?

\item[Question ($\scontrol{\electionsystem}{DCDV}$):] Is it possible
to by deleting at most $k$ voters ensure that $p$ is not a winner of
the resulting $\electionsystem$ election?
\end{description}

\subsubsection*{Control via Partition of Voters} 

In the case of control via partition of voters,  the following
two-stage election is performed.
First, the voter set $V$ is partitioned into two
subcommittees, $V_1$ and~$V_2$. The winners of election $(C,V_1)$ 
that survive the tie-handling rule compete against the winners of
$(C,V_2)$ that survive the tie-handling rule.
Again, our
tie-handling rules are TE and TP (ties-eliminate and
ties-promote).

\begin{description}
\item[Name:] $\scontrol{\electionsystem}{CCPV}$ and
$\scontrol{\electionsystem}{DCPV}$ (control via partition of voters).
\item[Given:] A set $C$ of candidates, a 
collection
$V$ of voters
  represented via preference lists over $C$, and a distinguished candidate
  $p \in C$.
\item[Question ($\scontrol{\electionsystem}{CCPV}$):] Is there
  a partition of $V$ into $V_1$ and $V_2$ such that $p$ is a winner of
  the two-stage election where the winners of election $(C,V_1)$ that
  survive the tie-handling rule compete against the winners of
  $(C,V_2)$ that survive the tie-handling rule?  Each subelection (in
  both stages) is conducted using election system~$\electionsystem$.

\item[Question ($\scontrol{\electionsystem}{DCPV}$):] Is there a
  partition of $V$ into $V_1$ and $V_2$ such that $p$ is not a winner
  of the two-stage election where the winners of election $(C,V_1)$
  that survive the tie-handling rule compete against the winners of
  $(C,V_2)$ that survive the tie-handling rule?  Each subelection (in
  both stages) is conducted using election system~$\electionsystem$.
\end{description}

\subsubsection*{Unique Winners and Irrationality}

Our bribery and control problems were each defined above only for rational
voters and in the \emph{nonunique-winner} model,
i.e., asking whether a given candidate can be made, or prevented from
being, \emph{a} winner.
Nonetheless, we have
proven all our control results both for the case of nonunique winners and
(to be able to fairly compare them with existing
control results, which 
except for the 
interesting 
``multi-winner'' 
model
of 
Procaccia, Rosenschein, and
Zohar~\cite{pro-ros-zoh:c:multiwinner}
are 
in the unique-winner model)
\emph{unique winners}
(a candidate is a unique winner if he or she
is a winner and is the only winner).
Similarly, all our bribery results are proven both in the unique-winner
model and (to be able to fairly compare them with existing bribery results
in the literature) in the nonunique-winner model.
In addition to the rational-voters case,
we also study these problems for the
case of voters who are allowed to be irrational. 
As mentioned earlier, in the 
case of irrational voters,
voters are represented via preference
tables rather than preference lists.

\subsection{Graphs}

An \emph{undirected graph} $G$ is a pair $(V(G),E(G))$, where $V(G)$ is
the set of vertices and $E(G)$ is the set of edges and each edge is an
unordered pair of distinct vertices.\footnote{In this paper, the
  symbols $E$ and $V$ are generally reserved for elections and voters,
  except the just introduced ``overloading'' of them to mean sets of
  edges and vertices in a given graph. The intended meaning of $E$ and
  $V$ will be clear from the context, even when our proofs involve multiple
  elections and graphs.}
A \emph{directed graph} is defined analogously, except
that the edges are represented as ordered pairs.  For example, if $u$
and $v$ are distinct vertices in an undirected graph $G$ then $G$ either has
an edge $e = \{u,v\}$ that connects $u$ and $v$ or it doesn't. On the
other hand, if $G$ is a directed graph then $G$ either has an edge $e' =
(u,v)$ from $u$ to~$v$, or an edge $e'' = (v,u)$ from $v$ to~$u$, or
both $e'$ and~$e''$, or neither $e'$ nor~$e''$.

For a directed graph $G$, the \emph{indegree} of a vertex $u \in V(G)$ 
is the number of $G$'s edges that enter~$u$ (i.e., the number of edges
of the form $(v,u)$ in $E(G)$).  Similarly, the \emph{outdegree} of $u
\in V(G)$ is the number of edges that leave~$u$ (i.e., the number of
edges of the form $(u,v)$ in $E(G)$).

\subsection{NP-Complete Problems and Reductions}

Without loss of generality, we assume that all problems that we
consider are encoded in a natural, efficient way over the
alphabet $\Sigma = \{0,1\}$. We use the standard notion of
$\np$-completeness, defined via polynomial-time 
many-one reductions.  We say that a
computational problem $A$ \emph{polynomial-time 
many-one reduces} to a problem $B$ if there
exists a polynomial-time computable function $f$ such that
\[ (\forall x \in \Sigma^*)[ x \in A \iff f(x) \in B]. \] 
A problem is
\emph{NP-hard} if all members of $\np$ polynomial-time 
many-one reduce to it.  Thus, if an $\np$-hard problem $A$
polynomial-time many-one reduces to a problem $B$, then $B$ is
$\np$-hard as well.
A problem is
\emph{NP-complete} if it is $\np$-hard and
is a member of $\np$.
When
clear from context  we will use ``reduce'' and ``reduction''
as shorthands for ``polynomial-time 
many-one reduce'' and ``polynomial-time many-one reduction.''

Our $\np$-hardness results typically follow via a reduction from either
the exact-cover-by-3-sets problem or from the vertex cover problem
(see, e.g.,~\cite{gar-joh:b:int}).
These are well-known $\np$-complete problems, but we define them here for
the sake of completeness.

\begin{description}
\item[Name:] X3C (exact cover by 3-sets). 
\item[Given:] A set $B = \{ b_1, \ldots, b_{3k}\}$, $k \geq 1$, and a family of
  sets $\mathcal{S} = \{S_1, \ldots, S_n\}$ such that for each $i$, $1
  \leq i \leq n$, it holds that $S_i \subseteq B$ and $\|S_i\| = 3$.
\item[Question:] Is there a set $A \subseteq \{1, \ldots, n\}$, $\|A\| = k$,
  such that $\bigcup_{i \in A}S_i = B$?
\end{description}

The set $A$ about which we ask in the above problem is called an
\emph{exact cover of~$B$}. It is a ``cover'' because every member of
$B$ belongs to some $S_i$ such that $i \in A$, and it is ``exact''
because each member of $B$ belongs to exactly one $S_i$ such that $i
\in A$.

Whenever we consider instances of the X3C problem, we assume that they are
well-formed, that is, we assume
that they follow the syntactic requirements stated in
the above
``Given'' field (e.g., the cardinality of the set $B$ is
indeed a multiple of three). We apply this convention of
considering only syntactically correct inputs to all other problems as
well. Let $A$ be some computational problem and let $x$ be an instance
of $A$. When we consider an algorithm for $A$, and input $x$ is
malformed, then we can immediately reject. When we are building a
reduction from $A$ to some problem $B$, then whenever we hit a
malformed input $x$ we can output a fixed $y$ not in $B$. (In our
reductions $B$ is never $\Sigma^*$, so this is always possible.)

$\copelandalpha$ elections can often be considered in terms of
appropriate graphs. This representation is particularly useful when we
face control problems that modify the structure of the candidate
set,
since in this case
operations on
an election directly translate into
suitable operations on
the corresponding graph. For
candidate control problems, we---instead 
of using
reductions from X3C---construct reductions from the vertex cover problem.
A vertex cover of an undirected graph $G$ is a subset
of $G$'s vertices such that each edge of $G$ is adjacent to at least
one vertex from that subset.

\begin{description}
\item[Name:] VertexCover. 
\item[Given:] An undirected graph $G$ and a nonnegative integer $k$.
\item[Question:] Is there a set $W$ such that $W \subseteq V(G)$, $\|W\| \leq k$,
  and for every edge $e \in E(G)$ it holds that $e \cap W \neq \emptyset$?
\end{description}

\subsection{Resistance and Vulnerability}
\label{sec:resistance-vulnerability}

Not all election systems can be affected by each control
type; if not, the system is said to be \emph{immune} to this type of control.
For example, 
if a candidate $c$ is not a Condorcet winner then it is
impossible to make him or her a Condorcet winner by
adding candidates (see~\cite{bar-tov-tri:j:control}
and~\cite{hem-hem-rot:j:destructive-control} for more such immunity
results). However, for
$\copelandalpha$ elections
it is easy to see that for each
type of control defined in Section~\ref{ss:problems}
there is a scenario in which the outcome of the election can indeed be changed
via conducting the corresponding control action.
If an election system is not immune to some type of control (as witnessed
by such a scenario), the election system is said to be \emph{susceptible} to that control type.

\begin{proposition}
\label{prop:susceptibility}
For each rational number~$\alpha$, $0 \leq \alpha \leq 1$,
$\copelandalpha$ is susceptible to each type of control defined in
Section~\ref{ss:problems}.
\end{proposition}

We say that an election system 
($\copelandalpha$ or $\copelandalphairrational$,
in our case) is \emph{resistant} to a particular
attack (be it a type of control or of bribery) if 
the appropriate
computational problem is $\np$-hard
and susceptibility holds.\footnote{It is true that for some
unnatural election systems immunity to bribery holds, e.g., the election
system ``Every candidate is a winner'' is immune to all types of
bribery. However, our $\copelandalpha$-type systems are all susceptible
to all the bribery types we look at in this paper, so we won't further
explicitly discuss or state susceptibility for the bribery cases.}
On the other hand, if the
computational problem is in $\p$ and
susceptibility holds,
then we say the system is \emph{vulnerable} to this attack. 
Because of how our bribery and control problems are defined,
the vulnerability definition merely requires that there exist
a polynomial-time algorithm that determines whether a successful
bribe or control action \emph{exists} on a given input. However,
in every single one of our vulnerability proofs we will provide
something far stronger. We will provide a polynomial-time algorithm
that actually \emph{finds} a successful bribe or control
action on each input for which a successful bribe or control action
exists, and on each input where no successful bribe or control action
exists will announce that fact.

The notions of
resistance and vulnerability (and of immunity and susceptibility) for
control problems in election systems were introduced by Bartholdi,
Tovey, and Trick~\cite{bar-tov-tri:j:control}, and we here follow
the definition alteration of~\cite{hem-hem-rot:c:hybrid}
of resistance from ``$\np$-complete'' to ``$\np$-hard,'' as
that change is compelling (because under the old definition, $\np$-completeness,
things could actually become nonresistant by being too hard, which is not
natural). However, for
all resistance claims in this paper $\np$-membership is 
clear, and so $\np$-completeness in fact does hold.

\section{Bribery}
\label{sec:bribery}

In this section we present our
results
on the complexity of bribery
for the
$\copelandalpha$ election systems, where
$\alpha$ is a rational number with $0 \leq \alpha \leq 1$. 
Our main result,
which will be presented in Section~\ref{sec:bribery:resistance},
is that
each such system is resistant to
bribery, regardless of voters' rationality and of our mode of operation
(constructive versus destructive).
In Section~\ref{sec:microbribery:vulnerability}, we will provide
vulnerability results for Llull and $\copelandzero$ with respect to
``microbribery.''

\subsection{Resistance to Bribery}
\label{sec:bribery:resistance}

\begin{theorem}
  \label{thm:bribery}
For each rational~$\alpha$, $0 \leq \alpha \leq 1$,
$\copelandalpha$ and $\copelandalphairrational$ 
are resistant to both constructive and destructive bribery,
in both the nonunique-winner model and the unique-winner model.
\end{theorem}

We prove Theorem~\ref{thm:bribery} via
Theorems~\ref{thm:bribery:dest-copelandalpha} and~\ref{thm:bribery:cons-copelandalpha}
and Corollary~\ref{cor:bribery:irrational} below.
Our proofs employ an approach that we call the UV technique.
For the constructive cases, this technique
proceeds by constructing
bribery
instances where the only briberies that could
possibly ensure that our favorite candidate $p$ is a winner 
involve
only voters who rank a group of special candidates (often the group will
contain exactly two candidates, $u$ and $v$) above~$p$.
The remaining voters, the bystanders so to speak, can be
used to create appropriate padding and structure within the election.
The destructive cases follow via a cute observation regarding the
dynamics of our constructive cases.

The remainder of this section is devoted to proving
Theorem~\ref{thm:bribery}.
We start with the case of rational voters in
Theorems~\ref{thm:bribery:dest-copelandalpha}
and~\ref{thm:bribery:cons-copelandalpha} below and then argue that the
analogous results for the case of irrational voters
follow via, essentially, the same proof.

\begin{theorem}
  \label{thm:bribery:dest-copelandalpha}
For each rational number~$\alpha$, $0 \leq \alpha \leq 1$, 
$\copelandalpha$ is resistant to constructive bribery in the unique-winner
model and to destructive bribery in the nonunique-winner model.
\end{theorem}

\begin{proofs}
  Fix an arbitrary rational number $\alpha$ with $0 \leq \alpha \leq
  1$.  Our proof provides reductions from the X3C problem to,
  respectively, the unique-winner variant of constructive bribery and
  to the nonunique-winner variant of destructive bribery. Our
  reductions will differ regarding only the specification of the goal
  (i.e., regarding which candidate we attempt to make a unique winner
  or which candidate we prevent from being a winner) and thus we
  describe them jointly as, essentially, a single reduction.

Our reduction will produce an instance of an appropriate bribery problem
with an odd number of voters, and so we will never have ties in head-to-head
contests. 
Thus our proof works regardless of which rational
number~$\alpha$ with $0 \leq \alpha \leq 1$ is chosen.

Let $(B,\mathcal{S})$ be an instance of X3C, where $B = \{b_1, b_2,
\ldots, b_{3k}\}$, $\mathcal{S}$ is a collection $\{S_1, S_2, \ldots,
S_n\}$ of
three-element subsets of $B$ with
$\bigcup_{j=1}^{n}S_j = B$, and
$k \geq 1$. If our input does not
meet these conditions then we output a fixed instance of
our bribery problem having a negative answer.

Construct a $\copelandalpha$ election $E = (C,V)$ as follows.  
The candidate set $C$ is $\{u,v,p\} \cup B$, where
none of~$u$, $v$, and $p$ is in~$B$.
The voter set $V$ contains $2n+4k+1$ voters of the following types.

\begin{enumerate}
\item For each $S_i$, we introduce
one voter of type~(i) and
one voter of type~(ii):
\[
\begin{array}{ll}
 \mbox{(i)}   &  u > v > S_i > p > B - S_i,  \\
 \mbox{(ii)}  &  \reversenotation{B - S_i} > p > u > v >
 \reversenotation{S_i}.
\end{array}
\]
\item We introduce
$k$ voters for each of the types~(iii)-1, (iii)-2, (iv)-1, and (iv)-2:
\[
\begin{array}{ll}
 \mbox{(iii)-1}     &  u > v > p > B,  \\
 \mbox{(iii)-2}     &  v > u > p > B,  \\
 \mbox{(iv)-1}      &  u > \reversenotation{B} >  p > v,  \\
 \mbox{(iv)-2}      &  v > \reversenotation{B} >  p > u.
\end{array}
\]
\item We introduce a single type (v) voter:
\[
\begin{array}{ll}
 \mbox{(v)}     &  B > p > u > v.
\end{array}
\]
\end{enumerate}

We have the following relative vote-scores:
\begin{enumerate}
  \item $\versus_E(u,v) = 2n + 1 \geq 2k + 1$, where the inequality
follows from our assumption $\bigcup_{j=1}^{n}S_j = B$ (which implies
$n \geq \|B\|/3 = k$),
  \item $\versus_E(u,p) = \versus_E(v,p) = 2k - 1$,
  \item for each $i \in \{1, 2, \ldots, 3k\}$,
 $\versus_E (u,b_i) = \versus_E(v,b_i) \geq 2k + 1$,
  \item for each $i \in \{1, 2, \ldots, 3k\}$, $\versus_E (b_i,p) = 1$, and
  \item for each $i,j \in \{1, 2, \ldots, 3k\}$ with $i \neq j$,
 $|\versus_E (b_i,b_j)| = 1$.
\end{enumerate}
For example, to see that $\versus_E (u,b_i) \geq 2k + 1$ for each $i
\in \{1, 2, \ldots, 3k\}$, note that each $b_i$ is in at least one
$S_j$ because of $\bigcup_{j=1}^{n}S_j = B$, so the voters of types
(i) and (ii) give $u$ an advantage of at least two votes over~$b_i$.
Furthermore, the voters of types~(iii)-1, (iii)-2, (iv)-1, and (iv)-2
give $u$ an advantage of $2k$ additional votes over each~$b_i$,
and the single type~(v) voter gives each~$b_i$ a one-vote advantage over~$u$.
Summing up, we obtain $\versus_E (u,b_i) \geq 2+2k-1 = 2k+1$.
The other relative vote-scores are similarly easy to verify.

These relative vote-scores yield the following 
$\copelandalpha$ scores or upper bounds on such scores:
\begin{enumerate}
  \item $\copelandalphascore_E(u) = 3k+2$,
  \item $\copelandalphascore_E(v) = 3k+1$,
  \item for each $i\in \{1, 2, \ldots, 3k\}$,
        $\copelandalphascore_E(b_i) \leq 3k$, and
  \item $\copelandalphascore_E(p) = 0$.
\end{enumerate}

To prove our theorem, we need the following claim.

\begin{claim}
\label{cla:bribery:copelandalpha}
The following three statements are equivalent:
\begin{enumerate}
\item\label{cla:bribery:copelandalpha-1}
 $(B,\mathcal{S}) \in \mbox{X3C}$.
\item\label{cla:bribery:copelandalpha-2}
 Candidate $u$ can be prevented from winning via bribing at most
$k$ voters of~$E$.
\item\label{cla:bribery:copelandalpha-3}
 Candidate $p$ can be made a unique winner via bribing at most $k$
voters of~$E$.
\end{enumerate}
\end{claim}

\sproofof{Claim~\ref{cla:bribery:copelandalpha}}
(\ref{cla:bribery:copelandalpha-1}) implies 
(\ref{cla:bribery:copelandalpha-2}):
It is easy to see that
if $(B,\mathcal{S}) \in \mbox{X3C}$ then
there is a bribe involving $k$ or fewer voters that prevents $u$
from being a winner:
It is enough to bribe those type~(i) 
voters that correspond to a cover of size $k$ to report $p$ as their top
choice (while not changing anything else in their preference lists):
$p > u > v > S_i > B - S_i$.  Call the resulting election~$E'$.
In $E'$ the following relative vote-scores change:
$\versus_{E'}(p,u) = \versus_{E'}(p,v) = n+k - (n-k) - 2k + 1 = 1$ and
$\versus_{E'}(p,b_i) \geq 1$
for each $i \in \{1, 2, \ldots, 3k\}$, while all other relative
vote-scores remain unchanged.
Thus $\copelandalphascore_{E'}(p) = 3k+2$,
$\copelandalphascore_{E'}(u) = 3k+1$, $\copelandalphascore_{E'}(v) =
3k$, and $\copelandalphascore_{E'}(b_i) < 3k$
for each $i \in \{1, 2, \ldots, 3k\}$, so $p$ defeats all
other candidates and is the unique winner.
In particular, this bribe (of at most $k$ voters in $E$) ensures 
that $u$ is not a winner.

(\ref{cla:bribery:copelandalpha-2}) implies 
(\ref{cla:bribery:copelandalpha-3}):
Suppose that there is a bribe involving $k$ or fewer voters that
prevents $u$ from being a winner.
Note that $u$ defeats
everyone except $p$ by more than $2k$ votes in~$E$.
This
means that via bribery of at most $k$ voters $u$'s score can decrease
by at most one.
Thus, to prevent $u$ from being a winner via such a bribery,
we need to ensure that $u$ receives  a $\copelandalpha$
score of $3k+1$ and some candidate other than $u$
gets a $\copelandalpha$ score of $3k+2$, that is, that candidate
defeats everyone. Neither $v$ nor any of the $b_i$'s can possibly
obtain a $\copelandalpha$ score of $3k+2$ via such a bribery,
since bribery of at most $k$
voters can affect only head-to-head contests where the relative
vote-scores of the participants are at most~$2k$. Thus, via such a bribery,
$u$ can be prevented from winning
only if $p$ can be made a (in fact, the
unique) winner of our election.

(\ref{cla:bribery:copelandalpha-3}) implies 
(\ref{cla:bribery:copelandalpha-1}):
Let $W$ be a set of at most $k$ voters whose bribery ensures that $p$
is a unique winner of our election. 
Thus we know that $\|W\| = k$ and
that $W$ contains only voters who rank both $u$ and $v$ above $p$
(as otherwise $p$ would not defeat both $u$ and~$v$),
which is the case only for voters of types~(i), (iii)-1, and~(iii)-2.
Furthermore, a bribery that makes $p$ the unique winner has to ensure that $p$
defeats all members of~$B$;
note that the type (iii)-1 and~(iii)-2 voters in $E$ already rank $p$ above
all of~$B$.  Thus, via a simple counting argument,
$W$ must contain exactly $k$ type~(i) voters that
correspond to a
size-$k$ cover of~$B$.~\eproofof{Claim~\ref{cla:bribery:copelandalpha}}

Since our reduction is computable in polynomial time,
Claim~\ref{cla:bribery:copelandalpha} completes the proof 
of Theorem~\ref{thm:bribery:dest-copelandalpha}.~\end{proofs}

\begin{theorem}
  \label{thm:bribery:cons-copelandalpha}
  For each rational~$\alpha$, $0 \leq \alpha \leq 1$, $\copelandalpha$
  is resistant to constructive bribery in the nonunique-winner model
  and to destructive bribery in the unique-winner model.
\end{theorem}

\begin{proofs}
  Fix an arbitrary rational number $\alpha$ with $0 \leq \alpha \leq
  1$. As in the proof of Theorem~\ref{thm:bribery:dest-copelandalpha},
  we handle the appropriate constructive and destructive cases jointly
  using essentially the same reduction for each of them, differing
  only in the specification of the goal of the briber. Thus we
  describe our reductions from X3C to the appropriate constructive and
  destructive bribery problems as a single reduction, separately
  specifying only the goals for each of the cases.

  Our reduction works as follows. We are given an X3C instance
  $(B,\mathcal{S})$, where $B = \{b_1, b_2, \ldots, b_{3k}\}$ is a
  set, $\mathcal{S}$ is a collection $\{S_1, S_2, \ldots, S_n\}$ of
  three-element subsets of $B$ with $\bigcup_{j=1}^{n}S_j = B$, and
  $k$ is a positive integer.  We form an election $E = (C,V)$, where
  $C = \{p,s,t,u,v\} \cup B$ and where $V$ is as specified below.  In the nonunique-winner constructive case we want to
  ensure that $p$ is a winner and in the unique-winner destructive
  case we want to prevent $s$ from being the unique winner. In each
  case we want to achieve our goal via bribing at most $k$ voters from
  $V$. $V$ contains $2n + 24k +1$ voters of the following types:
\begin{enumerate}
\item For each $S_i$, we introduce one voter of type~(i) and one voter
  of type~(ii):
\[
\begin{array}{ll}
  \mbox{(i)}   &
 s > t > u > v > S_i > p > B - S_i,  \\
  \mbox{(ii)}  & 
 \reversenotation{B - S_i} > p > v > u > t > s > \reversenotation{S_i}.
\end{array}
\]
\item We introduce $k$ voters for each of the types~(iii)-1, (iii)-2,
  (iv)-1, and (iv)-2:
\[
\begin{array}{ll}
 \mbox{(iii)-1}     &  s > t > u > v > p > B,  \\
 \mbox{(iii)-2}     &  s > t > v > u > p > B,  \\
 \mbox{(iv)-1}      &  u > \reversenotation{B} > p > s > v > t,  \\
 \mbox{(iv)-2}      &  v > \reversenotation{B} > p > s > u > t.
\end{array}
\]
\item We introduce $20k$ normalizing voters:
\[
\begin{array}{r@{\ }ll}
2k & \mbox{voters of type (v)-1} &                  u  > v > t > p > s > B, \\
2k & \mbox{voters of type (v)-2} &                  u  > v > s > t > p > B, \\
3k & \mbox{voters of type (v)-3} &                  s  > t > u > v > p > B, \\
3k & \mbox{voters of type (v)-4} &                  s  > v > t > u > p > B, \\
3k & \mbox{voters of type (v)-5} & t > \reversenotation{B} > p > u > s > v, \\
 k & \mbox{voters of type (v)-6} & \reversenotation{B} > p > s > u > v > t, \\
3k & \mbox{voters of type (v)-7} & s > \reversenotation{B} > p > u > v > t, \\
3k & \mbox{voters of type (v)-8} & \reversenotation{B} > p > s > v > t > u.
\end{array}
\]
\item Finally, we introduce a single type (vi) voter:
\[
\begin{array}{ll}
 \mbox{(vi)}     &  B > p > u > v > s > t.
\end{array}
\]
\end{enumerate}

In the nonunique-winner constructive case we want to ensure that $p$
is a winner and in the unique-winner destructive case we want to
prevent $s$ from being the unique winner. In each case we want to
achieve our goal via bribing at most $k$ voters. Thus within our
bribery we can affect the results of head-to-head contests between
only those candidates 
whose relative vote-scores are, in
absolute value, at most $2k$.
In~$E$, we have the following relative vote-scores:
\begin{enumerate}
\item
$\versus_E(s,t) > 2k$,
$\versus_E(s,u) > 2k$,
$\versus_E(s,v) > 2k$,
$\versus_E(t,p) > 2k$,
$\versus_E(t,u) > 2k$,
$\versus_E(v,t) > 2k$, and
$\versus_E(u,v) > 2k$,
\item 
$\versus_E(s,p) = \versus_E(u,p) = \versus_E(v,p) = 2k-1$,
\item for each $i \in \{1, 2, \ldots, 3k\}$,
 $\versus_E(b_i,p) = 1$,
 $\versus_E(s,b_i) > 2k$,
 $\versus_E(t,b_i) > 2k$,
 $\versus_E(u,b_i) > 2k$, and
 $\versus_E(v,b_i) > 2k$, and
\item for each $i,j \in \{1, 2, \ldots, 3k\}$ with $i \neq j$, we have
 $|\versus_E(b_i,b_j)| = 1$.
\end{enumerate}

To analyze~$E$, let $E'$ denote an arbitrary election resulting from
$E$ via bribing at most $k$ voters.  The relative vote-scores among
any two candidates in $E$ yield the following $\copelandalpha$ scores:
\begin{enumerate}
\item $\copelandalphascore_E(s) = 3k+4$, and since we have
  $\versus_E(s,c) > 2k$ for each candidate
  $c \in C$ with $p \neq c \neq s$, it follows that $3k+3 \leq
  \copelandalphascore_{E'}(s)$.
\item For each $x \in \{t,u,v\}$, $\copelandalphascore_E(x) = 3k+2$,
  and since we have $\versus_E(s,x) > 2k$, $\versus_E(t,u) > 2k$,
  $\versus_E(u,v) > 2k$, and $\versus_E(v,t) > 2k$, it follows that
  $\copelandalphascore_{E'}(x) \leq 3k+2$.
\item $\copelandalphascore_E(p) = 0$, and since we have
  $\versus_E(t,p) > 2k$, it follows that $\copelandalphascore_{E'}(p)
  \leq 3k+3$.
\item For each $i \in \{1, 2, \ldots, 3k\}$,
  $\copelandalphascore_E(b_i) \leq 3k$, and since we have
  $\versus_E(x,b_i) > 2k$ for each candidate $x \in \{s,t,u,v\}$,
  it follows that $\copelandalphascore_{E'}(b_i) \leq 3k$.
\end{enumerate}
Thus $s$ is the unique winner of~$E$, and the only candidate who is
able to prevent $s$ from being the unique winner via at most
$k$ voters being bribed is~$p$. 

We claim that $(B,\mathcal{S}) \in \mbox{X3C}$ if and only if 
there is a bribe involving at most $k$ voters that prevents $s$
from being the unique winner.
(Equivalently, $(B,\mathcal{S}) \in \mbox{X3C}$ if and only 
if there is a bribe of at most $k$ voters that ensures that
$p$ is a winner.)

From left to right, if $\mathcal{S}$ has an exact cover for~$B$, then,
via bribing the $k$ type (i) voters that correspond to this cover, $s$
can be prevented from being the unique winner.
In more detail,
if the $k$ bribed voters rank $p$ on top while leaving their
preferences otherwise unchanged (i.e., their votes are now $p > s > t
> u > v > S_i > B - S_i$), then the only relative vote-scores that
have changed in this new election, call it~$E'$, are:
$\versus_{E'}(p,s) = \versus_{E'}(p,u) = \versus_{E'}(p,v) = 1$,
$\versus_{E'}(t,p) = 4k-1$, and $\versus_{E'}(p,b_i) = 1$ for each $i
\in \{1, 2, \ldots, 3k\}$.  It follows that $p$ and $s$ tie for winner
in $E'$ with $\copelandalphascore_{E'}(p) =
\copelandalphascore_{E'}(s) = 3k+3$.

From right to left, suppose there is a bribe of at most $k$ voters
that prevents $s$ from being the unique winner.
By construction,
for each election $E'$ that results from $E$ via bribing at most $k$
voters, this is possible only if $\copelandalphascore_{E'}(p) =
\copelandalphascore_{E'}(s) = 3k+3$.  Let $W$ be a set of at most $k$
voters whose bribery ensures that $s$ is not the unique winner in the
resulting election.  Since $\versus_E(t,p) > 2k$, it is not possible
for $p$ to win the head-to-head contest with $t$ via such a bribery.
Thus, for $p$ to obtain a score of $3k+3$, $p$ must win the
head-to-head contests with each candidate in $\{s,u,v\} \cup B$.
However, since $\versus_E(s,p) = \versus_E(u,p) = \versus_E(v,p) =
2k-1$, we have $\|W\| = k$ and every voter in $W$ must rank each of
$s$, $u$, and $v$ ahead of~$p$.  Thus $W$ can contain only voters of
types~(i), (iii)-1, (iii)-2, (v)-2, (v)-3, and (v)-4.  However, since
$p$ also needs to defeat each member of $B$ and since all voters of
types (iii)-1, (iii)-2, (v)-2, (v)-3, and (v)-4 rank $p$ ahead of each
member of~$B$, $W$ must contain exactly $k$ type~(i) voters that
correspond to an exact cover for~$B$.

Since our reduction is computable in polynomial time, this completes
the proof of Theorem~\ref{thm:bribery:cons-copelandalpha}.~\end{proofs}

The proofs of the above theorems have an interesting feature.
When we discuss bribery,
we never rely
on the fact that the voters are rational.
Thus we can
allow the voters to be irrational and form 
$\sbribery{\copelandalphairrational}$ and
$\sdestbribery{\copelandalphairrational}$
instances simply by deriving the voters' 
preference
tables from the voters' preference lists
given in the above proofs.  It
is easy to see that the proofs remain valid after this change; 
in the proofs we assume that each bribed voter, after the bribery, 
prefers $p$ to all other candidates, but we do not make any further
assumptions (and, in particular, we do not use linearity of the preferences).
Thus we have the following corollary to the proofs of
Theorems~\ref{thm:bribery:dest-copelandalpha} and~\ref{thm:bribery:cons-copelandalpha}.

\begin{corollary}
  \label{cor:bribery:irrational}
For each rational number~$\alpha$, $0 \leq \alpha \leq 1$,
$\copelandalphairrational$ is resistant to both constructive bribery and
destructive bribery, in both the nonunique-winner model and the
unique-winner model.
\end{corollary}

Theorems~\ref{thm:bribery:dest-copelandalpha} and~\ref{thm:bribery:cons-copelandalpha}
and Corollary~\ref{cor:bribery:irrational}
together constitute a proof of Theorem~\ref{thm:bribery} and
establish that
for each rational $\alpha$, $0 \leq \alpha \leq 1$,
$\copelandalpha$ and $\copelandalphairrational$ possess broad---essentially 
perfect---resistance to bribery
regardless of
whether we are interested in
constructive or destructive results. However, the next section
shows that this perfect picture is, in fact, only near-perfect when
we consider microbribes, which don't allow changing
the complete preferences of
voters at once but rather change the results of head-to-head contests
between candidates in the voters' preferences.  We will show that
there is an efficient way of finding optimal microbriberies for the
case of irrational voters in $\copelandalpha$ elections.

\subsection{Vulnerability to Microbribery for Irrational Voters}
\label{sec:microbribery:vulnerability}

In this section we explore the problems related to microbribery of
irrational voters. In standard bribery problems, which were
considered in Section~\ref{sec:bribery:resistance}, we ask whether it is
possible to ensure that a designated candidate $p$ is a winner
(or, in the destructive case, to ensure that $p$ is not a winner)
via modifying the preference tables
of at most $k$ voters. That is, we can at unit cost completely redefine
the preference table of each voter bribed.
So in this model, we pay for a service (namely,
the modification of
the reported preference table) and we pay for it in bulk (when we buy a voter,
we have secured his or her total obedience). However,
sometimes it may be far more reasonable to adopt a more local approach in 
which we have to pay separately for each preference-table entry flip---to pay more 
the more we alter a vote. 

Throughout the remainder of this section we will use the term
\emph{microbribe} to refer to flipping an entry in a preference
table, and we will use the term
\emph{microbribery} to refer to bribing possibly irrational voters
via microbribes.  Recall that by ``irrational voters'' we simply
mean that they are allowed to have, but 
not that they must have, irrational preferences.

For each rational~$\alpha$, $0 \leq \alpha \leq 1$,
we define the following two problems.

\begin{description}
\item[Name:] $\spmicrobribery{\copelandalphairrational}$ and
             $\spdestmicrobribery{\copelandalphairrational}$.
\item[Given:] A set $C$ of candidates, a
collection
$V$ of voters 
specified via their preference tables over $C$,
a distinguished candidate $p \in C$, and a nonnegative integer $k$.
\item[Question (constructive):]
  Is it possible, by flipping at most $k$ entries in the preference tables of
  voters in $V$, to ensure that $p$ is a winner of the resulting election?

\item[Question (destructive):] 
  Is it possible, by flipping at most $k$ entries in the preference tables
  of voters in $V$, to guarantee that $p$ is not a winner of the 
  resulting election?
\end{description}

We can flip multiple entries in the preference table of the same voter,
but we have to pay separately for each flip.  The microbribery problems for
$\copelandalphairrational$
are very similar in flavor to the so-called bribery$'$
problems for approval voting that were studied by
Faliszewski, Hemaspaandra, and Hemaspaandra~\cite{fal-hem-hem:c:bribery}, 
where unit cost for flipping
approvals or disapprovals
of voters are paid.  However, the proofs
for
$\copelandalphairrational$ seem to be much more involved than their
counterparts for approval voting.  The reason
is that
$\copelandalphairrational$ elections
allow for very subtle and complicated interactions
between the candidates' scores.

Before we proceed with our results, let us define some notation that will
be useful throughout this section.
Let $E$
be an election with candidate set $C = \{c_1, c_2, \ldots, c_m\}$ and
voter collection
$V = \{v_1, v_2, \ldots, v_n\}$.
We define two functions, $\wincost_E$ and $\tiecost_E$, 
that describe the costs of 
ensuring a victory or a tie of a given candidate in a particular head-to-head
contest.

\begin{definition}
  Let $E = (C,V)$ be an election and let $c_i$ and $c_j$ be two
  distinct candidates in~$C$.
\begin{enumerate}
\item By $\wincost_E(c_i,c_j)$ we mean the minimum
  number of microbribes that ensure that $c_i$ defeats $c_j$ in their
  head-to-head contest. If $c_i$ already wins this contest then
  $\wincost_E(c_i,c_j) = 0$.

\item By $\tiecost_E(c_i,c_j)$ we mean the
  minimum number of microbribes that ensure that $c_i$ ties with
  $c_j$ in their head-to-head contest, or $\infty$ if $E$ has an odd
  number of voters and thus ties are impossible.
\end{enumerate}
\end{definition}

Our first result regarding microbribery is that destructive
microbribery is easy for
$\copelandalphairrational$. 
Since this is the paper's first vulnerability proof, we take this
opportunity to remind the reader (recall
Section~\ref{sec:resistance-vulnerability}) that although the
definition of vulnerability requires only that there be a
polynomial-time algorithm to determine whether a successful action (in
the present case, a destructive microbribery) \emph{exists}, we will in
each vulnerability proof provide something far stronger, namely a
polynomial-time algorithm that both determines whether a successful
action exists and that, when so, finds a successful action (e.g., for
our flow algorithms later on, the successful action will be implicit in the
flow computed).

\begin{theorem}
  \label{thm:bribery-prime-destructive}
For each rational~$\alpha$, $0 \leq \alpha \leq 1$,
$\copelandalphairrational$ is
vulnerable to destructive microbribery in both the nonunique-winner
model and the unique-winner model.
\end{theorem}

\begin{proofs}
Fix an arbitrary rational number $\alpha$ with $0 \leq \alpha \leq 1$.
We give an algorithm for $\copelandalphairrational$, for destructive
microbribery in the nonunique-winner model. (We omit the analogous algorithm
for the unique-winner case.)

Let $E = (C,V)$ be the input election where $C = \{p,c_1,c_2,
\ldots,c_m\}$ and $V = \{v_1, v_2, \ldots, v_n\}$, and let $k$ be the
number of microbribes that we are allowed to make.  We define
the predicate $M(E,p,c_i,k)$ to be true if and only if there is a microbribery
of cost at most $k$ that ensures that $c_i$'s score is higher than
that of $p$.  Our algorithm computes $M(E,p,c_i,k)$ for each $c_i \in
C$ and accepts if and only if it is true for at least one of them. We
now describe how to compute $M(E,p,c_i,k)$.\footnote{We stress that
  we have optimized our algorithm for simplicity rather than for
  performance.}

We set $E_1$, $E_2$, and $E_3$ to be elections identical to $E$ except
that
\begin{enumerate}
\item in $E_1$, $p$ defeats $c_i$ in their head-to-head contest,
\item in $E_2$, $p$ loses to $c_i$ in their head-to-head contest, and
\item in $E_3$, $p$ ties $c_i$ in their head-to-head contest (we disregard
  $E_3$ if the number of voters is odd and thus ties are impossible).
\end{enumerate}
Let $k_1$, $k_2$, and $k_3$ be the minimum costs of microbriberies
that transform $E$ to $E_1$, $E$ to $E_2$, and $E$ to $E_3$, respectively.
Such
microbriberies involve only the head-to-head contest between $p$ and
$c_i$.
We define the  predicate $M'(E',p,c_i,k')$, where $E' \in \{E_1,E_2,E_3\}$
and where $k'$ is an integer, to be true if and only if
there is a microbribery of cost at most $k'$ that does not involve the
head-to-head contest between $p$ and $c_i$ but that ensures that
$c_i$'s $\copelandalphairrational$
score is higher than $p$'s. It is easy to see that
\[ M(E,p,c_i,k) \iff \left( M'(E_1,p,c_i,k-k_1) \lor M'(E_2,p,c_i,k-k_2) \lor
M'(E_3,p,c_i,k-k_3) \right).  \]
Thus it is enough to focus on the problem of computing
$M'(E',p,c_i,k')$.

Let $(E',k')$ be one of $(E_1,k-k_1)$, $(E_2,k-k_2)$, and $(E_3,k-k_3)$. Define
$\promote_{E'}(c_i,w',w'',t)$, where $c_i \in C$ is a candidate and
$w'$, $w''$, and $t$ are nonnegative integers, to be the minimum cost
of a microbribery that, when applied to $E'$, increases
$c_i$'s $\copelandalphairrational$
score by $w' + (1-\alpha)w'' + \alpha t$ via
ensuring that
\begin{enumerate}
\item $c_i$ wins an additional $w'$ head-to-head contests against candidates
  in $C - \{p\}$ that used to defeat $c_i$ originally,
\item $c_i$ wins an additional $w''$ head-to-head contests against
  candidates in $C - \{p\}$ with whom $c_i$ used to tie originally, and
\item $c_i$ ties an additional $t$ head-to-head contests with candidates
  in $C-\{p\}$ that used to defeat $c_i$ originally.
\end{enumerate}
If such a microbribery does not exist then we set
$\promote_{E'}(c_i,w',w'',t)$ to be $\infty$. 
It is an easy exercise to see that $\promote_{E'}$ is computable in
polynomial time by a simple greedy algorithm.

We define $\demote_{E'}(c_i,\ell',\ell'',t)$ to be the minimum cost of
a microbribery that, when applied to election $E'$, decreases $p$'s
score by $\ell' + \alpha\ell'' + (1-\alpha)t$ via ensuring that
\begin{enumerate}
\item $p$ loses an additional $\ell'$ head-to-head contests to
  candidates in $C-\{c_i\}$ whom $p$ used to defeat originally,
\item $p$ loses an additional $\ell''$ head-to-head contests to
  candidates in $C-\{c_i\}$ with whom $p$ used to tie originally, and
\item $p$ ties an additional $t$ head-to-head contests with candidates in
  $C-\{c_i\}$ whom $p$ used to defeat originally.
\end{enumerate}
If such a microbribery does not exist then we set
$\demote_{E'}(c_i,\ell',\ell'',t)$ to be $\infty$. Note that $\demote_{E'}$ can be
computed in polynomial time using an algorithm similar to that for
$\promote_{E'}$.

Naturally, the microbriberies used implicitly within
$\promote_{E'}(c_i,w',w'',t')$, within $\demote_{E'}(c_i,\ell',\ell'',t'')$, and
within transforming $E$ to $E'$ are ``disjoint,'' i.e., they
never involve the same pair of candidates.  Thus $M'(E',p,c_i,k')$ is
true if and only if there exist
integers
$w',w'',\ell',\ell'',t',t'' \in \{0, 1, \ldots,m\}$ such that 
\[
\copelandalphascore_{E'}(c_i) + (w'+\ell'    + (1-\alpha)(t''+w'') + \alpha (t' + \ell'')) - \copelandalphascore_{E'}(p) > 0 \\
\]
and
\[
\promote_{E'}(c_i,w',w'',t') + \demote_{E'}(c_i,\ell',\ell'',t'') \leq k.
\]
There are only polynomially many combinations of such
$w',w'',\ell',\ell'',t'$, and $t''$, and we can try them all. Thus we
have given a polynomial-time algorithm for $M'(E',p,c_i,k')$. Via the
observations given at the beginning of our proof this implies that
$M(E,p,c_i,k)$ is computable in polynomial time and the proof
is complete.~\end{proofs}

The above destructive-case algorithm and approach is fairly straightforward; in the
destructive case we do not need to worry about any side effects of
promoting $c$ and demoting~$p$. The constructive case 
is
more complicated, but we still are able to obtain
polynomial-time algorithms via a fairly involved use of flow
networks to model how particular points shift between
candidates. In the remainder of this section we restrict ourselves to
the values $\alpha \in \{0,1\}$ or settings where the number of voters is odd and
so ties never happen. We remind the reader that $\copelandone$ and
$\copelandalphavarirrational{1}$, respectively, refer to Llull voting.

A flow network is a network of nodes with directed edges through which
we want to transport some amount of flow from the source to the sink
(these are two designated nodes). Each edge $e$ can carry up to $c(e)$
units of flow, and transporting each unit of flow through $e$ costs
$a(e)$. 
In the min-cost-flow problem we have a target flow value~$F$,
and the goal is to find a way
of transporting $F$ units of flow from the source to the sink, while 
minimizing the cost. (If there is no way of achieving target flow $F$, the
cost in effect is infinite.)

We now define the notions related to flow networks more formally.
Let $\naturals = \{0, 1, 2, \ldots \}$ and
$\integers = \{ \ldots, -2, -1, 0, 1, 2, \ldots \}$.

\begin{definition}
\label{def:flows}
\begin{enumerate}
\item A \emph{flow network} is a quintuple $(K,s,t,c,a)$,
where $K$ is a set of nodes
that includes the \emph{source} $s$ and the \emph{sink} $t$,
$c: K^2 \rightarrow \naturals$ is the \emph{capacity function},
and $a: K^2 \rightarrow \naturals$
is the \emph{cost function}. We assume that $c(u,u) = a(u,u) = 0$
for each node $u \in K$, and that at most one of $c(u,v)$ and
$c(v,u)$ is nonzero for each pair of distinct nodes $u, v \in K$.
We also assume that if $c(u,v) = 0$ then
$a(u,v) = 0$ as well.
\item Given a flow network $(K,s,t,c,a)$, a \emph{flow} is a
function $f: K^2 \rightarrow \integers$ that satisfies the following
conditions:
\begin{enumerate}
\item For each $u,v \in K$, we have $f(u,v) \leq c(u,v)$,
i.e., capacities limit the flow.
\item For each $u,v \in K$, we have $f(u,v) = -f(v,u)$.\footnote{Note that each flow is
fully defined via its nonnegative values.
Whenever we speak of a flow
(e.g., when defining some particular flows) we will just speak of its
nonnegative part.}

\item For each $u \in K - \{s,t\}$, we have $\sum_{v \in K}f(u,v) = 0$,
i.e., the flow is conserved in all nodes
  except the source and the sink.
\end{enumerate}

\item The \emph{value of flow $f$} is:
\[ 
   \flowvalue(f) = \sum_{v \in K}f(s,v). 
\]
The particular flow network we
have in mind will always be
clear from the context and so we will not indicate it explicitly
(we will not write it explicitly as a subscript to the function~$\flowvalue$).

\item The \emph{cost of flow $f$} is defined as:
\[ 
  \flowcost(f) = \sum_{u,v \in K} a(u,v)f(u,v). 
\] 
That is, we pay the price $a(u,v)$ for each unit of flow that passes
from node $u$ to node~$v$. 
\end{enumerate}
\end{definition}

Given a flow network $(K,s,t,c,a)$ we will often use 
the term \emph{edges} to refer to pairs of distinct
nodes $(u,v) \in K^2$ for which $c(u,v) > 0$.

Below we define the min-cost-flow problem,
which is well known from the literature.
The definition we employ here is not the most general one
but will suffice for our needs.
(Readers seeking a broader discussion of the problem may wish to
see, for example, the monograph by Ahuja, Magnanti, and
Orlin~\cite{ahu-mag-orl:b:flows}.)

\begin{definition}
  We define the \emph{min-cost-flow problem} as follows: Given a flow
  network $(K,s,t,c,a)$ and a target flow value~$F$, find a flow
  $f$ that has value $F$ (if one exists) and has minimum cost among
  all such flows, or otherwise indicate that no such flow $f$ exists.
\end{definition}

The min-cost-flow problem has a polynomial-time
algorithm.\footnote{%
  The min-cost-flow problem is often
  defined in terms of capacity and cost functions that are not
  necessarily limited to
  nonnegative integer values and so the corresponding flows
  are not restricted to integer values either.  However, crucially for
  us, it is known that if the capacity and cost functions have
  integral values (as we have assumed) then there exist optimal
  solutions to 
  the min-cost-flow problem
  that use only integer-valued flows and that can be found in
  polynomial time.  } There is a large body of work devoted to
flow problems and we will not even attempt to provide a complete list of
references here. Instead, we again point the reader to
the excellent
monograph by Ahuja, Magnanti, and Orlin~\cite{ahu-mag-orl:b:flows},
which provides descriptions of polynomial-time algorithms, theoretical
analysis, and numerous references to previous work on flow-related
problems.  We also mention that the issue of flows is so prevalent in
the study of algorithms that the textbook of Cormen et
al.~\cite[p.~787]{cor-lei-riv-ste:b:algorithms-second-edition} contains an
exposition of %
the min-cost-flow problem. %

Coming back to the study of constructive microbribery for Llull and
Copeland$^0$, with irrational voters allowed,
we now present the following result.

\begin{theorem}
  \label{thm:bribery-prime}
  For $\alpha \in \{0,1\}$, $\copelandalphairrational$ is
  vulnerable to constructive microbribery, in both
  the nonunique-winner model and the unique-winner model.
\end{theorem}

We prove Theorem~\ref{thm:bribery-prime} via
Lemmas~\ref{lem:bribery-prime-odd} through~\ref{lem:bribery-prime-llull-even}
below, which
cover three cases: 
(a)~an odd number of voters, where  all 
$\copelandalphairrational$ elections with $0 \leq \alpha \leq 1$ 
are identical due to the lack of ties, 
(b)~$\copelandalphavarirrational{1}$ with an even number of voters, and
(c)~$\copelandalphavarirrational{0}$ with an even number of voters.
These lemmas only discuss
the nonunique-winner model but in each case it is easy to see how to 
change the algorithms and proofs to make them work
for the unique-winner model.

\begin{lemma}
  \label{lem:bribery-prime-odd}
  For each rational $\alpha$ with  $0 \leq \alpha \leq 1$,
  there is a polynomial-time algorithm that
  solves the constructive microbribery problem for  
  $\copelandalphairrational$
  elections with an odd number of voters (in the nonunique-winner model).
\end{lemma}

\begin{proofs}
Our input is a nonnegative integer $k$ (the budget) and an election
$E = (C,V)$, where the candidate set $C$ is $\{c_0, c_1, \ldots,
c_m\}$, the number of voters is odd, and $p = c_0$ is the candidate whose
victory we want to ensure via at most $k$ microbribes.  Note that we
interchangeably use $p$ and $c_0$ to refer to the same candidate, since it
is sometimes convenient to be able to speak of $p$ and all other
candidates uniformly.  As the number of voters is odd, ties never
occur.  Thus any candidate $c_i$ has the same
$\copelandalphairrational$
score for each rational value of~$\alpha$, $0 \leq \alpha \leq 1$.
Fix an arbitrary such~$\alpha$.

We give a polynomial-time algorithm for the constructive microbribery
problem.  A high-level overview is that we try to find a
threshold value $T$ such that there is a microbribery of cost at most
$k$ that transforms $E$ into $E'$ such that (a) $p$ has $\copelandalphascore_{E'}$ exactly $T$, and (b) every
other candidate has $\copelandalphascore_{E'}$ at most~$T$.

Let $B$ be a number that is greater than the cost of any possible
microbribery within $E$ (e.g., $B = \|V\| \cdot \|C\|^2+1$).
For each possible threshold $T$, we consider a min-cost-flow instance $I(T)$
with node set $K = C \cup \{s,t\}$, where $s$ is the source and $t$ is
the sink,
the edge capacities and costs are specified in Figure~\ref{fig:min-cost-odd},
and the target flow value is
\[ 
  F = \sum_{c_i \in C} \copelandalphascore_E(c_i) = \frac{\|C\|(\|C\|-1)}{2}. 
\]
\begin{figure}[!tb]
\begin{center}
\begin{tabular}{|l|l|}
  \hline
  \multicolumn{1}{|c|}{Edge} & \multicolumn{1}{c|}{Parameters} \\
  \hline\hline
  \begin{tabular}{l}
  $e = (s,c_i)$, \\ where $c_i \in C$ 
  \end{tabular}
  & 
  $\begin{array}{l} c(e) = \copelandalphascore_E(c_i) \\  a(e) = 0\end{array}$ \\
  \hline

  \begin{tabular}{l}
  $e = (c_i,c_j)$, \\
  where $c_i, c_j \in C$ and $\versus_E(c_i,c_j) > 0$ 
  \end{tabular}
  &  
  $\begin{array}{l} c(e) = 1  \\ a(e) = \wincost_E(c_j,c_i) \end{array}$ \\

  \hline
    \begin{tabular}{l}
    $ e = (c_0,t)$ 
    \end{tabular}& 
  $\begin{array}{l} c(e) = T \\ a(e) = 0 \end{array}$ \\

  \hline
  \begin{tabular}{l}
  $ e = (c_i,t)$, \\ 
  where $i > 0$ and $c_i \in C$
  \end{tabular}
  & 
  $\begin{array}{l} c(e) = T \\  a(e) = B \end{array}$ \\
  \hline
  
  \begin{tabular}{l}
  every other edge $e$
  \end{tabular} 
   & $\begin{array}{l} c(e) = 0 \\ a(e) = 0 \end{array}$\\
  \hline
\end{tabular}
\end{center}
\caption{\label{fig:min-cost-odd}
Edge capacities and costs for min-cost-flow instance $I(T)$,
built from election $E$.
}
\end{figure}

\begin{example}
For illustration, consider the following example.  Suppose the given
election $E$ has four candidates and three voters, and the preference
tables of the voters (who each happen to be rational in this example)
can be obtained from their preference orders that are shown in
Figure~\ref{fig:constructive-microbribery-odd}, which also gives the
corresponding values of $\versus_E(c_i,c_j)$ for each pair of
candidates.  Thus we have $\copelandalphascore_E(c_0) = 2$,
$\copelandalphascore_E(c_1) = 0$, $\copelandalphascore_E(c_2) = 3$,
and $\copelandalphascore_E(c_3) = 1$.  Suppose further that we are
allowed to perform one microbribe, so $k=1$.  Clearly, one microbribe that
changes the preference of the third voter from $c_2 > c_0$ to $c_0 > c_2$ will
flip the outcome of their head-to-head contest from $c_2$ winning to
$c_0$ winning, which is enough to reach our goal of making $c_0$ win
the election, and this is of course the cheapest possible successful microbribery.
Finally, note that in this example we have $B = 49$.

\begin{figure}[!tb]
\begin{center}
\begin{tabular}{ll}
Voter~$1:$ & $c_0 > c_1 > c_2 > c_3$ \\ 
Voter~$2:$ & $c_3 > c_2 > c_1 > c_0$ \\ 
Voter~$3:$ & $c_2 > c_0 > c_3 > c_1$ \\ 
\end{tabular}
\quad\quad
\begin{tabular}{|c|rrrr|}
 \hline
$\versus_E(c_i,c_j)$ &
 \multicolumn{1}{c}{$c_0$} &
 \multicolumn{1}{c}{$c_1$} &
 \multicolumn{1}{c}{$c_2$} &
 \multicolumn{1}{c|}{$c_3$} \\
 \hline
       $c_0$         &  $0$  & $1$   & $-1$  &  $1$   \\
       $c_1$         & $-1$  & $0$   & $-1$  & $-1$   \\
       $c_2$         &  $1$  & $1$   &  $0$  &  $1$   \\
       $c_3$         & $-1$  & $1$   & $-1$  &  $0$   \\
 \hline
\end{tabular}
\end{center}
\caption{\label{fig:constructive-microbribery-odd}
Sample election $E$ for Example~\ref{exa:constructive-microbribery-odd}
in the proof of Lemma~\ref{lem:bribery-prime-odd}.}
\end{figure}

For each threshold $T$ with $0 \leq T \leq 3$, the flow network $I(T)$
corresponding to this instance $(E,c_0,k)$ of the constructive
microbribery problem is shown in
Figure~\ref{fig:constructive-microbribery-odd-flownetwork}, and we
have a target flow value of $F = 6$.  Every edge $e$ in this flow
network is labeled by the pair $(c(e),a(e))$ of numbers that give the
capacity and the cost of edge~$e$, respectively.

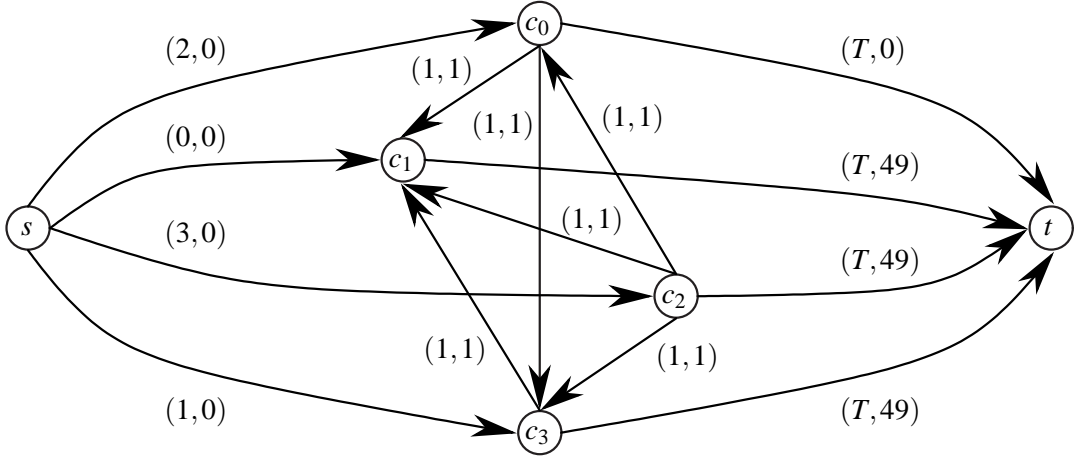
\begin{figure}[!tb]
\begin{center}
\input{constructive-microbribery-odd-flownetwork.eepic}
\end{center}
\caption{\label{fig:constructive-microbribery-odd-flownetwork}
Flow network $I(T)$ corresponding to the instance 
$(E,c_0,k)$ of Example~\ref{exa:constructive-microbribery-odd}.}
\end{figure}
\label{exa:constructive-microbribery-odd}
\end{example}

To continue the proof of Lemma~\ref{lem:bribery-prime-odd},
note that with an odd number of voters, constructive microbribery in
$\copelandalphairrational$
simply requires us to choose for which pairs of distinct candidates
we want to flip the outcome of their head-to-head contest
in order to ensure $p$'s victory.  Thus it is sufficient to represent
a microbribery $M$ as a collection of pairs $(c_i, c_j)$
of distinct candidates
for whom we need to flip the result of their head-to-head
contest from $c_i$ winning to $c_j$ winning.  Clearly, given such a
collection~$M$, the cheapest way to implement it costs
\[
  \sum_{(c_i,c_j)\in M} \wincost_E(c_j,c_i).
\]

A crucial observation for our algorithm is that we can directly
translate flows to microbriberies using the following interpretation.
Let $f$ be a flow (as per Definition~\ref{def:flows} with all edge flows being integers)
of value $F$ within instance $I(T)$.
The units of flow that travel
through the network correspond to
$\copelandalphairrational$ points. For each~$c_i \in C$,
we interpret the amount of flow that goes directly from $s$ to
$c_i$ as the number of
$\copelandalphairrational$ points that $c_i$ has before any
microbribery is attempted,\footnote{Note that for each $c_i \in C$
  any flow of value $F$ within $I(T)$ needs to send exactly
  $\copelandalphascore_E(c_i)$ units from $s$ to $c_i$.} and the amount of flow
that goes directly from $c_i$ to $t$ as the number of
$\copelandalphairrational$
points that $c_i$ has after the microbribery (defined by the flow).  The
units of flow that travel between distinct
$c_i$'s (i.e., through edges of the
form $(c_i,c_j)$, $i \neq j$)
correspond to the microbribes exerted: A unit
of flow traveling from node $c_i$ to $c_j$ corresponds to changing the
result of the head-to-head contest between $c_i$ and $c_j$ from $c_i$
winning to $c_j$ winning.
In this case, the
$\copelandalphairrational$ point moves from $c_i$ to
$c_j$ and the cost of the flow increases by $a(c_i,c_j) =
\wincost(c_j,c_i)$, exactly the minimum cost of a microbribery that
flips this contest's result. 
Let $M_f$ be the microbribery defined, as just described, by flow $f$.
It is easy to see that
\[
  \flowcost(f) =  B\cdot(F - f(c_0,t)) + \sum_{(c_i,c_j)\in M_f} \wincost_E(c_j,c_i).
\]
Thus we can easily extract the cost of microbribery $M_f$ from the cost
of flow $f$.

Our algorithm crucially depends on this correspondence
between flows and microbriberies. (Also, in the proofs of
Lemmas~\ref{lem:bribery-prime-copeland-even}
and~\ref{lem:bribery-prime-llull-even} that cover the case of an even
number of voters we simply show how to modify the instances $I(T)$ to
handle ties, and we show correspondences between the new networks and
microbriberies; the rest of these proofs is the same as here.)

Note that for small values of $T$ no flow of value $F$ exists for
$I(T)$.  The reason for this is that the edges coming into the sink
$t$ might not have enough capacity so as to hold a flow of value
$F$.  In such a situation it is impossible to guarantee
that every candidate gets at most $T$ points; there are too many
$\copelandalphairrational$ points to distribute.

Figure~\ref{fig:flow-alg} gives our algorithm for
constructive microbribery in $\copelandalphairrational$.
\begin{figure}[!tb]
\begin{tabbing}
123\=123\=123\=123\=123\=123\kill
\> \textbf{procedure}
$\copelandalphairrational$-odd-microbribery$(E = (C,V), k, p)$ \\
\> \textbf{begin} \\
\>\>   \textbf{if} $p$ is a winner of $E$ \textbf{then} %
\textbf{accept}; \\
\>\>    $F = \sum_{c_i \in C} \copelandalphascore_E(c_i) =
                                     \frac{\|C\|(\|C\|-1)}{2}$; \\
\>\>    \textbf{for} $T = 0$ to $\|C\|-1$ \textbf{do} \\
\>\>    \textbf{begin} \\
\>\>\>       build an instance $I(T)$ of min-cost-flow; \\
\>\>\>       \textbf{if} $I(T)$ has no flow of value $F$ \textbf{then} \\
\>\>\>\>          restart the for loop with the next value of $T$; \\
\>\>\>          $f = $ a minimum-cost flow for $I(T)$; \\ %
\>\>\>          \textbf{if} $f(c_0,t) < T$ \textbf{then} %
restart the loop; \\
\>\>\>          $\kappa = \flowcost(f) - B\cdot ( F - T)$; \\
\>\>\>          \textbf{if} $\kappa \leq k$ \textbf{then} %
\textbf{accept}; \\
\>\>    \textbf{end}; \\
\>\>    \textbf{reject}; \\
\> \textbf{end} \\
\end{tabbing}
\caption{\label{fig:flow-alg}The
constructive microbribery algorithm for $\copelandalphairrational$
elections with an odd number of voters.}
\end{figure}
This algorithm runs in polynomial time since, as we have already mentioned,
the min-cost-flow problem is solvable in polynomial time.  

Let us
now prove that this algorithm is correct.
We have presented above how a flow $f$ of value $F$ within the flow
network $I(T)$ (with $0 \leq T \leq F$) defines a microbribery. Based on
this, it is clear that if our algorithm accepts then there is a
microbribery of cost at most $k$ that ensures $p$'s victory.

On the other hand, suppose now that there exists a microbribery of
cost at most $k$ that ensures $p$'s victory in the election.  We will
show that our algorithm accepts in this case.

Let $M$ be a minimum-cost bribery (of cost at most $k$) that ensures
$p$'s victory. As pointed out above, $M$ can be represented as a
collection of pairs $(c_i,c_j)$ of distinct candidates for whom we
flip the result of the head-to-head contest from $c_i$ winning
to $c_j$ winning. The cost of $M$ is
\[ 
  \sum_{(c_i,c_j) \in M} \wincost_E(c_j,c_i). 
\]
Since applying microbribery $M$ ensures that $p$ is a winner, 
we have that each candidate among $c_1, c_2, \ldots, c_m$ has at most as many
$\copelandalphairrational$ points as $p$ does.  
Let $E'$ be the election that results from $E$ after applying microbribery
$M$ to $E$ (i.e., after flipping the results of the contests specified by
$M$ in an optimal way, as given by $\wincost_E$).  Let $T'$ be
$\copelandalphascore_{E'}(p)$, $p$'s
$\copelandalphairrational$ score after
implementing~$M$.
Clearly, $0 \leq T' \leq \|C\|-1$.

Consider instance $I(T')$ and let $f_M$ be the flow that corresponds to
the microbribery $M$. 
In this flow each edge of the form $(s,c_i)$
carries flow of its maximum capacity, $\copelandalphascore_E(c_i)$,
each edge of the
form $(c_i,c_j)$ 
carries one unit of flow exactly if $e$ is listed in~$M$
and carries zero units of flow otherwise, and each edge of the form
$(c_i,t)$ carries $\copelandalphascore_{E'}(c_i)$ units of flow.  It is easy to
see that this is a legal flow.  The cost of $f_M$ is
\[ 
  \flowcost(f_M) = B\cdot( F - T') + \sum_{(c_i,c_j) \in M} \wincost_E(c_j,c_i).
\] 
After applying $M$, $p$ gets $T'$
$\copelandalphairrational$
points that travel to the sink $t$ via edge $(c_0, t)$ with cost
$a(c_0,t) = 0$, and all the remaining $F - T'$ points travel via
edges $(c_i,t)$, $i \in \{1, 2, \ldots, m\}$, with cost $a(c_i,t) = B$.
The remaining part of $\flowcost(f_M)$ is the cost of the units of flow
traveling through the edges $(c_i,c_j)$ that directly correspond to the
cost of microbribery $M$.

Now consider some minimum-cost flow $f_{\min}$ for
$I(T')$. 
Since $f_M$ exists,
a minimum-cost flow must exist as well.
Clearly, we have 
\[
\flowcost(f_{\min}) \leq \flowcost(f_M).
\]

Let $T''$ be the number of
units of flow that $f_{\min}$ assigns to travel over the edge $(c_0,t)$,
i.e., $T'' = f_{\min}(c_0,t)$. 
The only edges with nonzero cost for sending flow through them
are those in the set $\{(c_i,c_j) \mid c_i,c_j \in C \land \versus_E(c_i,c_j) > 0\} \cup \{(c_i,t) \mid i \in \{1,\ldots,m\}\}$
and thus the cost of $f_{\min}$ can be expressed as (recall that
$\versus_E(c_i,c_j) > 0$ implies $i \neq j$)
\[ 
   \flowcost(f_{\min}) = B\cdot(F - T'') + \sum_{c_i,c_j \in C \land \versus_E(c_i,c_j)>0} f_{\min}(c_i,c_j)\cdot\wincost_E(c_j,c_i). 
\]
$B > \sum_{i,j, i \neq j} \wincost_E(c_i,c_j)$, for each
$c_i,c_j \in C$ such that $\versus_E(c_i,c_j) > 0$ we have
$f_{\min}(c_i,c_j) \in \{0,1\}$, and $\flowcost(f_{\min}) \leq
\flowcost(f_M)$, so it must hold that $T'' = T'$ and
\[
  \sum_{c_i,c_j \in C \land \versus_E(c_i,c_j)>0} f_{\min}(c_i,c_j)\cdot\wincost_E(c_j,c_i)
  \leq
  \sum_{(c_i,c_j) \in M} \wincost_E(c_j,c_i).
\]
Thus flow $f_{\min}$ corresponds to a microbribery that guarantees
$p$'s victory and has cost at most as high as that of $M$. Since $M$
was chosen to have minimum cost among all such microbriberies, flow
$f_{\min}$ corresponds to a microbribery of minimum cost and our
algorithm correctly accepts within the for loop of Figure~\ref{fig:flow-alg}, at the 
very latest when in the for loop $T$ is set to
$T'$.~\end{proofs}

We now turn to the algorithms showing that Llull and Copeland$^0$,
with irrational voters allowed, are
vulnerable to constructive microbribery when the number of voters is
even.  In this case we need to take into account
that it sometimes is more desirable to have some candidates tie each
other in a head-to-head contest than to have
one of them win the contest.

\begin{lemma}
  \label{lem:bribery-prime-even-copeland-ties}
  Let $E = (C,V)$ be an election with candidate set
  $C = \{c_0, c_1, \ldots, c_m\}$
  and with an even number of voters, specified via preference
  tables over $C$. If the election is conducted using
  $\copelandalphavarirrational{0}$ then
  no minimum-cost microbribery that ensures victory for $c_0$
  involves either (a)~flipping a result of a head-to-head contest
  between any two distinct candidates
  $c_i, c_j \in C - \{c_0\}$ from $c_i$ winning
  to $c_j$ winning, or (b)~changing a result of a head-to-head contest
  between
  any two distinct candidates in $C - \{c_0\}$ from a tie to one of 
  them winning.
\end{lemma}

\begin{proofs}
Our proof follows by way of contradiction.  Let $E = (C,V)$ be an election as
specified in the lemma.  For the sake of a contradiction
suppose there is a minimum-cost
microbribery $M$ that makes $c_0$ win and that there are two distinct
candidates, $c_i$ and $c_j$, in $C - \{c_0\}$
such that microbribery $M$ involves switching the result of the
head-to-head contest between these candidates
from $c_i$ winning to $c_j$ winning or from a tie to one of them
winning.  Consider the microbribery $M'$ that is identical to~$M$,
except that it makes $c_i$ tie with $c_j$ in a head-to-head
contest, either via an appropriate number of microbribes if $c_i$ and $c_j$ do
not tie originally, or via
leaving the corresponding preference-table entries untouched if they do tie
initially.
Clearly, this microbribery $M'$ has a lower cost than $M$ and it
still ensures $c_0$'s victory. This is a contradiction.~\end{proofs}

With Lemma~\ref{lem:bribery-prime-even-copeland-ties}
at hand, we can show that constructive microbribery is easy for
$\copelandalphavarirrational{0}$ for the case
of an even number of voters.

\begin{lemma}
  \label{lem:bribery-prime-copeland-even}
  There is a polynomial-time algorithm that solves the 
  constructive microbribery problem for
  $\copelandalphavarirrational{0}$ elections with an even number of voters
  (in the nonunique-winner model).
\end{lemma}

\begin{proofs}
Our input is election $E = (C,V)$, where $C = \{c_0, c_1, \ldots, c_m\}$,
$p = c_0$,
and $V$ is a collection of an even number of voters, each specified
via a preference table over $C$. Our algorithm is essentially the same
as that used in the proof of Lemma~\ref{lem:bribery-prime-odd},
except that instead of using
instances $I(T)$ we now use instances $J(T)$ defined below. 
In this proof we show how to
construct these instances and how they correspond to microbriberies within
$E$. The proof of Lemma~\ref{lem:bribery-prime-odd} shows how to
use such a correspondence to solve the microbribery problem at hand.

Let $T$ be a nonnegative integer, $0 \leq T \leq \|C\|-1$.  Instance
$J(T)$ is somewhat different from the instance $I(T)$ used in the proof of
Lemma~\ref{lem:bribery-prime-odd}. 
In particular, due to
Lemma~\ref{lem:bribery-prime-even-copeland-ties}
(and the fact that our goal is to
make $c_0$ a winner), we model only microbriberies
that have the following effects on our election:
\begin{enumerate}
  \item For any two distinct
     candidates $c_i$, $c_j$ in $C - \{c_0\}$, the result of the
     head-to-head contest between $c_i$ and $c_j$ may possibly turn into a tie.
   \item For each candidate $c_i$ in $C - \{c_0\}$, the result of a
     head-to-head contest between $c_0$ and $c_i$ may possibly turn into
     either a tie (from $c_i$ defeating $c_0$) or into $c_0$ defeating
     $c_i$ (from either a tie or from $c_i$ defeating $c_0$).
\end{enumerate}

Our instance $J(T)$ contains special nodes, namely the elements of
the sets $C'$ and $C''$ below, to handle these possible 
interactions. We define
\begin{eqnarray*}
  C' & = & \{ c_{ij} \condition i,j \in \{1, 2, \ldots, m\}
                      \land \versus_E(c_i,c_j) > 0 \} \mbox{ and} \\
  C''& = & \{ c_{i0} \condition i \in \{1, 2, \ldots, m\}
                      \land \versus_E(c_i,c_0) \geq 0 \}.
\end{eqnarray*}
For each possible threshold~$T$,
define $J(T)$ to be the flow network with node set
$K = C \cup C' \cup C'' \cup \{s,t\}$,
where $s$ is the source, $t$ is the sink, and the edge capacities and
costs are as stated in Figure~\ref{fig:min-cost-copeland}. 
(As before, we set $B$ to be a number that is greater than the cost of any possible
microbribery within $E$, e.g., $B = \|V\| \cdot \|C\|^2+1$.)
The target flow value is 
\[
  F = \sum_{v \in K}c(s,v).
\]

\begin{figure}[!tb]
\begin{center}
\begin{tabular}{|l|l|}
  \hline
  \multicolumn{1}{|c|}{Edge} & \multicolumn{1}{c|}{Parameters} \\
  \hline\hline
  \begin{tabular}{l}
  $e = (s,c_i)$, \\ where $c_i \in C$ 
  \end{tabular}
  & 
  $\begin{array}{l} c(e) = \score^0_E(c_i) \\  a(e) = 0\end{array}$ \\
  \hline

  \begin{tabular}{l}
  $e = (c_i,c_{ij})$, \\
  where $c_i,c_j \in C - \{c_0\}$ and $\versus_E(c_i,c_j) > 0$
  \end{tabular}
  &  
  $\begin{array}{l} c(e) = 1  \\ a(e) = \tiecost_E(c_j,c_i) \end{array}$ \\
  \hline

  \begin{tabular}{l}
  $ e = (c_i,t)$, \\ 
  where $c_i \in C - \{c_0\}$
  \end{tabular}
  & 
  $\begin{array}{l} c(e) = T \\  a(e) = B \end{array}$ \\
  \hline

  \begin{tabular}{l}
  $ e = (c_{ij},t)$, \\ 
  where $c_i,c_j \in C - \{c_0\}$ and $\versus_E(c_i,c_j) > 0$
  \end{tabular}
  & 
  $\begin{array}{l} c(e) = 1 \\  a(e) = B \end{array}$ \\
  \hline

  \begin{tabular}{l}
  $e = (c_i,c_{i0})$, \\
  where $c_i \in C - \{c_0\}$ %
  and $\versus_E(c_i,c_0) > 0$
  \end{tabular}
  &  
  $\begin{array}{l} c(e) = 1  \\ a(e) = \tiecost_E(c_0,c_i) \end{array}$ \\
  \hline

  \begin{tabular}{l}
  $e = (c_{i0},c_{0})$, \\
  where $c_i \in C - \{c_0\}$  and $\versus_E(c_i,c_0) \geq 0$
  \end{tabular}
  &  
  $\begin{array}{l} c(e) = 1  \\ a(e) = \wincost_E(c_0,c_i) - \tiecost_E(c_0,c_i) \end{array}$ \\
  \hline

  \begin{tabular}{l}
  $e = (s,c_{i0})$, \\
  where $c_i \in C - \{c_0\}$
  and $\versus_E(c_i,c_0) = 0$
  \end{tabular}
  &  
  $\begin{array}{l} c(e) = 1  \\ a(e) = 0 \end{array}$ \\
  \hline

  \begin{tabular}{l}
  $e = (c_{i0},t)$, \\
  where $c_i \in C - \{c_0\}$  and $\versus_E(c_i,c_0) \geq 0$
  \end{tabular}
  &  
  $\begin{array}{l} c(e) = 1 \\  a(e) = B \end{array}$ \\
  \hline

    \begin{tabular}{l}
    $ e = (c_0,t)$ 
    \end{tabular}& 
  $\begin{array}{l} c(e) = T \\ a(e) = 0 \end{array}$ \\

  \hline
  
  \begin{tabular}{l}
  every other edge $e$
  \end{tabular} 
   & $\begin{array}{l} c(e) = 0 \\ a(e) = 0 \end{array}$\\
  \hline
\end{tabular}
\end{center}
\caption{\label{fig:min-cost-copeland}Edge capacities and costs for
 min-cost-flow instance~$J(T)$, built from election~$E$.}
\end{figure}

Instance $J(T)$ is fairly complicated but it in fact does closely follow the
instance of microbribery that we have at hand. As in the
proof of Lemma~\ref{lem:bribery-prime-odd},
the units of flow that travel through the
network are interpreted as $\copelandalphavarirrational{0}$ points, and flows
are interpreted as specifying microbriberies.

\begin{example}
\label{exa:constructive-microbribery-even-copeland}
For illustration, we add a fourth voter to the election given in
Example~\ref{exa:constructive-microbribery-odd}; call the resulting
election~$E$.  Again, all voters happen to be rational, and their
preference tables can easily be derived from the preference orders
shown in Figure~\ref{fig:constructive-microbribery-even-copeland},
which also provides the corresponding values of $\versus_E(c_i,c_j)$
for each pair of candidates.  We now have $\copelandzeroscore_E(c_0) =
\copelandzeroscore_E(c_3) = 1$, $\copelandzeroscore_E(c_1) = 0$, and
$\copelandzeroscore_E(c_2) = 3$.  Suppose we are allowed to do two
microbribes and so have $k=2$.  Two microbribes that change, say, the
fourth voter's preference table to him or her now ranking $c_0$ above
of both $c_2$ and $c_3$ yields an election $E'$ with
$\copelandzeroscore_{E'}(c_0) = \copelandzeroscore_{E'}(c_2) = 2$,
$\copelandzeroscore_{E'}(c_3) = 1$, and $\copelandzeroscore_{E'}(c_1) = 0$,
thus reaching our goal of making $c_0$ a winner of the election.  It
is easy to see that this is a cheapest among all possible successful
microbribes.  Finally, note that in this example we have $B = 65$.

\begin{figure}[!tb]
\begin{center}
\begin{tabular}{ll}
Voter~$1:$ & $c_0 > c_1 > c_2 > c_3$ \\ 
Voter~$2:$ & $c_3 > c_2 > c_1 > c_0$ \\ 
Voter~$3:$ & $c_2 > c_0 > c_3 > c_1$ \\ 
Voter~$4:$ & $c_2 > c_3 > c_0 > c_1$ \\ 
\end{tabular}
\quad\quad
\begin{tabular}{|c|rrrr|}
 \hline
$\versus_E(c_i,c_j)$ &
 \multicolumn{1}{c}{$c_0$} &
 \multicolumn{1}{c}{$c_1$} &
 \multicolumn{1}{c}{$c_2$} &
 \multicolumn{1}{c|}{$c_3$} \\
 \hline
       $c_0$         &  $0$  & $2$   & $-2$  &  $0$   \\
       $c_1$         & $-2$  & $0$   & $-2$  & $-2$   \\
       $c_2$         &  $2$  & $2$   &  $0$  &  $2$   \\
       $c_3$         &  $0$  & $2$   & $-2$  &  $0$   \\
 \hline
\end{tabular}
\end{center}
\caption{\label{fig:constructive-microbribery-even-copeland} A sample
election $E$ for
Example~\ref{exa:constructive-microbribery-even-copeland} in the proof
of Lemma~\ref{lem:bribery-prime-copeland-even}.}
\end{figure}

For any threshold $T$ with $0 \leq T \leq 3$, the flow network $J(T)$
corresponding to this instance $(E,c_0,k)$ of the constructive
microbribery problem is shown in
Figure~\ref{fig:constructive-microbribery-even-copeland-flownetwork},
and we have a target flow value of $F = 6$.  Every edge $e$ in
this flow network is labeled by the pair $(c(e),a(e))$ of numbers that
give the capacity and the cost of edge~$e$, respectively.

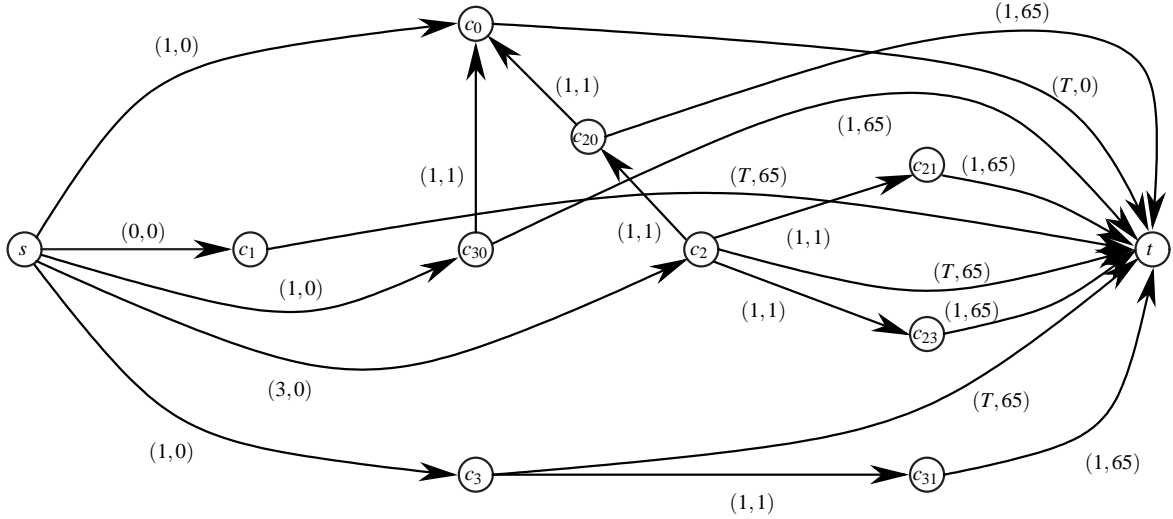
\begin{figure}[!tb]
\begin{center}
\input{constructive-microbribery-even-copeland-flownetwork.eepic}
\end{center}
\caption{\label{fig:constructive-microbribery-even-copeland-flownetwork}
Flow network $J(T)$ corresponding to the instance 
$(E,c_0,k)$ of Example~\ref{exa:constructive-microbribery-even-copeland}.}
\end{figure}
\end{example}

To continue the proof of Lemma~\ref{lem:bribery-prime-copeland-even},
let us now describe a bit more precisely how we interpret our flow network
$J(T)$.  In particular, we will argue that
each flow $f$ of value $F$ that travels through the network $J(T)$
corresponds to a microbribery
within $E$ that gives each candidate $c_i \in C$ exactly $f(c_i,t)$
$\copelandalphavarirrational{0}$ points:
\begin{enumerate}
\item For each $c_i \in C$, the units of flow that enter $c_i$ from $s$
  correspond to the number of $c_i$'s $\copelandalphavarirrational{0}$
points in $E$, prior to any microbribery.

\item For each $c_i \in C$, the units of flow that go directly from
  $c_i$ to $t$ correspond to the number of $c_i$'s
  $\copelandalphavarirrational{0}$ points after a microbribery as
  specified by the flow.

\item For each pair of distinct candidates
  $c_i, c_j \in C - \{c_0\}$ such that $c_i$ defeats
  $c_j$ in their
  head-to-head contest in~$E$, we have an edge $e = (c_i, c_{ij})$
  with capacity
  one and cost $\tiecost_E(c_j,c_i)$.  A unit of flow that travels
  through $e$ corresponds to a microbribe that makes $c_i$ tie with~$c_j$:
  $c_i$ loses the point, we pay
  $\tiecost_E(c_j,c_i)$, and then the unit of flow goes directly to~$t$.
  (From Lemma~\ref{lem:bribery-prime-even-copeland-ties}
  we know that we do not need to handle any other possible interactions between
  $c_i$ and $c_j$ in their head-to-head contest.)

\item For each candidate $c_i \in C - \{c_0\}$ such that $c_i$ defeats
  $c_0$ in a head-to-head contest, we need to allow for the possibility that a
  microbribery causes $c_0$ to either tie with $c_i$ or to defeat
  $c_i$. 
  Fix an arbitrary such~$c_i$.
  A unit of flow that travels directly from $c_i$ to $c_{i0}$ and then
  directly to $t$ corresponds to a microbribery after which $c_0$
  ties with~$c_i$: $c_i$ loses the point but $c_0$ does not receive it
  and the cost of the flow increases by $\tiecost_E(c_0,c_i)$.

  On the other hand, if that unit of flow travels from $c_i$ to $c_{i0}$
  and then
  directly to $c_0$, then this corresponds to a microbribery after which
  $c_0$ defeats~$c_i$.
  The point travels from $c_i$ to $c_0$ and the cost of the flow increases
  by $\wincost_E(c_0,c_i)$.

  If there is no flow entering node $c_{i0}$ then this means that our
  microbribery does not change the result of a head-to-head contest
  between $c_0$ and~$c_i$.

\item For each candidate $c_i \in C - \{c_0\}$ such that $c_i$ ties
  with $c_0$ in a head-to-head contest before any microbribery is
  attempted, we need to allow for $c_0$ defeating $c_i$ after the
  microbribery.  Fix any such~$c_i$.  A unit of
  flow that travels directly from $s$ to $c_{i0}$ and then directly to
  $c_0$ corresponds to a microbribery after which $c_0$ defeats $c_i$
  in a head-to-head contest: $c_0$ gets an additional point and the
  cost of the flow increases by $\wincost_E(c_0,c_i) -
  \tiecost_E(c_0,c_i) = \wincost_E(c_0,c_i)$.

  On the other hand, a unit of flow that travels from $s$
  directly to $c_{i0}$ and
  then directly to $t$ corresponds to a microbribery that does not change
  the result of a head-to-head contest between $c_0$ and~$c_i$.
 \end{enumerate}

Based on these comments, we can see the natural
correspondence between flows in $J(T)$ and microbriberies. 
In particular, each flow $f$ of value $F$ that travels through 
the network $J(T)$
corresponds to a microbribery 
within $E$ that gives each candidate $c_i \in C$ exactly $f(c_i,t)$
$\copelandalphavarirrational{0}$ points.

Now let $M_f$ be a microbribery defined by flow $f$ of value $F$ within
$J(T)$, $0 \leq T \leq \|C\|-1$, assuming that one exists.  Let
$\cost(M_f)$ be the minimum cost of implementing microbribery~$M_f$.
A close inspection
of instance $J(T)$ shows that the cost of such a flow $f$ is
\[
  \flowcost(f) = B\cdot(F - f(c_0,t)) + \cost(M_f).
\]
Because of the above equation, the fact that all units of flow that
are not accounted for as $c_0$'s points (i.e., the units of flow
that do not travel through the edge $(c_0,t)$)
impose cost $B$ on the flow,
and via arguments analogous to those in
Lemma~\ref{lem:bribery-prime-odd}, the following holds:
If for a given $J(T)$, $0 \leq T \leq \|C\|-1$, there exists a flow of
value $F$ then a minimum-cost flow $f_{\min}$ of value $F$ corresponds
to a minimum-cost microbribery that ensures that $c_0$ receives $T$
points and each other candidate receives at most $T$
points. (Intuitively, the reason for this is that $B$ is so large that
in a minimum-cost flow one would always send as few units of flow
through edges of cost $B$, but each node $c_i$ can
receive (or,
equivalently, send through some path to the sink) at most $T$ points.)
 Thus, to solve the constructive
microbribery problem for $\copelandalphavarirrational{0}$ elections
with an even number of voters it is enough to run the algorithm from
Figure~\ref{fig:flow-alg}, using the instances $J(T)$ instead of
$I(T)$ and using the new value of~$F$.~\end{proofs}

We now show that
Llull, with irrational voters allowed, is vulnerable to
constructive microbribery when there are an even number of voters.
The following lemma reduces the set of microbriberies we need to 
model in this case.

\begin{lemma}
  \label{lem:bribery-prime-even-llull-ties}
  Let $E = (C,V)$ be an election with candidate set
  $C = \{c_0, c_1, \ldots, c_m\}$
  and with an even number of voters, specified via  preference
  tables over $C$.  If the election is conducted using
  $\copelandalphavarirrational{1}$ then
  no minimum-cost microbribery that ensures victory for $c_0$
  involves obtaining a tie
  in a head-to-head contest between any two distinct
  candidates
  in $C - \{c_0\}$.
\end{lemma}

\begin{proofs}
Our proof is again by way of contradiction.  Let $E = (C,V)$ be an election
as specified in the lemma.  
Suppose there is a minimum-cost
microbribery that ensures $c_0$'s victory and that involves obtaining a
tie in a head-to-head contest between two distinct candidates in
$C - \{c_0\}$, say $c_i$ and~$c_j$.
That is, before this microbribery we have that either $c_i$
defeats $c_j$ or $c_j$ defeats $c_i$ in their head-to-head contest but
afterward they are tied.  Clearly, a microbribery
that is identical to this one except that does not change the
result of the head-to-head contest between $c_i$ and $c_j$ (i.e., one
that does not microbribe any voters to flip their preference-table entries
regarding $c_i$ versus~$c_j$) has a smaller cost and still ensures $c_0$'s
victory.
This is a contradiction.~\end{proofs}

\begin{lemma}
  \label{lem:bribery-prime-llull-even}
  There is a polynomial-time algorithm that solves the 
  constructive microbribery problem for
  $\copelandalphavarirrational{1}$ elections with an even number of voters
  (in the nonunique-winner model).
\end{lemma}

\begin{proofs}
We give a polynomial-time algorithm for constructive microbribery in $\copelandalphavarirrational{1}$
elections with an even number of voters.  Our input is a budget $k \in
\naturals$ and an election $E = (C,V)$, where $C = \{c_0, c_1, \ldots,
c_m\}$, $p = c_0$, and $V$ contains an even number of
voters specified via their preference tables over $C$.
Our goal is to ensure $p$'s victory via at most $k$ microbribes.

We use essentially the
algorithm
from the proof of
Lemma~\ref{lem:bribery-prime-odd}, except that instead of using
instances $I(T)$ we now employ instances $L(T)$ that are designed to
handle tie issues as appropriate for $\copelandalphavarirrational{1}$.
Lemma~\ref{lem:bribery-prime-even-llull-ties} tells us that our
min-cost-flow instances $L(T)$ do not need to model microbriberies
that incur ties between pairs of distinct candidates in $C - \{c_0\}$.
We also don't need to model microbriberies that change the outcome 
of the head-to-head contest between $c_0$ and any candidate in $C - \{c_0\}$
from $c_0$ winning to $c_0$ not winning
or from a tie to $c_0$ losing.
However, we do need to model possible microbribery-induced ties
between $c_0$ and each $c_i$ in $C - \{c_0\}$.

Define $B$ to be a number that is greater than the cost of any
possible microbribery within $E$ (e.g., $B = \|V\| \cdot \|C\|^2+1$ will
do).  Further, define the following three sets of nodes:
\begin{eqnarray*}
  C' & = & \{ c'_i \condition c_i \in C\}, \\
  C'' & = & \{ c_{ij} \condition i < j \land c_i,c_j \in C \land
 \versus_E(c_i,c_j) = 0 \}, \mbox{ and} \\
  C''' & = & \{ c_{0i} \condition c_i \in C \land  \versus_E(c_i,c_0) > 0 \}.
\end{eqnarray*}
Our flow network $L(T)$ has the node set
$K = C \cup C' \cup C'' \cup C''' \cup \{s,t\}$, where
$s$ is the source, $t$ is the sink, and the capacities and costs
of edges are defined in Figure~\ref{fig:min-cost-llull}.
Each instance $L(T)$ asks for a minimum-cost flow of value
\[
  F = \sum_{c_i \in C} c(s,c_i).
\]

\begin{figure}[!tb]
\begin{center}
\begin{tabular}{|l|l|}
  \hline
  \multicolumn{1}{|c|}{Edge} & \multicolumn{1}{c|}{Parameters} \\
  \hline\hline

  \begin{tabular}{l}
  $e = (s,c_i)$, \\ where $c_i \in C$ 
  \end{tabular}
  & 
  $\begin{array}{l} c(e) = \llullscore_E(c_i) \\  a(e) = 0\end{array}$ \\
  \hline

  \begin{tabular}{l}
  $e = (c_i,c_j)$, \\
  where $c_i, c_j \in C - \{c_0\}$ and $\versus_E(c_i,c_j) > 0$ 
  \end{tabular}
  &  
  $\begin{array}{l} c(e) = 1  \\ a(e) = \wincost_E(c_j,c_i) \end{array}$ \\
  \hline

  \begin{tabular}{l}
  $e = (c_i,c_{ij})$, \\
  where $i < j$, $c_i, c_j \in C$  and $\versus_E(c_i,c_j) = 0$ 
  \end{tabular}
  &  
  $\begin{array}{l} c(e) = 1  \\ a(e) = \wincost_E(c_j,c_i) \end{array}$ \\
  \hline

  \begin{tabular}{l}
  $e = (c_j,c_{ij})$, \\
  where $i < j$, $c_i, c_j \in C$  and $\versus_E(c_i,c_j) = 0$ 
  \end{tabular}
  &  
  $\begin{array}{l} c(e) = 1  \\ a(e) = \wincost_E(c_i,c_j) \end{array}$ \\
  \hline

  \begin{tabular}{l}
  $e = (c_{ij},t)$, \\
  where $i < j$, $c_i, c_j \in C$  and $\versus_E(c_i,c_j) = 0$ 
  \end{tabular}
  &  
  $\begin{array}{l} c(e) = 1  \\ a(e) = B \end{array}$ \\
  \hline

  \begin{tabular}{l}
  $e = (c_i,c_{0i})$, \\
  where $c_i \in C - \{c_0\}$  and $\versus_E(c_i,c_0) > 0$ 
  \end{tabular}
  &  
  $\begin{array}{l} c(e) = 1  \\ a(e) = \wincost(c_0,c_i) \end{array}$ \\
  \hline

  \begin{tabular}{l}
  $e = (c'_i,c_{0i})$, \\
  where $c_i \in C - \{c_0\}$  and $\versus_E(c_i,c_0) > 0$ 
  \end{tabular}
  &  
  $\begin{array}{l} c(e) = 1  \\ a(e) = \tiecost(c_0,c_i) \end{array}$ \\
  \hline

  \begin{tabular}{l}
  $e = (c_{0i},c_0)$, \\
  where $c_i \in C - \{c_0\}$  and $\versus_E(c_i,c_0) > 0$ 
  \end{tabular}
  &  
  $\begin{array}{l} c(e) = 1  \\ a(e) = 0 \end{array}$ \\
  \hline

  \begin{tabular}{l}
  $ e = (c_i,c'_i)$, \\
  where $c_i \in C$ 
  \end{tabular}& 
  $\begin{array}{l} c(e) = T \\ a(e) = 0 \end{array}$ \\
  \hline

  \begin{tabular}{l}
  $ e = (c'_i,t)$, \\
  where $c_i \in C - \{c_0\}$ 
  \end{tabular}& 
  $\begin{array}{l} c(e) = T \\ a(e) = B \end{array}$ \\
  \hline

  \begin{tabular}{l}
  $ e = (c'_0,t)$ \\
  \end{tabular}& 
  $\begin{array}{l} c(e) = T \\ a(e) = 0 \end{array}$ \\
  \hline

  \begin{tabular}{l}
  every other edge $e$
  \end{tabular} 
   & $\begin{array}{l} c(e) = 0 \\ a(e) = 0 \end{array}$\\
  \hline
\end{tabular}
\end{center}
\caption{\label{fig:min-cost-llull}Edge capacities and costs for
min-cost-flow instance $L(T)$, built from election $E$.}
\end{figure}

\begin{example}
\label{exa:constructive-microbribery-even-llull}
Consider the sample election $E$ from
Example~\ref{exa:constructive-microbribery-even-copeland} again, which
is given in Figure~\ref{fig:constructive-microbribery-even-copeland}.
Since we are using
$\copelandalphavarirrational{1}$ now, we have the following scores:
$\copelandonescore_E(c_0) = \copelandonescore_E(c_3) = 2$, 
$\copelandonescore_E(c_1) = 0$, and $\copelandonescore_E(c_2) = 3$.
Let $k = 1$.  We can reach our goal
of making $c_0$ win via one microbribe that switches the preference
table of, say, the third voter to now prefer $c_0$ over~$c_2$.  This
microbribe gives $c_0$ one more $\copelandalphavarirrational{1}$ point (i.e.,
$\copelandonescore_{E'}(c_0) = 3$ in the modified election~$E'$), but it
doesn't change the score of any of the other candidates within
$\copelandalphavarirrational{1}$.
Again, we have $B = 65$.

For any threshold $T$ with $0 \leq T \leq 3$, the flow network $L(T)$
corresponding to this instance $(E,c_0,k)$ of the constructive
microbribery problem is shown in
Figure~\ref{fig:constructive-microbribery-even-llull-flownetwork}.
We now have a target flow value of $F = 7$, and every edge $e$ in
this flow network is labeled by the pair $(c(e),a(e))$ of numbers that
give the capacity and the cost of edge~$e$, respectively.

\begin{figure}[!tb]
\begin{center}
\input{constructive-microbribery-even-llull-flownetwork.eepic}
\end{center}
\caption{\label{fig:constructive-microbribery-even-llull-flownetwork}
Flow network  $L(T)$ corresponding to the instance 
$(E,c_0,k)$ of Example~\ref{exa:constructive-microbribery-even-llull}.}
\end{figure}
\end{example}

To continue the proof of Lemma~\ref{lem:bribery-prime-llull-even},
note that a flow $f$ of value $F$ within~$L(T)$, $0 \leq T \leq \|C\|-1$,
corresponds to a microbribery $M_f$ within election $E$ that leaves
each candidate $c_i$ with exactly $f(c_i,c'_i)$
$\copelandalphavarirrational{1}$ points. 
As in the proofs of Lemmas~\ref{lem:bribery-prime-odd}
and~\ref{lem:bribery-prime-copeland-even}, points traveling through
the network $L(T)$ are here interpreted as
$\copelandalphavarirrational{1}$ points
and flows are interpreted as specifying microbriberies.
More specifically, we interpret the units of flow traveling through
$L(T)$ as follows:
\begin{enumerate}
\item For each $c_i \in C$, the units of flow that enter $c_i$ from
  $s$ are interpreted as the
  $\copelandalphavarirrational{1}$ points that $c_i$ has before any
  microbribery is attempted.

\item For each $c_i \in C$, the units of flow that travel directly
  from $c_i$ to $c'_i$ are interpreted as the
  $\copelandalphavarirrational{1}$ points that $c_i$
  has after the microbribery defined by $f$ has been performed.

\item If a candidate $c_i \in C - \{c_0\}$ originally
  defeats $c_0$ but our flow models a microbribery in which $c_0$ and
  $c_i$ end up tied in their head-to-head contest, then we have a
  single
  $\copelandalphavarirrational{1}$ point that travels from $c_i$ to $c'_i$,
  then to $c_0$ through
  $c_{0i}$, at cost $\tiecost_E(c_0,c_i)$, and then to~$c'_0$.
  This way the same unit of flow is accounted both for the
  score of $c_0$ and for the score of $c_i$. Note that such a unit of
  flow then travels to $t$ through edge $(c_0,t)$ at zero
  cost.

\item For any two distinct candidates $c_i$ and $c_j$ such that
  $c_i,c_j \in C - \{c_0\}$
  where $c_i$ defeats $c_j$ in a head-to-head
  contest in $E$, a unit of flow traveling from $c_i$ to $c_j$
  corresponds to a microbribery that flips the result of their
  head-to-head contest. Thus $c_j$ receives the
  $\copelandalphavarirrational{1}$ point and the cost of
  the flow increases by $\wincost_E(c_j,c_i)$. 

\item For any $c_i \in C - \{c_0\}$ where $c_i$ defeats $c_0$ in a
  head-to-head contest in $E$, a unit of flow traveling from $c_i$ to
  $c_0$ via $c_{0i}$ corresponds to a microbribery that flips the
  result of their head-to-head contest. Thus $c_i$ receives the
  $\copelandalphavarirrational{1}$
  point and the cost of the flow increases by $\wincost_E(c_0,c_i)$.
  Note that since the edge $(c_{0i},c_0)$ has capacity
  one we ensure
  that at most one unit of flow travels from $c_i$ to $c_0$ (either on
  a path $c_i, c_{0i},c_0$ (modeling a microbribery that flips the
  result of the head-to-head contest between $c_i$ and $c_0$ to $c_0$
  winning) or on a path $c_i,c'_i,c_{0i},c_0$ (modeling a microbribery
  that enforces a tie between $c_0$ and $c_i$)).

\item For each $c_i,c_j \in C$, we have to take into account the possibility
  that $c_i$ and $c_j$ are tied in their head-to-head
  contest within $E$, but via microbribery we want to change the
  result of this contest. Let $c_i, c_j \in C$ be two such candidates
  and let $i < j$.  Here, a unit of flow traveling from $c_i$ to
  $c_{ij}$ (or, analogously, from $c_j$ to $c_{ij}$) is interpreted as
  a microbribery that ensures $c_j$'s ($c_i$'s) victory in the
  head-to-head contest. Since $c_i$ and $c_j$ were already tied, $c_j$
  ($c_i$) already has his or her point for the victory and $c_i$
  ($c_j$)
  gets rid of his or her point through the node $c_{ij}$. The
  cost of the flow increases by $\wincost_E(c_j,c_i)$
  (respectively, by $\wincost_E(c_i,c_j)$).  Also, we point out that via the
  introduction of node $c_{ij}$ we ensure that only one of $c_i$ and~$c_j$, 
  let us call him or her $c_k$, can lose a point by sending it from $c_k$
  to $c_{ij}$ and then to $t$; the capacity of edge $(c_{ij},t)$ is
  only one.
\end{enumerate}

Through the above description, we can see the natural correspondence
between flows in $L(T)$ and microbriberies. In particular, each flow
$f$ of value $F$ corresponds to a microbribery $M_f$ within $E$ that
gives each candidate $c_i$ exactly $f(c_i,c'_i)$ points.

Let us now analyze the cost of~$f$.  It
is easy to see that each unit of flow that is not accounted for as a
$\copelandalphavarirrational{1}$ 
point of $c_0$ reaches the sink $t$ via an edge of cost $B$.
Also, the only other edges through which units of flow travel at
nonzero cost are those that define the microbribery $M_f$.
Thus the cost of our flow $f$ can be expressed as
\[
  \flowcost(f) = B \cdot (F - f(c_0, c'_0)) + \cost(M_f).
\]

Fix $T$ such that $0 \leq T \leq \|C\|-1$.
Given the above properties of $L(T)$ and by the
arguments presented in the proof of Lemma~\ref{lem:bribery-prime-odd},
if a flow of value $F$ exists within the flow network of
instance~$L(T)$, then each minimum-cost flow in $L(T)$
corresponds to a microbribery that ensures that $c_0$ has exactly $T$
$\copelandalphavarirrational{1}$ 
points and every other candidate has at most $T$ 
$\copelandalphavarirrational{1}$ points.
Thus if there exists a value $T'$ such that
\begin{enumerate}
\item there is a flow of value $F$ in $L(T')$ and
\item the cost of a minimum-cost flow $f$ of value $F$ in $L(T')$ is $K$,
\end{enumerate}
then there is a microbribery of cost $K - B\cdot(F-T')$ that ensures
$c_0$'s victory.

On the other hand, via Lemma~\ref{lem:bribery-prime-even-llull-ties}
and our correspondence between flows for $L(T)$ and
microbriberies in $E$, there is a value $T''$ such that a minimum-cost
flow in $L(T'')$ corresponds to a minimum-cost microbribery that
ensures $p$'s victory.  Thus the algorithm from
Figure~\ref{fig:flow-alg}, used with the instances $L(T)$ instead of
$I(T)$ and with our new value of~$F$, solves in polynomial time the constructive
microbribery problem for
$\copelandalphavarirrational{1}$ with an even number of voters.~\end{proofs}

Together, Theorem~\ref{thm:bribery-prime-destructive} and
Lemmas~\ref{lem:bribery-prime-odd}, \ref{lem:bribery-prime-copeland-even},
and~\ref{lem:bribery-prime-llull-even} show that, in particular, both 
$\copelandalphavarirrational{1}$ and
$\copelandalphavarirrational{0}$ are vulnerable to
microbribery, both in the constructive and the destructive
settings. It is interesting to note that all our microbribery proofs
above would work just as well if we considered a slight twist on the
definition of the microbribery problem, namely, if instead of saying
that each flip in a voter's preference table has unit cost we would
allow each voter to have a possibly different price for flipping each
separate entry in his or her preference table. 
This change would affect only the computing of the values of the functions
$\wincost$ and $\tiecost$. (Technically, we would also have to modify
Lemmas~\ref{lem:bribery-prime-even-copeland-ties} 
and~\ref{lem:bribery-prime-even-llull-ties},
which in our unit-cost setting say that an optimal
microbribery never involves certain specified pairs of candidates,
whereas in the priced setting we would need to rephrase them to state
that there \emph{exist} optimal microbriberies that do not involve
those specified pairs of candidates.)

An interesting direction for further study of the complexity of
bribery within
$\copelandalpha$ systems is to consider a version
of the microbribery problem for the case of rational voters. There,
one would pay unit cost for a switch of two adjacent candidates on a
given voter's preference list.

For $\copelandalphairrational$, we would also like to know
the complexity of constructive microbribery when
$\alpha$ is a rational number strictly between $0$ and $1$. 
Our network-flow-based approach does not seem to generalize easily to
values of $\alpha$ strictly between $0$ and $1$ (when the number of voters is
even) because in a flow network it is hard to ``split'' a unit of flow
in a tie. A
promising approach would be to have several units of flow
model one $\copelandalphairrational$ point (e.g., for the case of $\alpha =
\frac{1}{2}$ we could try to use two units of flow to model a single
Copeland$^{0.5}$ point), but then it seems very difficult
(if not impossible) to find edge costs that appropriately model the
microbribery. (It is possible to do so in a very restricted
setting, namely where $\alpha = \frac{1}{2}$ and there are exactly two voters
that can be bribed.) Also, the results regarding hardness of
manipulation of Faliszewski, Hemaspaandra, and
Schnoor~\cite{fal-hem-sch:c:copeland-ties-matter} suggest that
microbribery for $\alpha$ strictly between $0$ and $1$ might be
$\np$-hard. However, again, it is nontrivial to translate their
reduction to the world of microbribery.

On a related note,
Kern and Paulusma~\cite{ker-pau:j:fifa} have shown that the following 
problem, which they call $\mbox{SC}(0,\alpha,1)$, is $\np$-complete.
Let $\alpha$ be a rational number such that $0 < \alpha < 1$ and $\alpha 
\neq \frac{1}{2}$.  We are given
an undirected graph $G = (V(G),E(G))$, where each vertex $u \in V(G)$ is 
assigned a rational value $c_u$ of the form $i+j\alpha$, for
nonnegative integers $i$ and~$j$.  
The question, which we have rephrased
to state in terms of (a variant of) our notion of $\copelandalpha$,
is whether it is possible to
(possibly partially) orient the edges of $G$ such that for each vertex
$u \in V(G)$ it holds that $u$'s $\copelandalpha$ score is at most
$c_u$. Here, by ``$\copelandalpha$ score of a vertex $u$'' we mean, as is
natural, the number of vertices $u$ ``defeats'' (i.e., the number of
vertices $v$ such that there is a directed edge from $u$ to $v$) plus
$\alpha$ times the number of vertices that $u$ ``ties'' with (i.e., the
number of vertices such that there is an undirected edge between $u$ and
$v$).

Problem $\mbox{SC}(0,\alpha,1)$ is very closely 
related to our microbribery problem.  However, 
we do not see an immediate reduction from $\mbox{SC}(0,\alpha,1)$ to 
microbribery.  A natural approach would be to embed graph $G$ into an 
election (in the sense that will be explored in Section~\ref{sec:control})
in such a way that 
our preferred candidate $p$ can become a winner, via a microbribery, 
if and only if it is possible to orient the edges of $G$ in a way 
respecting the constraints defined by the values $c_u$ (for each $u$ in 
$V(G)$).  We would, of course, have to set the budget of our microbribery 
high enough to allow modifying each of the edges in $G$ and none of the 
edges outside of $G$.  However, this is difficult.  The proof of Kern and 
Paulusma uses values $c_u$ that can be implemented only via using tied 
head-to-head contests.  The agent performing microbribery could, 
potentially, affect those head-to-head contests, thus spoiling our reduction.

\newcommand{\copelandalphaonescore}{\mathit{score}^1}
\newcommand{\copelandalphaone}{\mbox{\rm{}Copeland$^{1}$}}
\newtheorem{construction}[theorem]{Construction}

\section{Control}
\label{sec:control}

In this section we focus on the complexity of control in
$\copelandalpha$ elections. In control problems we are
trying to ensure that our preferred candidate $p$ is a winner (or, in
the destructive case, that our despised candidate is not a winner) of
a given election via affecting this election's structure (namely, via
adding, deleting, or partitioning either candidates or voters).  
In contrast with bribery problems, in control problems we
are never allowed to change any of the votes and, consequently, the
issues that we encounter and the proof
techniques we use are quite different from those presented in the
previous section. For the same reason as previously for each standard type of control a
resistance result in the rational-voters case implies an analogous resistance
result in the irrational-voters case, and a vulnerability result in the
irrational-voters case implies an analogous vulnerability result in the
rational-voters case.

The literature regarding the complexity of
control problems is not large. To the best of our knowledge, the
only election systems for which a comprehensive analysis has been conducted
previously are plurality, Condorcet, and (variants of) approval voting
(see~\cite{bar-tov-tri:j:control,hem-hem-rot:j:destructive-control,hem-hem-rot:c:hybrid,bet-uhl:c:parameterized-complecity-candidate-control,erd-now-rot:t-With-MFCS08-Ptr:sp-av};
see also~\cite{pro-ros-zoh:c:multiwinner} for some results on
(variants of) approval voting, single nontransferable vote, and
cumulative voting with respect to constructive control via adding
voters).  Among plurality, Condorcet, and (the standard variant of)
approval voting, 
plurality appears to be the least
vulnerable to control and so it is natural to compare our new results
with those for plurality. However, we mention in passing that 
Hemaspaandra, Hemaspaandra, and Rothe~\cite{hem-hem-rot:c:hybrid}
show how to construct hybrid election systems
that are resistant to all standard types of control
(including both AC and AC$_{\mathrm{u}}$; AC is not discussed or proven
in~\cite{hem-hem-rot:c:hybrid}---the ``AC'' there is our ``AC$_{\mathrm{u}}$''---but
we mention that the techniques clearly can handle it without any problem).
(It should also be noted that these hybrid systems were not designed
as ``natural'' systems to be applied in real-world elections but
rather their purpose was to prove a certain impossibility theorem
impossible.)

Our main result in this section is
Theorem~\ref{thm:control}.

\begin{theorem}
  \label{thm:control}
  Let $\alpha$ be a rational number with $0 \leq \alpha \leq 1$.
  $\copelandalpha$ elections
  are resistant and vulnerable
  to control types as indicated in Table~\ref{tab:control}
  in both the nonunique-winner model and the unique-winner model,
  for both the rational and the irrational voter model.
\end{theorem}

\begin{table}[t]
\begin{center}
\begin{tabular}{|l|c|c|c|c|c|c||c|c|}
\hline 
\multicolumn{1}{|c|}{}
 & \multicolumn{6}{|c||}{$\copelandalpha$}
 & \multicolumn{2}{c|}{Plurality} \\ \cline{2-7}
\multicolumn{1}{|c|}{}
 & \multicolumn{2}{c|}{$\alpha = 0$}
 & \multicolumn{2}{c|}{$0 < \alpha < 1$}
 & \multicolumn{2}{c||}{$\alpha = 1$}
 & \multicolumn{2}{c|}{} \\
\hline 
Control type         & CC & DC & CC & DC & CC & DC & CC & DC \\
\hline
AC$_{\rm u}$  & V  & V  & R  & V  & V  & V  & R & R \\
AC            & R  & V  & R  & V  & R  & V  & R & R \\
DC            & R  & V  & R  & V  & R  & V  & R & R \\
RPC-TP        & R  & V  & R  & V  & R  & V  & R & R \\
RPC-TE        & R  & V  & R  & V  & R  & V  & R & R \\
PC-TP         & R  & V  & R  & V  & R  & V  & R & R \\
PC-TE         & R  & V  & R  & V  & R  & V  & R & R \\
\hline
PV-TE         & R  & R  & R  & R  & R  & R  & V & V \\
PV-TP         & R  & R  & R  & R  & R  & R  & R & R \\
AV            & R  & R  & R  & R  & R  & R  & V & V \\
DV            & R  & R  & R  & R  & R  & R  & V & V \\

\hline
\end{tabular}
\caption{\label{tab:control}Comparison of control results 
for
$\copelandalpha$ elections, where $\alpha$ with $0 \leq \alpha \leq 1$
is a rational number, and for plurality-rule elections.  R means
resistance to a particular control type and V means vulnerability. The
results regarding plurality
are due to
Bartholdi, Tovey, 
and Trick~\cite{bar-tov-tri:j:control} and 
Hemaspaandra, Hemaspaandra, and
Rothe~\cite{hem-hem-rot:j:destructive-control}.
(Note that CCAC and DCAC resistance results for plurality, not handled explicitly
in~\cite{bar-tov-tri:j:control,hem-hem-rot:j:destructive-control}, follow immediately from
the respective CCAC$_{\rm u}$ and DCAC$_{\rm u}$ results.)}
\end{center}
\end{table}
In particular, we will prove in this section
that the notion widely referred to in the literature simply
as ``Copeland elections,'' which we here for clarity call Copeland$^{0.5}$,
possesses all ten of our basic types (see Table~\ref{tab:control}) of
constructive resistance (and in addition, even has constructive AC$_{\rm u}$
resistance).  And we will establish that the other notion
that in the literature is occasionally referred to as ``Copeland
elections,'' namely Copeland$^0$, 
as well as Llull elections, which are here denoted by Copeland$^1$,
both possess all ten of our basic types of
constructive resistance.  However,
we will show that Copeland$^0$ and Copeland$^1$ are vulnerable
to an eleventh type of constructive control, the
incongruous but historically resonant notion of constructive 
control by adding an unlimited number of candidates
(i.e., CCAC$_{\rm u}$).

Note that
Copeland$^{0.5}$ has a higher number of constructive resistances,
by three,
than even plurality, which was 
before this paper the reigning champ among natural election systems
with a polynomial-time winner-determination procedure.  
(Although the  results regarding plurality in Table~\ref{tab:control} 
are stated for the unique-winner version of control, for all the table's
$\copelandalpha$ cases, $0 \leq \alpha \leq 1$,
our results hold both in the cases of unique winners and of nonunique
winners, so that regardless of which of the two winner models one finds
more natural, one will know what holds in that model.)
Admittedly, plurality
does perform better with respect to destructive candidate control problems,
but still our study of
$\copelandalpha$
makes  significant steps forward
in the quest for a fully control-resistant
natural election system with an easy winner problem.

Among the systems with a polynomial-time winner problem,
Copeland$^{0.5}$---and indeed all Copeland$^{\alpha}$, $0 < \alpha <
1$---have the most resistances currently known for any natural
election system whose voters vote by giving preference lists.
We mention that after our work, Erd\'{e}lyi, Nowak, and
Rothe~\cite{erd-now-rot:t-With-MFCS08-Ptr:sp-av} have shown that a
variant of approval voting proposed by Brams and
Sanver~\cite{bra-san:j:critical-strategies-under-approval}---a certain
rather subtle election system with a richer voter preference type
(each voter specifies both a permutation and a set) that combines
approval with preference-based voting---has nineteen (out of a
possible twenty-two) control resistances.

This section is organized as follows. The next two sections are
devoted to proving Theorem~\ref{thm:control},
and Section~\ref{sec:control-fpt} considers the case of control in 
elections with a bounded number of candidates
or voters.  In particular, Section~\ref{sec:control-candidate} focuses on
the upper part of Table~\ref{tab:control} and studies control problems
that affect the candidate structure. Section~\ref{sec:control-voter}
is devoted to voter control and covers the lower part of
Table~\ref{tab:control}. Finally, in Section~\ref{sec:control-fpt} we
study the fixed-parameter complexity of control problems. In particular, we
take the role of someone who tries to solve in-general-resistant
control problems and we devise some efficient algorithms for the case
where the number of candidates or the number of voters is bounded.

All our resistance results regarding
candidate control follow via reductions from 
vertex cover
and all our vulnerability results follow via 
greedy
algorithms.  
Our resistance results for the case of control by modifying voter
structure follow from reductions from 
the X3C
problem.

\subsection{Candidate Control}
\label{sec:control-candidate}

We start our discussion of candidate control for
$\copelandalpha$ with our results on destructive control. It is
somewhat disappointing that for each rational $\alpha$, $0 \leq \alpha
\leq 1$, $\copelandalpha$ is vulnerable to each type of destructive
candidate control. On the positive side, our vulnerability proofs
follow via natural greedy algorithms and will allow us to smoothly get
into the spirit of candidate-control problems.

\subsubsection{Destructive Candidate Control}

The results for destructive control by adding and deleting candidates
use the following observation.

\begin{observation}
\label{obs:scores-new}
Let $(C,V)$ be an election, and let $\alpha$ be a rational
number such that $0 \leq \alpha \leq 1$.
For every candidate $c \in C$ it holds that
\[\copelandalphascore_{(C,V)}(c) = \sum_{d \in C - \{c\}}
\copelandalphascore_{(\{c,d\},V)}(c).\]
\end{observation}

\begin{theorem}
\label{thm-dcac-dc}
For each rational number $\alpha$ with $0 \leq \alpha \leq 1$,
 $\copelandalpha$
  is vulnerable to destructive control via
  adding candidates (both limited and unlimited,
  i.e., DCAC and DCAC$_{\rm u}$),
  in both the nonunique-winner model and the unique-winner model,
  for both the rational and the irrational voter model.
\end{theorem}

\begin{proofs}
Our input is a set $C$ of
candidates, a set $D$ of spoiler candidates, a collection $V$ of
voters with preferences (either preference lists or preference
tables) over $C \cup D$, a candidate $p \in C$, and a 
nonnegative integer $k$
(for the unlimited version of the problem we let $k = \|D\|$).
We ask whether there is a subset $D'$ of $D$ such that
$\|D'\| \leq k$ and $p$ is not a winner (is not a unique winner) of
$\copelandalpha$ election $E' = (C \cup D',V)$.
Note that if  $k = 0$, this amounts to determining whether
$p$ is not a winner (is not a unique winner) of election $E$,
which can easily be done in polynomial time.

For the remainder of this proof we will assume that $k > 0$.
Let $c$ be any candidate in $(C \cup D) - \{p\}$. We define
$a(c)$ to be the maximum value of the expression
\[
  \copelandalphascore_{(C\cup D',V)}(c) -
   \copelandalphascore_{(C\cup D',V)}(p)
\]
under the conditions that $D' \subseteq D$, $c \in  C \cup D'$,
and $\|D'\| \leq k$.
From Observation~\ref{obs:scores-new} it follows that $a(c)$ is the
maximum value of
\[
  \copelandalphascore_{(C\cup \{c\},V)}(c) -
   \copelandalphascore_{(C\cup \{c\},V)}(p) +
\sum_{d \in D'-\{c\}} \left(\copelandalphascore_{(\{c,d\},V)}(c) -
              \copelandalphascore_{(\{p,d\},V)}(p)\right)
\]
under the conditions that $D' \subseteq D$, $c \in  C \cup D'$,
and $\|D'\| \leq k$.

Clearly, $p$ can be prevented from being a winner (a unique winner)
if and only if there
exists a candidate $c \in (C \cup D) - \{p\}$ such that $a(c) > 0$
(such that $a(c) \geq 0$).

Given a candidate $c \in (C \cup D) - \{p\}$,
it is easy to construct in polynomial time a
set $D' \subseteq D$, $\|D'\| \leq k$,
that yields the value $a(c)$. We start with $D' = \emptyset$.
If $c \in D$, we add $c$ to $D'$. Then we add those candidates
$d \in D - D'$ to $D'$ such that
$\copelandalphascore_{(\{c,d\},V)}(c) - \copelandalphascore_{(\{p,d\},V)}(p)$
is positive, starting with those for whom this value is highest,
until $\|D'\| = k$ or no
more such candidates exist.~\end{proofs}

\begin{theorem}
  \label{thm:dcdc}
  For each rational number $\alpha$ with $0 \leq \alpha \leq 1$,
  $\copelandalpha$ is vulnerable to destructive control via
  deleting candidates (DCDC), 
  in both the nonunique-winner model and the unique-winner model,
  for both the rational and the irrational voter model.
\end{theorem}

\begin{proofs}
Our approach is similar to that of the previous theorem.
We are given an election $E = (C,V)$, a candidate $p \in
C$, and a nonnegative integer $k$.
Our goal is to check whether $p$ can be prevented from being a winner (a unique winner) via
deleting at most $k$ candidates in $C - \{p\}$.

Let $c$ be any candidate in $C - \{p\}$.  Define
$d(c)$ to be the maximum value of the expression
\[
  \copelandalphascore_{(C - D,V)}(c) -
   \copelandalphascore_{(C - D,V)}(p)
\]
under the conditions that $D \subseteq C$, $\{p,c\} \cap D = \emptyset$,
and $\|D\| \leq k$.

From Observation~\ref{obs:scores-new} it follows that $d(c)$ is the
maximum value of
\[
  \copelandalphascore_{(C,V)}(c) -
   \copelandalphascore_{(C,V)}(p) +
\sum_{d \in D} \left(\copelandalphascore_{(\{c,d\},V)}(p) -
              \copelandalphascore_{(\{p,d\},V)}(c)\right)
\]
under the conditions that $D \subseteq C$, $\{p,c\} \cap D = \emptyset$,
and $\|D\| \leq k$.

Clearly, $p$ can be prevented from being a winner (a unique winner)
if and only if there
exists a candidate $c \in C - \{p\}$ such that $d(c) > 0$
(such that $d(c) \geq 0$).

Given a candidate $c \in C - \{p\}$, it is easy to construct in polynomial
time a set $D$, $\|D\| \leq k$, that yields the value $d(c)$.
Simply let $D$ consist of as many as possible but at most $k$
candidates $d \not \in \{c,p\}$ such that $\copelandalphascore_{(\{p,d\},V)}(p) - \copelandalphascore_{(\{c,d\},V)}(c)$
is positive, starting with those for whom this value is highest.~\end{proofs}

Destructive control via partitioning of candidates (with or without
run-off) is also easy.

\begin{theorem}
  \label{thm:partition-destructive}
  For each rational number $\alpha$ with $0 \leq \alpha \leq 1$,
  $\copelandalpha$ is vulnerable to destructive control via
  partitioning of candidates and via partitioning of candidates with
  run-off (in both the TP and TE model, i.e., DCPC-TP, DCPC-TE, DCRPC-TP,
  and DCRPC-TE),
  in both the nonunique-winner model and the unique-winner model,
  for both the rational and the irrational voter model.
\end{theorem}

\begin{proofs}
It is easy to see that in the TP model, $p$ can be prevented from
being a winner via partitioning of candidates (with or without
run-off) if and only if there is a set 
$C' \subseteq C$ such that $p \in C'$ and
$p$ is not a winner of $(C',V)$.  It follows that
$p$ can be prevented from being a winner
if and only if $p$ can be prevented from being a winner
by deleting at most $\|C\| - 1$ candidates, which can
be determined in polynomial time by
Theorem~\ref{thm:dcdc}.

For
the TE model, it is easy to see that if there is a set
$C' \subseteq C$ such that $p \in C'$ and $p$ is not a unique winner
of $(C',V)$ then $p$ can be prevented from being a unique winner via
partitioning of candidates (with or without run-off).  One simply
partitions the candidates into $C'$ and $C - C'$ and thus $p$ fails to
advance to the final stage.  On the other hand, if $p$ can be
prevented from being a winner (a unique winner) via partitioning of
candidates (with or without run-off) in the TE model, then there
exists a set $C' \subseteq C$ such that $p \in C'$ and $p$ is not a unique winner
of $(C',V)$. This is so because then either $p$ does not advance to
the final stage (and this means that $p$ is not a unique winner of his
or her first-stage election) or $p$ is not a winner (not a unique
winner) of the final stage (note that not being a winner implies not being
a unique winner).

Thus, $p$ can be prevented from being a winner (a unique winner) via
partitioning of candidates (with or without run-off) in the TE model
if and only if there is a set $C' \subseteq C$ such that $p \in C'$ and
$p$ is not a unique winner of $(C',V)$. Clearly, such a set exists if
and only if $p$ can be prevented from being a unique winner via
deleting at most $\|C\|-1$ candidates, which
by Theorem~\ref{thm:dcdc} can be tested in
polynomial time.

It remains to show that
$\copelandalpha$ is vulnerable to destructive control 
via partitioning of candidates (with or without
run-off), in both the rational and the irrational voter model,
in the unique-winner model with the TP tie-handling rule.
In the argument below we focus on
the DCRPC-TP case but it is easy to see that essentially the same
reasoning works for DCPC-TP.

First we determine whether $p$ can be precluded from being a winner
in our current control scenario.
This can be done in polynomial time as explained above.
If $p$ can be precluded from being a winner, $p$ can
certainly be precluded from being a unique winner, and
we are done.
For the remainder of the proof, suppose that $p$ cannot be
precluded from being a winner
in our current control scenario, i.e., for every set $D \subseteq C$
such that $p \in D$, $p$ is a winner of $(D,V)$.
Let
\[
D_1 = \{c \in C - \{p\} \condition
\mbox{$p$  defeats $c$ in a head-to-head contest}\}
\]
and let $D_2 = D - (D_1  \cup \{p\})$.  Note that for
all $c \in D_2$, $p$ ties $c$ in a head-to-head contest,
since otherwise $p$ would not be a winner of $(\{c,p\},V)$.
If $D_2 = \emptyset$,
then $p$ is a Condorcet winner and no partition (with or without
run-off) can prevent $p$ from being a unique
winner~\cite{hem-hem-rot:j:destructive-control}.
For the remainder of the proof, we assume that $D_2 \neq \emptyset$.
We will show that $p$ can be precluded from  being
a unique winner in our current control scenario.

If $\alpha < 1$, we let the first subelection be
$(D_1 \cup \{p\}, V)$.  Note that $p$ is the unique winner of this
subelection.  The final stage of the election involves
$p$ and one or more candidates from $D_2$.  Note that
every pair of candidates in $D_2 \cup \{p\}$ is tied
in a head-to-head election (since if $c$ would
defeat $d$ in a head-to-head election, $c$ would be the
unique winner of $(\{c,d,p\}, V)$, which contradicts the assumption
that $p$ is a winner of every subelection it participates in).
It follows that all candidates that participate in the final stage of
the election are winners, and so $p$ is not a unique winner.

Finally, consider the case that $\alpha = 1$.
Then $\copelandalphascore_{(C,V)}(p) = \|C\| - 1$.
If there is a candidate $d \in C - \{p\}$ such that
$\copelandalphascore_{(C,V)}(d) = \|C\| - 1$, then $d$ will
always (i.e., in every subelection containing $d$) be a winner, 
and thus $p$ will not 
be a unique winner of the final stage of the election,
regardless of which partition of $C$ was chosen.  Now suppose that
$\copelandalphascore_{(C,V)}(d) < \|C\| - 1$ for
all $d \in C - \{p\}$. Then
$\copelandalphascore_{(C,V)}(d) \leq \|C\| - 2$ for
all $d \in C - \{p\}$. Let $c$ be a candidate in $D_2$
and let the first subelection be $(C - \{c\},V)$.
Let $C'$ be the set of winners
of $(C - \{c\},V)$.  Since 
$\copelandalphascore_{(C - \{c\},V)}(p) = \|C\| - 2$,
it holds that $p \in C'$ and for every $d \in C' - \{p\}$, 
$\copelandalphascore_{(C - \{c\},V)}(d) = \|C\| - 2$.
Since $\copelandalphascore_{(C,V)}(d) \leq \|C\| - 2$,
it follows that $c$ defeats $d$ in a head-to-head election.
The final stage of the election involves candidates $C' \cup \{c\}$.
Note that 
$\copelandalphascore_{(C' \cup \{c\},V)}(c) = \|C'\|$,
and thus $c$ is a winner of the election, and we have precluded
$p$ from being a unique winner.~\end{proofs}

The above vulnerability results for the case of destructive candidate
control should be contrasted with the essentially perfect resistance
to constructive candidate control (with the exception of
constructive control via adding an unlimited number of candidates for
$\copelandalpha$ with $\alpha \in \{0,1\}$) that will be 
shown in Section~\ref{sec:constructive-candidate-control}.
But first, in Section~\ref{sect:constructing}, we will provide some technical
prerequisites.

\subsubsection{Constructing Instances of Elections}
\label{sect:constructing}

Many of our proofs in the next section require constructing fairly
involved instances of $\copelandalpha$ elections. In this section we
provide several lemmas and observations that simplify building such
instances.

We first note that each election $E = (C,V)$ induces a directed graph
$G(E)$ whose vertices are $E$'s candidates and whose edges correspond
to the results of the head-to-head contests in $E$. That is, for each two
distinct vertices of $G(E)$ (i.e., for each two distinct candidates), $a$ and
$b$, there is an edge from $a$ to $b$ if and only if $a$ defeats $b$ in
their head-to-head contest (i.e., if and only if $\versus_E(a,b) >
0$).  Clearly, $G(E)$ does not depend on the value of $\alpha$.  The
following fundamental result is due to McGarvey.
This result allows us to basically identify elections with their election
graphs in the proofs of resistance for candidate control. 

\begin{lemma}[\cite{mcg:j:election-graph}]
  \label{thm:election-graph}
  There is a polynomial-time algorithm that given as input
  an antisymmetric directed graph~$G$ outputs an election
  $E$ such that $G = G(E)$.
\end{lemma}
\begin{proofs}
  For the sake of completeness, we give the algorithm.
  Let $G$ be an antisymmetric directed graph.  The algorithm
  computes the election $E = (C,V)$,
  where $C = V(G)$ and for each edge $(a,b)$ in $G$ there
  are exactly two voters, one with preference list $a > b > C -
  \{a,b\}$ and one with preference list $\reversenotation{C -
    \{a,b\}} > a > b$. Since $G$ is antisymmetric, it is easy to see
  that $G = G(E)$.~\end{proofs}

The above basic construction of McGarvey was improved upon by
Stearns~\cite{ste:j:election-graph}. While McGarvey's construction
requires twice as many voters as there are edges in $G$, the
construction of Stearns needs at most $\|V(G)\|+2$ voters. Stearns also
provides a lower bound on the number of voters that are needed to
represent an arbitrary graph via an election. (It is easy to see that
any such graph can be modeled via two irrational voters but the lower
bound for the case of rational votes is somewhat harder.)

We will often construct complicated elections via combining
simpler ones.  Whenever we speak of \emph{combining} two elections,
say $E_1 = (C_1,V_1)$ and $E_2 = (C_2,V_2)$, we mean building, via the
algorithm from Lemma~\ref{thm:election-graph}, an election $E = (C,V)$
whose election graph is a disjoint union of the election graphs of
$E_1$ and $E_2$ with, possibly, some edges added between the vertices
of $G(E_1)$ and $G(E_2)$ (in each case we will explicitly state which
edges, if any, are added).  In particular, we will often want to add
some padding candidates to an election, without affecting the original
election much. In order to do so, we will typically combine our main
election with one of the following  ``padding'' elections.
Note that this construction, which we originally developed for use in
the study of control for $\copelandalpha$ voting, has also proven useful
in the study of manipulation
for $\copelandalpha$~\cite{fal-hem-sch:c:copeland-ties-matter}.

\begin{lemma}
  \label{thm:pad-election}
  Let $\alpha$ be a rational number such that $0 \leq \alpha \leq 1$.
  For each positive integer~$n$, there is 
  a polynomial-time (in $n$) computable 
  election $\pad_n = (C,V)$
  such that $\|C\| = 2n+1$ and for each candidate $c_i \in C$ it holds
  that $\copelandalphascore_{\pad_n}(c) = n$.
\end{lemma}
\begin{proofs}
  Fix a positive integer~$n$.  By Lemma~\ref{thm:election-graph} it is
  enough to construct (in polynomial time in $n$) a directed, antisymmetric graph $G$ with $2n+1$
  vertices, each with its indegree and outdegree equal to $n$.  We set
  $G$'s vertex set to be $\{0,1, \ldots, 2n\}$ and we put an edge from
  vertex $i$ to vertex $j$ ($i \neq j$) if and only if $(j - i) \bmod
  (2n+1) \leq n$.  As a result there is exactly one directed edge between
  every two distinct vertices and for each vertex $i$ we have edges
  going out from $i$ to exactly the vertices $(i+1) \bmod (2n+1), (i+2)
  \bmod (2n+1), \ldots, (i+n) \bmod (2n+1)$.  Thus, both the indegree and the
  outdegree of each vertex is equal to~$n$ and the proof is
  complete.~\end{proofs}

Lemma~\ref{thm:election-graph} is very useful when building an
election in which we need direct control over the results of all
head-to-head contests. However, in many cases explicitly specifying
the results of all head-to-head contests would be very tedious.
Instead it would be easier to specify the results of only the
important head-to-head contests and require all candidates to have
certain suitable scores. In the next lemma we show how to construct
elections specified in such a way via combining a ``small'' election
containing the important head-to-head contest with a ``large'' padding
election. We mention that a generalized version of this lemma has since
been used to study manipulation for $\copelandalpha$~\cite{fal-hem-sch:c:copeland-ties-matter}.

\begin{lemma}
  \label{thm:construction-lemma}
  Let $E = (C,V)$ be an election where $C = \{c_1, \ldots, c_{n'}\}$,
  let $\alpha$ be a rational number such that $0 \leq \alpha \leq 1$, and
  let $n \geq n'$ be an integer.  For each
  candidate~$c_i$ we denote the number of head-to-head ties of $c_i$ in
  $E$ by~$t_i$.
  Let $k_1, \ldots , k_{n'}$
  be a sequence of $n'$ nonnegative integers such that for each $k_i$ we
  have $0 \leq k_i \leq n$. There is an algorithm that in polynomial time in $n$
  outputs an election $E' = (C',V')$ such that:
  \begin{enumerate}
  \item $C' = C \cup D$, where $D = \{d_1, \ldots, d_{2n^2}\}$,

  \item $E'$ restricted to $C$ is $E$,
  \item the only ties in head-to-head contests in $E'$ are between candidates in $C$,

  \item for each $i$, $1 \leq i \leq n'$,
   $\copelandalphascore_{E'}(c_i) = 2n^2 - k_i + t_i \alpha$, and
  \item for each $i$, $1 \leq i \leq 2n^2$,
   $\copelandalphascore_{E'}(d_{i}) \leq n^2+1$.
  \end{enumerate}
\end{lemma}

\begin{proofs}
  We build $E'$ via combining $E$ with a padding election $F$ (see
  Lemma~\ref{thm:pad-election} and the paragraph just before it).  $F
  = (D,W)$, where $D = \{d_1, \ldots, d_{2n^2}\}$, is essentially the
  election $\pad_{n^2}$ with one arbitrary candidate removed.
  We partition the candidates in $D$ into $n$ groups, $D_1, \ldots,
  D_n$, each with exactly $2n$ candidates and we set the results 
  of head-to-head contests between each $c_i \in C$ and the candidates
  in $D$ according to the following scheme. For each $j \in \{1, \ldots, n'\}$
  such that $i \neq j$, $c_i$ defeats all members of $D_j$ and
  $c_i$ defeats  exactly as many candidates in $D_i$ (and loses to
  all the remaining ones) as needed to ensure that
  \[
  \copelandalphascore_{E'}(c_i) = 2n^2 - k_i + t_i \alpha.
  \]
  It is easy to see that this is possible: 
  $c_i$'s score in $(C'-D_i,V')$ is 
  $2n^2-2n+k'+t_i\alpha$ for some $k'$ such that $0 \leq k' \leq n'-t_i$.
  There are $2n$ candidates in $D_i$
  and so $c_i$ can reach any score of the form $2n^2 - k + t_i\alpha$,
  where $k$ is an integer between $0$ and $n$, via defeating in
  head-to-head contests an appropriate number of candidates in $D_i$
  and losing to all the remaining ones.

  Finally, since $F$ is $\pad_{n^2}$ with one candidate removed, each
  $d_i$ gets at most $n^2$ points from defeating other members of $D$
  and at most one point from possibly defeating some member of
  $C$. Thus, for each $d_i \in D$, it holds that
  $\copelandalphascore_{E'}(d_i) \leq n^2 + 1$. This completes the
  proof.~\end{proofs}

Instead of invoking Lemma~\ref{thm:construction-lemma} directly, we
will often simply describe an election in terms of the results of
important head-to-head contests and the scores of the important
candidates and then mention that such an election can be built, 
possibly with adding extra padding candidates that do not affect the
general structure of the election, using
Lemma~\ref{thm:construction-lemma}.  In each such case it will be
clear that Lemma~\ref{thm:construction-lemma} can indeed be used to
build the election we describe.

\subsubsection{Constructive Candidate Control}
\label{sec:constructive-candidate-control}

Let us now turn to the case of constructive candidate control.  Here
we show that resistance holds for $\copelandalpha$ in all cases (i.e.,
for all rational values of $\alpha$ with $0 \leq \alpha \leq 1$ and
for all constructive candidate control scenarios), except for
CCAC$_{\rm u}$ for $\alpha \in \{0,1\}$ where vulnerability holds (see
Theorem~\ref{thm:unlimited-adding-constructive}).

All our resistance proofs in this section follow via reductions from the
vertex cover problem. 
Recall that in the vertex cover problem our input is $(G,k)$ where $G$
is an undirected graph and $k$ a nonnegative integer and we accept if
and only if $G$ has a vertex cover of size at most~$k$. Without the loss
of generality, we assume that $V(G) = \{1, \ldots, n\}$ and $E(G) =
\{e_1, \ldots, e_m\}$.
Note that if either $m = 0$, $n
= 0$, or $k \geq \min(n,m)$ then the instance has a trivial solution
and so in our proofs we will always assume that both $n$ and $m$ are
nonzero and that $k$ is less than $\min(n,m)$. In each case, if the
input to our reduction does not meet these requirements (or is
otherwise malformed) the reduction outputs a fixed ``yes'' instance or
a fixed ``no'' instance depending on the (easily obtained) 
solution to $(G,k)$ or the malformation of the input.
Also note that for every input $(G,k)$ that meets our requirements, 
$G$ has a vertex cover of size less than or equal to $k$ if and only if $G$ has a vertex
cover of size $k$.

\begin{theorem}
  \label{thm:ccac}
  Let $\alpha$ be a rational number such that $0 \leq \alpha \leq 1$.
  $\copelandalpha$ is resistant to constructive control via adding
  candidates (CCAC),
  in both the nonunique-winner model and the unique-winner model,
  for both the rational and the irrational voter model.
\end{theorem}
\begin{proofs}
  We give a reduction from the vertex cover problem.  Let $(G,k)$ be
  an instance of the vertex cover problem, where $G$ is an undirected
  graph, $k$ is a nonnegative integer, $V(G) = \{1, \ldots, n\}$,
  $E(G) = \{e_1, \ldots, e_m\}$, $n \neq 0$, $m \neq 0$, and $k <
  \min(n,m)$.
  We construct an instance of
  CCAC for $\copelandalpha$ such that a designated candidate $p$ can
  become a winner after adding at most $k$ candidates if and only if
  $G$ has a vertex cover of size at most~$k$.  

  Our reduction works as follows.  Via
  Lemma~\ref{thm:construction-lemma}, we build an election $E' =
  (C',V')$ such that:
  \begin{enumerate}
  \item $\{p, e_1, \ldots, e_m\} \subseteq C'$,
  \item $\copelandalphascore_{E'}(p) = 2\ell^2 - 1$ in the
    nonunique-winner case ($\copelandalphascore_{E'}(p)
    = 2\ell^2$ in the unique-winner case),
  \item for each $e_i \in C'$, $\copelandalphascore_{E'}(e_i) =
    2\ell^2$, and
  \item the scores of all candidates in $C' - \{p, e_1, \ldots, e_m\}$ 
    are at most $2\ell^2-n-2$.
  \end{enumerate}

  We form election $E = (C, V)$ by combining $E'$ with candidates $D
  = \{1, \ldots, n\}$ (corresponding to the vertices of $G$).  The
  results of the head-to-head contests within $D$ are set arbitrarily,
  and the head-to-head contests between the members of $C$ and the
  members of $D$ are set as follows: All candidates in $C - \{e_1,
  \ldots, e_m\}$ defeat all members of $D$, and for each $i \in D$ and
  each $e_j \in \{e_1, \ldots, e_m\}$, candidate $i$ defeats $e_j$ if
  $e_j$ is an edge incident to $i$ and loses otherwise.  Our reduction
  outputs an instance $(C,D,V,p,k)$ of CCAC and the question is whether
  it is possible to choose a subset $D' \subseteq D$, $\|D'\| \leq k$,
  such that $p$ is a winner (the unique winner)
  of $\copelandalpha$ election $(C \cup D',
  V)$.  It is clear that this reduction is computable in polynomial
  time. We will now show that it is correct.

If $G$ does have a vertex cover of size $k$ then add the
candidates in $D$ that correspond to the cover. Adding these
candidates increases the score of $p$ by~$k$, while the scores of the $e_i$'s
can increase only by $k-1$ each, since each edge is incident with at least
one member of the vertex cover. 
Clearly, candidates in $C - \{p, e_1, \ldots, e_m\}$ can never become winners
by adding at most $k$ candidates from~$D$,
and thus $p$ becomes a winner (the unique winner).

For the converse,
assume that $p$ can become a winner (the unique winner) via adding at
most $k$ candidates from the set $D$. In order for $p$ to become a
winner (the unique winner), it must be the case that via adding
candidates each $e_i$ gets at least one point less than~$p$. However,
this is possible only if we add candidates that correspond to a 
cover.~\end{proofs}

Interestingly, when the parameter $\alpha$ is strictly between $0$ and
$1$ (i.e., $0 < \alpha < 1$) then $\copelandalpha$ is resistant to
constructive control via adding candidates even if we allow adding an
unlimited number of candidates (the CCAC$_\mathrm{u}$ case).  The
reason for this is that for each rational $\alpha$ strictly between
$0$ and~$1$ our construction will ensure, via its structure,
that we can add at most $k$ candidates. On the other hand, 
both $\copelandzero$ and $\copelandone$ are vulnerable to
constructive control via adding an unlimited number of candidates
(CCAC$_\mathrm{u}$, see Theorem~\ref{thm:unlimited-adding-constructive}).

\begin{theorem}
  \label{thm:ccacu}
  Let $\alpha$ be a rational number such that $0 < \alpha < 1$.
  $\copelandalpha$ is resistant to constructive control via adding an
  unlimited number of candidates (CCAC$_{\mathrm u}$), 
  in both the nonunique-winner model and the unique-winner model,
  for both the rational and the irrational voter model.
\end{theorem}
\begin{proofs}
  We give a reduction from the vertex cover problem. Our reduction
  follows the same general structure as that in the proof of
  Theorem~\ref{thm:ccac}.

  For the unique-winner case, we will need to specify one of the
  candidates' scores in terms of a number $\varepsilon > 0$ such that
  $1-\varepsilon \geq \alpha$.  Let $t_1$ and $t_2$ be two positive
  integers such that $\alpha = \frac{t_1}{t_2}$ and such that their
  greatest common divisor is $1$.  Clearly, two such numbers exist
  because $\alpha$ is rational and greater than $0$. We set
  $\varepsilon$ to be $\frac{1}{t_2}$.  By elementary number-theoretic 
  arguments, there are two positive integer constants,
  $k_1$ and $k_2$, such that $k_1\alpha = k_2 - \varepsilon$.

Let $(G,k)$ be an instance of the vertex cover problem, where $G$ is
an undirected graph and $k$ is 
a nonnegative integer. Let $\{e_1, \ldots,e_m\}$ be $G$'s
edges and let $\{1, \ldots, n\}$ be $G$'s vertices. As before,
we assume that both $n$ and $m$ are nonzero and that $k <
\min(n,m)$. Using Lemma~\ref{thm:construction-lemma}, we can build an
election $E' = (C,V')$ with the following properties:
\begin{enumerate}
  \item $\{p, r, e_1, \ldots, e_m\} \subseteq C$ (the remaining
    candidates in $C$ are used for padding),
  \item $\copelandalphascore_{E'}(p) = 2\ell^2 -1 $,
  \item $\copelandalphascore_{E'}(r) = 2\ell^2 -1 -k + k\alpha$
    in the nonunique-winner case ($\copelandalphascore_{E'}(r) = 2\ell^2 -1 -k + k\alpha - \varepsilon$
    in the unique-winner case\footnote{Note that via the second paragraph of the proof it is easy
    to build an election where $r$ has a score of this form. To obtain the $-\varepsilon$ part
    of $r$'s score we could, for example, have $r$ tie with $k_1$ padding candidates to obtain $k_2 - \varepsilon$ points.
    The $k_2$ points could be accounted for as part of $2\ell^2-1$.}),
  \item for each $e_i \in C$,
    $\copelandalphascore_{E'}(e_i) = 2\ell^2 -1 + \alpha$
    in the nonunique-winner case ($\copelandalphascore_{E'}(e_i) = 2\ell^2 - 1$ in the unique-winner case), and
  \item the scores of all candidates in $C - \{p,r, e_1, \ldots, e_m\}$ 
    are at most $2\ell^2-n-2$.
\end{enumerate}

We form election $E = (C \cup D,V)$ via combining $E'$ with candidates
$D = \{1, \ldots, n\}$ and appropriate voters such that the results of
the head-to-head contests are:
\begin{enumerate}
  \item $p$ ties with all candidates in $D$,
  \item for each $e_j$, if $e_j$ is incident with some $i \in D$ then candidate
    $i$ defeats candidate
 $e_j$,
 and otherwise they tie, and
  \item all other candidates in $C$ defeat each of the candidates in~$D$.
\end{enumerate}

We will now show that $G$ contains a vertex cover of size at most $k$
if and only if there is a set $D' \subseteq D$ such that $p$ is a
winner (the unique winner) of $\copelandalpha$ election $(C \cup
D',V)$.  It is easy to see that if $D'$ corresponds to a vertex cover
of size at most $k$ then $p$ is a winner (the unique winner) of
$\copelandalpha$ election $(C \cup D',V)$. The reason is that adding
any member of $D'$ increases $p$'s score by $\alpha$ and increases $r$'s
score by one, and for each $e_j$, adding $i \in D'$ increases $e_j$'s
score by $\alpha$ if and only if $e_j$ is not incident with $i$. Thus,
via a simple calculation of the scores of the candidates, it is easy to see
that $p$ is a winner (the unique winner) of this election.

On the other hand, assume that $p$ can become a winner (the unique winner)
of $\copelandalpha$ election $(C \cup D',V)$
via adding some subset $D'$ of candidates from $D$. First, note that
$\|D'\| \leq k$, since otherwise $r$ would end up with more points than
(at least as many points as) $p$ and so $p$ would not be
a winner (would not be a unique winner).  We claim
that $D'$ corresponds to a vertex cover of $G$. For the sake of
contradiction, assume that there is some edge $e_j$ incident to
vertices $u$ and $v$ such that neither $u$ nor $v$ is in $D'$.
However, if this were the case then candidate $e_j$ would have more
points than (at least as many points as)  $p$ and so $p$ would
not be a winner (would not be a unique winner).  Thus,
$D'$ must form a vertex cover of size at most~$k$.~\end{proofs}

Note that in the above proof it is crucial that $\alpha$ is neither
$0$ nor $1$.  If $\alpha$ were $0$ then the proof would fall apart because
we would not be able to ensure that $D'$ is a vertex cover, and if
$\alpha$ were $1$ then we would not be able to limit the size of~$D'$. 
In fact, we will now show, as
Theorem~\ref{thm:unlimited-adding-constructive},
that both $\copelandzero$ and $\copelandone$ are vulnerable to
control via adding an unlimited number of candidates (CCAC$_{\mathrm{u}}$).

\begin{theorem}
  \label{thm:unlimited-adding-constructive}
Let $\alpha \in \{0,1\}$.
  $\copelandalpha$ is
vulnerable to constructive control via
  adding an unlimited number of candidates (CCAC$_\mathrm{u}$), 
  in both the nonunique-winner model and the unique-winner model,
  for both the rational and the irrational voter model.
\end{theorem}

\noindent{\bf Proof.}\quad
Our input is candidate set $C$, spoiler candidate set $D$, a
collection of voters with preferences (either preference lists or
preference tables) over $C \cup D$, and a candidate $p \in C$. Our goal is
to check whether there is some subset $D' \subseteq D$ such that $p$ is a
winner (the unique winner) of $(C \cup D',V)$ within $\copelandalpha$.
We will show that we can find such a set $D'$, if it exists,
by the following simple algorithm.
\begin{quote}
Let $D_1 = \{d \in D \ | \ \copelandalphascore_{(\{p,d\},V)}(p) = 1\}.$
Initialize $D'$ to be $D_1$, and delete every $d \in D'$ for
which $\copelandalphascore_{(C \cup D',V)}(p) <
\copelandalphascore_{(C \cup D',V)}(d)$.
For the unique-winner problem, delete
every $d \in D'$ for which $\copelandalphascore_{(C \cup D',V)}(p) \leq
\copelandalphascore_{(C \cup D',V)}(d)$.
\end{quote}

Clearly, this algorithm runs in polynomial time.
To show that the algorithm works, first note that for 
all $\widehat{D} \subseteq D$, if $p$ is a winner (the unique winner) of
$(C \cup \widehat{D}, V)$, then
$p$ is a winner (the unique winner) of
$(C \cup (\widehat{D} \cap D_1), V)$.  This is so because,
by Observation~\ref{obs:scores-new},
\begin{eqnarray*}
\copelandalphascore_{(C \cup \widehat{D},V)}(p) & = & 
\copelandalphascore_{(C \cup (\widehat{D} \cap D_1),V)}(p) +
\sum_{d \in \widehat{D} - D_1} \copelandalphascore_{(\{p,d\},V)}(p)\\
& =  & \copelandalphascore_{(C \cup (\widehat{D} \cap D_1),V)}(p).
\end{eqnarray*}

Now suppose that for some $\widehat{D} \subseteq D_1$,
$p$ is a winner (the unique winner) of  $(C \cup \widehat{D},V)$,
but that the algorithm computes a set $D'$ such that $p$
is not a winner (not a unique winner) of $(C \cup D',V)$.
We first consider the case that $\widehat{D} \subseteq D'$.
Since $p$ is not a winner (not a unique winner) of  $(C \cup D',V)$,
it follows by the construction of $D'$ that
there exists a candidate $d \in C - \{p\}$ such that
$\copelandalphascore_{(C \cup D',V)}(p) <
\copelandalphascore_{(C \cup D',V)}(d)$ 
(such that $\copelandalphascore_{(C \cup D',V)}(p) \leq
\copelandalphascore_{(C \cup D',V)}(d)$).
However, in the nonunique-winner model we then have
\begin{eqnarray*}
\copelandalphascore_{(C \cup D',V)}(p) &=&
\copelandalphascore_{(C \cup \widehat{D},V)}(p)  + \|D'\| - \|\widehat{D}\| \\
&\geq&
\copelandalphascore_{(C \cup \widehat{D},V)}(d)  + \|D'\| - \|\widehat{D}\|
\geq
\copelandalphascore_{(C \cup D',V)}(d),
\end{eqnarray*}
which is a contradiction.
In the unique-winner model, the first
``$\geq$'' in the above inequality
becomes a ``$>$'' and we reach a contradiction as well.

Finally, consider the case that $\widehat{D} \not \subseteq D'$.
Let $d$ be the first candidate in $\widehat{D}$ that is deleted from
$D'$ in the algorithm.  Then there is a set $D''$ such
that $\widehat{D} \subseteq D'' \subseteq D_1$ and 
$\copelandalphascore_{(C \cup D'',V)}(p) < 
\copelandalphascore_{(C \cup D'',V)}(d)$ in the nonunique-winner case
($\copelandalphascore_{(C \cup D'',V)}(p) \leq 
\copelandalphascore_{(C \cup D'',V)}(d)$
in the unique-winner case).
Since $\widehat{D} \subseteq D'' \subseteq D_1$, we have
\begin{enumerate}
\item $\copelandalphascore_{(C \cup \widehat{D},V)}(p) =
\copelandalphascore_{(C \cup D'',V)}(p)  - (\|D''\| - \|\widehat{D}\|) < 
\copelandalphascore_{(C \cup D'',V)}(d)  - (\|D''\| - \|\widehat{D}\|) \leq
\copelandalphascore_{(C \cup \widehat{D},V)}(d)$
in the nonunique-winner case, and
\item $\copelandalphascore_{(C \cup \widehat{D},V)}(p) =
\copelandalphascore_{(C \cup D'',V)}(p)  - (\|D''\| - \|\widehat{D}\|) \leq
\copelandalphascore_{(C \cup D'',V)}(d)  - (\|D''\| - \|\widehat{D}\|) \leq
\copelandalphascore_{(C \cup \widehat{D},V)}(d)$
in the unique-winner case.
\end{enumerate}
It follows that $p$ is not a winner (not a unique winner) of
$(C \cup \widehat{D},V)$. This is again a contradiction.~\qedsymbol

The remainder of this section is dedicated to showing that for any
rational $\alpha$ such that $0 \leq \alpha \leq 1$, $\copelandalpha$
is resistant to constructive control via deleting candidates and
to constructive control
via partitioning candidates (with or without run-off and in both the
TE and the TP model). We first handle the case of constructive control
via deleting candidates (CCDC) and then, using our proof for the CCDC
case as a building block, we handle the constructive
partition-of-candidates cases. The constructions in the proof of
Theorem~\ref{thm:ccdc} are crucial for the remaining proofs, so we
encourage the reader to read that proof particularly carefully, as it
is very difficult to understand the remainder of the section without
understanding the
constructions and arguments in the proof of Theorem~\ref{thm:ccdc}.

\begin{theorem}
  \label{thm:ccdc}
  Let $\alpha$ be a rational number such that $0 \leq \alpha \leq 1$.
  $\copelandalpha$
  is resistant to constructive control via deleting candidates (CCDC), 
  in both the nonunique-winner model and the unique-winner model,
  for both the rational and the irrational voter model.
\end{theorem}
\begin{proofs}
The proof follows via a reduction from the vertex cover problem.  We
first handle the nonunique-winner case.

Let $(G,k)$ be a given input instance of the vertex cover problem, where
$G$ is an undirected graph and $k$ is a nonnegative integer. Let $V(G) = \{1, \ldots, n\}$ and let
$E(G) = \{e_1, \ldots, e_m\}$.  
Again, we assume that $n$ and $m$ are nonzero and that $k < \min(n,m)$.
We build election $E' = (C',V')$,
where $C' = \{p, z, e_1, \ldots, e_m, 1, \ldots, n\}$ and the voter
set $V'$ yields the following results of head-to-head contests (see
Lemma~\ref{thm:election-graph}):
\begin{enumerate}
  \item $p$ defeats $z$,
  \item $z$ defeats each candidate $e_i \in C'$,
  \item each candidate $e_i \in C$ defeats exactly those two candidates $u, v \in \{1, \ldots, n\}$ 
    that the edge $e_i$ is incident with,
  \item each candidate $u \in \{1, \ldots, n\}$ defeats $p$ and
    all candidates $e_i \in C'$ such that vertex $u$ is not incident with
    $e_i$, and
  \item all the remaining contests result in a tie.
\end{enumerate}
Let $\ell = n+m$. We form an election $E = (C,V)$ via combining
election $E'$ with election $\pad_\ell = (C'',V'')$, where $C'' =
\{t_0, \ldots, t_{2\ell}\}$ and the set $V''$ of voters is set as in
Lemma~\ref{thm:pad-election}.  We select the following results of
head-to-head contests between the candidates in $C'$ and the candidates
in $C''$: $p$ and all candidates $e_i \in C'$ defeat everyone in $C''$,
and each candidate in $C''$ defeats all candidates in $C' - \{p, e_1,
\ldots , e_m\}$.
It is easy to verify that election $E$ yields the following $\copelandalpha$
scores:
\begin{enumerate}
  \item $\copelandalphascore_E(p) = m\alpha + 1+ 2\ell+1$,
  \item $\copelandalphascore_E(z) = m + n\alpha$,
  \item for each $e_i \in C$, $\copelandalphascore_E(e_i) = m\alpha + 2+ 2\ell+1$,
  \item for each
  $u \in \{1, \ldots , n\}$,
  $\copelandalphascore_E(u) \leq 1 + m +n\alpha$, and
  \item for each $t_i \in C$, $\copelandalphascore_E(t_i) = \ell + n + 1$.
\end{enumerate}

The set of winners of $E$ is $W = \{e_1, \ldots, e_m\}$.  We claim
that $p$ can become a winner of $\copelandalpha$ election $E$ via
deleting at most $k$ candidates if and only if the graph $G$ has a
vertex cover of size at most $k$. Note that $p$ is the only nonwinner
of $E$ that can become a winner after deleting up to $k$ candidates.
All other candidates lose by more than $k$ points to the members of $W$.

We now show that if $p$ can become a winner via deleting at most $k$
candidates then there is a set $D \subseteq \{1, \ldots, n\}$ such
that $\|D\| \leq k$ and $p$ is a winner of election $(C-D,V)$.  Let
$D'$ be a smallest (in terms of cardinality) subset of $C$ such that $p$ is a
winner of election $(C-D',V)$. 
Clearly, no candidate in $C''$ belongs
to $D'$ because each member of $C''$ loses his or her head-to-head
contests with each member of $\{p\} \cup W$ so deleting any such
candidate from $C$ does not bring $p$ any closer to being a winner.
Similarly, $z$ wins all head-to-head contests with the members of $W$
and so $D'$ does not contain $z$.  Thus, $D' \subseteq \{1, \ldots, n,
e_1, \ldots, e_m\}$. 
Choosing a subset $D'$ of $C$
that ensures that $p$ is a winner means, in essence,
choosing a $D'$ such that each candidate in $\{e_1, \ldots, e_m\}$ is either
in $D'$ or loses at least one point due to some candidate from 
$\{1, \ldots, n\}$ being
in $D'$. (Keep in mind that deleting any $e_i$ affects the scores of
the remaining members of $\{e_1, \ldots, e_m\}$ and $p$ equally.)
Let us assume that there is some $e_i \in D'$
such that $e_i$ is an edge incident to vertices $u$ and $v$ in
$G$.
$D'$ does not contain $u$ and does not contain $v$, since
$e_i$ defeats both $u$ and $v$ in their head-to-head contests;
so if $D'$ included
either $u$ or $v$ then $p$ would be a winner of election 
$(C - (D' - \{e_i\}),V)$, contradicting the fact that $D'$ is a
smallest set such that deleting $D'$ from $C$ ensures $p$'s victory. 
However, if $D'$ contains neither $u$ nor $v$ then $p$ is a winner of
election $(C - ((D' \cup \{u\}) - \{e_i\})$.  Thus, by removing all
members of $\{e_1, \ldots, e_m\}$ from $D'$ and replacing each of them
with one of the vertices they are incident with, we can build a set $D
\subseteq \{1, \ldots, n\}$ such that $\|D\| \leq k$ and $p$ is a
winner of election $(C-D,V)$.

We will now argue that the set $D$ from the previous paragraph
corresponds to a vertex cover of $G$.  In election~$E$, each of $e_1,
\ldots, e_m$ has exactly one point of advantage over
$p$. Deleting any candidate $u$ corresponding to a vertex of $G$ does
not affect $p$'s score but it does decrease by one the scores of all
the candidates $e_1, \ldots, e_m$ that correspond to the edges
incident with $u$. Since deleting the candidates in $D$ makes $p$ a
winner and since $D$ contains at most $k$ candidates that
correspond to vertices of $G$, it must be the case that $D$
corresponds to a vertex cover of $G$.  On the other hand, it is easy
to see that if $G$ has a vertex cover of size at most $k$ then
deleting the candidates that correspond to this vertex cover
guarantees $p$'s victory. Thus our reduction is correct and, as is easily
seen, computable in polynomial time. The proof for the nonunique-winner 
case is complete.

For the proof in the unique-winner case, we need to add one
more candidate, $\hat{z}$, that is
a ``clone'' of $z$ (i.e., $\hat{z}$ ties in
the head-to-head contest with $z$ and has the same results as $z$ in all
other head-to-head contests). In such a modified election, $p$
has the same $\copelandalpha$ score as each of the $e_i$'s and, 
since a candidate ties $p$ if and only if he or she ties each of the $e_i$'s,
$p$ has to
gain at least one point over each of them to become the unique
winner. The rest of the argument remains the same.~\end{proofs}

We will use the above construction in the resistance proofs
for the cases of control via partition of candidates
(with or without run-off, in TP and TE)
below. In particular, we will need the fact that in this construction
the only candidates that can be winners
after deleting at most $k$ candidates are members of the set 
$\{p\} \cup W$ (recall that $W = \{e_1, \ldots, e_m\}$).

Often, when proving things about
candidate control problems we perform 
operations on the candidate set (e.g., deleting candidates, partitioning
the candidate set, etc.)
while the voter set remains unchanged.  
To simplify our notation, in the proofs below we thus will
often use the candidate set as if it were the whole election and leave
$V$ implicit from context. For
example, if we had two candidate sets $C$ and $D$, where $D \subseteq
C$, and a voter collection $V$ with voter preferences over the
candidates from $C$, we may use $C - D$ to mean the set of
candidates that belong to $C$ but not to $D$, but may also use $C - D$ to mean the
election $(C-D,V)$ (where, as usual, the preferences of voters in $V$
as used there are implicitly restricted to the candidates in $C-D$).
In the second case, we typically write
\emph{election} $C-D$ rather than \emph{set} $C-D$.  The intended
meaning is always clear from context.

\begin{theorem}
\label{thm:ccrpc}
Let $\alpha$ be a rational number such that $0 \leq \alpha \leq 1$.
$\copelandalpha$ is resistant to constructive control via run-off
partition of candidates in both the ties-promote model (CCRPC-TP) and the
ties-eliminate model (CCRPC-TE), 
  in both the nonunique-winner model and the unique-winner model,
  for both the rational and the irrational voter model.
\end{theorem}

\sproof
The proof follows via reductions from the vertex cover
problem to appropriate variants of the CCRPC problem for $\copelandalpha$
(i.e., to CCRPC-TP and CCRPC-TE in both the nonunique-winner model and 
the unique-winner model).

Let $(G,k)$ be an instance of the  vertex cover problem, where $G$ is an undirected
graph and $k$ is a nonnegative integer. 
As before, we let $V(G) = \{1,\ldots, n\}$ and $E(G) =
\{e_1,\ldots,e_m\}$, where $n \neq 0, m \neq 0$ and $k < \min(n,m)$.
Our goal is to build an election $E$ in which our
favorite candidate can become a winner (the unique winner) via 
run-off partitioning of candidates (with either the TP or the TE model)
if and only if $G$ has a vertex cover of size at most $k$, and  we do so
by combining suitable 
elections $F$ and $H$ using Construction~\ref{con:ghr} below.
We will later specify  $F$ and $H$ separately for each of the
two variants of the $\copelandalpha$ CCRPC problem (TP and TE), but before doing so 
we will outline our proof in more detail and prove a useful property 
of Construction~\ref{con:ghr} (see Lemma~\ref{thm:ccrpc-structure}).

\begin{construction}
  \label{con:ghr}
  Let $F$ and $H$ be two elections, $F$ with candidates $f_1 = p,f_2,
  \ldots, f_n$ and $H$ with candidates $r, h_1, \ldots, h_q$, $q
  \geq 2$.  We form election $E = (C,V)$, where $C = \{r, f_1, \ldots,
  f_n, h_1, \ldots, h_q\}$, by combining $F$ and $H$ and setting the
  results of head-to-head contests between candidates of $F$ and $H$
  as follows:
  \begin{enumerate}
    \item For each $f_i \in C$, $f_i$ defeats $r$.
    \item For each $h_i, f_j \in C$, $h_i$ defeats $f_j$.
  \end{enumerate}
\end{construction}

In the next lemma we will show that the only partitions $(C_1,C_2)$ of
$C$ such that $p$ is a winner (the unique winner) of the resulting
$\copelandalpha$ run-off election are of the form $C_1 = F - D$, $C_2
= H \cup D$, where $D$ is a subset of $F - \{p\}$ (without loss of generality,
we will assume that $p \in C_1$).
Next we will specify two variants of the
election $H$, one for the TP case and one for the TE case, such that
the only partitions of the form presented above that may possibly lead
to $p$ being a winner (the unique winner) have $\|D\| \leq k$. We will
conclude the construction by choosing $F$ to be one of the
elections from the proof of Theorem~\ref{thm:ccdc}, so that $p$ can
become a winner (the unique winner) of $F$ by deleting at most $k$
candidates if and only if $G$ has a vertex cover of size at most $k$.

Let $E$, $F$, and $H$ be elections as in Construction~\ref{con:ghr}.
We further assume that there are no ties in the head-to-head contests among the candidates of $H$
(and that thus there are no ties in the head-to-head contests among the corresponding candidates in
$E$) and that in the case of CCRPC-TE, $H - \{r\}$ has a unique
winner.  Let us assume that $p$ can become a winner (the unique
winner) of election $E$ via run-off partition of candidates and let
$(C_1,C_2)$ be a partition of candidates such that $p \in C_1$ is a
winner (the unique winner) of the thus-formed $\copelandalpha$ run-off
election (performed using either the TP or the TE tie-handling rule).

\begin{lemma}
  \label{thm:ccrpc-structure}
  Let $E$, $F$, and $H$ be elections as in Construction~\ref{con:ghr},
  and let $(C_1,C_2)$ be a partition of $C$ that makes $p$ a winner
  (the unique winner) in either the CCRPC-TP or the CCRPC-TE scenario,
  where we adopt the assumptions explained in the previous two paragraphs.
  There is a set $D \subseteq F - \{p\}$ such that 
  $C_1 = F - D$ and $C_2 = H \cup D$.

\end{lemma}
\sproof We will handle the TP and TE cases in parallel.
For the sake of seeking a contradiction, let us assume that $C_1 \cap H \neq
\emptyset$.  We consider three cases.
\begin{description}
\item[Case~1:] $C_1$ contains at least two candidates from $H -
  \{r\}$. Let $h_i$ and $h_j$ be two such candidates such that $h_i$
  wins his or her head-to-head contest with $h_j$.  Note that $p$ is
  not a winner of $(C_1,V)$ because
  $\copelandalphascore_{(C_1,V)}(h_i) \geq \|C_1 \cap F\| + 1$ whereas
  $\copelandalphascore_{(C_1,V)}(p) \leq \|C_1 \cap F\|$ (since $p$
  gets at most $\|C_1 \cap F\|-1$ points from defeating members of $C_1
  \cap F$ and one additional point from defeating $r$ if $r
  \in C_1$). Thus, in this case, $p$ does not advance to the final stage,
  irrespective of the tie-handling model used.

\item[Case~2:] $C_1$ contains exactly one member
  of $H - \{r\}$,
  say $h_i$. By an analysis similar to the one above, if $C_1$ does
  not contain $r$ then $\copelandalphascore_{(C_1,V)}(p) <
  \copelandalpha_{(C_1,V)}(h_i)$ and
  irrespective of the tie-handling model $p$ does not advance to the final
  stage.  If $r \in C_1$ then $\copelandalphascore_{(C_1,V)}(p) \leq
  \copelandalpha_{(C_1,V)}(h_i)$.
  Thus, in the TE model, $p$ certainly does not advance to
  the final stage. In the TP model, the set of candidates
  that advance from $(C_1,V)$ to the final stage includes
  $h_i$ and it might include $p$. However, the final stage includes at least
  one member of $H - \{r,h_i\}$, namely
  a winner of $(C_2,V)$ (it is easy to verify that in our current
  case $(C_2,V)$ has at least one winner that belongs to $H-\{r\}$).
  Thus, either $p$ does not
  participate in the final stage or, via the same argument as in
  Case~1, $p$ is not a winner of the final stage because he or she
  meets at least two members of $H-\{r\}$ there.

\item[Case~3:] $C_1 \cap H = \{r\}$. Since $r$ loses the head-to-head contests
  with all members of $F$, $r$ certainly does not advance to
  the final stage of the election. Let us assume that $p$ participates in
  the final stage. However, at least one winner (the only winner, in
  the TE model)\footnote{Recall that we assumed that in the TE case
    $H -\{r\}$ has a unique winner and that each member of $H-\{r\}$ defeats each
    member of $F$ in their head-to-head contests.} of $(C_2,V)$ is a member
  of $H - \{r\}$. Then, via the same argument as in the first subcase
  of Case~2 above, $p$ is not a winner of the final
  stage.
\end{description}
Thus the lemma holds.~\eproofof{Lemma~\ref{thm:ccrpc-structure}}

We now define variants of election $H$ appropriate for the TE and TP models, in
the nonunique-winner model, such that the set $D$ in
Lemma~\ref{thm:ccrpc-structure} is forced to have at most $k$
elements. (We will handle the unique-winner cases at the end of this
proof.)  

For the TP case, we set $H'$ to be an election with candidate
set $\{r, h_1, \ldots, h_q\}$, $q \geq 3$, such that there exists a nonnegative
integer $\ell$ for which we have the following scores:
\begin{enumerate}
  \item $\copelandalphascore_{H'}(r) = \ell$,
  \item for $i \in \{1,2\}$,
 $\copelandalphascore_{H'}(h_i) = \ell - k - 1$, and
  \item for each $i \in \{3, \ldots, q\}$,
 $\copelandalphascore_{H'}(h_i) \leq \ell - k - 1$.
\end{enumerate}
Such an election is easy to build in polynomial time using
Lemma~\ref{thm:construction-lemma}. 

For the TE case, we set $H''$ to have candidate set
 $\{r, h_1, \ldots, h_q\}$, $q \geq 2$,
with the following scores:
\begin{enumerate}
  \item $\copelandalphascore_{H''}(r) = \ell$,
  \item $\copelandalphascore_{H''}(h_1) = \ell - k$, and
  \item for each $i \in \{2, \ldots, q\}$, 
 $\copelandalphascore_{H''}(h_i) < \ell - k - 1$.
\end{enumerate}

Note that both $H'$ and $H''$ satisfy the assumptions that
we made regarding $H$ before Lemma~\ref{thm:ccrpc-structure}.

\begin{lemma}
  \label{thm:ccrpc-k-bound}
  For the TP case with $H=H'$ and for the TE case with $H = H''$, 
  the set $D$ in Lemma~\ref{thm:ccrpc-structure} has the additional
  property that $\|D\| \leq k$.
\end{lemma}
\sproof Recall that in $E$ each candidate $h_i \in H - \{r\}$ wins
each of his or her head-to-head contests with candidates in $F$ and
that each candidate $f_i \in F$ wins his or her head-to-head contest
with $r$. From Lemma~\ref{thm:ccrpc-structure} we know that 
if $p$ is a winner in either the CCRPC-TP or the CCRPC-TE scenario
with the partition $(C_1,C_2)$ of~$C$, then $C_1 = F -
D$ and $C_2 = H \cup D$.  If $\|D\| > k$ were to hold
in the TP case, then at least
$h_1$ and $h_2$ would be
winners of $(C_2,V)$, and so even if $p$ were promoted to the
final stage, he or she would meet two members of $H - \{r\}$ there and
would not become a global winner.  If $\|D\| > k$ were to hold
in the TE case, then
$h_1$ would be the unique winner of $(C_2,V)$ and the final stage
would involve, at best, $p$ and $h_1$. Naturally, $p$ would
lose.~\eproofof{Lemma~\ref{thm:ccrpc-k-bound}}

For the TP case (TE case), we set $F$ to be the election built in the
proof of Theorem~\ref{thm:ccdc} for the nonunique-winner model (for
the unique-winner model).
In particular, candidates $e_1, \ldots, e_m$ correspond to
the edges of graph $G$.  Since $p$ can become a winner of his or her
subcommittee in the TP model (in the TE model) if and only if $p$ can
become a winner (the unique winner) of election $F - D$, where $D
\subseteq F - \{p\}$ and $\|D\| \leq k$, it follows by our choice of
$F$, that $p$ can advance to the final stage only if $G$ has a vertex
cover of size at most $k$.  On the other hand, if $G$ has a vertex
cover of size at most $k$ then it is easy to see that if we partition
the candidates in $C$ as in Lemma~\ref{thm:ccrpc-structure} with $D$
containing the candidates corresponding to an at-most-size-$k$ vertex
cover of $G$ then $p$ advances to the final stage of the election and
is a winner there.  Why? In the TP case,
the subcommittee $H \cup D$ has $r$ as the unique winner and the
subcommittee $F - D$'s winner set contains $p$ and some subset
of $\{e_1,\ldots,e_m\}$ (see the note below the proof of
Theorem~\ref{thm:ccdc}).  Since $p$ and all candidates in $\{e_1,
\ldots, e_m\}$ tie in their head-to-head contests and since they all
defeat $r$, they all are winners of the final stage.  Similarly, in
the TE case, $r$ is the only candidate that can be a unique winner
of subcommittee $H \cup D$,
and the subcommittee $F-D$ has $p$ as the unique
winner. Since $p$ defeats $r$, $p$ is the winner of the final stage.
This completes the proof for the nonunique-winner case.

The proof for the TE case in the nonunique-winner model works
just as well in the unique-winner model and so it remains to handle
the TP case in the unique-winner model. To do so,  we form the
election $E$ using Construction~\ref{con:ghr} with $F$ set to the
unique-winner version of the election from Theorem~\ref{thm:ccdc} and with
$H$ set to $H'$. Via Lemmas~\ref{thm:ccrpc-structure}
and~\ref{thm:ccrpc-k-bound} and the subsequent discussion we have that
any partition of $E$'s candidate set $C$ into $C_1$ and $C_2$ such that
$p \in C_1$ is a winner of the final-stage election requires
$C_1 = F - D$ and $C_2 = H \cup D$, where $D$ corresponds to an
at-most-size-$k$ vertex cover of $G$. On the other hand, by the choice
of $F$, it is easy to see that using such a $D$ guarantees that $p$ is
the unique winner of the final stage election.  The proof is
complete.~\eproofof{Theorem~\ref{thm:ccrpc}}

$\copelandalpha$ is also resistant to constructive control via
partition of candidates (without run-off) for each rational value of
$\alpha$ between (and including) $0$ and $1$. However, the proofs for
the TP and TE cases are not as uniform as in the CCRPC scenario and so we
treat these cases separately.

\begin{theorem}
  \label{thm:ccpc-tp}
  Let $\alpha$ be a rational number such that $0 \leq \alpha \leq 1$.
  $\copelandalpha$ is resistant to constructive control via partition
  of candidates with the ties-promote tie-handling rule (CCPC-TP),
  in both the nonunique-winner model and the unique-winner model,
  for both the rational and the irrational voter model.
\end{theorem}
\sproof The proof is very similar to that of Theorem~\ref{thm:ccrpc}
and we maintain similar notation. In particular, $(G,k)$ is the given
instance of the vertex cover problem we reduce from (where
$V(G)=\{1,\ldots,n\}$, $E(G) = \{e_1, \ldots, e_m\}$, $n \neq 0, m \neq 0$,
and $k < \min(n,m)$), we set
$F$ to be the nonunique-winner variant of the election built in the
proof of Theorem~\ref{thm:ccdc} when we handle the nonunique-winner
case for $\copelandalpha$-CCPC-TP, and we set $F$ to be the unique-winner
variant of that election when we handle the unique-winner case for
$\copelandalpha$-CCPC-TP.

We define $H$ to be an election with candidate set $\{r, h_1, \ldots, h_q\}$,
$q \geq 3$, such that there exists a nonnegative integer $\ell$ for which
we have the following scores:
\begin{enumerate}
  \item $\copelandalphascore_H(r) = \ell$,
  \item for $i \in \{1,2,3\}$, $\copelandalphascore_H(h_i) = \ell -k-1$,
  \item for each $i \in \{4, \ldots, q\}$, $\copelandalphascore_H(h_i) \leq \ell - k -1$, and
  \item the results of the head-to-head contests between $r$, $h_1$,
    $h_2$, and $h_3$ are that $h_1$ defeats $h_2$, $h_2$ defeats
    $h_3$, $h_3$ defeats $h_1$, and $r$ defeats each of $h_1$, $h_2$,
    and $h_3$.\footnote{The reason for such a cycle of head-to-head
      contest results is that if we delete at most one candidate from
      $H$ then there still will be at least one candidate with score
      $\ell - k - 1$.}
\end{enumerate}

Election $H$ is easy to build using
Lemma~\ref{thm:construction-lemma}.
Election $E = (C,V)$ is formed via applying Construction~\ref{con:ghr} to
elections $F$ and $H$.

In constructive control via partition of candidates (CCPC), the first
stage of the election is held among some subset $C' \subseteq C$ of
candidates.  In the TP model, the winners of
the first stage compete with the candidates in $C - C'$. The
following two lemmas describe what properties $C'$ needs to satisfy for
$p$ to become a winner (the unique winner) of the final stage.

\begin{lemma}
  \label{thm:ccpc-tp-p}
  Let $C'$ be a subcommittee such that $p$ is a winner (the unique winner) of the final stage in $E$ whose first stage involves
  subcommittee~$C'$ in the CCPC-TP model.
  Then $p$ is not a member of~$C'$.
\end{lemma}
\sproof
If $p$ were in $C'$ together with at least two members of $H$ then $p$
would not be a winner of this subelection. If $p$ were in $C'$ with
less than two members of $H$ then $p$ would either meet at least two members
of $H$ in the final stage or $p$ would not make it to the final
stage.  In either case, $p$ would not be a winner of the final
stage.~\eproofof{Lemma~\ref{thm:ccpc-tp-p}}

\begin{lemma}
  \label{thm:ccpc-tp-structure}
  Let $C'$ be a subcommittee such that $p$ is a winner (the unique winner)
  of the final stage of $E$ whose first stage involves subcommittee~$C'$
  in the CCPC-TP model.
  Then $C'$ is of the form $H \cup D$, where $D \subseteq F - \{p\}$ and
  $\|D\| \leq k$.
\end{lemma}
\sproof
From Lemma~\ref{thm:ccpc-tp-p} we know that $p$ is not in $C'$.
If more than two members of $H$ were not in $C'$ then $p$ would meet
at least two members of $H$ in the final stage and so would not be a
winner (would not be a unique winner).

Assume that exactly one member of $H$,
say $h$, is not in $C'$.  If $r$ is not in $C'$ then the set of
winners of $(C',V)$ includes at least one member of $H - \{r\}$ and so
$p$ meets at least two members of $H - \{r\}$ in the final stage and thus is
not a winner of that stage. So we additionally assume that $r$ is in
$C'$.  Since $H - \{h\} \subseteq C'$, at least two of $h_1$, $h_2$, and $h_3$
are in $C'$. If $C'$ contains more than $k$ candidates from $F$ then
at least one of $h_1$, $h_2$, and $h_3$ is a winner of $(C',V)$ and
$p$, again, competes (and loses) against at least two members of $H$ in the
final subelection.  Thus let us assume that $C'$ contains at most $k$
members of $F$, call them $d_1, d_2, \ldots, d_j$, where $j \leq k$.
In such a case, $p$ loses to at least one candidate $s$ in $F - \{d_1,
\ldots, d_j\}$ (recall that $k < n$ because otherwise the input vertex
cover instance is trivial and so our $s$ can be, e.g., one of the candidates
in $F - \{d_1, \ldots, d_j\}$ corresponding to a vertex in $G$) and so in the final stage $p$ has a score
lower than $h$'s score: $h$ wins the head-to-head contests with
everyone with whom $p$ wins his or her head-to-head contests except
for~$r$, but $h$ also wins his or her head-to-head contests with both
$p$ and~$s$.  As a result, $p$ does not win the final stage. This
completes the proof.~\eproofof{Lemma~\ref{thm:ccpc-tp-structure}}

Since $r$ is the unique winner of any subelection of the form $H \cup
D$, where $D$ contains at most $k$ candidates from $F - \{p\}$ and all
members of $F$ defeat $r$ in their head-to-head contests, via
Lemma~\ref{thm:ccpc-tp-structure}, we have that $p$ can become a
winner (the unique winner) in our current control scenario (CCPC-TP)
if and only if we can select
a set $D$ of at most $k$ candidates from $F$ such that $p$ is a winner
(the unique winner) of $F - D$. By the choice of $F$ and 
Theorem~\ref{thm:ccdc}, this is possible only if $G$ has a vertex
cover of size at most~$k$. This completes the proof.~\eproofof{Theorem~\ref{thm:ccpc-tp}}

We now turn to the TE variant of constructive control via
partitioning candidates. We start by showing resistance
for Llull's system (i.e., for $\copelandalphaone$).

\begin{theorem}
  \label{thm:ccpc-te-1}
  $\copelandalphaone$ is resistant to constructive control via
  partition of candidates with the ties-eliminate tie-handling rule
  (CCPC-TE),
  in both the nonunique-winner model and the unique-winner model,
  for both the rational and the irrational voter model.
\end{theorem}
\begin{proofs}
  We use the same reduction as that in the proof of
  Theorem~\ref{thm:ccpc-tp}, except that we now use a slightly 
  different variant of the election $H$. 

  Let $(G,k)$ be our input instance of the vertex cover problem. 
  As in the proof of Theorem~\ref{thm:ccpc-tp}, we form an election $E$
  via combining elections $F$ and $H$, where $F$ and $H$ are as
  follows. If we are in the nonunique-winner model
  then $F$ is the nonunique-winner variant of the election
  from the proof Theorem~\ref{thm:ccdc}, and otherwise it is the unique-winner
  variant of that election. $H$ is an election whose
  candidate set is $\{r,h_1, \ldots,h_\ell\}$ and whose voter set is such 
  that  $r$ ties all head-to-head
  contests with the candidates $h_1, \ldots, h_\ell$ and the
  scores of the candidates satisfy:
  \begin{enumerate}
  \item $\copelandalphaonescore_{H}(r) = \ell$,
  \item $\copelandalphaonescore_{H}(h_1) = \ell - k$, and
  \item for each $h \in H - \{r,h_1\}$, $\copelandalphaonescore_{H}(h) < \ell - k$.
  \end{enumerate}
  It is easy to see that such an election can be built in polynomial
  time in $k$ (thus ensuring that $\ell$ is polynomially bounded in
  $k$).  A simple variant of the construction from
  Lemma~\ref{thm:construction-lemma} can be used for this purpose. (We
  cannot use Lemma~\ref{thm:construction-lemma} directly because it
  doesn't allow us to have $r$ tie all its head-to-head contests, but
  clearly a similar construction works.)
  As usual, we assume
  $k < \min(n,m)$, where $n$ is the number of vertices in $G$ and
  $m$ is the number of $G$'s edges.

  It is easy to see that if $G$ does have a vertex cover of size $k$
  then using the subcommittee $H \cup D$, where $D$ is a
  subset of candidates from $F$ that corresponds to this vertex cover
  does make $p$ a winner (the unique winner) of the
  final stage of our two-stage election.

  For the converse, we show that if $p$ can be made a winner
  (the unique winner) via partition of candidates then
  $G$ does have a vertex cover of size at most $k$.  First, we note
  that $p$ is not a winner (and so certainly not a unique winner) of
  any subelection that involves any of the candidates $h_1, \ldots,
  h_\ell$. (This is because each of $h_1, \ldots, h_\ell$ wins all
  his or her head-to-head contests with members of $F$ and ties the
  head-to-head contest with $r$.) Thus if $p$ is a winner (the unique
  winner) of our two-stage election then all candidates $h_1, \ldots,
  h_\ell$ participate in the first-stage subcommittee and $p$ does not.
  In this case, the subcommittee also contains $r$ because otherwise 
  $h_1$ would be the unique winner of the subcommittee and, via the
  previous argument, $p$ would not be a winner of the final stage. 
  Similarly, if the subcommittee contained more than $k$ members of
  $F$ then, again, $h_1$ would be its unique winner preventing $p$
  from being a winner of the final stage.

  Let $C'$ be a subcommittee (i.e., the set of the candidates that
  participate in the first stage of the election) such that $p$ is a
  winner (the unique winner) of our two-stage
  election. Via the previous paragraph it holds that $C' = H \cup D$,
  where $D \subseteq F$ and $\|D\| \leq k$.  Either $r$ is the unique
  winner of $C'$ (if $\|D\| < k$) or $C'$ does not have a unique
  winner. Thus the winner set of the final stage is the same as that
  of $F-D$ because all candidates in $F-D$ defeat $r$ in their
  head-to-head contests.  As a result, via Theorem~\ref{thm:ccdc}, we have
  that $p$ is a winner (the unique winner) if and only
  if $D$ corresponds to a vertex cover of
  $G$ of size at most $k$.~\end{proofs}

\begin{theorem}
  \label{thm:ccpc-te-smaller-than-1}
  Let $\alpha$ be a rational number, $0 \leq \alpha <
  1$. $\copelandalpha$ is resistant to constructive control via
  partition of candidates with the ties-eliminate tie-handling rule
  (CCPC-TE), 
  in both the nonunique-winner model and the unique-winner model,
  for both the rational and the irrational voter model.
\end{theorem}
\begin{proofs}
  The proof follows via a reduction from the vertex cover problem. Let
  $(G,k)$ be our input instance, where $G$ is an undirected graph and
  $k$ is a nonnegative integer. By combining two subelections, $F$ and $H$, we will build an
  election $E$ such that a candidate $p$ in $E$ can become a winner
  (the unique winner) via partitioning of candidates if and only if
  $G$ has a vertex cover of size at most~$k$.
  
  In the nonunique-winner case, we take $F$ to be the nonunique-winner
  variant of the election built in the proof of
  Theorem~\ref{thm:ccdc}.  In the unique-winner case, we set $F$ to be
  this election's unique-winner variant. Election $H$ has candidate set
  $\{r,\hat{r}, h_1, \ldots,h_k\}$ and a voter set that yields the
  following results of head-to-head contests within~$H$:
  \begin{enumerate}
  \item $r$ and $\hat{r}$ are tied,
  \item $r$ ties with every candidate $h_i$, $i \in \{1,\ldots, k\}$,
  \item $\hat{r}$ defeats every candidate $h_i$, $i \in \{1,\ldots, k\}$, and
  \item The results of head-to-head contests between candidates
        $h_1, \ldots, h_k$ are set arbitrarily.
  \end{enumerate}
  Within election $H$ the candidates have the following
  $\copelandalpha$ scores:
  \begin{enumerate}
  \item $\copelandalphascore_H(r) = k\alpha + \alpha$,
  \item $\copelandalphascore_H(\hat{r}) = k + \alpha$, and
  \item for each $i \in \{1, \ldots, k\}$, $\copelandalphascore_H(h_i)
    \leq k-1 + \alpha$.
  \end{enumerate}
  We form election $E = (C,V)$ via combining elections $F$ and $H$ in
  such a way that $r$ defeats all the candidates in $F$ and for all
  the other head-to-head contests between the candidates from $F$ and
  the candidates from $H$ the result is a tie.

  It is easy to see that if $G$ does have a vertex cover of size 
  $k$ then we can make $p$ a winner (the unique winner) via
  partitioning of candidates. To do so we hold the first-stage election
  among the candidates in $H \cup D$, where $D$ is a set of candidates in
  $F$ that correspond to a vertex cover within $G$ of size 
  $k$. 
  Since $r$ and $\hat{r}$ are the winners of this subelection, no candidates
  from this subelection proceed to the final stage of the election. By
  Theorem~\ref{thm:ccdc}, it follows that $p$ is a winner (the unique
  winner) of the final stage.

  For the converse, let us assume that $p$ can become a winner (the
  unique winner) via partitioning of candidates and let $C'$ be a
  subset of candidates such that if the first-stage election is
  $(C',V)$ then $p$ is a winner (the unique winner) of the
  \emph{final} stage.  If $p$ and $r$ participate in the same
  subelection (be it the first stage or the final stage) then it is
  easy to see that $p$ is not a winner of that subelection.  (This is
  true because $r$ wins his or her head-to-head contests with all members of $F$ and
  ties with everyone else, whereas $p$ ties with all candidates
  $\hat{r},h_1, \ldots, h_k$ but loses to $r$.)  Thus $r
  \in C'$, $p \notin C'$, and $r$ is not the unique winner of $(C',V)$

  We will now show that $C'$ contains at most $k$ members of $F$. For
  the sake of seeking a contradiction let us assume that this is not the case,
  and let $d$ be the number of members of $F$ in $C'$, $d > k$, and let
  $k'$ be the number of members of $\{h_1, \ldots, h_k\}$ in $C'$. Let us rename the
  candidates in $F$ so that $f_1, \ldots, f_d$ are those that belong
  to $C'$.  Finally, we set $b = 0$ if $\hat{r} \notin C'$, and $b =
  1$ otherwise.  We have the following $\copelandalpha$ scores within
  $(C',V)$:
  \begin{enumerate}
  \item $\copelandalphascore_{(C',V)}(r) = d + k'\alpha + b\alpha$,
  \item $\copelandalphascore_{(C',V)}(\hat{r}) = k' + d\alpha +\alpha$,
    provided that $\hat{r} \in C'$,
  \item for each $h_i \in C'$, $\copelandalphascore_{(C',V)}(h_i) \leq
    k'-1 + d\alpha + \alpha$, and
  \item for each $f_i \in C'$, $\copelandalphascore_{(C',V)}(f_i) \leq d-1
    + k'\alpha + b\alpha$.
  \end{enumerate}
  Since $k' \leq k$ and $d > k$, it is easy to see that irrespective
  of whether $\hat{r}$ participates in~$C'$, $r$ is the unique
  winner of $(C',V)$, which is a contradiction. Thus $C'$ contains
  at most $k$ members of $F$.

  Let $D$ be the set of candidates from $F$ that are in $C'$
  and that do not participate in the final stage of the election.
  Since
  $p$ is a winner (the unique winner) of the final stage, $r$ does not
  participate in that stage. Thus the final stage is held with
  candidate set $(F - D) \cup H'$, where $H'$ is some subset of $H -
  \{r\}$. It is easy to see that the winners of $F-D$ have such high
  scores in addition to tieing their head-to-head contests with all members
  of $H'$ that the winner set of $((F-D) \cup H',V)$ is the same as
  that of $(F-D,V)$. 

  To see this, we note the following: Let $c$ be a winner of $F -
  D$.  In $F-D$, $c$'s score is at least $m\alpha + 2 + 2\ell + 1
  -\|D\|$, where $\ell = n+m$, $n$ is the number of vertices of $G$,
  and $m$ is the number of edges of $G$. (See Theorem~\ref{thm:ccdc}
  and the scores of candidates $p$ and $e_i$.)  Since, due to our
  usual convention, $k < \min(m,n)$, it holds that
  the lowest score a winner of $F-D$ might have is at least $m\alpha +
  2 +2\ell + 1 - \|D\|$. The highest score a candidate in $H'$ might
  have is $k + \alpha$. Since all members of $H'$ tie in their
  head-to-head contents with members of $F - D$ (and since $F$ contains
  $m+n+2$ candidates), the highest score a
  member of $H'$ can have in $(F-D) \cup H'$ is 
\[
 k + \alpha + (m + n + 2)\alpha - \|D\| <
 k + \alpha + \ell + 2 - \|D\| <
 m\alpha + 2 +2\ell + 1 - \|D\| \leq
 \copelandalphascore_{(F-D) \cup H'}(c).
\]
  Thus any winner of $F-D$ is still a winner of $(F-D) \cup H'$.

Thus, via Theorem~\ref{thm:ccdc}, if $p$ is a
  winner (the unique winner) of $F-D$, $\|D\| \leq k$,
  then $D$ corresponds to a vertex cover of size at most $k$ in
  $G$. This completes the proof.~\end{proofs}

\subsection{Voter Control}
\label{sec:control-voter}

In this section, we show that for each rational $\alpha$, $0 \leq
\alpha \leq 1$, $\copelandalpha$ is resistant to all types of voter
control. Table~\ref{tab:voter-control} lists for each type of voter
control, each rational $\alpha$, $0 \leq \alpha \leq 1$, and each
winner model (i.e., the nonunique-winner model and the unique-winner model)
the theorem in which each given case is handled.
We start with control via adding voters.

\newcommand{\thm}[1]{\multicolumn{1}{|c|}{Thm.~\ref{#1}}}

\begin{table}
\begin{center}
\begin{tabular}{|l|cccccc|}
\hline
                  & \multicolumn{2}{|c|}{$\alpha = 0$} & \multicolumn{2}{|c|}{$0 < \alpha < 1$} & \multicolumn{2}{|c|}{$\alpha = 1$} \\
\cline{2-7}
                  & unique & \multicolumn{1}{|c|}{nonunique}   & unique & \multicolumn{1}{|c|}{nonunique}  & unique & \multicolumn{1}{|c|}{nonunique} \\
\hline
CCAV              & \multicolumn{6}{|c|}{\multirow{2}{*}{Thm.~\ref{thm:av}}} \\
\cline{0-0}
DCAV              & & & & & & \\ 
\hline
CCDV              &\thm{thm:ucdv}         & \thm{thm:copeland-cdv} & \thm{thm:ucdv}         & \thm{thm:copeland-cdv} & \thm{thm:ucdv}      & \thm{thm:llull-cdv}\\ 
\hline
DCDV              &\thm{thm:copeland-cdv} & \thm{thm:ucdv}         & \thm{thm:copeland-cdv} & \thm{thm:ucdv}         & \thm{thm:llull-cdv} & \thm{thm:ucdv} \\ 
\hline
CCPV-TP           & \multicolumn{6}{|c|}{\multirow{2}{*}{Thm.~\ref{thm:cpv-tp}}} \\
\cline{0-0}
DCPV-TP           & & & & & & \\ 
\hline
CCPV-TE           & \multicolumn{4}{|c|}{Thm.~\ref{thm:cpv-te}} & \multicolumn{2}{|c|}{Thm.~\ref{thm:cpv-one-con}} \\
\hline
DCPV-TE           & \multicolumn{4}{|c|}{Thm.~\ref{thm:cpv-lessone}} & \multicolumn{2}{|c|}{Thm.~\ref{thm:cpv-one-dest}} \\
\hline
\end{tabular}
\end{center}
\caption{\label{tab:voter-control}Table of theorems covering all resistance results for voter control for $\copelandalpha$. Each
theorem covers both the case of rational voters and the case of irrational voters.}
\end{table}

\begin{theorem}
  \label{thm:av}
  Let $\alpha$ be a rational number such that $0 \leq \alpha \leq
  1$. $\copelandalpha$ is resistant to both constructive and
  destructive control via adding voters (CCAV and DCAV),
  in both the nonunique-winner model and the unique-winner model,
  for both the rational and the irrational voter model.
\end{theorem}
\begin{proofs}
Our result follows via reductions from the X3C problem. We
will first show how to handle the nonunique-winner constructive case
and later we will argue that the construction can be easily modified
for each of the remaining cases.

Let $(B,\calS)$ be an X3C instance where $B = \{b_1, \ldots, b_{3k}\}$
and $\calS = \{S_1, \ldots, S_n\}$ is
a finite collection of
three-element subsets of $B$. 
Without loss of generality, we assume that $k$ is odd (if it is even,
we simply add $b_{3k+1}, b_{3k+2}, b_{3(k+1)}$ to $B$ and $S_{n+1} =
\{b_{3k+1}, b_{3k+2}, b_{3(k+1)}\}$ to $\calS$, and add $1$ to $k$).
The question is whether one can pick $k$ sets
$S_{a_1}, \ldots, S_{a_k}$ such that $B = \bigcup_{j=1}^k S_{a_j}$.

We build a $\copelandalpha$ election $E = (C,V)$ as follows. The
candidate set $C$ contains candidates $p$
(the preferred candidate), $r$ ($p$'s rival), $s$, all members of
$B$, and some number of padding candidates. We select the voter collection $V$
such that in their head-to-head contests, $s$ defeats $p$, $r$ defeats each $b_i$,
and such that 
we have the following $\copelandalpha$ scores for these
candidates, where $\ell$ is some sufficiently large (but polynomially bounded in $n$) nonnegative integer:
\begin{enumerate}
  \item $\copelandalphascore_E(p) = \ell - 1$,
  \item $\copelandalphascore_E(r) = \ell + 3k$, and
  \item all other candidates have $\copelandalpha$ scores below $\ell-1$.
\end{enumerate}
It is easy to see that $E$ can be constructed in polynomial time by
Lemma~\ref{thm:construction-lemma}.
In addition, we ensure that we have the following results of head-to-head contests between
the candidates in $C$:
\begin{enumerate}
  \item $\versus_E(s,p) = k-1$,
  \item for each $i \in \{1, \ldots, k\}$, $\versus_E(r, b_i) = k-3$, and
  \item for all other pairs of candidates $c$, $d$, we have
    $|\versus_E(c,d)| \geq k+1$.
\end{enumerate}
This can be done since we can add $2$ to $\versus_E(c,d)$ and leave
all other relative vote scores the same by adding two voters, $c > d >
C - \{c,d\}$ and $C - \{c,d\} > c > d$ (see Lemma~\ref{thm:election-graph}). Since $k$
is odd and the number of voters is even (see Lemma~\ref{thm:construction-lemma}), it is
easy to see that we can fulfill these requirements.

We also specify the set $W$ of voters that the chair can potentially
add. For each set $S_i \in \calS$ we have a single voter $w_i \in W$ with
preference list \[ p > B - S_i > r > S_i > \cdots \] (all unmentioned
candidates follow in any fixed arbitrary order).  We claim that
$\calS$
contains a $k$-element cover of $B$ if and only if $p$ can become a winner of
the above election via adding at most $k$ voters selected from $W$.

If $\calS$ contains a $k$-element cover of $B$, say $S_{a_1}, \ldots ,  S_{a_k}$, then we can make
$p$ a winner via adding the voters from $U = \{ w_{a_1}, \ldots, w_{a_k}\}$. Adding
these $k$ voters increases $p$'s score by one, since $p$ now
defeats $s$ in their head-to-head contest.
Since voters in $U$
correspond to a cover, the score of $r$ goes
down by $3k$ points.  Why?
For each $b_i \in B$, adding
the $k-1$ voters in $U$ that correspond to the sets
in the cover not containing $b_i$ increases the relative performance
of $b_i$ versus $r$ by $k-1$ votes, thus giving $b_i$ two votes of
advantage over $r$. Adding the remaining voter from $U$
decreases this advantage to $1$, but still $b_i$ wins the
head-to-head contest with $r$. 

We now show that if we can make $p$ a winner by adding at most $k$
voters then $\calS$ contains a $k$-element cover of $B$.
Note that $p$ is the only candidate
that can possibly become a winner by adding at most $k$ voters,
that $p$ can at best obtain $\copelandalpha$ score $\ell$, that $p$
will obtain this score only if we
add exactly $k$ voters, and that $r$ can lose at most $3k$ points via
losing his or her head-to-head contests with each of the $b_i$'s.
Thus the only way for $p$ to become a winner by adding at most $k$
voters from $W$ is that we add exactly $k$ voters such that $r$ loses
his or her head-to-head contest with each~$b_i$. Assume that $U \subseteq
W$ is such a set of voters that does not correspond to a cover of
$B$. This means that there is some candidate $b_i$ such that at least
two voters in $U$ prefer $r$ to $b_i$. However, if this is the case
then $b_i$ cannot defeat $r$ in their head-to-head contest and $p$ is
not a winner. $U$ corresponds to a cover. This
completes the proof of the nonunique-winner constructive case of
the theorem.

For the constructive unique-winner case, we modify election $E$ so that
$\copelandalphascore_E(p) = \ell$.  All other listed properties of the 
relative vote scores and absolute $\copelandalpha$ scores are unchanged.
As in the previous case, it is easy
to see that $p$ can become the unique winner via adding $k$ voters
that correspond to a cover of $B$.  For the converse, we will
show that we
still need to add exactly $k$ voters if $p$ is to become the unique
winner. 

If we added fewer than $k-1$ voters then $p$ would not get any extra points
and so it would be impossible for $p$ to become the unique winner.
Let us now show that adding exactly $k-1$ voters cannot make $p$ the
unique winner.  If we added exactly $k-1$ voters then $p$ would get
$\alpha$ points extra from the tie with $s$.  Now consider some
candidate $b_i \in S_j$, where $S_j$ corresponds to one of the added
voters, $w_j$. Since $w_j$ prefers $r$ to $b_i$, adding $w_j$ to the
election increases the relative performance of $r$ versus $b_i$ to
$k-2$. Thus adding the remaining $k-2$ voters can result in $b_i$
either tieing or losing his or her head-to-head contest with~$r$. In
either case $p$ would not have a high enough score to become the unique
winner. Thus we know that exactly $k$ candidates must be added if we
want $p$ to become the unique winner and, via the same argument as in
the previous case, we know that they have to correspond to a cover.

For the destructive cases it suffices to note that the proof for the
constructive nonunique-winner case
works also as a proof for the destructive
unique-winner case (where we are preventing $r$ from being the unique
winner) and the constructive unique-winner case
works also as a proof
for the destructive nonunique-winner case (where we are preventing $r$ from
being a winner).~\end{proofs}

Let us now turn to the case of control via deleting voters.
Unfortunately, the proofs here are not as uniform as before and we need
in some cases
to handle $\alpha = 1$ separately from the case where $0 \leq \alpha <
1$. Also, we cannot use the construction lemma
(Lemma~\ref{thm:construction-lemma}) anymore to so conveniently build
our elections. In the case of deleting voters (or partitioning voters)
we need to have a very clear understanding of how each voter affects
the election and the whole point of introducing the construction lemma
was to abstract away from such low-level details.

Analogously to the case of candidate control, we will later reuse the
resistance proofs for deleting voters
within the resistance proofs for
partitioning voters.

\begin{theorem}
\label{thm:llull-cdv}
  $\copelandone$ is resistant to constructive control via
  deleting voters (CCDV) in the nonunique-winner model  and to
  destructive control via deleting voters (DCDV) in the unique-winner
  model,
  for both the rational and the irrational voter model.
\end{theorem}

\begin{proofs}
Let $(B,\calS)$ be an instance of X3C, where $B = \{b_1, \ldots,
b_{3k}\}$ and $\calS = \{S_1, \ldots, S_n\}$ is a finite family of
three-element subsets of $B$.  Without loss of generality, we assume
that $n \geq k$ and that $k > 2$ (if $n < k$ then $\calS$ does not 
contain a cover of $B$, and if $k \leq 2$ then we can solve the problem
by brute force).
We build an election $E = (C,V)$ such
that the preferred candidate $p$ can become a $\copelandone$ winner of
$E$ by deleting at most $k$ voters if and only if $\calS$ contains a
$k$-element cover of $B$.

We let the candidate set $C$ be $\{p,r,b_1, \ldots, b_{3k}\}$ and
we let $V$ be the following collection of $4n-k$ voters:
\begin{enumerate}
  \item  We have $n-1$ voters with preference $ B > p > r$, 
  \item  we have $n - k + 1$ voters with preference $ p > r > B$, and
  \item  for each $S_i \in \calS$, we have two voters, $v_i$ and $v'_i$, such 
    that 
   \begin{enumerate}
     \item $v_i$ has preference $r > B - S_i > p > S_i$, and
     \item $v'_i$ has preference $r > S_i > p > B - S_i$.
   \end{enumerate}
\end{enumerate}
It is easy to see that for all $b_i \in B$,
$\versus_E(r,b_i) = 2n - k + 2$,
$\versus_E(b_i,p) = k - 2$, and
$\versus_E(r,p) = k$.

If $\calS$ contains a $k$-element cover of $B$, say $\{S_{a_1},
\ldots, S_{a_k}\}$, then we delete voters $v_{a_1}, \ldots, v_{a_k}$.
In the resulting election, $p$ ties every other candidate
in their head-to-head contests, and
thus $p$ is a winner.

For the converse, suppose that there is a subset $W$ of at most $k$
voters such that $p$ is a winner of $\widehat{E} = (C,V - W)$.
It is easy to see that $\copelandalphavarscore{1}_{\widehat{E}}(r) = 3k+1$.
Since $p$ is a winner of $\widehat{E}$,
$p$ must tie-or-defeat
every other candidate in their head-to-head contests.  By deleting at most
$k$ voters,  $p$ can at best tie $r$ in their head-to-head contest.
And $p$ will tie $r$ only if $\|W\| = k$ and every voter in 
$W$ prefers $r$ to $p$.  It follows that $W$ is a size
$k$ subset of $\{v_1,v'_1, \ldots, v_n,v'_n\}$.

Let $b_i \in B$.  Recall that $\versus_E(b_i,p) = k - 2$ and
that $p$ needs to at least tie $b_i$ in their head-to-head contest in
$\widehat{E}$.  Since $\|W\| = k$, it follows that $W$ can
contain at most one voter that prefers $p$ to $b_i$. Since $k > 2$,
it follows that $W$ contains only voters from the set
$\{v_1, \ldots, v_n\}$ and that the voters in $W$ correspond to a
$k$-element cover of $B$.

This completes the proof for the nonunique-winner constructive
case.
This proof also handles the unique-winner destructive case,
since $r$ is always a winner after deleting
at most $k$ voters from $E$ and $b_i$ is never a winner after
deleting at most $k$ voters from $E$. And so $r$ can be made to not uniquely
win by deleting at most $k$ voters if and only if $p$ can be 
made a winner by deleting at most $k$ voters.
\end{proofs}

\begin{theorem}
\label{thm:ucdv}
  Let $\alpha$ be a rational number such that $0 \leq \alpha \leq
  1$. $\copelandalpha$ is resistant to
  constructive control via deleting voters (CCDV) in the unique-winner
  model and to destructive control via deleting voters (DCDV) in the 
  nonunique-winner model,
  for both the rational and the irrational voter model.
\end{theorem}

\begin{proofs}
As in the proof of the previous theorem, 
let $(B,\calS)$ be an instance of X3C, where $B = \{b_1, \ldots,
b_{3k}\}$ and $\calS = \{S_1, \ldots, S_n\}$ is a finite family of
three-element subsets of $B$.  Without loss of generality, we assume
that $n \geq k$ and that $k > 2$ (if $n < k$ then $\calS$ does not 
contain a cover of $B$, and if $k \leq 2$ then we can solve the problem
by brute force).
We build an election $E = (C,V)$ such
that:
\begin{enumerate}
\item If $\calS$ contains a $k$-element cover of $B$, then
the preferred candidate $p$ can become the unique $\copelandalpha$
winner of $E$ by deleting at most $k$ voters, and
\item if $r$ can become a nonwinner by deleting at most $k$ voters,
then $\calS$ contains a $k$-element cover of $B$.
\end{enumerate}

We use the election from the proof of Theorem~\ref{thm:llull-cdv} with
one extra voter with preference $p > r > B$.  That is, we
let the candidate set $C$ be $\{p,r,b_1, \ldots, b_{3k}\}$ and
we let $V$ be the following collection of $4n-k+1$ voters:
\begin{enumerate}
  \item  We have $n-1$ voters with preference $ B > p > r$,
  \item  we have $n - k + 2$ voters with preference $ p > r > B$, and
  \item  for each $S_i \in \calS$ we have two voters, $v_i$ and $v'_i$, such 
    that 
   \begin{enumerate}
     \item $v_i$ has preference $r > B - S_i > p > S_i$, and
     \item $v'_i$ has preference $r > S_i > p > B - S_i$.
   \end{enumerate}
\end{enumerate}
It is easy to see that for all $b_i \in B$,
$\versus_E(r,b_i) = 2n - k + 3$,
$\versus_E(b_i,p) = k - 3$, and
$\versus_E(r,p) = k -1$.

If $\calS$ contains a $k$-element cover of $B$, say $\{S_{a_1},
\ldots, S_{a_k}\}$, then we delete voters $v_{a_1}, \ldots, v_{a_k}$.
In the resulting election, $p$ defeats every other candidate
in their head-to-head contests, and thus
$p$ is the unique winner.

To prove the second statement,
suppose that there is a subset $W$ of at most $k$
voters such that $r$ is not a winner of $\widehat{E} = (C,V - W)$.
Since $\versus_E(r,b_i) = 2n - k + 3$ and $n \geq k$, it
is immediate that $r$ defeats every $b_i \in B$ in their
head-to-head contests in $\widehat{E}$.
In order for $r$ not to be a winner of $\widehat{E}$,
$p$ must certainly defeat $r$ and tie-or-defeat every $b_i \in B$ in their
head-to-head contests.
But $p$ can defeat $r$ in their head-to-head contest
only if $\|W\| = k$ and every voter in 
$W$ prefers $r$ to $p$.  It follows that $W$ is a size-$k$
subset of $\{v_1,v'_1, \ldots, v_n,v'_n\}$.

Let $b_i \in B$.  Recall that $\versus_E(b_i,p) = k - 3$ and
that $p$ needs to at least tie $b_i$ in their head-to-head contest in
$\widehat{E}$.  Since $\|W\| = k$, it follows that $W$ can
contain at most one voter that prefers $p$ to $b_i$. Since $k > 2$,
it follows that $W$ contains only voters from the set
$\{v_1, \ldots, v_n\}$ and that the voters in $W$ correspond to a
$k$-element cover of $B$.~\end{proofs}

\begin{theorem}
\label{thm:copeland-cdv}
  Let $\alpha$ be a rational number such that $0 \leq \alpha < 1$.
  $\copelandalpha$ is resistant to constructive control via
  deleting voters (CCDV)  in the nonunique-winner model and to
  destructive control via deleting voters (DCDV) in the unique-winner
  model,
  for both the rational and the irrational voter model.
\end{theorem}

\begin{proofs}
Let $(B,\calS)$ be an instance of X3C, where $B = \{b_1, \ldots,
b_{3k}\}$ and $\calS = \{S_1, \ldots, S_n\}$ is a finite family of
three-element subsets of $B$.  Without loss of generality, we assume
that $n \geq k$ and that $k > 2$ (if $n < k$ then $\calS$ does not
contain a cover of $B$, and if $k \leq 2$ then we can solve the problem
by brute force).
We build an election $E = (C,V)$ such
that:
\begin{enumerate}
\item If $\calS$ contains a $k$-element cover of $B$, then
the preferred candidate $p$ can become a $\copelandalpha$
winner of $E$ by deleting at most $k$ voters, and
\item if $r$ can be made to not uniquely win the election
by deleting at most $k$ voters, then $\calS$ contains a $k$-element
cover of $B$.
\end{enumerate}

Our election is similar to the elections from the proofs of
Theorems~\ref{thm:llull-cdv}
and~\ref{thm:ucdv}.  To avoid problems when $\alpha = 0$,
we introduce a new candidate $\hat{r}$ to ensure that $p$ and $r$ are
the only possible winners after deleting at most $k$ candidates.
We let the candidate set $C$ be $\{p,r, \hat{r}, b_1, \ldots, b_{3k}\}$ and
we let $V$ be the following collection of $4n-k+2$ voters:

\begin{enumerate}
\item We have $n-2$ voters with preference
$B > p > r > \hat{r}$,
\item  we have $n - k + 2$ voters with preference
$p > r > \hat{r} > B$,
\item  for each $S_i \in \calS$, we have two voters, $v_i$ and $v'_i$, such
    that
   \begin{enumerate}
     \item $v_i$ has preference $r > \hat{r} >  B - S_i > p > S_i$, and
     \item $v'_i$ has preference $r > \hat{r} >  S_i > p > B - S_i$,
\end{enumerate}
\item we have one voter with preference
$ r > p > \hat{r} > B $, and
\item we have one voter with preference
$ B > p > r > \hat{r}$.
\end{enumerate}
It is easy to see that for all $b_i \in B$,
$\versus_E(r,b_i) =
\versus_E(\hat{r},b_i) = 2n - k + 4$,
$\versus_E(r,\hat{r}) = 4n - k + 2$,
$\versus_E(b_i,p) = k - 4$,
$\versus_E(r,p) = k$, and
$\versus_E(\hat{r},p) = k-2$.

If $\calS$ contains a $k$-element cover of $B$, say $\{S_{a_1},
\ldots, S_{a_k}\}$, then we delete voters $v_{a_1}, \ldots, v_{a_k}$.
In the resulting election, $p$ ties $r$ in their head-to-head contest
and $p$ defeats every other candidate in their head-to-head contests.
It follows that $p$ is a winner. 

To prove the second statement,
suppose that there is a subset $W$ of at most $k$
voters such that $r$ is not a unique winner of $\widehat{E} = (C,V - W)$.
It is easy to see that $r$ defeats every candidate
in  $\{\hat{r}, b_1, \ldots, b_{3k}\}$ in their head-to-head contests in 
$\widehat{E}$.
So it certainly cannot be the case that $r$ defeats $p$ in
their head-to-head contest
in $\widehat{E}$.  It follows that
$\|W\| = k$ and that every voter in $W$ prefers $r$ to $p$.
Note that both $r$ and $p$ defeat $\hat{r}$ in their
head-to-head contest in $\widehat{E}$ and that
both $r$ and $\hat{r}$ defeat every $b_i \in B$ in their
head-to-head contests in $\widehat{E}$.  It follows that the
only possible winners in $\widehat{E}$ are $r$ and $p$.
(Note that without $\hat{r}$, it would be possible that
after deleting $k$ voters, some $b_i$ defeats all candidates other than $r$ in
their head-to-head contests.  If $\alpha = 0$, this
could prevent $r$ from being the unique winner without necessarily
making $p$ a winner.)

Let $b_i \in B$.  Recall that $\versus_E(b_i,p) = k - 4$ and
that $p$ needs to defeat $b_i$ in their head-to-head contest in
$\widehat{E}$.  Since $\|W\| = k$, it follows that $W$ can
contain at most one voter that prefers $p$ to $b_i$. Since $k > 2$
and every voter in $W$ prefers $r$ to $p$,
it follows that $W$ contains only voters from the set
$\{v_1, \ldots, v_n\}$ and that the voters in $W$ correspond to a
$k$-element cover of $B$.~\end{proofs}

\begin{theorem}
\label{thm:cpv-tp}
  Let $\alpha$ be a rational number such that $0 \leq \alpha \leq 1$.
  $\copelandalpha$ is resistant to both constructive and destructive
  control via partitioning voters in the TP model (CCPV-TP and DCPV-TP),
  in both the nonunique-winner model and the unique-winner model,
  for both the rational and the irrational voter model.
\end{theorem}

\begin{proofs}
Let $(B,\calS)$ be an instance of X3C, where $B = \{b_1, \ldots,
b_{3k}\}$ and $\calS = \{S_1, \ldots, S_n\}$ is a finite family of
three-element subsets of $B$.  Without loss of generality, we assume
that $n \geq k$ and that $k > 2$ (if $n < k$ then $S$ does not 
contain a cover of $B$, and if $k \leq 2$ then we can solve the problem
by brute force).
We build an election $E = (C,V)$ such
that:
\begin{enumerate}
\item If $\calS$ contains a $k$-element cover of $B$, then
the preferred candidate $p$ can become the unique $\copelandalpha$
winner of $E$ via partitioning voters in the TP model, and
\item if $r$ can be made to not uniquely win $E$ via partitioning
voters in the TP model,
then $\calS$ contains a $k$-element cover of $B$.
\end{enumerate}
Note that this implies that 
$\copelandalpha$ is resistant to both constructive and destructive
control via partitioning voters in the TP model,
in both the nonunique-winner model and the unique-winner model.

Our construction is an extension of the construction from
Theorem~\ref{thm:ucdv}. 
We let the candidate set $C$ be $\{p,r,s,b_1, \ldots, b_{3k}\}$ and
we let $V$ be the following collection of voters:
\begin{enumerate}
  \item  We have $k + 1$ voters with preference $s > r > B > p$,
  \item  we have $n-1$ voters with preference $B > p > r > s$, 
  \item  we have $n - k + 2$ voters with preference $ p > r > B > s$, and
  \item  for each $S_i \in \calS$ we have two voters, $v_i$ and $v'_i$, such 
    that 
   \begin{enumerate}
     \item $v_i$ has preference $r > B - S_i > p > S_i > s$, and 
     \item $v'_i$ has preference $r > S_i > p > B - S_i > s$.
   \end{enumerate}
\end{enumerate}
Let $\widehat{V} \subseteq V$ be the collection of all the voters in $V$ except
for the $k+1$ voters with preference $s > r > B > p$.  Note that 
$\widehat{V}$ is exactly the voter collection used in the proof
of 
Theorem~\ref{thm:ucdv} with
candidate $s$ added as the least desirable candidate.
Since $s$ does not influence the differences
between the scores of the other candidates,
the following claim follows immediately
from the proof of Theorem~\ref{thm:ucdv}.

\begin{claim}
\label{cl:llull-dv}
If $r$ can become a nonwinner of $(C,\widehat{V})$
by deleting at most $k$ voters, then
$\calS$ contains a $k$-element cover of $B$.
\end{claim}

Recall that we need to prove that 
if $\calS$ contains a $k$-element cover of $B$, then
$p$ can be made the unique $\copelandalpha$
winner of $E$ via partitioning voters in the TP model, and
that if $r$ can be made to not uniquely win $E$ via partitioning
voters in the TP model, then $\calS$ contains a $k$-element cover of $B$.

If $\calS$ contains a $k$-element cover of $B$, say $\{S_{a_1},
\ldots, S_{a_k}\}$, then we let the second subelection consist
of the $k+1$ voters with preference $s > r > B > p$ and
voters $v_{a_1}, \ldots, v_{a_k}$.  Then $p$ is the unique
winner of the first subelection, $s$ is the unique winner
of the second subelection, and $p$ uniquely wins the final run-off
between $p$ and $s$.

To prove the second statement, suppose there is a partition of voters such 
that $r$ is not a unique winner of the resulting election in model TP\@. Note 
that in at least one of the subelections, without loss of generality say the second
subelection, a majority of the voters
prefers $r$ to all candidates in $\{p, b_1, \ldots, b_{3k}\}$.
Since $r$ is the unique winner of every run-off he or
she participates in, $r$ cannot be a winner of either
subelection.   Since $r$ defeats every candidate in $\{p, b_1, \ldots, b_{3k}\}$
in their head-to-head contests in the second subelection,
in order for $r$ not to be a winner of the second subelection,
it must certainly be the case that $s$ defeats $r$ in their
head-to-head contest in the second subelection.
This implies that
at most $k$ voters from $\widehat{V}$ can be part of the second subelection.

Now consider the first subelection.  Note that $r$ cannot be a winner
of the first subelection.  Then, clearly, $r$ cannot be a winner
of the first subelection restricted to voters in $\widehat{V}$.
By Claim~\ref{cl:llull-dv} it follows
that $\calS$ contains a $k$-element cover of $B$.~\end{proofs}

\begin{theorem}
\label{thm:cpv-te}
  Let $\alpha$ be a rational number such that $0 \leq \alpha < 1$.
  $\copelandalpha$ is resistant to constructive 
  control via partitioning voters in the TE model (CCPV-TE),
  in both the nonunique-winner model and the unique-winner model,
  for both the rational and the irrational voter model.
\end{theorem}

\begin{proofs}
We use the exact same construction as in the proof of 
Theorem~\ref{thm:cpv-tp}.  We will show that
if $\calS$ contains a $k$-element cover of $B$
then $p$ can be made the unique $\copelandalpha$
winner of $E$ via partitioning voters
in the TE model, and that if $p$ can be made a winner
by partitioning voters in the TE model then
$\calS$ contains a $k$-element cover of $B$.

If $\calS$ contains a $k$-element cover of $B$, say $\{S_{a_1},
\ldots, S_{a_k}\}$, then we let the second subelection consist
of the $k+1$ voters with preference $s > r > B > p$ and
voters $v_{a_1}, \ldots, v_{a_k}$.  Then $p$ is the unique
winner of the first subelection, $s$ is the unique winner
of the second subelection, and $p$ uniquely wins the final run-off
between $p$ and $s$.

To prove the second statement, suppose there is a partition of voters such 
that $p$ is a $\copelandalpha$ winner
of the resulting election in model TE\@. Note 
that in at least one of the subelections, without loss of generality say the second
subelection, a majority of the voters
prefers $r$ to all candidates in $\{p, b_1, \ldots, b_{3k}\}$.
Since $r$ is the unique winner of every run-off he or
she participates in, $r$ can certainly not be the unique winner of the
second subelection.
Since $r$ defeats every candidate in $\{p, b_1, \ldots, b_{3k}\}$
in their head-to-head contests in the second subelection,
and since $s$ does not influence the relative vote scores of
the candidates in  $\{p, r, b_1, \ldots, b_{3k}\}$,
no candidate in $\{p, b_1, \ldots, b_{3k}\}$ is a winner of the
second subelection.  It follows that $s$ is a winner of the second
subelection.  
If $s$ were to tie $r$ in their head-to-head contest in the second subelection,
then $s$ would tie all candidates in their head-to-head contests in the second
subelection, and $r$ would be the unique winner of the second subelection (since $\alpha < 1$).
It follows that $s$ defeats $r$
in their head-to-head contest in the second subelection.
This implies that
at most $k$ voters from $\widehat{V}$ can be part of the second subelection.

Now consider the first subelection. Note that $p$ must be the unique winner
of the first subelection.  So, certainly, 
$r$ cannot be a winner
of the first subelection.  Then, clearly, $r$ cannot be a winner
of the first subelection restricted to voters in $\widehat{V}$.
By Claim~\ref{cl:llull-dv} it follows
that $\calS$ contains a $k$-element cover of $B$.~\end{proofs}

\begin{theorem}
\label{thm:cpv-one-con}
  $\copelandone$ is resistant to constructive
  control via partitioning voters in the TE model (CCPV-TE),
  in both the nonunique-winner model and the unique-winner model,
  for both the rational and the irrational voter model.
\end{theorem}
\begin{proofs}
We use the same construction as in the proof of 
Theorem~\ref{thm:cpv-tp}, except that
we have one fewer voter with preference $s > r > B > p$.
We will show that if $\calS$ contains a $k$-element cover of $B$
then $p$ can be made the unique $\copelandone$
winner of $E$ via partitioning voters
in the TE model,
and that if $p$ can be made a winner by
partitioning
voters in the TE model then
$\calS$ contains a $k$-element cover of $B$.

If $\calS$ contains a $k$-element cover of $B$, say $\{S_{a_1},
\ldots, S_{a_k}\}$, then we let the second subelection consist
of the $k$ voters with preference $s > r > B > p$ and
voters $v_{a_1}, \ldots, v_{a_k}$.  Then $p$ is the unique
winner of the first subelection and proceeds to the run-off,
and $r$ and $s$ are winners of the second subelection, and 
so no candidate from the second election proceeds to the run-off.
It follows that $p$ is the only candidate participating in the final
run-off, and so $p$ is the unique winner of the election.

To prove the second statement, suppose there is a partition of voters such 
that $p$ is a $\copelandone$ winner of the resulting election in model TE\@. Note 
that in at least one of the subelections, without loss of generality say the second
subelection, a majority of the voters
prefers $r$ to all candidates in $\{p, b_1, \ldots, b_{3k}\}$.
Since $r$ is the unique winner of every run-off he or
she participates in, $r$ should not participate in the final
run-off.  In particular, $r$ cannot be the unique winner of 
the second subelection.  The only way to avoid this is
if $r$ does not defeat $s$ in their head-to-head contest in the
second subelection.
This implies that
at most $k$ voters from $\widehat{V}$ can be part of the second subelection.

Now consider the first subelection. Note that 
$p$ must be the unique winner of the first subelection. So, certainly,
$r$ cannot be a winner
of this subelection.
Then, clearly, $r$ cannot be a winner of this subelection
restricted to voters in $\widehat{V}$.
By Claim~\ref{cl:llull-dv} it follows
that $\calS$ contains a $k$-element cover of $B$.
\end{proofs}

\begin{theorem}
\label{thm:cpv-one-dest}
$\copelandone$ is resistant to destructive control via partitioning voters
in the TE model (DCPV-TE), in both the nonunique-winner model and the unique winner model,
for both the rational and the irrational voter model.
\end{theorem}
\begin{proofs}
  We use the same construction as in the proof of
  Theorem~\ref{thm:cpv-one-con}, except that we have one fewer voter
  with preference $p > r > B > s$.  We will show that if $\calS$
  contains a $k$-element cover of $B$ then $r$ can become a
  nonwinner of $E$ via partitioning of voters in the TE model, and
  that if $r$ can be made to not uniquely win $E$ via partitioning of
  voters in the TE model then $\calS$ contains a $k$-element cover of
  $B$. 

  Let $\widehat{V} \subset V$ be the collection of all voters except for the $k$
  voters with preference $s > r > B > p$. Note that $\widehat{V}$ is exactly
  the voter collection used in the proof of
  Theorem~\ref{thm:llull-cdv} with candidate
  $s$ added as the least desirable candidate. Since $s$ does not influence the
  differences between the scores of the other candidates, the following 
  claim follows immediately from the proof of Theorem~\ref{thm:llull-cdv}. 

  \begin{claim}
  \label{thm:cpv-one-dest-claim}
  If $r$ can be made to not uniquely win $(C,\widehat{V})$ by deleting at most $k$
  voters, then $\calS$ contains a $k$-element cover of $B$.
  \end{claim}

  If $\calS$ contains a $k$-element cover of $B$, say $\{S_{a_1}, \ldots, S_{a_k}\}$,
  then we let the second subelection consist of the $k$ voters with preference
  $s > r > B > p$ and voters $v_{a_1},\ldots, v_{a_k}$. Then $p$ is a winner of
  the first subelection, $s$ is a winner of the second subelection, and it
  follows that $r$ does not participate in the run-off. 

  For the second statement, suppose there is a partition of voters such that
  $r$ is not a unique winner of the resulting election in model TE\@. Since $r$
  uniquely wins any run-off he or she participates in, it follows that 
  $r$ does not uniquely win either subelection. 
  Note that in at least one of the subelections, without loss of generality say the second subelection,
  a majority of the voters prefers $r$ to all candidates in
  $\{p, b_1, \ldots, b_{3k}\}$.
  It follows that $r$ cannot defeat $s$ in their head-to-head contest in the
  second subelection. This implies that at most $k$ voters from $\widehat{V}$ can be
  part of the second subelection.

  Now consider the first subelection. Note that $r$ cannot be a unique
  winner of the first subelection. Then, clearly, $r$ cannot be a
  unique winner of the first subelection restricted to voters in
  $\widehat{V}$. By Claim~\ref{thm:cpv-one-dest-claim}, it follows that $\calS$ 
  contains a $k$-element cover of $B$.
\end{proofs}

\begin{theorem}
\label{thm:cpv-lessone}
  Let $\alpha$ be a rational number such that $0 \leq \alpha < 1$.
  $\copelandalpha$ is resistant to destructive
  control via partitioning voters in the TE model (DCPV-TE),
  in both the nonunique-winner model and the unique-winner model,
  for both the rational and the irrational voter model.
\end{theorem}

\begin{proofs}
Let $(B,\calS)$ be an instance of X3C, where $B = \{b_1, \ldots,
b_{3k}\}$ and $\calS = \{S_1, \ldots, S_n\}$ is a finite family of
three-element subsets of $B$.  Without loss of generality, we assume
that $n \geq k$ and that $k > 2$ (if $n < k$ then $S$ does not 
contain a cover of $B$, and if $k \leq 2$ then we can solve the problem
by brute force).
We build an election $E = (C,V)$ such
that:
\begin{enumerate}
\item If $\calS$ contains a $k$-element cover of $B$, then
$r$ can become a nonwinner
of $E$ via partitioning voters in the TE model, and
\item if $r$ can be made to not uniquely win $E$ via partitioning
voters in the TE model,
then $\calS$ contains a $k$-element cover of $B$.
\end{enumerate}
Note that this implies that 
$\copelandalpha$ is resistant to destructive
control via partitioning voters in the TE model,
in both the nonunique-winner model and the unique-winner model.

In the proof of Theorem~\ref{thm:cpv-tp},
we extended the construction from the proof of Theorem~\ref{thm:ucdv}. 
In the proof of the present theorem, we extend the construction
from Theorem~\ref{thm:copeland-cdv} in the same way. 

We let the candidate set $C$ be $\{p,r,\hat{r},s,b_1, \ldots, b_{3k}\}$ and
we let $V$ be the following collection of voters:
\begin{enumerate}
  \item  We have $k + 1$ voters with preference $s > r > \hat{r} > B > p$,
\item we have $n-2$ voters with preference
$B > p > r > \hat{r} > s$,
\item  we have $n - k + 2$ voters with preference
$p > r > \hat{r} > B > s$,
\item  for each $S_i \in \calS$, we have two voters, $v_i$ and $v'_i$, such
    that
   \begin{enumerate}
     \item $v_i$ has preference $r > \hat{r} >  B - S_i > p > S_i > s$, and
     \item $v'_i$ has preference $r > \hat{r} >  S_i > p > B - S_i > s$,
\end{enumerate}
\item we have one voter with preference
$ r > p > \hat{r} > B > s $, and
\item we have one voter with preference
$ B > p > r > \hat{r} > s$.
\end{enumerate}
Let $\widehat{V} \subseteq V$ be the collection of all the voters in $V$ except
for the $k+1$ voters with preference $s > r > \hat{r} > B > p$.  Note that 
$\widehat{V}$ is exactly the voter collection used in the proof
of 
Theorem~\ref{thm:copeland-cdv} with
candidate $s$ added as the least desirable candidate.
Since $s$ does not influence the differences
between the scores of the other candidates,
the following claim follows immediately
from the proof of Theorem~\ref{thm:copeland-cdv}.

\begin{claim}
\label{cl:copeland-dv}
If $r$ can be made to not uniquely win $(C,\widehat{V})$
by deleting at most $k$ voters, then
$\calS$ contains a $k$-element cover of $B$.
\end{claim}

If $\calS$ contains a $k$-element cover of $B$, say $\{S_{a_1},
\ldots, S_{a_k}\}$, then we let the second subelection consist
of the $k+1$ voters with preference $s > r > \hat{r} >  B > p$ and
voters $v_{a_1}, \ldots, v_{a_k}$. Then $p$ is a 
winner of the first subelection, $s$ is the unique winner
of the second subelection, and it follows that
$r$ does not participate in the run-off.

For the second statement, suppose there is a partition of voters such 
that $r$ is not a unique winner of the resulting election in model TE\@. Since 
$r$ uniquely wins any run-off he or she participates in,
it follows that $r$ does
not uniquely win either subelection.
Note that in at least one of the subelections, without loss of generality say the second
subelection, a majority of the voters
prefers $r$ to $\hat{r}$ and both
$r$ and $\hat{r}$ to all candidates in $\{p, b_1, \ldots, b_{3k}\}$.
If $s$ were to tie $r$ in their head-to-head contest in
the second subelection, then $s$ would tie all candidates
in the second subelection in their head-to-head contests, and $r$ would be the unique winner
of the second subelection.  It follows that $s$ 
defeats $r$ in their
head-to-head contest in the second subelection.
This implies that
at most $k$ voters from $\widehat{V}$ can be part of the second subelection.

Now consider the first subelection.  Note that $r$ cannot be the unique winner
of the first subelection.  Then, clearly, $r$ cannot be the unique
winner of the first subelection restricted to voters in $\widehat{V}$.
By Claim~\ref{cl:copeland-dv} it follows
that $\calS$ contains a $k$-element cover of $B$.
\end{proofs}

\subsection{FPT Algorithm Schemes for Bounded-Case Control}\label{sec:control-fpt}

\newcommand{\fpt}{\ensuremath{\mathrm{FPT}}}
\newcommand{\bigo}{{\protect\cal O}}
\newcommand{\bcj}{\mbox{\rm{}BC$_{j}$}}
\newcommand{\bvj}{\mbox{\rm{}BV$_{j}$}}
\newcommand{\card}[1]{{ \mathopen\parallel {#1} \mathclose\parallel }}

Resistance to control is generally viewed as a desirable property in
system design.  However, suppose one is trying
to solve resistant control problems.  Is there any hope?  

In their seminal paper on NP-hard winner-determination problems,
Bartholdi, Tovey, and Trick~\cite{bar-tov-tri:j:who-won} suggested
considering hard election problems for the cases of a bounded number
of candidates or a bounded number of voters, and they obtained
efficient-algorithm results for such cases.  Within the study of
elections, this same approach---seeking efficient
fixed-parameter algorithms---has, for example,
also been used (although somewhat tacitly---see 
the coming discussion in the second paragraph of 
Footnote~\ref{f:not-family}) within the study of
bribery~\cite{fal-hem-hem:c:bribery,fal-hem-hem:tRevisedOutByConfToAppear:bribery}.
To the best of our knowledge, this bounded-case approach to 
finding the limits of resistance results has not been previously
used to study control problems.  In this section
we do precisely that.

In particular, we obtain 
for
resistant-in-general control problems
a broad range of efficient algorithms 
for the case when the number of
candidates or voters is bounded.  
Our algorithms are not merely polynomial time. 
Rather, 
we give algorithms that prove membership in
FPT (fixed-parameter
tractability, i.e., the problem is not merely individually in
P for each fixed value of the
parameter of interest (voters or candidates),
but indeed has a single P algorithm having degree that is 
bounded independently of 
the value of the fixed number of voters or candidates) when the
number of candidates is bounded, and also when the number of voters is
bounded.    And we prove that 
our FPT claims hold even under the succinct input model---in
which the voters are input 
via ``(preference-list,
binary-integer-giving-frequency-of-that-preference-list)'' 
pairs---and even in the case of irrational voters.

We obtain such algorithms
for all the voter-control cases,
both for bounded candidates and for bounded voters, and for all the
candidate-control cases with bounded candidates.  On the other hand,
we show that for the resistant-in-general
irrational-voter, candidate-control cases,
resistance still holds even if the number of voters is limited to
being at most two.

We structure this section as follows.  We first start by briefly
stating our notions and  notations.  We next state, and then
prove, our fixed-parameter tractability results.  
Regarding those, we first address FPT results for the (standard) 
constructive and destructive cases.  We then show that 
in many cases we can assert FPT results 
that are more general still---in particular, we
will look at ``extended control'': 
completely pinpointing
whether 
under a given type of control 
we can ensure that at least one of a specified collection of 
``Copeland Outcome Tables'' (to be defined later) can be
obtained.
Finally, we give our
resistance results.

\subsubsection*{Notions and Notations}

The study of fixed-parameter complexity 
(see, e.g.,~\cite{nie:b:invitation-fpt})
has been expanding explosively
since it was parented as a field by Downey, Fellows, and others in the
late 1980s and the 1990s.  Although the area has built a rich variety
of complexity classes regarding parameterized problems, for the
purpose of the current paper we need focus only on one very
important class, namely, the
class $\fpt$.  Briefly put, a problem parameterized by some value $j$
is said to be \emph{fixed-parameter tractable} (equivalently, to
belong to the class $\fpt$) 
if there is an algorithm for the problem whose running time is
$f(j)n^{O(1)}$.  (Note in particular that there is some particular constant 
for the ``big-oh'' that holds for all inputs, regardless of 
what $j$ value the particular input has.)

In our context, we will consider two parameterizations: bounding the
number of candidates and bounding the number of voters.  We will use
the same notations used throughout this paper to describe problems,
except we will postpend a ``-BV${}_j$'' to a problem name to state
that the number of voters may be at most $j$, and we will postpend a
``-BC${}_j$'' to a problem name to state that the number of candidates
may be at most $j$.  In each case, the bound applies to the full
number of such items involved in the problem.  For example, in the
case of control by adding voters, the $j$ must bound the total of the
number of voters in the election added together with the number of
voters in the pool of voters available for adding.

Typically, we have been viewing input votes as coming in each on a
ballot.  However, one can also consider the case of succinct inputs,
in which our algorithm is given the votes as ``(preference-list,
binary-integer-giving-frequency-of-that-preference-list)'' pairs.
(We mention in passing that for the ``adding voter'' cases, 
when we speak of succinctness we require that not just the 
always-voting voters be specified succinctly but also that 
the pool of voters-available-to-be-added be specified succinctly.)
Succinct inputs have been studied extensively in the case of
bribery~\cite{fal-hem-hem:c:bribery,fal-hem-hem:tRevisedOutByConfToAppear:bribery},
and speaking more broadly, 
succinctness-of-input issues are often very germane to
complexity classification (see, e.g.,~\cite{wag:j:succinct}).
Note that proving an FPT result for the succinct case of a problem
immediately implies an FPT result for the same problem (without the
requirement of succinct inputs being in place), and indeed is a 
stronger result, since succinctness can potentially exponentially
compress the input.

Finally, we would like to be able to concisely express many results in
a single statement.  To do so, we borrow a notational approach from
transformational grammar, and use square brackets as an ``independent
choice'' notation.  So, for example, the claim
\mbox{$\scriptsize
\begin{bmatrix} \textrm{\rm{}It} \\ \textrm{\rm{}She} \\ \textrm{\rm{}He} \end{bmatrix}
\begin{bmatrix} \textrm{\rm{}runs} \\ \textrm{\rm{}walks} \end{bmatrix}
$} %
is a shorthand for six assertions: It runs; She runs; He runs;
It
walks; She walks; and He 
walks.  
A special case is the symbol
``$\emptyset$'' which, when it appears in such a bracket, means that
when unwound it should be viewed as no text at all.  For example,
``$\bigl[\begin{smallmatrix} \textrm{\rm{}Succinct} \\ \emptyset
\end{smallmatrix}\bigr]$ Copeland is fun'' asserts both ``Succinct 
Copeland is fun'' and ``Copeland is fun.''

\subsubsection*{Fixed-Parameter Tractability Results}

We immediately state our main results, which show that for all the
voter-control cases FPT schemes hold for both the bounded-voter and
bounded-candidate cases, and for all the candidate-control cases FPT
schemes hold for the bounded-candidate cases.

\begin{theorem}\label{t:v}
For each rational $\alpha$, $0\leq \alpha \leq 1$, and each 
choice from the independent choice brackets below, 
the specified parameterized (as $j$ varies over $\naturals$) 
problem 
is in $\fpt$:
\begin{center}
$\begin{bmatrix}
\textrm{\rm{}succinct}\\\textrm{\rm{}$\emptyset$}
\end{bmatrix}
\hbox{-}
\begin{bmatrix}
\textrm{\rm{}\copelandalpha}\\\textrm{\rm{}\copelandalphairrational}
\end{bmatrix}
\hbox{-}
\begin{bmatrix}
\textrm{\rm{}C}\\\textrm{\rm{}D}
\end{bmatrix}
\hbox{\rm{}C}
\begin{bmatrix}
\textrm{\rm{}AV}\\
\textrm{\rm{}DV}\\
\textrm{\rm{}PV-TE}\\
\textrm{\rm{}PV-TP}
\end{bmatrix}
\hbox{-}
\begin{bmatrix}
\textrm{\rm{}\bvj}\\\textrm{\rm{}\bcj}
\end{bmatrix}$.
\end{center}
\end{theorem}

\begin{theorem}\label{t:c-bc}
For each rational $\alpha$, $0\leq \alpha \leq 1$, and each 
choice from the independent choice brackets below, the specified
parameterized (as $j$ varies over $\naturals$) 
problem 
is in $\fpt$:
\begin{center}
$
\begin{bmatrix}
\textrm{\rm{}succinct}\\\textrm{\rm{}$\emptyset$}
\end{bmatrix}
\hbox{-}
\begin{bmatrix}
\textrm{\rm{}\copelandalpha}\\\textrm{\rm{}\copelandalphairrational}
\end{bmatrix}
\hbox{-}
\begin{bmatrix}
\textrm{\rm{}C}\\\textrm{\rm{}D}
\end{bmatrix}
\hbox{\rm{}C}
\begin{bmatrix}
\textrm{\rm{}AC${}_{\rm u}$}\\
\textrm{\rm{}AC}\\
\textrm{\rm{}DC}\\
\textrm{\rm{}PC-TE}\\
\textrm{\rm{}PC-TP}\\
\textrm{\rm{}RPC-TE}\\
\textrm{\rm{}RPC-TP}
\end{bmatrix}
\hbox{-$\bcj$}$.
\end{center}
\end{theorem}

Readers not interested in a discussion of those results and their
proofs
can at this point safely skip to the next 
labeled section header.

Before proving the above theorems, let us first make a few
observations about them.  First, for cases where under a particular
set of choices that same case is known (e.g., due to 
the results of 
Sections~\ref{sec:control-candidate} and~\ref{sec:control-voter}) 
to be in P even for the unbounded case, the above results
are uninteresting as they follow from the earlier results (such
cases
do not include any of the ``succinct'' cases, since those 
were not treated earlier).  However,
that is a small minority of the cases.  Also, for clarity as to what
cases are covered, we have included some items that are not formally
needed.  For example, since FPT for the succinct case implies FPT for
the no-succinctness-restriction case, and since FPT for the
irrationality-allowed case implies FPT for the rational-only case, the
first two choice brackets in each of the theorems could, 
without decreasing the results' strength, be removed by
eliminating their ``$\emptyset$'' and ``$\copelandalpha$'' choices.

We now turn to the proofs.  
Since proving every case would be uninterestingly repetitive, we will
at times 
(after carefully warning the reader) 
prove the cases of one or two control types when that is enough to
make clear how the omitted cases' proofs go.

Let us start with those cases that can be done simply by appropriately
applied brute force.  

We first prove Theorem~\ref{t:c-bc}.  

\medskip

\sproofof{Theorem~\ref{t:c-bc}}
If we are limited to having at most $j$ candidates, 
then for each of the cases mentioned, the total number of ways 
of adding/deleting/partitioning candidates is simply a (large)
constant.    For example, there will be at most (``at most'' rather
than ``exactly'' since $j$ is merely an upper bound on
the number of candidates) $2^j$ possible run-off partitions 
and there will be at most 
$2^{j-1}$ 
relevant
ways of deleting
candidates (since we can't (destructive case) or 
would never (constructive case) delete the distinguished 
candidate).  So we can brute-force try all ways of 
adding/deleting/partitioning candidates, 
and for each such way can see whether 
we get the desired outcome.  This works in polynomial time 
(with a fixed
degree independent of $j$ and $\alpha$) even in the succinct 
case, and even with irrationality allowed.~\eproofof{Theorem~\ref{t:c-bc}}

A brute-force approach similarly works for the case of voter control
when the number of voters is fixed.  In particular, we prove the following
subcase of Theorem~\ref{t:v}.
\begin{lemma}\label{l:short}
For each rational $\alpha$, $0\leq \alpha \leq 1$, and each 
choice from the independent choice brackets below, the 
specified 
parameterized (as $j$ varies over $\naturals$) 
problem 
is in $\fpt$:
\begin{center}
$\begin{bmatrix}
\textrm{\rm{}succinct}\\\textrm{\rm{}$\emptyset$}
\end{bmatrix}
\hbox{-}
\begin{bmatrix}
\textrm{\rm{}\copelandalpha}\\\textrm{\rm{}\copelandalphairrational}
\end{bmatrix}
\hbox{-}
\begin{bmatrix}
\textrm{\rm{}C}\\\textrm{\rm{}D}
\end{bmatrix}
\hbox{\rm{}C}
\begin{bmatrix}
\textrm{\rm{}AV}\\
\textrm{\rm{}DV}\\
\textrm{\rm{}PV-TE}\\
\textrm{\rm{}PV-TP}
\end{bmatrix}
\hbox{-\textrm{\rm{}\bvj}}$.
\end{center}
\end{lemma}

When considering ``$\bvj$'' cases---namely in this proof and in the
resistance section starting on page~\pageref{ss-star:resistance}---we
will not 
even discuss succinctness.  
The reason is that if the number of voters is bounded, say by
$j$, then succinctness doesn't asymptotically change the input sizes
interestingly, since succinctness at very best would compress the vote
description by a factor of about $j$---which in this case is a fixed
constant (relative to the value of the parameterization, which itself
is $j$).

\medskip

\sproofof{Lemma~\ref{l:short}}
If we are limited to having at most $j$ voters, note that we can, for
each of these four types of control, brute-force check all possible
approaches to that type of control.  For example, for the case of
control by deleting voters, 
we clearly
have no more than $2^j$ possible vote deletion choices, and for the
case of control 
by partitioning of voters, we again have at most $2^j$ partitions
(into $V_1$ and $V-V_1$) to consider.  And $2^j$ is just a (large)
constant.  So a direct brute-force check yields a polynomial-time
algorithm, and by inspection one can see that its run-time's degree is
bounded above independently of $j$.~\eproofof{Lemma~\ref{l:short}}

We now come to the interesting cluster of $\fpt$ cases: the
voter-control cases when the number of candidates is bounded.  Now, at
first, one might think that we can handle this, just as the above
cases, via a brute-force approach.  And that is almost
correct: One can get polynomial-time algorithms for these
cases via a brute-force approach.  However, for the succinct cases, the
degrees of these algorithms will be huge, and \emph{will not be
  independent of the bound, $j$, on the number of candidates}.  For
example, even in the rational case, one would from this approach
obtain run-times with terms such as $n^{\card{C}!}$.  
That is, one
would obtain a family of P-time algorithms, but one 
would not have an FPT algorithm.

To overcome this obstacle, we will employ
Lenstra's~\cite{len:j:integer-fixed}
algorithm for bounded-variable-cardinality integer programming.
Although Lenstra's algorithm is truly amazing in its power, even it
will not be enough to accomplish our goal.  Rather, we will use a
scheme that involves a fixed (though very large) number of Lenstra-type
programs each being focused on a different resolution path
regarding the 
given
problem.  

What we need to prove, to complete the proof of 
Theorem~\ref{t:v}, is the following lemma.

\begin{lemma}\label{t:leftover}
For each rational $\alpha$, $0\leq \alpha \leq 1$, and each 
choice from the independent choice brackets below, the 
specified 
parameterized (as $j$ varies over $\naturals$) 
problem 
is in $\fpt$:
\begin{center}
$\begin{bmatrix}
\textrm{\rm{}succinct}\\\textrm{\rm{}$\emptyset$}
\end{bmatrix}
\hbox{-}
\begin{bmatrix}
\textrm{\rm{}\copelandalpha}\\\textrm{\rm{}\copelandalphairrational}
\end{bmatrix}
\hbox{-}
\begin{bmatrix}
\textrm{\rm{}C}\\\textrm{\rm{}D}
\end{bmatrix}
\hbox{\rm{}C}
\begin{bmatrix}
\textrm{\rm{}AV}\\
\textrm{\rm{}DV}\\
\textrm{\rm{}PV-TE}\\
\textrm{\rm{}PV-TP}
\end{bmatrix}
\hbox{-\textrm{\rm{}\bcj}}$.
\end{center}
\end{lemma}

Let us start by recalling that, regarding the first choice bracket, the
``succinct'' case implies the ``$\emptyset$'' case, so we need only
address the succinct case.  Recall also that, regarding the second
choice bracket, for each rational $\alpha$, $0 \leq \alpha \leq 1$,
the ``\copelandalphairrational'' case implies the ``\copelandalpha''
case, so we need only address the $\copelandalphairrational$ case.

So all that remains is to handle each pair of choices from the third
and forth choice brackets.  To prove every case would be very
repetitive.  So we will simply prove in detail a difficult,
relatively representative
case, and then will for the other cases either 
mention the
type of adjustment needed to obtain their proofs, or will simply leave
it 
as a simple but tedious exercise that will be clear, as
to how to do, to anyone who reads this section.

So, in particular, let us prove the following result.

\begin{lemma}\label{t:first-example-case}
For each rational $\alpha$, $0\leq \alpha \leq 1$, 
the following 
parameterized (as $j$ varies over $\naturals$) 
problem 
is in $\fpt$:
$
\textrm{\rm{}succinct}
\hbox{-}
\textrm{\rm{}\copelandalphairrational}
\hbox{-}
\textrm{\rm{}C}
\hbox{\rm{}C}
\textrm{\rm{}PV-TP}
\hbox{-\textrm{\rm{}\bcj}}$.
\end{lemma}

\begin{proofs}
Let $\alpha$, $0\leq \alpha \leq 1$, be some arbitrary, fixed
rational number.  In particular, suppose that $\alpha$ can be
expressed as $b/d$, where $b \in \naturals$, $d \in \naturals^+$, $b$
and $d$ share no common integer divisor greater than 1, and if $b=0$
then $d=1$.  We won't explicitly invoke $b$ and $d$ in our algorithm,
but each time we speak of evaluating a certain set of pairwise
outcomes ``with respect to $\alpha$,'' one can think of it as
evaluating that with respect to a strict pairwise win giving $d$
points, a pairwise tie giving $b$ points, and a strict pairwise loss
giving $0$ points.

We need a method of specifying the pairwise outcomes among a set
of candidates.  To do this, we will use the notion of a \emph{Copeland
outcome table} over a set of candidates.  This will not actually be
a table, but rather will be a function (a symmetric one---it will not
be affected by the order of its two arguments) that, when given a pair
of distinct candidates as inputs, will say which of the three possible
outcomes allegedly happened: 
Either there is a tie, or one candidate won, or the
other candidate won. Note that a COT is simply a representation
of an election graph (see Section~\ref{sect:constructing}).  So, in a $j$-candidate election, there are
exactly $3^{\binom{j}{2}}$ such functions.  (We will not care about
the names of the candidates, and so will assume that the tables simply
use the names $1$ through $j$, and that we match the names of the
actual candidates with those integers by linking them
lexicographically, i.e., the lexicographically first candidate
will be associated with the integer 1 and so on.)  Let us call
a $j$-candidates Copeland outcome table a $j$-COT.

We need to build our algorithm 
that shows that 
the problem 
$
\textrm{\rm{}succinct}
\hbox{-}
\textrm{\rm{}\copelandalphairrational}
\hbox{-}
\textrm{\rm{}C}
\hbox{\rm{}C}
\textrm{\rm{}PV-TP}
\hbox{-\textrm{\rm{}\bcj}}$, $j\in\naturals$,
is in $\fpt$.  So, let $j$ be some fixed integer bound on the number
of candidates.\footnote{\label{f:not-family}We will now seem to
specify the algorithm merely for this bound.  However, it 
is important to note that we do enough to 
establish that there 
exists a single 
algorithm that fulfills the requirements
of the definition of $\fpt$.  In particular, the specification
we are about to give is sufficiently uniform that one can 
simply consider 
a single algorithm that, on a given input, notes the value of 
$j$, the number of candidates, and then does what the ``$j$''
algorithm we are about to specify does.

We take this moment to mention in passing that our earlier work,
\cite{fal-hem-hem:c:bribery}
and (this is an expanded, full version of that)
\cite{fal-hem-hem:tRevisedOutByConfToAppear:bribery},
that gives P-time algorithms for the fixed parameter (fixed candidate
and fixed voters) cases in fact, in all such claims we 
have in that work, implicitly is giving
FPT algorithms, even though those papers don't explicitly note 
that.  The reason is generally the same as why that is true
in this paper---namely, the Lenstra technique is not just 
powerful but is also ideally suited for 
FPT algorithms and for being used inside algorithms that 
are FPT algorithms.  Most interestingly, the Lenstra approach
tends to work even on succinct inputs, and so the FPT comment we 
made applies even to those results in our abovementioned earlier papers 
that are about the succinct-inputs case of fixed-number-of-candidates and 
fixed-number-of-voters claims.  (The fixed-number-of-candidates and 
fixed-number-of-voters
Dodgson winner/score
work 
of Bartholdi, Tovey, and Trick~\cite{bar-tov-tri:j:who-won}
is
known to be about FPT algorithms (see~\cite{bet-guo-nie:c:dodgson-parametrized}).  
Although the 
paper of
Bartholdi, Tovey, and Trick~\cite{bar-tov-tri:j:who-won}
doesn't address the succinct input model, 
\cite{fal-hem-hem:c:bribery}
notes that their approach works fine even in the succinct
cases of the winner problem.  That is true
not just for the P-ness of their algorithms even in the
succinct case, but also for 
the FPT-ness of their algorithms even in the succinct case.)}

\begin{figure}[!htpb]
\small
For each $j'$-COT, $T_1$, \\
For each $j'$-COT, $T_2$, \\
Do
\begin{quote}
If
\begin{quote}
 when we have a $\copelandalphairrational$ election (involving all the
  input voters), with respect to $\alpha$, between all the candidates
  who win under $T_1$ with respect to $\alpha$, and all the candidates 
  who win under $T_2$ with respect to $\alpha$, the preferred candidate of the 
  input problem is a winner,
\end{quote}
then
\begin{quote}
  create and run 
  the integer 
  linear program constraint feasibility problem that checks
  whether there exists a partition of the voters such that 
  the first subelection has $j'$-COT $T_1$ and the second subelection
  has $j'$-COT $T_2$, and if so, then accept. 
\end{quote}
  \end{quote}
\caption{\label{f:top-level-code}The top-level code for the case
  $ \textrm{\rm{}succinct} \hbox{-} \textrm{\rm{}\copelandalphairrational}
  \hbox{-} \textrm{\rm{}C} \hbox{C} \textrm{\rm{}PV-TP} \hbox{-\textrm{\rm{}\bcj}}$.}
\end{figure}

Let us suppose we are given an input instance.  Let $j' \leq j$ be the
number of candidates in this instance (recall that $j$ is not the
number of candidates, but rather is an upper bound on the number of
candidates).

The top level of our algorithm is specified by the pseudocode in
Figure~\ref{f:top-level-code}.  (Although this algorithm seemingly is
just trying to tell whether the given control is possible for the
given case, rather than telling how to partition to achieve that control, note
that which iteration through the double loop accepts and the precise
values of the variables inside the integer linear program constraint
feasibility problem that made that iteration be satisfied will in fact
tell us precisely what the partition is that makes the preferred
candidate win.)

Now, note that the total number of $j'$-COTs that exist (we do not
need to care whether all can be realized via actual votes) is
$3^{\binom{j'}{2}}$.  So the code inside the two loops executes at
most $9^{\binom{j'}{2}}$ times, which is constant-bounded since $j'
\leq j$, and we have fixed $j$.

So all that remains is to give the 
integer linear program 
constraint
feasibility problem mentioned inside the inner loop.  
The setting here can sometimes be confusing, e.g.,
when we speak of constants that can grow without limit.
It is important to keep in mind that in this integer linear program
constraint feasibility problem, the number of variables and
constraints is constant (over all inputs), and the integer linear
program constraint feasibility problem's ``constants'' are the only
things that change with respect to the input. 
This is the 
framework that allows us to invoke Lenstra's powerful algorithm.  

We first specify the set of constants of the 
integer linear program
constraint
feasibility problem. 
In particular, 
for each $i$, $1 \leq i \leq 2^{\binom{j'}{2}}$, we
will have a constant, $n_i$, that is the number of input voters whose
vote is of the $i$th type (among the $2^{\binom{j'}{2}}$ possible vote
possibilities; keep in mind that voters are allowed to be irrational,
thus the value $2^{\binom{j'}{2}}$ is correct).  Note that the number
of these constants that we have is itself constant-bounded (for fixed
$j$), though of course the values that these constants 
(of the integer linear program
constraint
feasibility problem)
take on can grow
without limit.

In addition, let us define 
some constants that will not vary with the input but
rather are simply a notational shorthand that we will use to describe
how the integer linear program constraint feasibility problem is
defined (what constraints occur in it).  In particular, 
for each $i$ and $\ell$ such that 
$1 \leq i \leq j'$,
$1 \leq \ell \leq j'$, and $ i \neq \ell$, let $val1_{i,\ell}$
be $1$ if $T_1$ asserts that
(in their head-to-head contest)
$i$ ties or defeats $\ell$, 
and let it be
$0$ if $T_1$ asserts that
(in their head-to-head contest) $i$ loses to $\ell$.
Let $val2_{i,\ell}$ be identically defined, except 
with respect to $T_2$.
Informally put, these values will be used to let 
our integer linear program constraint feasibility problem
seek to enforce such a win/loss/tie pattern 
with respect to the given input vote numbers
and the given type of allowed control action.

The integer linear program constraint feasibility problem's variables,
which of course are all \emph{integer} variables, are the following 
$2^{\binom{j'}{2}}$ variables.  
For each $i$, $1 \leq i \leq 2^{\binom{j'}{2}}$, we
will have a variable, $m_i$, that represents 
how many of the $n_i$ voters having the $i$th among the 
$2^{\binom{j'}{2}}$ possible vote types go into the first subelection.

Finally, we must specify the constraints of our integer linear program
constraint feasibility problem.  We will have three groups of constraints.

The first constraint group 
is enforcing that 
plausible numbers are put
in the first partition.  In particular,
for each $i$, $1 \leq i \leq 2^{\binom{j'}{2}}$, 
we have the constraints $0 \leq m_i$ and
$m_i \leq n_i$.

The second constraint group 
is enforcing that 
after the partitioning
we really do have in the first subelection a situation in which all
the pairwise contests come out exactly as specified by $T_1$.  In particular,
for each $i$ and $\ell$ such that 
$1 \leq i \leq j'$,
$1 \leq \ell  \leq j'$, and $i \neq \ell$, 
we do the following.  Consider the equation
\begin{equation}\label{e:enforce}
(\sum_{\{a~\mid~ 
\parbox[t]{1.5in}{\scriptsize$
1 \leq a \leq 2^{\binom{j'}{2}}$
and in votes of type $a$ it holds that $i$~is preferred to $\ell\}
$} %
} m_a )
{\rm~~~~OP~~~~}
(\sum_{\{a~\mid~
\parbox[t]{1.5in}{\scriptsize$
1 \leq a \leq 2^{\binom{j'}{2}}$
and in votes of type $a$ it holds that $\ell$~is preferred to $i\}
$} %
} m_a),
\end{equation}
where $a$ in each sum varies over the $2^{\binom{j'}{2}}$ possible
preferences.  If $val1(i,\ell)=1$ we will have a constraint of the
above form with OP set to ``$\geq$''.  If $val1(\ell,i)=1$ we will
have a constraint of the above form with OP set to ``$\leq$''.  Note
that this means that if $val1(i,\ell)=val1(\ell,i)=1$, i.e., those
two voters are purported to tie, we will add two constraints.

The third constraint group has the same function as the second
constraint group, except it regards the second subelection rather than
the first subelection.  In particular, for each $i$ and $\ell$ such
that $1 \leq i \leq j'$, $1 \leq \ell \leq j'$, and $i \neq \ell$, we
do the following.  Consider again equation~\eqref{e:enforce} from above,
except with each of the two occurrences of $m_a$ replaced by $n_a -
m_a$.  If $val2(i,\ell)=1$ we will have a constraint of that form
with OP set to ``$\geq$''.  If $val2(\ell,i)=1$ we will have a
constraint of that form with OP set to ``$\leq$''.  As above,
this means that if $val2(i,\ell)= val2(\ell,i)=1$, 
we will add two constraints.

This completes the specification of the 
integer linear programming constraint feasibility problem.

Note that our top-level code, from Figure~\ref{f:top-level-code},
clearly runs within polynomial time relative to even 
the
succinct-case input to the original $\rm CCPV\hbox{-}TP$ problem, and
that that polynomial's degree is bounded above independently of 
$j$.  Note in particular that our 
algorithm constructs at most a large constant
(for $j$ fixed) number of 
integer linear programming constraint feasibility problems, and each of those 
is itself polynomial-sized relative to even 
the
succinct-case input to the original $\rm CCPV\hbox{-}TP$ problem, 
and  that polynomial size's degree is bounded above independently of 
$j$.  Further, note that the 
integer linear programming constraint feasibility problems 
clearly do test what they are supposed to test---most 
importantly, they test
that the subelections match the pairwise outcomes 
specified by $j'$-COTs $T_1$ and $T_2$.
Finally and crucially, by 
Lenstra's algorithm~(\cite{len:j:integer-fixed},
see also~\cite{dow:c:parameterized-survey,nie:thesis-habilition:fixed-param}
which are very clear regarding the ``linear''s later in this sentence),
since this 
integer linear programming constraint feasibility problem 
has a fixed number of constraints (and in our case in fact also
has a fixed number of variables), it can be solved---relative to its 
size (which includes the filled-in constants, such as our $n_i$ for 
example, which are in effect inputs to the integer program's 
specification)---via 
a linear number of arithmetic operations on linear-sized 
integers.  So, overall, we are in polynomial time 
even relative to succinctly specified input, and the polynomial's degree
is bounded above independently of $j$.  Thus we have established
membership in the class $\fpt$.~\end{proofs}

We now describe very briefly how the above proof of 
Lemma~\ref{t:first-example-case} can be adjusted 
to handle all the 
partition cases 
from Lemma~\ref{t:leftover}, namely, the cases
$\begin{bmatrix}
\textrm{\rm{}succinct}\\\textrm{\rm{}$\emptyset$}
\end{bmatrix}
\hbox{-}
\begin{bmatrix}
\textrm{\rm{}\copelandalpha}\\\textrm{\rm{}\copelandalphairrational}
\end{bmatrix}
\hbox{-}
\begin{bmatrix}
\textrm{\rm{}C}\\\textrm{\rm{}D}
\end{bmatrix}
\hbox{\rm{}C}
\begin{bmatrix}
\textrm{\rm{}PV-TE}\\
\textrm{\rm{}PV-TP}
\end{bmatrix}
\hbox{-\textrm{\rm{}\bcj}}$.  
As noted before, the first two brackets can
be ignored, as we have chosen the more demanding choice for each.  Let
us discuss the other variations.  Regarding changing from constructive
to destructive, in Figure~\ref{f:top-level-code} change ``is a
winner'' to ``is not a winner.''  Regarding changing from PV-TP to
PV-TE, in the ``if'' block in Figure~\ref{f:top-level-code} change each
``all the candidates who win'' to ``the candidate who wins (if there
is a unique candidate who wins).''

The only remaining cases are the cases
$\begin{bmatrix}
\textrm{\rm{}succinct}\\\textrm{\rm{}$\emptyset$}
\end{bmatrix}
\hbox{-}
\begin{bmatrix}
\textrm{\rm{}\copelandalpha}\\\textrm{\rm{}\copelandalphairrational}
\end{bmatrix}
\hbox{-}
\begin{bmatrix}
\textrm{\rm{}C}\\\textrm{\rm{}D}
\end{bmatrix}
\hbox{\rm{}C}
\begin{bmatrix}
\textrm{\rm{}AV}\\
\textrm{\rm{}DV}
\end{bmatrix}
\hbox{-\textrm{\rm{}\bcj}}$.  
However, these cases are even more straightforward than the 
partition cases we just covered, so for space reasons we will not 
write them out, but rather will briefly comment on 
these cases.  Basically, one's top-level code for these cases
loops over all 
$j'$-COTs, and for each 
(there are $3^{\binom{j'}{2}}$)
checks whether the right outcome happens under that $j'$-COT
(i.e., the distinguished candidate either is (constructive case) or 
is not (destructive case) a winner), and if so, it runs Lenstra's
algorithm on an 
integer linear programming constraint feasibility problem 
to see whether we can by the allowed action (adding/deleting)
get to a state where that particular $j'$-COT matches our 
(after addition or deletion of voters) election.  In the integer 
program, the variables will be the obvious ones, namely, 
for each $i$, $1 \leq i \leq 2^{\binom{j'}{2}}$, we will
have a variable, $m_i$, that describes how many voters of type $i$
to add/delete.  As our key constants
(of the 
integer linear program 
constraint
feasibility problem), we will have,
for each $i$, 
$1 \leq i \leq 2^{\binom{j'}{2}}$,
a value, $n_i$,
for the number of type $i$ voters in the input.
Also, if this is a problem about addition of voters, 
we will have additional constants, $\widehat{n_i}$,
$1 \leq i \leq 2^{\binom{j'}{2}}$,
representing 
the number of type $i$ voters among the pool, $W$, of voters
available for addition.  And if our problem has an internal ``$k$''
(a limit on the number of additions or deletions), we enforce 
that with the natural constraints, as do we also 
with the natural constraints 
enforce the obvious relationships between the $m_i$, $n_i$,
$\widehat{n_i}$, and so on.  Most critically, we have 
constraints ensuring that after the additions/deletions specified by
the $m_i$, each pairwise outcome specified by the 
$j'$-COT is realized.

Finally, although everything in
Section~\ref{sec:control-fpt} (both the part so
far and the part to come)
is written 
for the case of the nonunique-winner model, all the results 
hold analogously in the unique-winner model, with the 
natural, minor proof modifications.  (Also, we mention in 
passing that due to the connection, found in
Footnote~5 
of~\cite{hem-hem-rot:j:destructive-control}, between unique-winner
destructive control and nonunique-winner constructive control, one
could use some of our nonunique-winner constructive-case results to
indirectly prove some of the unique-winner destructive-case results.)

\subsubsection*{FPT and Extended Control}
In this section, we look at extended control.  By that we do not mean
changing the ten standard control notions of
adding/deleting/partitioning candidates/voters.  Rather, we mean
generalizing past merely looking at the constructive (make a
distinguished candidate a winner) and the destructive (prevent a
distinguished candidate from being a winner) cases.  In particular, we
are interested in control where the goal can be far
more flexibly specified, for example (though in the partition cases
we will be even more flexible than this), we will allow 
as our goal region any
(reasonable---there are some time-related conditions) subcollection of
``Copeland outcome tables'' (specifications of who won/lost/tied each
head-to-head contest).  Since from a Copeland outcome
table, in concert with
the current $\alpha$, one can read off the $\copelandalphairrational$ scores of
the candidates, this allows us a tremendous range of descriptive
flexibility in specifying our control goals, e.g., we can specify a
linear order desired for the candidates with respect to their
$\copelandalphairrational$ scores, we can specify a linear-order-with-ties
desired for the candidates with respect to their $\copelandalphairrational$
scores, we can specify the exact desired $\copelandalphairrational$ scores for
one or more candidates, we can specify that we want to ensure that
no candidate from a certain subgroup has a $\copelandalphairrational$ score that
ties or defeats the $\copelandalphairrational$ score of any candidate from a
certain other subgroup, etc.\footnote{\label{f:tricky-caveat}We 
mention up front that that initial
example list applies with some additional
minor technical caveats.  Those examples
were speaking as if in the final election we have all the candidates
receiving $\copelandalphairrational$ scores 
in the final election.  But in fact
in the partition cases this is not (necessarily)
so, and so in those cases we will
focus on the Copeland outcome tables most natural to the given case.
For example, in control by partition of voters, we will focus on
subcollections of pairs of Copeland outcome tables for the two
subelections.  Also, though our Copeland outcome tables as defined
below are not explicitly labeled with candidate names, but rather use
a lexicographical correspondence with the involved candidates, in some
cases we would---though we don't repeat this in
the discussion below---need to allow 
the inclusion in the goal specification of the names of the candidates
who are in play in a given table or tables, most particularly, in the cases of
addition and deletion of candidates, and in 
some 
partition cases.}
Later in this section we will give a 
list repeating some of these examples and adding some 
new examples.

All the FPT algorithms given in the previous section regard, on their
surface, the standard control problem, which tests whether a given
candidate can be made a winner (constructive case) or can be precluded
from being a winner (destructive case).  We now note that the
general approaches used in that section in fact yield 
FPT schemes even for the far more
flexible notions of 
control mentioned above.  In fact, one gets, for
all the FPT cases covered in the previous section, FPT algorithms 
for the extended-control problem
for those cases---very loosely put, FPT algorithms
that test, for virtually any natural collection of outcome tables (as
long as that collection itself can be recognized in a way that doesn't
take too much running time, i.e., the checking time is 
polynomial and of a degree that is bounded 
independently of $j$), whether by the given type of control one
can reach one of those outcome tables.

Let us discuss this in a bit more detail.  A key concept used inside
the proof of Lemma~\ref{t:first-example-case} was that of a Copeland
outcome table---a function that for each distinct pair of candidates
specifies either a tie or specifies who is the (not tied) winner in
their pairwise contest.  Let us consider the control algorithm given
in the proof of that lemma, and in particular let us consider the
top-level code specified in Figure~\ref{f:top-level-code}.  That code
double-loops over size $j'$ Copeland outcome tables (a.k.a.~$j'$-COTs),
regarding the subpartitions, and for each case when the outcome tables'
subelection cases, followed by the final election that they imply,
correspond to the desired type of constructive (the
distinguished person wins) or destructive (the distinguished
person does not win) outcome, we check whether those two $j'$-COTs
can be made to hold via the current type of control (for the 
case being discussed,
PV-TP).  

However, note that simply by easily varying that top-level code
we can
obtain a natural FPT 
algorithm (a single algorithm, see Footnote~\ref{f:not-family}
the analogue of which applies here)
for any question of whether
via the allowed type of control one can reach any 
run-time-quick-to-recognize collection of pairs of $j'$-COTs (in the
subelection), or even whether a given candidate collection and one of
a given (run-time-quick-to-recognize) $j''$-COT collection
over that candidate collection 
($j''$ being the size of that final-round candidate collection)
can be reached in the final election.
This is true not just for the partition cases 
(where, informally put, 
we would do this by, in Figure~\ref{f:top-level-code},
changing the condition inside the ``if'' to instead look
for membership in that collection of 
$j'$-COTs\footnote{Let us discuss this 
a bit more formally, again using 
PV-TP as an example.   Consider 
any family of 
 boolean
functions $F_j$, $j \in \naturals$, 
such that each $F_j$ is computable, even when its first 
argument is succinctly specified, in polynomial time 
with the polynomial degree bounded independently of $j$.
Now, consider 
changing Figure~\ref{f:top-level-code}'s  code to:
\begin{quote}
For each $j'$-COT, $T_1$, \\
For each $j'$-COT, $T_2$, \\
If ($F_{j'}(input, T_1, T_2)$)  \\
then $\cdots$.
\end{quote}
Note that this change gives 
an FPT control scheme
for a certain
extended control problem.  In particular, it 
does so for the 
extended control problem whose goal is to ensure 
that we can 
realize at least one of the set of $(T_1,T_2)$
such that $F_{j'}$ ($j'$ being the number of candidates 
in the particular input), given as its inputs the problem's input, $T_1$,
and $T_2$ evaluates to true.
That is, the $F_j$ functions are recognizing (viewed a bit 
differently, are defining) the goal
set of the extended control problem.

{}From the input, $T_1$, and $T_2$ 
we can easily tell the scores in the final election. 
So this approach can be used to choose as our extended-control goals 
natural features of the final election.
}%
)
but also for all 
the cases we attacked via Lenstra's method (though for 
the nonpartition cases
we will typically single-loop over Copeland outcome tables that 
may represent the outcome after control is exerted; also, for 
some of these cases, the caveat at the end of 
Footnote~\ref{f:tricky-caveat} will apply).
And it is even easier to 
notice that for those cases we attacked by direct brute force this 
also holds.

So, as just a few examples (some echoing the start of this section, and
some new), all the following have (with the caveats mentioned above
about needed names attached, e.g., in cases of candidate
addition/deletion/partition, and regarding the partition cases
focusing not necessarily directly on the final table) FPT 
extended control
algorithms 
for all the types of control and boundedness cases for
which the FPT results of the previous section are stated.
\begin{enumerate}
\item Asking whether under the stated action one can obtain in the
  final election (simply in the election in the case when there is no
  partitioning) the outcome that all the $\copelandalphairrational$-system
  scores of the candidates precisely match the relations of the
  lexicographic names of the candidates.
\item More generally than that, asking whether under the stated action
  one can obtain in the final election (simply in the election in the
  case when there is no partitioning) a certain
  linear-order-without-ties regarding the $\copelandalphairrational$-system
  scores of the candidates.
\item More generally still, asking whether under the stated action one
  can obtain in the final election (simply in the election in the case
  when there is no partitioning) a certain linear-order-with-ties
  regarding the $\copelandalphairrational$-system scores of the candidates.
\item Asking whether under the stated action one
  can obtain in the final election (simply in the election in the case
  when there is no partitioning) the situation that exactly 1492
  candidates tie as winner regarding their 
  $\copelandalphairrational$-system scores.
\item Asking whether under the stated action one can obtain in the
  final election (simply in the election in the case when there is no
  partitioning) the situation that no two candidates have the same
  $\copelandalphairrational$-system scores as each other.
\end{enumerate}
Again, these are just a very few examples.  Our point is that the
previous section is flexible enough to address not just
constructive/destructive control, but also to address far more general
control issues.

\subsubsection*{Resistance Results}\label{ss-star:resistance}
Theorems~\ref{t:v} and~\ref{t:c-bc} give $\fpt$ schemes for all 
voter-control
cases with bounded voters, for all voter-control cases with
bounded candidates, and for all candidate-control cases with bounded
candidates.  This might lead one to hope that all the cases admit
$\fpt$ schemes.  However, the remaining type of case, 
the candidate-control 
cases with bounded voters, does not follow this pattern.  In fact, 
we note that for $\copelandalphairrational$ all the 
candidate-control cases 
that we showed earlier in this paper 
(i.e., without bounds on the number of voters) 
to be resistant remain resistant 
even for the case of bounded voters.  
This resistance holds even when the input is not in succinct 
format, and so it certainly also holds when the input is 
in succinct format.

The reason for this is that, for the case of irrational voters, with
just \emph{two} voters (with preferences over $j$ candidates) any
given $j$-COT can be achieved.  To do this, for each distinct pair of
candidates $i$ and $\ell$, to have $i$ preferred in their pairwise
contest have both voters prefer $i$ to $\ell$, to have $\ell$
preferred in their pairwise contest have both voters prefer $\ell$ to
$i$, and to have a tie in the pairwise contest have one voter prefer
$\ell$ to $i$ and one voter prefer $i$ to $\ell$.  
Since in the proofs of resistance for candidate control, we identified
elections with their election graphs, i.e., with their COTs, it is not
hard to see that all these resistance proofs carry over even to the
case of two irrational voters.

The only open cases remaining regard the rational-voter,
candidate-control, bounded-voter cases.  We note that Betzler and
Uhlmann~\cite{bet-uhl:c:parameterized-complecity-candidate-control}
have recently resolved some of these open issues.

\section{Control in Condorcet Elections}\label{sec:control-condorcet}

In this section we show that Condorcet elections are resistant to
constructive control via deleting voters (CCDV) and via partition of
voters (CCPV). These results were originally claimed in the seminal
paper of Bartholdi, Tovey, and Trick~\cite{bar-tov-tri:j:control}, but
the proofs there were based on the assumption that a voter can be
indifferent between several candidates.  Their
model of elections did not allow that (and neither does ours). Here we
show how one can obtain these results in the case when the voters'
preference lists are linear orders---which is both their model and ours.

Recall that a candidate $c$ of election $E = (C,V)$ is a Condorcet winner of $E$
if he or she defeats all other candidates in their head-to-head
contests. Alternatively, one could say that a candidate $c$ is a
Condorcet winner of election $E$ if and only if he or she has
Copeland$^0$ score of $\|C\|-1$. Since each election can have at
most one Condorcet winner, it doesn't make sense here to
differentiate between the unique-winner and the nonunique-winner models.

\begin{theorem}
  \label{thm:ccdv-condorcet}
  Condorcet elections are resistant to constructive control via
  deleting voters.
\end{theorem}

\begin{proofs}
This follows immediately from the proof of Theorem~\ref{thm:ucdv}. Note
that a Condorcet winner is always a unique $\copelandalpha$ winner,
for each rational $\alpha$ with $0 \leq \alpha \leq 1$,
and note that in the proof of Theorem~\ref{thm:ucdv}, if $\calS$ contains
a $k$-element cover of $B$, then we can delete $k$ voters such that
in the resulting election $p$ defeats every other candidate in their
head-to-head, contest, i.e., $p$ is a Condorcet winner in the resulting
election.~\end{proofs}%

Before we proceed with our proof of resistance for the case of
constructive control via partition of voters (CCPV), we have to
mention a slight quirk of Bartholdi, Tovey, and Trick's model of voter
partition. If one reads their paper carefully, it becomes apparent
that they have a quiet assumption that each given set of voters can
only be partitioned into subelections that each elect exactly one
winner, thus severely restricting the chair's partitioning possibilities.
That was why Hemaspaandra, Hemaspaandra, and
Rothe~\cite{hem-hem-rot:j:destructive-control} replaced Bartholdi,
Tovey, and Trick's convention with the more natural ties-promote and ties-eliminate
rules (see the discussion in~\cite{hem-hem-rot:j:destructive-control}),
but for this current section of our paper we go back to Bartholdi, Tovey, and
Trick's model, since our goal here is to reprove their results without
breaking \emph{their} model.

\begin{theorem}
  \label{thm:ccpv-condorcet}
  Condorcet elections are resistant to constructive control via
  partitioning voters (CCPV) in Bartholdi, Tovey, and Trick's
  model (see the paragraph above).
\end{theorem}

\begin{proofs}
The proof follows via a reduction from the X3C problem. In fact, we
use exactly the construction from the proof of Theorem~\ref{thm:cpv-tp}.
Let $E = (C,V)$ be the election constructed in that proof.

Since $s$ is the only candidate that $p$ defeats in a head-to-head
contest, the only way for $p$ to become a winner via
partitioning voters is to guarantee that $p$ wins within his or her
subelection and that $s$ wins within the other one. (Note that since
$p$ is not a Condorcet winner, $p$ cannot win in both
subelections.)

If $\calS$ contains a $k$-element cover, say, $\{S_{a_1}, \ldots,
S_{a_k}\}$, then letting $V_p = \widehat{V} - \{v_{a_1}, \ldots,
v_{a_k}\}$ and $V_s = V - V_p$ will make $p$ the Condorcet winner
in this CCPV scenario.

For the converse, let $(V_p,V_s)$ be a partition of the collection of
voters such that $p$ is the global Condorcet winner in the CCPV scenario
where we use two subelections,
one with voters $V_p$ and one with voters $V_s$. Via the above
paragraph we can assume, without loss of generality, that $p$ is the
Condorcet winner in $(C,V_p)$ and that $s$ is the Condorcet winner in
$(C,V_s)$. 
We can assume that $V_s$
contains the $k+1$ voters in $V - \widehat{V}$ 
(i.e., the voters with preference $s > r > B > p$).
Also, $V_s$
contains at most $k$ voters from $\widehat{V}$, as
otherwise $s$ would certainly not be a Condorcet winner in $(C,V_s)$.

As a result, $p$ can be made the Condorcet winner of $(C,\widehat{V})$ by
deleting at most $k$ voters.
It follows from Claim~\ref{cl:llull-dv} that $\calS$
contains a $k$-element cover of $B$.~\end{proofs}

\section{Conclusions}\label{sec:conclusions}
We have shown that from the computational point of view the election
systems of Llull and Copeland (i.e., Copeland$^{0.5}$) are broadly resistant to bribery and
procedural
control, regardless of whether the voters are required to have
rational preferences.  It is rather charming that Llull's 700-year-old
system shows 
perfect resistance to bribery and 
more resistances to (constructive)
control than any other natural 
system (even far more modern ones) with an easy winner-determination
procedure---other than
\copelandalpha, $0 < \alpha < 1$---is known to possess, and this is even more
remarkable when one considers that Llull's system was defined long
before control of elections was even explicitly studied.  
Copeland$^{0.5}$ voting matches
Llull's 
perfect resistance to 
bribery and in addition has perfect resistance to (constructive) control.

A natural open direction would be to study the complexity 
of control for additional election systems.
Particularly interesting would be to
find existing, 
natural voting systems that have polynomial-time winner determination
procedures but that are resistant to all standard types of both
constructive \emph{and destructive} control.  Also 
extremely  interesting would be to
find 
single 
results 
that classify, for broad families of election systems, precisely what it is
that makes control easy or hard, i.e., to obtain  dichotomy 
meta-results for control (see Hemaspaandra and 
Hemaspaandra~\cite{hem-hem:j:dichotomy} for some 
discussion regarding work of that flavor for manipulation).

\addcontentsline{toc}{section}{Acknowledgments}\section*{Acknowledgments}
We thank Felix Brandt, Frieder Stolzenburg, and the anonymous AAAI-07, AAIM-08, and
COMSOC-08 referees for helpful comments.

\addcontentsline{toc}{section}{References}

%
%
\bibliographystyle{alpha}
\end{document}

%% file: constructive-microbribery-odd-flownetwork.eepic
\setlength{\unitlength}{0.00078740in}
\begingroup\makeatletter\ifx\SetFigFont\undefined%
\gdef\SetFigFont#1#2#3#4#5{%
  \reset@font\fontsize{#1}{#2pt}%
  \fontfamily{#3}\fontseries{#4}\fontshape{#5}%
  \selectfont}%
\fi\endgroup%
{\renewcommand{\dashlinestretch}{30}
\begin{picture}(7064,3027)(0,-10)
\thicklines
\put(157,1506){\ellipse{284}{284}}
\put(6907,1506){\ellipse{284}{284}}
\put(3532,156){\ellipse{284}{284}}
\put(4432,1056){\ellipse{284}{284}}
\put(2632,1956){\ellipse{284}{284}}
\put(3532,2856){\ellipse{284}{284}}
\path(3532,2710)(3532,336)
\blacken\path(3472.000,576.000)(3532.000,336.000)(3592.000,576.000)(3532.000,504.000)(3472.000,576.000)
\path(3532,302)(2632,1776)
\blacken\path(2808.278,1602.432)(2632.000,1776.000)(2705.860,1539.897)(2719.549,1632.615)(2808.278,1602.432)
\path(4432,1202)(3555,2676)
\blacken\path(3729.280,2500.425)(3555.000,2676.000)(3626.153,2439.067)(3640.902,2531.622)(3729.280,2500.425)
\path(3532,2710)(2632,2114)
\blacken\path(2798.974,2296.537)(2632.000,2114.000)(2865.230,2196.486)(2772.071,2206.758)(2798.974,2296.537)
\path(4432,1202)(2655,1789)
\blacken\path(2901.708,1770.693)(2655.000,1789.000)(2864.069,1656.749)(2814.522,1736.305)(2901.708,1770.693)
\path(4432,910)(3559,318)
\blacken\path(3723.961,502.358)(3559.000,318.000)(3791.310,403.040)(3698.045,412.289)(3723.961,502.358)
\path(157,1648)(158,1649)(160,1653)
	(165,1659)(171,1668)(181,1681)
	(194,1698)(209,1719)(228,1743)
	(250,1771)(274,1801)(300,1833)
	(328,1867)(357,1901)(388,1936)
	(419,1969)(451,2003)(483,2035)
	(515,2066)(547,2095)(579,2123)
	(612,2150)(645,2175)(679,2198)
	(713,2221)(748,2242)(784,2263)
	(822,2282)(861,2301)(902,2319)
	(945,2337)(990,2355)(1020,2366)
	(1052,2378)(1085,2389)(1120,2400)
	(1155,2411)(1192,2422)(1231,2433)
	(1271,2445)(1313,2456)(1357,2467)
	(1403,2479)(1452,2491)(1502,2503)
	(1555,2515)(1610,2527)(1668,2540)
	(1729,2553)(1792,2566)(1858,2579)
	(1926,2593)(1997,2607)(2070,2622)
	(2145,2636)(2222,2651)(2300,2666)
	(2379,2681)(2459,2696)(2539,2711)
	(2619,2725)(2697,2739)(2773,2753)
	(2846,2767)(2917,2779)(2983,2791)
	(3044,2802)(3101,2812)(3151,2821)
	(3196,2829)(3234,2836)(3267,2841)
	(3293,2846)(3314,2849)(3329,2852)(3352,2856)
\blacken\path(3125.830,2755.765)(3352.000,2856.000)(3105.269,2873.991)(3186.484,2827.215)(3125.830,2755.765)
\path(303,1506)(304,1505)(308,1504)
	(314,1502)(324,1499)(338,1494)
	(356,1488)(380,1480)(408,1470)
	(441,1459)(478,1447)(520,1433)
	(564,1419)(612,1403)(662,1388)
	(713,1371)(766,1355)(819,1339)
	(872,1323)(925,1307)(977,1292)
	(1028,1278)(1078,1264)(1127,1251)
	(1176,1238)(1223,1227)(1269,1216)
	(1314,1205)(1359,1195)(1404,1186)
	(1448,1178)(1492,1170)(1536,1162)
	(1580,1155)(1625,1149)(1670,1142)
	(1717,1137)(1764,1131)(1802,1127)
	(1841,1123)(1880,1119)(1920,1115)
	(1962,1112)(2005,1108)(2048,1105)
	(2094,1102)(2141,1099)(2189,1097)
	(2240,1094)(2292,1091)(2347,1089)
	(2404,1087)(2463,1085)(2524,1082)
	(2588,1080)(2654,1079)(2723,1077)
	(2794,1075)(2868,1073)(2943,1072)
	(3021,1070)(3100,1069)(3181,1068)
	(3262,1066)(3344,1065)(3426,1064)
	(3507,1063)(3587,1062)(3664,1061)
	(3739,1060)(3811,1060)(3878,1059)
	(3940,1058)(3997,1058)(4049,1057)
	(4094,1057)(4133,1057)(4166,1057)
	(4193,1056)(4214,1056)(4229,1056)(4252,1056)
\blacken\path(4012.000,996.000)(4252.000,1056.000)(4012.000,1116.000)(4084.000,1056.000)(4012.000,996.000)
\path(157,1360)(158,1359)(160,1355)
	(165,1349)(171,1340)(181,1327)
	(194,1310)(209,1290)(228,1265)
	(250,1238)(274,1208)(300,1176)
	(328,1142)(357,1108)(388,1074)
	(419,1040)(451,1007)(483,975)
	(515,944)(547,915)(579,887)
	(612,861)(645,836)(679,812)
	(713,790)(748,769)(784,748)
	(822,729)(861,710)(902,692)
	(945,674)(990,657)(1020,645)
	(1052,634)(1085,623)(1120,611)
	(1155,600)(1192,589)(1231,578)
	(1271,567)(1313,556)(1357,544)
	(1403,533)(1452,521)(1502,509)
	(1555,497)(1610,485)(1668,472)
	(1729,459)(1792,446)(1858,432)
	(1926,419)(1997,404)(2070,390)
	(2145,376)(2222,361)(2300,346)
	(2379,331)(2459,316)(2539,301)
	(2619,287)(2697,272)(2773,259)
	(2846,245)(2917,233)(2983,221)
	(3044,210)(3101,200)(3151,191)
	(3196,183)(3234,176)(3267,171)
	(3293,166)(3314,163)(3329,160)(3352,156)
\blacken\path(3105.269,138.009)(3352.000,156.000)(3125.830,256.235)(3186.484,184.785)(3105.269,138.009)
\path(3678,2856)(3680,2856)(3683,2855)
	(3690,2854)(3701,2852)(3716,2849)
	(3737,2846)(3764,2841)(3797,2836)
	(3836,2829)(3881,2821)(3933,2812)
	(3989,2802)(4052,2791)(4119,2779)
	(4190,2767)(4264,2754)(4341,2740)
	(4420,2726)(4501,2711)(4582,2697)
	(4663,2682)(4743,2667)(4822,2652)
	(4900,2637)(4976,2623)(5050,2609)
	(5121,2595)(5191,2581)(5257,2568)
	(5321,2555)(5382,2542)(5441,2529)
	(5497,2517)(5550,2505)(5602,2494)
	(5650,2482)(5697,2471)(5742,2460)
	(5784,2449)(5825,2438)(5864,2427)
	(5901,2416)(5937,2405)(5972,2394)
	(6005,2383)(6038,2372)(6069,2361)
	(6114,2344)(6157,2327)(6199,2309)
	(6238,2291)(6276,2272)(6313,2253)
	(6348,2232)(6383,2211)(6417,2188)
	(6450,2164)(6483,2139)(6515,2112)
	(6548,2084)(6580,2054)(6612,2024)
	(6644,1992)(6675,1960)(6706,1927)
	(6735,1894)(6764,1862)(6790,1832)
	(6814,1803)(6836,1777)(6855,1753)
	(6870,1734)(6883,1717)(6893,1705)(6907,1686)
\blacken\path(6716.329,1843.621)(6907.000,1686.000)(6812.936,1914.805)(6807.343,1821.249)(6716.329,1843.621)
\path(2778,1956)(2779,1956)(2783,1956)
	(2789,1955)(2799,1954)(2814,1953)
	(2833,1952)(2858,1950)(2889,1948)
	(2926,1945)(2970,1942)(3019,1939)
	(3075,1935)(3136,1930)(3202,1925)
	(3273,1920)(3349,1915)(3428,1909)
	(3509,1903)(3594,1897)(3680,1890)
	(3767,1884)(3854,1877)(3941,1871)
	(4028,1864)(4113,1858)(4197,1852)
	(4280,1845)(4360,1839)(4438,1833)
	(4514,1827)(4587,1822)(4658,1816)
	(4726,1810)(4791,1805)(4854,1800)
	(4915,1795)(4973,1790)(5029,1785)
	(5082,1781)(5133,1776)(5183,1772)
	(5230,1768)(5276,1763)(5319,1759)
	(5362,1755)(5403,1751)(5442,1747)
	(5480,1743)(5517,1739)(5554,1735)
	(5589,1731)(5641,1725)(5691,1719)
	(5740,1713)(5787,1706)(5832,1700)
	(5877,1694)(5920,1687)(5963,1680)
	(6005,1673)(6047,1666)(6088,1658)
	(6130,1650)(6172,1641)(6214,1632)
	(6256,1623)(6299,1614)(6341,1604)
	(6383,1594)(6425,1584)(6465,1574)
	(6504,1565)(6541,1555)(6576,1546)
	(6607,1538)(6636,1531)(6660,1524)
	(6680,1519)(6696,1514)(6709,1511)(6727,1506)
\blacken\path(6479.697,1512.423)(6727.000,1506.000)(6511.814,1628.046)(6565.129,1550.964)(6479.697,1512.423)
\path(3678,156)(3680,156)(3683,157)
	(3690,158)(3701,160)(3716,163)
	(3737,166)(3764,171)(3797,176)
	(3836,183)(3881,191)(3933,200)
	(3989,210)(4052,221)(4119,233)
	(4190,245)(4264,258)(4341,272)
	(4420,286)(4501,301)(4582,315)
	(4663,330)(4743,345)(4822,360)
	(4900,375)(4976,389)(5050,403)
	(5121,417)(5191,431)(5257,444)
	(5321,457)(5382,470)(5441,483)
	(5497,495)(5550,507)(5602,518)
	(5650,530)(5697,541)(5742,552)
	(5784,563)(5825,574)(5864,585)
	(5901,596)(5937,607)(5972,618)
	(6005,629)(6038,640)(6069,651)
	(6114,668)(6157,685)(6199,703)
	(6238,721)(6276,740)(6313,759)
	(6348,780)(6383,801)(6417,824)
	(6450,848)(6483,873)(6515,900)
	(6548,928)(6580,958)(6612,988)
	(6644,1020)(6675,1052)(6706,1085)
	(6735,1118)(6764,1150)(6790,1180)
	(6814,1209)(6836,1235)(6855,1259)
	(6870,1278)(6883,1295)(6893,1307)(6907,1326)
\blacken\path(6812.936,1097.195)(6907.000,1326.000)(6716.329,1168.379)(6807.343,1190.751)(6812.936,1097.195)
\path(4578,1056)(4580,1056)(4584,1056)
	(4592,1056)(4605,1056)(4622,1056)
	(4646,1057)(4675,1057)(4711,1057)
	(4752,1058)(4798,1059)(4849,1059)
	(4903,1060)(4961,1061)(5021,1063)
	(5082,1064)(5144,1065)(5205,1067)
	(5266,1068)(5325,1070)(5382,1072)
	(5438,1074)(5491,1076)(5542,1078)
	(5590,1080)(5636,1083)(5680,1085)
	(5721,1088)(5760,1091)(5797,1094)
	(5832,1097)(5866,1101)(5898,1105)
	(5928,1109)(5957,1113)(5985,1117)
	(6013,1122)(6039,1127)(6076,1135)
	(6112,1143)(6146,1153)(6179,1163)
	(6212,1174)(6245,1186)(6277,1199)
	(6309,1214)(6342,1230)(6375,1248)
	(6409,1267)(6443,1287)(6478,1309)
	(6513,1331)(6547,1353)(6580,1376)
	(6611,1397)(6639,1417)(6664,1435)
	(6685,1450)(6701,1463)(6727,1482)
\blacken\path(6568.627,1291.953)(6727.000,1482.000)(6497.825,1388.839)(6591.358,1382.877)(6568.627,1291.953)
\path(303,1506)(305,1507)(308,1510)
	(315,1515)(325,1523)(339,1533)
	(358,1547)(380,1563)(406,1582)
	(434,1602)(465,1623)(498,1645)
	(531,1667)(565,1689)(599,1710)
	(632,1730)(664,1749)(697,1767)
	(728,1783)(759,1798)(790,1812)
	(820,1824)(851,1836)(882,1847)
	(914,1856)(946,1865)(979,1873)
	(1014,1881)(1042,1887)(1070,1892)
	(1100,1897)(1131,1901)(1163,1906)
	(1197,1910)(1233,1914)(1270,1917)
	(1310,1921)(1352,1924)(1396,1927)
	(1443,1930)(1492,1932)(1545,1935)
	(1599,1937)(1656,1939)(1716,1941)
	(1777,1943)(1840,1945)(1904,1946)
	(1968,1948)(2031,1949)(2092,1951)
	(2151,1952)(2206,1953)(2257,1953)
	(2303,1954)(2342,1955)(2375,1955)
	(2401,1955)(2422,1956)(2452,1956)
\blacken\path(2212.000,1896.000)(2452.000,1956.000)(2212.000,2016.000)(2284.000,1956.000)(2212.000,1896.000)
\put(2537,1911){\makebox(0,0)[lb]{\smash{{\SetFigFont{11}{13.2}{\rmdefault}{\mddefault}{\updefault}$c_1$}}}}
\put(4337,1011){\makebox(0,0)[lb]{\smash{{\SetFigFont{11}{13.2}{\rmdefault}{\mddefault}{\updefault}$c_2$}}}}
\put(3437,111){\makebox(0,0)[lb]{\smash{{\SetFigFont{11}{13.2}{\rmdefault}{\mddefault}{\updefault}$c_3$}}}}
\put(6862,1461){\makebox(0,0)[lb]{\smash{{\SetFigFont{11}{13.2}{\rmdefault}{\mddefault}{\updefault}$t$}}}}
\put(112,1461){\makebox(0,0)[lb]{\smash{{\SetFigFont{11}{13.2}{\rmdefault}{\mddefault}{\updefault}$s$}}}}
\put(3437,2811){\makebox(0,0)[lb]{\smash{{\SetFigFont{11}{13.2}{\rmdefault}{\mddefault}{\updefault}$c_0$}}}}
\put(1057,2631){\makebox(0,0)[lb]{\smash{{\SetFigFont{11}{13.2}{\rmdefault}{\mddefault}{\updefault}$(2,0)$}}}}
\put(1057,2046){\makebox(0,0)[lb]{\smash{{\SetFigFont{11}{13.2}{\rmdefault}{\mddefault}{\updefault}$(0,0)$}}}}
\put(1057,246){\makebox(0,0)[lb]{\smash{{\SetFigFont{11}{13.2}{\rmdefault}{\mddefault}{\updefault}$(1,0)$}}}}
\put(1057,1416){\makebox(0,0)[lb]{\smash{{\SetFigFont{11}{13.2}{\rmdefault}{\mddefault}{\updefault}$(3,0)$}}}}
\put(3082,2136){\makebox(0,0)[lb]{\smash{{\SetFigFont{11}{13.2}{\rmdefault}{\mddefault}{\updefault}$(1,1)$}}}}
\put(3667,1506){\makebox(0,0)[lb]{\smash{{\SetFigFont{11}{13.2}{\rmdefault}{\mddefault}{\updefault}$(1,1)$}}}}
\put(4297,606){\makebox(0,0)[lb]{\smash{{\SetFigFont{11}{13.2}{\rmdefault}{\mddefault}{\updefault}$(1,1)$}}}}
\put(5512,2631){\makebox(0,0)[lb]{\smash{{\SetFigFont{11}{13.2}{\rmdefault}{\mddefault}{\updefault}$(T,0)$}}}}
\put(5512,1866){\makebox(0,0)[lb]{\smash{{\SetFigFont{11}{13.2}{\rmdefault}{\mddefault}{\updefault}$(T,49)$}}}}
\put(5512,1236){\makebox(0,0)[lb]{\smash{{\SetFigFont{11}{13.2}{\rmdefault}{\mddefault}{\updefault}$(T,49)$}}}}
\put(5512,246){\makebox(0,0)[lb]{\smash{{\SetFigFont{11}{13.2}{\rmdefault}{\mddefault}{\updefault}$(T,49)$}}}}
\put(2677,2496){\makebox(0,0)[lb]{\smash{{\SetFigFont{11}{13.2}{\rmdefault}{\mddefault}{\updefault}$(1,1)$}}}}
\put(3937,2181){\makebox(0,0)[lb]{\smash{{\SetFigFont{11}{13.2}{\rmdefault}{\mddefault}{\updefault}$(1,1)$}}}}
\put(2767,651){\makebox(0,0)[lb]{\smash{{\SetFigFont{11}{13.2}{\rmdefault}{\mddefault}{\updefault}$(1,1)$}}}}
\end{picture}
}

%% file: constructive-microbribery-even-copeland-flownetwork.eepic
\setlength{\unitlength}{0.00065242in}
\begingroup\makeatletter\ifx\SetFigFont\undefined%
\gdef\SetFigFont#1#2#3#4#5{%
  \reset@font\fontsize{#1}{#2pt}%
  \fontfamily{#3}\fontseries{#4}\fontshape{#5}%
  \selectfont}%
\fi\endgroup%
{\renewcommand{\dashlinestretch}{30}
\begin{picture}(9300,4174)(0,-10)
\thicklines
\put(3750,3943){\ellipse{270}{270}}
\put(4650,3043){\ellipse{270}{270}}
\put(5550,2143){\ellipse{270}{270}}
\put(3750,2143){\ellipse{270}{270}}
\put(3750,343){\ellipse{270}{270}}
\put(1950,2143){\ellipse{270}{270}}
\put(150,2143){\ellipse{270}{270}}
\put(9150,2143){\ellipse{270}{270}}
\put(7350,343){\ellipse{270}{270}}
\put(7350,1468){\ellipse{270}{270}}
\put(7350,2818){\ellipse{270}{270}}
\path(285,2143)(1770,2143)
\blacken\path(1530.000,2083.000)(1770.000,2143.000)(1530.000,2203.000)(1602.000,2143.000)(1530.000,2083.000)
\path(3750,2278)(3750,3763)
\blacken\path(3810.000,3523.000)(3750.000,3763.000)(3690.000,3523.000)(3750.000,3595.000)(3810.000,3523.000)
\path(2085,2143)(2086,2143)(2089,2144)
	(2094,2145)(2102,2146)(2114,2148)
	(2130,2151)(2150,2155)(2176,2159)
	(2207,2165)(2243,2171)(2285,2179)
	(2333,2187)(2386,2197)(2444,2207)
	(2507,2218)(2575,2230)(2648,2242)
	(2724,2255)(2804,2269)(2887,2283)
	(2972,2297)(3058,2312)(3147,2326)
	(3236,2341)(3326,2356)(3416,2370)
	(3505,2384)(3594,2398)(3682,2412)
	(3768,2425)(3854,2438)(3937,2450)
	(4019,2462)(4100,2474)(4178,2485)
	(4255,2495)(4329,2505)(4402,2514)
	(4473,2522)(4542,2531)(4610,2538)
	(4676,2545)(4740,2552)(4803,2558)
	(4865,2563)(4925,2568)(4985,2573)
	(5043,2577)(5100,2580)(5157,2583)
	(5213,2586)(5269,2588)(5324,2590)
	(5379,2591)(5434,2592)(5488,2593)
	(5542,2593)(5543,2593)(5597,2593)
	(5651,2592)(5706,2591)(5761,2590)
	(5816,2588)(5871,2586)(5927,2583)
	(5984,2580)(6041,2577)(6099,2573)
	(6158,2568)(6218,2563)(6279,2558)
	(6342,2552)(6406,2545)(6471,2538)
	(6538,2531)(6607,2522)(6678,2514)
	(6750,2505)(6824,2495)(6900,2485)
	(6977,2474)(7057,2462)(7138,2450)
	(7221,2438)(7305,2425)(7391,2412)
	(7478,2398)(7566,2384)(7655,2370)
	(7743,2356)(7832,2341)(7920,2326)
	(8008,2312)(8094,2297)(8178,2283)
	(8260,2269)(8338,2255)(8414,2242)
	(8485,2230)(8553,2218)(8615,2207)
	(8673,2197)(8725,2187)(8772,2179)
	(8814,2171)(8850,2165)(8880,2159)
	(8905,2155)(8926,2151)(8942,2148)
	(8953,2146)(8970,2143)
\blacken\path(8723.225,2125.621)(8970.000,2143.000)(8744.079,2243.795)(8804.556,2172.196)(8723.225,2125.621)
\path(3885,343)(3886,343)(3889,343)
	(3895,344)(3903,344)(3916,345)
	(3934,347)(3956,349)(3983,351)
	(4017,354)(4055,357)(4100,361)
	(4151,365)(4207,370)(4268,375)
	(4335,381)(4406,387)(4481,394)
	(4560,401)(4642,408)(4727,416)
	(4813,424)(4901,433)(4990,441)
	(5079,450)(5168,459)(5256,468)
	(5343,477)(5430,487)(5514,496)
	(5597,505)(5679,515)(5758,524)
	(5835,534)(5909,544)(5982,553)
	(6052,563)(6120,573)(6186,583)
	(6249,592)(6311,602)(6370,613)
	(6428,623)(6484,633)(6538,644)
	(6590,655)(6641,666)(6691,677)
	(6739,689)(6786,700)(6832,712)
	(6877,725)(6921,738)(6965,751)
	(7008,764)(7050,778)(7099,795)
	(7147,812)(7194,829)(7241,848)
	(7287,867)(7334,887)(7380,907)
	(7425,929)(7471,952)(7518,975)
	(7564,1000)(7611,1026)(7659,1054)
	(7707,1082)(7756,1112)(7806,1143)
	(7857,1176)(7909,1210)(7961,1245)
	(8015,1282)(8069,1321)(8125,1360)
	(8181,1400)(8238,1442)(8295,1484)
	(8352,1527)(8409,1570)(8465,1614)
	(8521,1657)(8575,1699)(8628,1740)
	(8678,1780)(8726,1818)(8770,1853)
	(8812,1887)(8849,1917)(8883,1944)
	(8913,1969)(8938,1989)(8959,2007)
	(8977,2021)(8990,2032)(9000,2041)(9015,2053)
\blacken\path(8865.073,1856.221)(9015.000,2053.000)(8790.110,1949.925)(8883.814,1948.051)(8865.073,1856.221)
\path(3885,343)(7170,343)
\blacken\path(6930.000,283.000)(7170.000,343.000)(6930.000,403.000)(7002.000,343.000)(6930.000,283.000)
\path(7485,343)(7487,343)(7491,344)
	(7499,346)(7512,348)(7530,352)
	(7553,356)(7582,362)(7617,369)
	(7657,378)(7701,387)(7750,397)
	(7802,409)(7856,421)(7911,434)
	(7967,447)(8023,461)(8078,475)
	(8131,489)(8183,503)(8233,517)
	(8280,532)(8325,546)(8368,561)
	(8408,576)(8446,590)(8481,606)
	(8515,621)(8546,637)(8575,653)
	(8603,670)(8630,687)(8655,705)
	(8678,723)(8701,743)(8723,763)
	(8744,785)(8766,809)(8786,833)
	(8805,859)(8824,887)(8841,916)
	(8859,947)(8876,980)(8892,1015)
	(8908,1053)(8924,1094)(8940,1136)
	(8956,1182)(8971,1230)(8987,1281)
	(9002,1334)(9017,1389)(9032,1445)
	(9047,1502)(9061,1560)(9074,1616)
	(9087,1671)(9099,1723)(9110,1772)
	(9119,1815)(9128,1854)(9134,1886)
	(9140,1912)(9144,1932)(9150,1963)
\blacken\path(9163.302,1715.972)(9150.000,1963.000)(9045.488,1738.774)(9118.076,1798.061)(9163.302,1715.972)
\path(244,2049)(245,2048)(248,2047)
	(253,2045)(262,2041)(274,2035)
	(290,2028)(311,2018)(337,2006)
	(368,1992)(404,1976)(445,1957)
	(492,1936)(543,1913)(599,1888)
	(659,1862)(722,1834)(789,1804)
	(859,1774)(931,1743)(1004,1712)
	(1079,1680)(1154,1649)(1229,1618)
	(1304,1587)(1379,1557)(1452,1528)
	(1525,1499)(1596,1472)(1665,1446)
	(1733,1421)(1800,1397)(1864,1375)
	(1927,1354)(1988,1334)(2048,1315)
	(2105,1298)(2162,1282)(2216,1268)
	(2270,1254)(2322,1242)(2373,1231)
	(2422,1221)(2471,1213)(2519,1205)
	(2567,1199)(2614,1194)(2660,1190)
	(2706,1186)(2752,1184)(2798,1183)
	(2843,1182)(2889,1183)(2934,1184)
	(2980,1186)(3026,1190)(3072,1194)
	(3119,1199)(3166,1206)(3214,1213)
	(3263,1222)(3312,1232)(3363,1243)
	(3414,1255)(3467,1269)(3522,1283)
	(3577,1299)(3634,1317)(3693,1335)
	(3754,1355)(3816,1376)(3880,1399)
	(3945,1423)(4013,1448)(4081,1474)
	(4152,1501)(4223,1530)(4296,1559)
	(4370,1589)(4444,1620)(4518,1651)
	(4592,1683)(4666,1715)(4738,1746)
	(4809,1777)(4877,1807)(4943,1837)
	(5006,1865)(5065,1892)(5120,1917)
	(5171,1940)(5217,1961)(5258,1980)
	(5293,1996)(5324,2010)(5349,2022)
	(5370,2032)(5386,2039)(5398,2045)(5415,2053)
\blacken\path(5223.391,1896.520)(5415.000,2053.000)(5172.296,2005.098)(5262.990,1981.466)(5223.391,1896.520)
\path(244,2237)(245,2238)(247,2242)
	(251,2248)(256,2257)(265,2271)
	(276,2289)(290,2312)(307,2340)
	(328,2373)(351,2409)(377,2450)
	(405,2494)(436,2541)(468,2590)
	(502,2641)(537,2692)(572,2744)
	(608,2796)(644,2846)(681,2896)
	(716,2944)(752,2991)(787,3035)
	(822,3078)(856,3118)(890,3157)
	(924,3194)(957,3228)(990,3261)
	(1023,3292)(1056,3321)(1089,3349)
	(1122,3375)(1156,3400)(1190,3424)
	(1225,3446)(1261,3468)(1298,3489)
	(1336,3509)(1369,3525)(1404,3542)
	(1439,3557)(1475,3572)(1513,3587)
	(1551,3601)(1592,3615)(1633,3629)
	(1677,3642)(1722,3655)(1770,3667)
	(1819,3680)(1871,3692)(1925,3704)
	(1982,3716)(2041,3728)(2102,3740)
	(2166,3751)(2233,3763)(2302,3775)
	(2373,3786)(2446,3798)(2521,3809)
	(2597,3820)(2674,3831)(2752,3842)
	(2830,3852)(2907,3863)(2982,3872)
	(3055,3882)(3125,3891)(3192,3899)
	(3254,3906)(3311,3913)(3363,3919)
	(3408,3925)(3448,3929)(3482,3933)
	(3509,3936)(3530,3939)(3546,3940)(3570,3943)
\blacken\path(3339.295,3853.695)(3570.000,3943.000)(3324.411,3972.768)(3403.297,3922.162)(3339.295,3853.695)
\path(285,2098)(287,2098)(290,2096)
	(297,2094)(308,2091)(323,2087)
	(344,2080)(370,2073)(401,2063)
	(439,2052)(482,2039)(531,2025)
	(584,2009)(642,1993)(703,1975)
	(768,1956)(835,1937)(903,1918)
	(972,1898)(1042,1879)(1111,1860)
	(1180,1842)(1247,1824)(1313,1806)
	(1377,1790)(1439,1774)(1500,1759)
	(1557,1745)(1613,1732)(1667,1720)
	(1718,1709)(1767,1699)(1815,1690)
	(1860,1681)(1904,1674)(1946,1667)
	(1987,1661)(2027,1656)(2065,1652)
	(2102,1649)(2139,1646)(2174,1644)
	(2209,1643)(2244,1642)(2286,1642)
	(2327,1643)(2367,1645)(2408,1648)
	(2448,1652)(2488,1658)(2529,1664)
	(2570,1672)(2612,1681)(2654,1691)
	(2698,1703)(2743,1716)(2789,1730)
	(2836,1746)(2885,1763)(2935,1781)
	(2986,1801)(3038,1821)(3091,1843)
	(3144,1864)(3196,1887)(3247,1909)
	(3296,1931)(3344,1952)(3387,1972)
	(3427,1990)(3463,2007)(3494,2021)
	(3520,2033)(3540,2043)(3556,2051)(3579,2062)
\blacken\path(3388.375,1904.323)(3579.000,2062.000)(3336.600,2012.579)(3427.441,1989.515)(3388.375,1904.323)
\path(5652,2233)(7215,2728)
\blacken\path(7004.315,2598.339)(7215.000,2728.000)(6968.085,2712.739)(7054.840,2677.278)(7004.315,2598.339)
\path(7475,2732)(7477,2732)(7482,2731)
	(7492,2729)(7507,2726)(7527,2722)
	(7552,2718)(7583,2712)(7619,2705)
	(7657,2698)(7699,2690)(7742,2681)
	(7785,2673)(7828,2664)(7870,2655)
	(7911,2646)(7949,2638)(7986,2630)
	(8021,2622)(8054,2614)(8085,2606)
	(8115,2598)(8143,2590)(8170,2582)
	(8196,2574)(8221,2566)(8246,2557)
	(8271,2549)(8295,2540)(8320,2530)
	(8344,2520)(8369,2510)(8394,2499)
	(8420,2487)(8447,2474)(8475,2460)
	(8504,2446)(8535,2430)(8567,2413)
	(8601,2395)(8636,2377)(8672,2357)
	(8708,2337)(8745,2316)(8782,2296)
	(8817,2276)(8850,2257)(8879,2240)
	(8905,2225)(8927,2213)(8944,2203)(8970,2188)
\blacken\path(8732.132,2255.962)(8970.000,2188.000)(8792.099,2359.905)(8824.481,2271.953)(8732.132,2255.962)
\path(5649,2044)(7186,1468)
\blacken\path(6940.208,1496.037)(7186.000,1468.000)(6982.318,1608.406)(7028.684,1526.955)(6940.208,1496.037)
\path(5424,2224)(4767,2926)
\blacken\path(4974.804,2791.770)(4767.000,2926.000)(4887.189,2709.772)(4881.797,2803.340)(4974.804,2791.770)
\path(4555,3137)(3865,3828)
\blacken\path(4077.040,3700.567)(3865.000,3828.000)(3992.126,3615.776)(3983.708,3709.120)(4077.040,3700.567)
\path(3885,2188)(3886,2189)(3889,2190)
	(3895,2193)(3904,2197)(3917,2203)
	(3935,2212)(3957,2223)(3985,2236)
	(4018,2252)(4057,2270)(4101,2291)
	(4150,2315)(4204,2340)(4262,2368)
	(4325,2397)(4391,2429)(4460,2461)
	(4531,2494)(4603,2529)(4677,2563)
	(4752,2598)(4827,2633)(4901,2667)
	(4974,2701)(5047,2734)(5118,2767)
	(5188,2798)(5255,2829)(5321,2858)
	(5385,2887)(5447,2914)(5507,2940)
	(5564,2965)(5620,2989)(5674,3012)
	(5726,3034)(5777,3054)(5825,3074)
	(5872,3093)(5918,3111)(5963,3128)
	(6006,3144)(6048,3159)(6089,3174)
	(6130,3188)(6169,3201)(6208,3214)
	(6247,3226)(6285,3238)(6332,3252)
	(6379,3266)(6425,3278)(6472,3290)
	(6518,3302)(6564,3313)(6610,3323)
	(6656,3333)(6702,3341)(6748,3350)
	(6794,3357)(6840,3364)(6886,3370)
	(6931,3375)(6977,3380)(7023,3384)
	(7068,3387)(7113,3389)(7157,3390)
	(7201,3391)(7245,3391)(7288,3390)
	(7330,3388)(7371,3385)(7412,3382)
	(7452,3377)(7492,3372)(7530,3367)
	(7567,3360)(7604,3353)(7640,3345)
	(7675,3337)(7709,3327)(7742,3318)
	(7775,3307)(7806,3296)(7837,3284)
	(7868,3272)(7898,3259)(7928,3245)
	(7958,3230)(7989,3215)(8019,3198)
	(8049,3180)(8079,3161)(8109,3141)
	(8140,3119)(8171,3096)(8203,3071)
	(8235,3045)(8268,3017)(8302,2987)
	(8336,2956)(8372,2922)(8409,2887)
	(8447,2849)(8486,2810)(8526,2770)
	(8567,2728)(8608,2684)(8649,2641)
	(8690,2597)(8730,2553)(8770,2510)
	(8807,2468)(8843,2429)(8875,2392)
	(8905,2359)(8931,2329)(8954,2304)
	(8972,2282)(8987,2265)(8998,2252)(9015,2233)
\blacken\path(8810.255,2371.850)(9015.000,2233.000)(8899.684,2451.866)(8902.978,2358.201)(8810.255,2371.850)
\path(3885,3943)(3886,3943)(3889,3943)
	(3894,3942)(3902,3942)(3914,3941)
	(3930,3940)(3951,3939)(3978,3938)
	(4009,3936)(4047,3934)(4090,3931)
	(4140,3928)(4196,3925)(4257,3921)
	(4324,3917)(4397,3912)(4475,3907)
	(4558,3902)(4646,3896)(4737,3891)
	(4832,3884)(4929,3878)(5029,3871)
	(5131,3864)(5235,3857)(5339,3850)
	(5443,3842)(5548,3835)(5652,3827)
	(5755,3819)(5857,3811)(5957,3803)
	(6056,3795)(6153,3787)(6247,3779)
	(6340,3771)(6430,3763)(6517,3755)
	(6602,3747)(6684,3739)(6764,3730)
	(6841,3722)(6916,3714)(6988,3705)
	(7057,3697)(7124,3689)(7188,3680)
	(7251,3671)(7311,3663)(7369,3654)
	(7424,3645)(7478,3636)(7530,3626)
	(7580,3617)(7629,3607)(7675,3598)
	(7721,3587)(7765,3577)(7807,3567)
	(7848,3556)(7888,3545)(7927,3534)
	(7965,3522)(8024,3503)(8080,3484)
	(8133,3463)(8185,3441)(8234,3419)
	(8281,3395)(8326,3370)(8369,3344)
	(8411,3316)(8451,3287)(8490,3256)
	(8528,3223)(8565,3188)(8601,3152)
	(8636,3113)(8671,3073)(8705,3031)
	(8738,2987)(8770,2941)(8802,2894)
	(8833,2846)(8863,2798)(8891,2749)
	(8919,2700)(8945,2652)(8970,2606)
	(8992,2561)(9013,2519)(9032,2481)
	(9048,2446)(9062,2416)(9074,2389)
	(9084,2368)(9091,2350)(9097,2337)(9105,2318)
\blacken\path(8956.568,2515.909)(9105.000,2318.000)(9067.164,2562.476)(9039.806,2472.835)(8956.568,2515.909)
\path(7485,1468)(7488,1469)(7493,1470)
	(7504,1472)(7519,1475)(7541,1480)
	(7569,1486)(7603,1494)(7641,1502)
	(7684,1512)(7729,1522)(7775,1533)
	(7822,1544)(7869,1555)(7914,1566)
	(7958,1577)(8000,1588)(8039,1598)
	(8077,1608)(8112,1618)(8145,1628)
	(8176,1637)(8206,1647)(8234,1657)
	(8261,1667)(8287,1677)(8313,1687)
	(8338,1697)(8362,1708)(8387,1720)
	(8411,1732)(8435,1744)(8460,1757)
	(8485,1772)(8511,1787)(8538,1803)
	(8566,1820)(8595,1839)(8626,1859)
	(8657,1880)(8690,1902)(8724,1926)
	(8758,1950)(8792,1974)(8826,1998)
	(8859,2021)(8889,2043)(8916,2063)
	(8940,2081)(8960,2096)(8976,2107)(9000,2125)
\blacken\path(8844.000,1933.000)(9000.000,2125.000)(8772.000,2029.000)(8865.600,2024.200)(8844.000,1933.000)
\path(5685,2143)(5687,2143)(5690,2141)
	(5698,2140)(5709,2136)(5725,2132)
	(5746,2126)(5773,2119)(5805,2110)
	(5843,2100)(5886,2088)(5934,2076)
	(5986,2062)(6042,2047)(6100,2032)
	(6161,2017)(6223,2001)(6285,1986)
	(6347,1970)(6409,1955)(6470,1941)
	(6530,1927)(6588,1914)(6644,1902)
	(6699,1890)(6751,1879)(6802,1870)
	(6851,1860)(6898,1852)(6944,1845)
	(6988,1838)(7030,1832)(7072,1827)
	(7112,1823)(7152,1820)(7191,1817)
	(7229,1814)(7267,1813)(7305,1812)
	(7343,1811)(7380,1812)(7418,1812)
	(7456,1814)(7494,1816)(7533,1818)
	(7572,1822)(7612,1826)(7653,1831)
	(7695,1836)(7738,1843)(7783,1850)
	(7830,1858)(7878,1867)(7927,1876)
	(7979,1887)(8033,1898)(8088,1910)
	(8145,1923)(8203,1936)(8263,1950)
	(8323,1965)(8384,1980)(8445,1995)
	(8506,2010)(8565,2026)(8622,2040)
	(8676,2055)(8727,2068)(8774,2081)
	(8816,2092)(8853,2102)(8884,2110)
	(8911,2118)(8931,2123)(8947,2128)(8970,2134)
\blacken\path(8752.917,2015.362)(8970.000,2134.000)(8722.627,2131.476)(8807.440,2091.593)(8752.917,2015.362)
\path(229,2019)(230,2018)(232,2015)
	(236,2009)(243,2001)(252,1989)
	(264,1973)(280,1952)(299,1927)
	(321,1898)(347,1865)(376,1828)
	(408,1788)(442,1744)(478,1698)
	(517,1651)(556,1602)(597,1552)
	(638,1503)(679,1453)(721,1405)
	(762,1357)(802,1311)(843,1266)
	(882,1223)(921,1182)(958,1143)
	(995,1105)(1032,1070)(1068,1036)
	(1103,1004)(1137,973)(1172,945)
	(1206,917)(1240,892)(1274,867)
	(1308,844)(1342,822)(1376,801)
	(1411,780)(1447,761)(1483,742)
	(1517,726)(1551,710)(1586,695)
	(1622,680)(1658,666)(1696,652)
	(1735,639)(1775,626)(1817,613)
	(1860,601)(1905,589)(1952,577)
	(2001,566)(2052,554)(2105,543)
	(2160,532)(2217,522)(2277,511)
	(2339,500)(2403,490)(2469,480)
	(2536,469)(2606,459)(2676,449)
	(2747,440)(2819,430)(2890,421)
	(2961,412)(3030,403)(3098,395)
	(3162,388)(3223,381)(3280,374)
	(3333,368)(3380,363)(3422,358)
	(3458,355)(3489,351)(3514,349)
	(3534,347)(3548,345)(3570,343)
\blacken\path(3325.553,304.975)(3570.000,343.000)(3336.418,424.482)(3402.690,358.210)(3325.553,304.975)
\path(4785,3043)(4786,3043)(4789,3044)
	(4795,3046)(4804,3050)(4817,3054)
	(4835,3060)(4858,3068)(4886,3077)
	(4919,3089)(4958,3102)(5002,3117)
	(5052,3133)(5106,3152)(5166,3172)
	(5229,3193)(5295,3215)(5365,3238)
	(5437,3262)(5511,3286)(5586,3311)
	(5661,3336)(5737,3361)(5813,3385)
	(5887,3409)(5961,3433)(6034,3456)
	(6105,3479)(6174,3501)(6242,3522)
	(6307,3542)(6371,3562)(6433,3580)
	(6492,3598)(6550,3615)(6606,3632)
	(6660,3647)(6713,3662)(6764,3676)
	(6814,3689)(6862,3702)(6910,3714)
	(6956,3726)(7001,3737)(7046,3747)
	(7089,3757)(7133,3767)(7175,3776)
	(7218,3785)(7260,3793)(7312,3803)
	(7365,3813)(7417,3822)(7469,3830)
	(7521,3838)(7574,3846)(7626,3853)
	(7679,3859)(7731,3865)(7784,3871)
	(7837,3875)(7889,3879)(7942,3883)
	(7994,3885)(8046,3887)(8098,3888)
	(8149,3889)(8199,3888)(8249,3887)
	(8298,3885)(8346,3882)(8393,3878)
	(8439,3874)(8484,3869)(8527,3862)
	(8569,3855)(8610,3848)(8649,3839)
	(8687,3829)(8723,3819)(8758,3808)
	(8790,3796)(8822,3784)(8852,3771)
	(8880,3757)(8907,3742)(8932,3726)
	(8956,3710)(8979,3693)(9000,3675)
	(9022,3654)(9043,3631)(9062,3608)
	(9079,3583)(9095,3556)(9109,3527)
	(9122,3496)(9133,3463)(9144,3427)
	(9153,3389)(9160,3349)(9167,3305)
	(9172,3259)(9176,3210)(9180,3159)
	(9182,3104)(9183,3048)(9183,2989)
	(9183,2929)(9182,2867)(9180,2806)
	(9178,2745)(9175,2685)(9172,2628)
	(9168,2574)(9165,2525)(9162,2480)
	(9159,2442)(9157,2410)(9154,2384)
	(9153,2364)(9150,2334)
\blacken\path(9114.179,2578.779)(9150.000,2334.000)(9233.583,2566.839)(9166.717,2501.166)(9114.179,2578.779)
\put(105,2098){\makebox(0,0)[lb]{\smash{{\SetFigFont{8}{9.6}{\rmdefault}{\mddefault}{\updefault}$s$}}}}
\put(9105,2098){\makebox(0,0)[lb]{\smash{{\SetFigFont{8}{9.6}{\rmdefault}{\mddefault}{\updefault}$t$}}}}
\put(3660,3898){\makebox(0,0)[lb]{\smash{{\SetFigFont{8}{9.6}{\rmdefault}{\mddefault}{\updefault}$c_0$}}}}
\put(3640,2098){\makebox(0,0)[lb]{\smash{{\SetFigFont{8}{9.6}{\rmdefault}{\mddefault}{\updefault}$c_{30}$}}}}
\put(3660,298){\makebox(0,0)[lb]{\smash{{\SetFigFont{8}{9.6}{\rmdefault}{\mddefault}{\updefault}$c_3$}}}}
\put(1860,2098){\makebox(0,0)[lb]{\smash{{\SetFigFont{8}{9.6}{\rmdefault}{\mddefault}{\updefault}$c_1$}}}}
\put(5460,2098){\makebox(0,0)[lb]{\smash{{\SetFigFont{8}{9.6}{\rmdefault}{\mddefault}{\updefault}$c_2$}}}}
\put(4525,2998){\makebox(0,0)[lb]{\smash{{\SetFigFont{8}{9.6}{\rmdefault}{\mddefault}{\updefault}$c_{20}$}}}}
\put(7235,2773){\makebox(0,0)[lb]{\smash{{\SetFigFont{8}{9.6}{\rmdefault}{\mddefault}{\updefault}$c_{21}$}}}}
\put(7235,1423){\makebox(0,0)[lb]{\smash{{\SetFigFont{8}{9.6}{\rmdefault}{\mddefault}{\updefault}$c_{23}$}}}}
\put(7235,298){\makebox(0,0)[lb]{\smash{{\SetFigFont{8}{9.6}{\rmdefault}{\mddefault}{\updefault}$c_{31}$}}}}
\put(2175,1783){\makebox(0,0)[lb]{\smash{{\SetFigFont{8}{9.6}{\rmdefault}{\mddefault}{\updefault}$(1,0)$}}}}
\put(1140,478){\makebox(0,0)[lb]{\smash{{\SetFigFont{8}{9.6}{\rmdefault}{\mddefault}{\updefault}$(1,0)$}}}}
\put(1185,3718){\makebox(0,0)[lb]{\smash{{\SetFigFont{8}{9.6}{\rmdefault}{\mddefault}{\updefault}$(1,0)$}}}}
\put(915,2233){\makebox(0,0)[lb]{\smash{{\SetFigFont{8}{9.6}{\rmdefault}{\mddefault}{\updefault}$(0,0)$}}}}
\put(2085,973){\makebox(0,0)[lb]{\smash{{\SetFigFont{8}{9.6}{\rmdefault}{\mddefault}{\updefault}$(3,0)$}}}}
\put(3300,2683){\makebox(0,0)[lb]{\smash{{\SetFigFont{8}{9.6}{\rmdefault}{\mddefault}{\updefault}$(1,1)$}}}}
\put(4875,2233){\makebox(0,0)[lb]{\smash{{\SetFigFont{8}{9.6}{\rmdefault}{\mddefault}{\updefault}$(1,1)$}}}}
\put(4380,3403){\makebox(0,0)[lb]{\smash{{\SetFigFont{8}{9.6}{\rmdefault}{\mddefault}{\updefault}$(1,1)$}}}}
\put(5865,1603){\makebox(0,0)[lb]{\smash{{\SetFigFont{8}{9.6}{\rmdefault}{\mddefault}{\updefault}$(1,1)$}}}}
\put(6225,2188){\makebox(0,0)[lb]{\smash{{\SetFigFont{8}{9.6}{\rmdefault}{\mddefault}{\updefault}$(1,1)$}}}}
\put(5775,73){\makebox(0,0)[lb]{\smash{{\SetFigFont{8}{9.6}{\rmdefault}{\mddefault}{\updefault}$(1,1)$}}}}
\put(7710,883){\makebox(0,0)[lb]{\smash{{\SetFigFont{8}{9.6}{\rmdefault}{\mddefault}{\updefault}$(T,65)$}}}}
\put(7395,1918){\makebox(0,0)[lb]{\smash{{\SetFigFont{8}{9.6}{\rmdefault}{\mddefault}{\updefault}$(T,65)$}}}}
\put(5775,2683){\makebox(0,0)[lb]{\smash{{\SetFigFont{8}{9.6}{\rmdefault}{\mddefault}{\updefault}$(T,65)$}}}}
\put(8340,3403){\makebox(0,0)[lb]{\smash{{\SetFigFont{8}{9.6}{\rmdefault}{\mddefault}{\updefault}$(T,0)$}}}}
\put(7890,3988){\makebox(0,0)[lb]{\smash{{\SetFigFont{8}{9.6}{\rmdefault}{\mddefault}{\updefault}$(1,65)$}}}}
\put(6630,3088){\makebox(0,0)[lb]{\smash{{\SetFigFont{8}{9.6}{\rmdefault}{\mddefault}{\updefault}$(1,65)$}}}}
\put(7620,2773){\makebox(0,0)[lb]{\smash{{\SetFigFont{8}{9.6}{\rmdefault}{\mddefault}{\updefault}$(1,65)$}}}}
\put(7485,1603){\makebox(0,0)[lb]{\smash{{\SetFigFont{8}{9.6}{\rmdefault}{\mddefault}{\updefault}$(1,65)$}}}}
\put(8610,388){\makebox(0,0)[lb]{\smash{{\SetFigFont{8}{9.6}{\rmdefault}{\mddefault}{\updefault}$(1,65)$}}}}
\end{picture}
}

%% file: constructive-microbribery-even-llull-flownetwork.eepic
\setlength{\unitlength}{0.00087489in}
\begingroup\makeatletter\ifx\SetFigFont\undefined%
\gdef\SetFigFont#1#2#3#4#5{%
  \reset@font\fontsize{#1}{#2pt}%
  \fontfamily{#3}\fontseries{#4}\fontshape{#5}%
  \selectfont}%
\fi\endgroup%
{\renewcommand{\dashlinestretch}{30}
\begin{picture}(6150,4026)(0,-10)
\thicklines
\put(150,1950){\ellipse{270}{270}}
\put(1500,2850){\ellipse{270}{270}}
\put(2400,1950){\ellipse{270}{270}}
\put(2400,3750){\ellipse{270}{270}}
\put(2400,150){\ellipse{270}{270}}
\put(4650,1050){\ellipse{270}{270}}
\put(4650,150){\ellipse{270}{270}}
\put(4650,2850){\ellipse{270}{270}}
\put(4650,3750){\ellipse{270}{270}}
\put(6000,1950){\ellipse{270}{270}}
\put(3300,1050){\ellipse{270}{270}}
\put(3980,2405){\ellipse{270}{270}}
\path(150,1815)(151,1813)(153,1810)
	(156,1803)(162,1792)(169,1777)
	(180,1756)(193,1731)(208,1700)
	(227,1665)(247,1625)(270,1581)
	(295,1534)(321,1484)(349,1433)
	(377,1380)(406,1327)(436,1274)
	(465,1222)(494,1172)(522,1122)
	(550,1075)(578,1030)(605,987)
	(631,946)(657,907)(682,870)
	(706,835)(730,803)(754,772)
	(778,743)(801,716)(825,690)
	(848,665)(872,642)(896,619)
	(920,598)(945,577)(972,557)
	(999,537)(1027,518)(1056,499)
	(1086,482)(1117,464)(1149,448)
	(1183,432)(1219,416)(1257,401)
	(1296,386)(1338,371)(1382,356)
	(1428,341)(1476,326)(1527,312)
	(1579,297)(1633,283)(1688,269)
	(1743,255)(1799,242)(1854,229)
	(1908,216)(1959,205)(2007,194)
	(2051,185)(2090,176)(2125,169)
	(2153,163)(2176,159)(2194,155)(2220,150)
\blacken\path(1972.988,136.403)(2220.000,150.000)(1995.649,254.244)(2055.023,181.726)(1972.988,136.403)
\path(2400,285)(2400,1770)
\blacken\path(2460.000,1530.000)(2400.000,1770.000)(2340.000,1530.000)(2400.000,1602.000)(2460.000,1530.000)
\path(2400,3615)(2400,2130)
\blacken\path(2340.000,2370.000)(2400.000,2130.000)(2460.000,2370.000)(2400.000,2298.000)(2340.000,2370.000)
\path(2535,3750)(4470,3750)
\blacken\path(4230.000,3690.000)(4470.000,3750.000)(4230.000,3810.000)(4302.000,3750.000)(4230.000,3690.000)
\path(3435,1050)(4470,1050)
\blacken\path(4230.000,990.000)(4470.000,1050.000)(4230.000,1110.000)(4302.000,1050.000)(4230.000,990.000)
\path(2535,150)(4470,150)
\blacken\path(4230.000,90.000)(4470.000,150.000)(4230.000,210.000)(4302.000,150.000)(4230.000,90.000)
\path(249,1860)(3120,1050)
\blacken\path(2872.725,1057.422)(3120.000,1050.000)(2905.309,1172.913)(2958.312,1095.617)(2872.725,1057.422)
\path(249,2040)(1383,2719)
\blacken\path(1207.913,2544.230)(1383.000,2719.000)(1146.267,2647.186)(1238.863,2632.696)(1207.913,2544.230)
\path(1599,2940)(1601,2940)(1605,2941)
	(1612,2943)(1623,2946)(1640,2949)
	(1661,2954)(1688,2960)(1721,2967)
	(1759,2976)(1802,2985)(1849,2995)
	(1900,3006)(1954,3017)(2010,3029)
	(2067,3041)(2125,3053)(2184,3065)
	(2241,3076)(2298,3087)(2354,3098)
	(2408,3108)(2460,3117)(2511,3126)
	(2560,3134)(2606,3141)(2651,3148)
	(2695,3154)(2737,3159)(2777,3164)
	(2816,3168)(2854,3171)(2892,3174)
	(2928,3176)(2964,3178)(3000,3179)
	(3035,3180)(3071,3180)(3106,3180)
	(3141,3179)(3177,3178)(3213,3176)
	(3249,3174)(3286,3171)(3324,3168)
	(3363,3164)(3403,3159)(3444,3154)
	(3487,3148)(3532,3141)(3578,3134)
	(3626,3126)(3676,3117)(3728,3108)
	(3781,3098)(3836,3087)(3892,3076)
	(3949,3065)(4006,3053)(4063,3041)
	(4120,3029)(4175,3017)(4228,3006)
	(4278,2995)(4325,2985)(4367,2976)
	(4404,2967)(4436,2960)(4463,2954)
	(4484,2949)(4500,2946)(4524,2940)
\blacken\path(4276.614,2940.000)(4524.000,2940.000)(4305.718,3056.417)(4361.016,2980.746)(4276.614,2940.000)
\path(4767,2760)(5865,2040)
\blacken\path(5631.400,2121.431)(5865.000,2040.000)(5697.203,2221.780)(5724.511,2132.124)(5631.400,2121.431)
\path(4771,1136)(5865,1860)
\blacken\path(5697.972,1677.513)(5865.000,1860.000)(5631.746,1777.584)(5724.901,1767.284)(5697.972,1677.513)
\path(3205,954)(2526,276)
\blacken\path(2653.436,488.038)(2526.000,276.000)(2738.226,403.123)(2644.881,394.706)(2653.436,488.038)
\path(4785,3750)(4787,3749)(4792,3748)
	(4801,3745)(4816,3741)(4836,3736)
	(4861,3728)(4892,3719)(4928,3708)
	(4968,3696)(5011,3682)(5056,3667)
	(5102,3651)(5149,3635)(5194,3618)
	(5239,3601)(5281,3584)(5322,3567)
	(5360,3549)(5396,3532)(5429,3514)
	(5461,3496)(5490,3478)(5517,3459)
	(5542,3439)(5566,3419)(5588,3398)
	(5609,3376)(5629,3353)(5648,3329)
	(5664,3306)(5679,3283)(5694,3258)
	(5709,3232)(5723,3205)(5736,3175)
	(5750,3144)(5763,3111)(5776,3075)
	(5789,3037)(5802,2997)(5815,2954)
	(5828,2908)(5841,2859)(5854,2809)
	(5867,2755)(5880,2700)(5893,2644)
	(5906,2586)(5919,2528)(5931,2472)
	(5942,2416)(5953,2364)(5963,2315)
	(5972,2271)(5979,2233)(5985,2200)
	(5990,2174)(5994,2154)(6000,2123)
\blacken\path(5895.488,2347.226)(6000.000,2123.000)(6013.302,2370.028)(5968.076,2287.939)(5895.488,2347.226)
\path(4785,150)(4787,151)(4792,152)
	(4801,155)(4816,159)(4836,164)
	(4861,172)(4892,181)(4928,192)
	(4968,204)(5011,218)(5056,233)
	(5102,249)(5149,265)(5194,282)
	(5239,299)(5281,316)(5322,333)
	(5360,351)(5396,368)(5429,386)
	(5461,404)(5490,422)(5517,441)
	(5542,461)(5566,481)(5588,502)
	(5609,524)(5629,547)(5648,571)
	(5664,594)(5679,617)(5694,642)
	(5709,668)(5723,695)(5736,725)
	(5750,756)(5763,789)(5776,825)
	(5789,863)(5802,903)(5815,946)
	(5828,992)(5841,1041)(5854,1091)
	(5867,1145)(5880,1200)(5893,1256)
	(5906,1314)(5919,1372)(5931,1428)
	(5942,1484)(5953,1536)(5963,1585)
	(5972,1629)(5979,1667)(5985,1700)
	(5990,1726)(5994,1746)(6000,1777)
\blacken\path(6013.302,1529.972)(6000.000,1777.000)(5895.488,1552.774)(5968.076,1612.061)(6013.302,1529.972)
\path(2310,248)(1545,2679)
\blacken\path(1674.275,2468.078)(1545.000,2679.000)(1559.809,2432.057)(1595.429,2518.747)(1674.275,2468.078)
\path(150,2085)(151,2087)(152,2091)
	(155,2099)(159,2112)(166,2130)
	(174,2153)(185,2182)(198,2217)
	(212,2257)(229,2301)(248,2350)
	(268,2402)(289,2456)(311,2511)
	(333,2567)(356,2623)(380,2678)
	(403,2731)(426,2783)(449,2833)
	(472,2880)(495,2925)(517,2968)
	(539,3008)(561,3046)(583,3081)
	(605,3115)(627,3146)(650,3175)
	(672,3203)(696,3230)(719,3255)
	(744,3278)(769,3301)(795,3322)
	(820,3342)(847,3361)(874,3379)
	(902,3397)(931,3414)(962,3430)
	(995,3446)(1029,3461)(1065,3477)
	(1103,3491)(1143,3506)(1186,3520)
	(1231,3534)(1278,3548)(1328,3562)
	(1379,3576)(1434,3590)(1490,3603)
	(1548,3617)(1607,3630)(1667,3644)
	(1727,3656)(1787,3669)(1845,3681)
	(1902,3692)(1955,3702)(2005,3711)
	(2050,3720)(2090,3727)(2125,3733)
	(2154,3739)(2177,3743)(2194,3746)(2220,3750)
\blacken\path(1991.914,3654.204)(2220.000,3750.000)(1973.667,3772.809)(2053.954,3724.454)(1991.914,3654.204)
\path(2535,1950)(5820,1950)
\blacken\path(5580.000,1890.000)(5820.000,1950.000)(5580.000,2010.000)(5652.000,1950.000)(5580.000,1890.000)
\path(3229,1162)(3228,1164)(3227,1168)
	(3224,1176)(3219,1189)(3212,1207)
	(3203,1230)(3192,1260)(3178,1296)
	(3162,1337)(3144,1383)(3124,1434)
	(3102,1489)(3079,1546)(3055,1606)
	(3031,1667)(3006,1728)(2981,1789)
	(2956,1849)(2931,1908)(2907,1964)
	(2883,2018)(2860,2070)(2837,2119)
	(2815,2166)(2794,2210)(2774,2252)
	(2754,2290)(2734,2327)(2715,2361)
	(2696,2393)(2677,2423)(2659,2451)
	(2641,2478)(2622,2502)(2604,2526)
	(2586,2548)(2567,2569)(2543,2594)
	(2518,2617)(2493,2638)(2467,2658)
	(2440,2677)(2412,2694)(2382,2710)
	(2351,2724)(2318,2738)(2283,2751)
	(2246,2762)(2206,2773)(2165,2783)
	(2122,2793)(2077,2802)(2032,2809)
	(1985,2817)(1939,2823)(1894,2829)
	(1852,2834)(1812,2838)(1777,2842)
	(1747,2844)(1722,2846)(1702,2848)(1672,2850)
\blacken\path(1915.460,2893.903)(1672.000,2850.000)(1907.477,2774.168)(1839.628,2838.825)(1915.460,2893.903)
\path(3390,1162)(3390,1165)(3392,1171)
	(3394,1182)(3397,1199)(3401,1221)
	(3407,1250)(3413,1284)(3420,1322)
	(3429,1363)(3437,1406)(3446,1450)
	(3455,1492)(3463,1534)(3472,1574)
	(3481,1611)(3489,1646)(3497,1679)
	(3504,1709)(3512,1737)(3520,1764)
	(3527,1788)(3535,1811)(3543,1833)
	(3551,1854)(3559,1874)(3569,1897)
	(3579,1919)(3591,1941)(3603,1962)
	(3616,1984)(3630,2006)(3646,2028)
	(3662,2051)(3681,2075)(3701,2100)
	(3722,2125)(3743,2150)(3765,2176)
	(3786,2200)(3806,2222)(3824,2241)
	(3839,2257)(3863,2283)
\blacken\path(3744.301,2065.950)(3863.000,2283.000)(3656.125,2147.344)(3749.049,2159.553)(3744.301,2065.950)
\path(4555,1162)(4555,1165)(4553,1171)
	(4551,1182)(4548,1199)(4544,1221)
	(4539,1250)(4533,1284)(4526,1322)
	(4518,1363)(4510,1406)(4502,1450)
	(4493,1492)(4485,1534)(4476,1574)
	(4468,1611)(4461,1646)(4453,1679)
	(4446,1709)(4438,1737)(4431,1764)
	(4424,1788)(4417,1811)(4409,1833)
	(4402,1854)(4394,1874)(4385,1897)
	(4375,1919)(4364,1941)(4353,1962)
	(4341,1984)(4327,2006)(4313,2028)
	(4297,2051)(4280,2075)(4261,2100)
	(4242,2125)(4222,2150)(4201,2176)
	(4181,2200)(4163,2222)(4146,2241)
	(4133,2257)(4110,2283)
\blacken\path(4313.958,2142.995)(4110.000,2283.000)(4224.078,2063.486)(4221.312,2157.169)(4313.958,2142.995)
\path(3880,2494)(2535,3615)
\blacken\path(2757.776,3507.433)(2535.000,3615.000)(2680.947,3415.252)(2664.053,3507.440)(2757.776,3507.433)
\put(105,1905){\makebox(0,0)[lb]{\smash{{\SetFigFont{12}{14.4}{\rmdefault}{\mddefault}{\updefault}$s$}}}}
\put(5955,1905){\makebox(0,0)[lb]{\smash{{\SetFigFont{12}{14.4}{\rmdefault}{\mddefault}{\updefault}$t$}}}}
\put(2310,3705){\makebox(0,0)[lb]{\smash{{\SetFigFont{12}{14.4}{\rmdefault}{\mddefault}{\updefault}$c_0$}}}}
\put(1410,2805){\makebox(0,0)[lb]{\smash{{\SetFigFont{12}{14.4}{\rmdefault}{\mddefault}{\updefault}$c_1$}}}}
\put(2285,1905){\makebox(0,0)[lb]{\smash{{\SetFigFont{12}{14.4}{\rmdefault}{\mddefault}{\updefault}$c_{03}$}}}}
\put(3210,1005){\makebox(0,0)[lb]{\smash{{\SetFigFont{12}{14.4}{\rmdefault}{\mddefault}{\updefault}$c_2$}}}}
\put(2310,105){\makebox(0,0)[lb]{\smash{{\SetFigFont{12}{14.4}{\rmdefault}{\mddefault}{\updefault}$c_3$}}}}
\put(4560,3705){\makebox(0,0)[lb]{\smash{{\SetFigFont{12}{14.4}{\rmdefault}{\mddefault}{\updefault}$c'_0$}}}}
\put(4560,2805){\makebox(0,0)[lb]{\smash{{\SetFigFont{12}{14.4}{\rmdefault}{\mddefault}{\updefault}$c'_1$}}}}
\put(4560,1005){\makebox(0,0)[lb]{\smash{{\SetFigFont{12}{14.4}{\rmdefault}{\mddefault}{\updefault}$c'_2$}}}}
\put(4560,105){\makebox(0,0)[lb]{\smash{{\SetFigFont{12}{14.4}{\rmdefault}{\mddefault}{\updefault}$c'_3$}}}}
\put(735,2085){\makebox(0,0)[lb]{\smash{{\SetFigFont{12}{14.4}{\rmdefault}{\mddefault}{\updefault}$(0,0)$}}}}
\put(645,285){\makebox(0,0)[lb]{\smash{{\SetFigFont{12}{14.4}{\rmdefault}{\mddefault}{\updefault}$(2,0)$}}}}
\put(1635,825){\makebox(0,0)[lb]{\smash{{\SetFigFont{12}{14.4}{\rmdefault}{\mddefault}{\updefault}$(1,2)$}}}}
\put(2445,735){\makebox(0,0)[lb]{\smash{{\SetFigFont{12}{14.4}{\rmdefault}{\mddefault}{\updefault}$(1,1)$}}}}
\put(2445,2850){\makebox(0,0)[lb]{\smash{{\SetFigFont{12}{14.4}{\rmdefault}{\mddefault}{\updefault}$(1,1)$}}}}
\put(1635,3075){\makebox(0,0)[lb]{\smash{{\SetFigFont{12}{14.4}{\rmdefault}{\mddefault}{\updefault}$(T,0)$}}}}
\put(3075,3840){\makebox(0,0)[lb]{\smash{{\SetFigFont{12}{14.4}{\rmdefault}{\mddefault}{\updefault}$(T,0)$}}}}
\put(3525,1140){\makebox(0,0)[lb]{\smash{{\SetFigFont{12}{14.4}{\rmdefault}{\mddefault}{\updefault}$(T,0)$}}}}
\put(3165,240){\makebox(0,0)[lb]{\smash{{\SetFigFont{12}{14.4}{\rmdefault}{\mddefault}{\updefault}$(T,0)$}}}}
\put(3075,600){\makebox(0,0)[lb]{\smash{{\SetFigFont{12}{14.4}{\rmdefault}{\mddefault}{\updefault}$(1,2)$}}}}
\put(5145,3165){\makebox(0,0)[lb]{\smash{{\SetFigFont{12}{14.4}{\rmdefault}{\mddefault}{\updefault}$(T,0)$}}}}
\put(5010,2670){\makebox(0,0)[lb]{\smash{{\SetFigFont{12}{14.4}{\rmdefault}{\mddefault}{\updefault}$(T,65)$}}}}
\put(5010,1095){\makebox(0,0)[lb]{\smash{{\SetFigFont{12}{14.4}{\rmdefault}{\mddefault}{\updefault}$(T,65)$}}}}
\put(4920,465){\makebox(0,0)[lb]{\smash{{\SetFigFont{12}{14.4}{\rmdefault}{\mddefault}{\updefault}$(T,65)$}}}}
\put(465,3480){\makebox(0,0)[lb]{\smash{{\SetFigFont{12}{14.4}{\rmdefault}{\mddefault}{\updefault}$(2,0)$}}}}
\put(735,1410){\makebox(0,0)[lb]{\smash{{\SetFigFont{12}{14.4}{\rmdefault}{\mddefault}{\updefault}$(3,0)$}}}}
\put(2715,2490){\makebox(0,0)[lb]{\smash{{\SetFigFont{12}{14.4}{\rmdefault}{\mddefault}{\updefault}$(1,2)$}}}}
\put(4695,2040){\makebox(0,0)[lb]{\smash{{\SetFigFont{12}{14.4}{\rmdefault}{\mddefault}{\updefault}$(1,65)$}}}}
\put(3570,1500){\makebox(0,0)[lb]{\smash{{\SetFigFont{12}{14.4}{\rmdefault}{\mddefault}{\updefault}$(1,2)$}}}}
\put(3705,2670){\makebox(0,0)[lb]{\smash{{\SetFigFont{12}{14.4}{\rmdefault}{\mddefault}{\updefault}$(1,0)$}}}}
\put(4560,1500){\makebox(0,0)[lb]{\smash{{\SetFigFont{12}{14.4}{\rmdefault}{\mddefault}{\updefault}$(1,1)$}}}}
\put(3860,2355){\makebox(0,0)[lb]{\smash{{\SetFigFont{12}{14.4}{\rmdefault}{\mddefault}{\updefault}$c_{02}$}}}}
\end{picture}
}

%% file: main-corr-withoutcomments.bbl
\begin{thebibliography}{GMHS99}

\bibitem[AB00]{aus-ban:b:positive-political-I}
D.~{Austen-Smith} and J.~Banks.
\newblock {\em Positive Political Theory {I}: {Collective} Preference}.
\newblock Univ.~of Mich.~Press, 2000.

\bibitem[AMO93]{ahu-mag-orl:b:flows}
R.~Ahuja, T.~Magnanti, and J.~Orlin.
\newblock {\em Network Flows: Theory, Algorithms, and Applications}.
\newblock Prentice-Hall, 1993.

\bibitem[Arr63]{arr:b:polsci:social-choice}
K.~Arrow.
\newblock {\em Social Choice and Individual Values}.
\newblock John Wiley and Sons, 1951 (revised editon, 1963).

\bibitem[AT07]{alt-ten:c:axiomatic-personalized-ranking}
A.~Altman and M.~Tennenholtz.
\newblock An axiomatic approach to personalized ranking systems.
\newblock In {\em Proceedings of the 20th International Joint Conference on
  Artificial Intelligence}, pages 1187--1192. AAAI Press, January 2007.

\bibitem[BGN08]{bet-guo-nie:c:dodgson-parametrized}
N.~Betzler, J.~Guo, and R.~Niedermeier.
\newblock Parameterized computational complexity of {D}odgson and {Y}oung
  elections.
\newblock In {\em Proceedings of the 11th Scandinavian Workshop on Algorithm
  Theory}, pages 402--413. Springer-Verlag {\it Lecture Notes in Computer
  Science \#5124}, 2008.

\bibitem[BO91]{bar-oli:j:polsci:strategic-voting}
J.~{{Bartholdi}}, III and J.~Orlin.
\newblock Single transferable vote resists strategic voting.
\newblock {\em Social Choice and Welfare}, 8(4):341--354, 1991.

\bibitem[BS06]{bra-san:j:critical-strategies-under-approval}
S.~Brams and R.~Sanver.
\newblock Critical strategies under approval voting: {W}ho gets ruled in and
  ruled out.
\newblock {\em Electoral Studies}, 25(2):287--305, 2006.

\bibitem[BTT89a]{bar-tov-tri:j:manipulating}
J.~{{Bartholdi}}, III, C.~Tovey, and M.~Trick.
\newblock The computational difficulty of manipulating an election.
\newblock {\em Social Choice and Welfare}, 6(3):227--241, 1989.

\bibitem[BTT89b]{bar-tov-tri:j:who-won}
J.~{{Bartholdi}}, III, C.~Tovey, and M.~Trick.
\newblock Voting schemes for which it can be difficult to tell who won the
  election.
\newblock {\em Social Choice and Welfare}, 6(2):157--165, 1989.

\bibitem[BTT92]{bar-tov-tri:j:control}
J.~{{Bartholdi}}, III, C.~Tovey, and M.~Trick.
\newblock How hard is it to control an election?
\newblock {\em Mathematical and Computer Modeling}, 16(8/9):27--40, 1992.

\bibitem[BU08]{bet-uhl:c:parameterized-complecity-candidate-control}
N.~Betzler and J.~Uhlmann.
\newblock Parameterized complexity of candidate control in elections and
  related digraph problems.
\newblock In {\em Proceedings of the 2nd Annual International Conference on
  Combinatorial Optimization and Applications}, pages 43--53. Springer-Verlag
  {\it Lecture Notes in Computer Science \#5156}, July 2008.

\bibitem[CLRS01]{cor-lei-riv-ste:b:algorithms-second-edition}
T.~Cormen, C.~Leiserson, R.~Rivest, and C.~Stein.
\newblock {\em Introduction to Algorithms}.
\newblock MIT Press/McGraw Hill, second edition, 2001.

\bibitem[Con85]{con:b:condorcet-paradox}
{J.-A.-N.~de Caritat, Marquis de} Condorcet.
\newblock {\em Essai sur l'Application de L'Analyse \`{a} la Probabilit\'{e}
  des D\'{e}cisions Rendues \`{a} la Pluralit\'{e} des Voix}.
\newblock 1785.
\newblock Facsimile reprint of original published in Paris, 1972, by the
  Imprimerie Royale.

\bibitem[Cop51]{cop:m:copeland}
A.~Copeland.
\newblock A ``reasonable'' social welfare function.
\newblock Mimeographed notes from a Seminar on Applications of Mathematics to
  the Social Sciences, University of Michigan, 1951.

\bibitem[CS03]{con-san:c:voting-tweaks}
V.~Conitzer and T.~Sandholm.
\newblock Universal voting protocol tweaks to make manipulation hard.
\newblock In {\em Proceedings of the 18th International Joint Conference on
  Artificial Intelligence}, pages 781--788. Morgan Kaufmann, August 2003.

\bibitem[CS06]{con-san:c:nonexistence}
V.~Conitzer and T.~Sandholm.
\newblock Nonexistence of voting rules that are usually hard to
  manipulate\typeout{MINOR PANIC: MISSING PAGE NUMBERS CON-SAN}.
\newblock In {\em Proceedings of the 21st National Conference on Artificial
  Intelligence}, pages 627--634. AAAI Press, July 2006.

\bibitem[CSL07]{con-lan-san:j:when-hard-to-manipulate}
V.~Conitzer, T.~Sandholm, and J.~Lang.
\newblock When are elections with few candidates hard to manipulate?
\newblock {\em Journal of the ACM}, 54(3):Article~14, 2007.

\bibitem[DKNS01]{dwo-kum-nao-siv:c:rank-aggregation}
C.~Dwork, R.~Kumar, M.~Naor, and D.~Sivakumar.
\newblock Rank aggregation methods for the web.
\newblock In {\em Proceedings of the 10th International World Wide Web
  Conference}, pages 613--622. ACM Press, March 2001.

\bibitem[Dow03]{dow:c:parameterized-survey}
R.~Downey.
\newblock Parameterized complexity for the skeptic.
\newblock In {\em Proceedings of the 18th Annual IEEE Conference on
  Computational Complexity}, pages 147--168. IEEE Computer Society Press, July
  2003.

\bibitem[DS00]{dug-sch:j:polsci:gibbard}
J.~Duggan and T.~Schwartz.
\newblock Strategic manipulability without resoluteness or shared beliefs:
  {Gibbard}--{Satterthwaite} generalized.
\newblock {\em Social Choice and Welfare}, 17(1):85--93, 2000.

\bibitem[EHRS07]{erd-hem-rot-spa:c:lobbying}
G.~Erd\'{e}lyi, L.~Hemaspaandra, J.~Rothe, and H.~Spakowski.
\newblock On approximating optimal weighted lobbying, and frequency of
  correctness versus average-case polynomial time.
\newblock In {\em Proceedings of the 16th International Symposium on
  Fundamentals of Computation Theory}, pages 300--311. Springer-Verlag {\it
  Lecture Notes in Computer Science \#4639}, August 2007.

\bibitem[EL05]{elk-lip:c:polsci:universal-tweaks-coalitions}
E.~Elkind and H.~Lipmaa.
\newblock Small coalitions cannot manipulate voting\typeout{MINOR PANIC:
  elk-lip: missing dates of publication, pages, LNCS fvolume and remove to
  appeare}.
\newblock In {\em Proceedings of the 9th International Conference on Financial
  Cryptography and Data Security}, pages 285--297. Springer-Verlag {\it Lecture
  Notes in Computer Science \#3570}, February/March 2005.

\bibitem[ENR08a]{erd-now-rot:c:sp-av}
G.~Erd\'{e}lyi, M.~Nowak, and J.~Rothe.
\newblock Sincere-strategy preference-based approval voting broadly resists
  control.
\newblock In {\em Proceedings of the 33rd International Symposium on
  Mathematical Foundations of Computer Science}, pages 311--322.
  Springer-Verlag {\it Lecture Notes in Computer Science \#5162}, August 2008.

\bibitem[ENR08b]{erd-now-rot:t-With-MFCS08-Ptr:sp-av}
G.~Erd\'{e}lyi, M.~Nowak, and J.~Rothe.
\newblock Sincere-strategy preference-based approval voting fully resists
  constructive control and broadly resists destructive control.
\newblock Technical Report cs.GT/0806.0535, ACM Computing Research Repository
  (CoRR), June 2008.
\newblock A precursor appears as \cite{erd-now-rot:c:sp-av}.

\bibitem[ER97]{eph-ros:j:multiagent-planning}
E.~Ephrati and J.~Rosenschein.
\newblock A heuristic technique for multi-agent planning.
\newblock {\em Annals of Mathematics and Artificial Intelligence},
  20(1--4):13--67, 1997.

\bibitem[Fal08]{fal:c:nonuniform-bribery}
P.~Faliszewski.
\newblock Nonuniform bribery (short paper).
\newblock In {\em Proceedings of the 7th International Conference on Autonomous
  Agents and Multiagent Systems}, pages 1569--1572, May 2008.

\bibitem[FHH06a]{fal-hem-hem:c:bribery}
P.~Faliszewski, E.~Hemaspaandra, and L.~Hemaspaandra.
\newblock The complexity of bribery in elections.
\newblock In {\em Proceedings of the 21st National Conference on Artificial
  Intelligence}, pages 641--646. AAAI Press, July 2006.

\bibitem[FHH06b]{fal-hem-hem:tRevisedOutByConfToAppear:bribery}
P.~Faliszewski, E.~Hemaspaandra, and L.~Hemaspaandra.
\newblock How hard is bribery in elections?
\newblock Technical Report TR-895, Department of Computer Science, University
  of Rochester, Rochester, NY, April 2006.
\newblock Revised, September 2006.

\bibitem[FHHR]{fal-hem-hem-rot:btoappearWithTrPtr:richer}
P.~Faliszewski, E.~Hemaspaandra, L.~Hemaspaandra, and J.~Rothe.
\newblock A richer understanding of the complexity of election
  systems\typeout{Minor Panic: fal-hem-hem-rot--book: Missing pages, year,
  EXACT publisher name and book name, etc}.
\newblock In S.~Ravi and S.~Shukla, editors, {\em Fundamental Problems in
  Computing: {Essays} in Honor of {Professor} {Daniel} {J.} {Rosenkrantz}}.
  Springer.
\newblock To appear. Preliminary version available
  as~\cite{fal-hem-hem-rot:t:richer}.

\bibitem[FHHR06]{fal-hem-hem-rot:t:richer}
P.~Faliszewski, E.~Hemaspaandra, L.~Hemaspaandra, and J.~Rothe.
\newblock A richer understanding of the complexity of election systems.
\newblock Technical Report TR-903, Department of Computer Science, University
  of Rochester, Rochester, NY, September 2006.

\bibitem[FHHR07]{fal-hem-hem-rot:c:llull}
P.~Faliszewski, E.~Hemaspaandra, L.~Hemaspaandra, and J.~Rothe.
\newblock Llull and {C}opeland voting broadly resist bribery and control.
\newblock In {\em Proceedings of the 22nd AAAI Conference on Artificial
  Intelligence}, pages 724--730. AAAI Press, July 2007.

\bibitem[FHHR08]{fal-hem-hem-rot:c:llull-aaim}
P.~Faliszewski, E.~Hemaspaandra, L.~Hemaspaandra, and J.~Rothe.
\newblock Copeland voting fully resists constructive control.
\newblock In {\em Proceedings of the 4th International Conference on
  Algorithmic Aspects in Information and Management}, pages 165--176.
  Springer-Verlag {\it Lecture Notes in Computer Science \#5034}, June 2008.

\bibitem[FHS08]{fal-hem-sch:c:copeland-ties-matter}
P.~Faliszewski, E.~Hemaspaandra, and H.~Schnoor.
\newblock Copeland voting: Ties matter.
\newblock In {\em Proceedings of the 7th International Conference on Autonomous
  Agents and Multiagent Systems}, pages 983--990, May 2008.

\bibitem[Gib73]{gib:j:polsci:manipulation}
A.~Gibbard.
\newblock Manipulation of voting schemes.
\newblock {\em Econometrica}, 41(4):587--601, 1973.

\bibitem[GJ79]{gar-joh:b:int}
M.~Garey and D.~Johnson.
\newblock {\em Computers and Intractability: {A} Guide to the Theory of
  {NP}-Completeness}.
\newblock {W. H. Freeman and Company}, 1979.

\bibitem[GMHS99]{gho-mun-her-sen:c:voting-for-movies}
S.~Ghosh, M.~Mundhe, K.~Hernandez, and S.~Sen.
\newblock Voting for movies: {T}he anatomy of recommender systems.
\newblock In {\em Proceedings of the 3rd Annual Conference on Autonomous
  Agents}, pages 434--435. ACM Press, 1999.

\bibitem[HH]{hem-hom:jtoappearWithPtr:dodgson-greedy}
C.~Homan and L.~Hemaspaandra.
\newblock Guarantees for the success frequency of an algorithm for finding
  {Dodgson}-election winners\typeout{MINOR PANIC: hem-hom: missing issue, year,
  volume, number and remove the To-appear note}.
\newblock {\em Journal of Heuristics}.
\newblock To appear. Full version available
  as~\cite{hem-hom:t3OutByJour:dodgson-greedy}.

\bibitem[HH05]{hem-hom:t3OutByJour:dodgson-greedy}
C.~Homan and L.~Hemaspaandra.
\newblock Guarantees for the success frequency of an algorithm for finding
  {Dodgson}-election winners.
\newblock Technical Report TR-881, Department of Computer Science, University
  of Rochester, Rochester, NY, September 2005.
\newblock Revised, June 2007.

\bibitem[HH07]{hem-hem:j:dichotomy}
E.~Hemaspaandra and L.~Hemaspaandra.
\newblock Dichotomy for voting systems.
\newblock {\em Journal of Computer and System Sciences}, 73(1):73--83, 2007.

\bibitem[HHR07a]{hem-hem-rot:j:destructive-control}
E.~Hemaspaandra, L.~Hemaspaandra, and J.~Rothe.
\newblock Anyone but him: {The} complexity of precluding an alternative.
\newblock {\em Artificial Intelligence}, 171(5-6):255--285, April 2007.

\bibitem[HHR07b]{hem-hem-rot:c:hybrid}
E.~Hemaspaandra, L.~Hemaspaandra, and J.~Rothe.
\newblock Hybrid elections broaden complexity-theoretic resistance to control.
\newblock In {\em Proceedings of the 20th International Joint Conference on
  Artificial Intelligence}, pages 1308--1314. AAAI Press, January 2007.

\bibitem[HP01]{hae-puk:j:electoral-writings-ramon-llull}
G.~H{\"{a}}gele and F.~Pukelsheim.
\newblock The electoral writings of {Ramon} {Llull}.
\newblock {\em Studia Lulliana}, 41(97):3--38, 2001.

\bibitem[KP01]{ker-pau:j:fifa}
W.~Kern and D.~Paulusma.
\newblock The new {FIFA} rules are hard: {Complexity} aspects of sports
  competitions.
\newblock {\em Discrete Applied Mathematics}, 108(3):317--323, 2001.

\bibitem[{Len}83]{len:j:integer-fixed}
H.~{Lenstra, Jr.}
\newblock Integer programming with a fixed number of variables.
\newblock {\em Mathematics of Operations Research}, 8(4):538--548, 1983.

\bibitem[LN95]{lev-nal:j:voting-rules}
J.~Levin and B.~Nalebuff.
\newblock An introduction to vote-counting schemes.
\newblock {\em The Journal of Economic Perspectives}, 9(1):3--26, 1995.

\bibitem[McG53]{mcg:j:election-graph}
D.~McGarvey.
\newblock A theorem on the construction of voting paradoxes.
\newblock {\em Econometrica}, 21(4):608--610, 1953.

\bibitem[MPS08]{mcc-pri-sli:j:dodgeson}
J.~{McCabe-Dansted}, G.~Pritchard, and A.~Slinko.
\newblock Approximability of {D}odgson's rule.
\newblock {\em Social Choice and Welfare}, 31(2):311--330, 2008.

\bibitem[MS97]{mer-saa:j:copeland2}
V.~Merlin and D.~Saari.
\newblock {Copeland} method {II}: {M}anipulation, monotonicity, and paradoxes.
\newblock {\em Journal of Economic Theory}, 72(1):148--172, 1997.

\bibitem[MU95]{mcl-urk:b:polsci:classics}
I.~McLean and A.~Urken.
\newblock {\em Classics of Social Choice}.
\newblock University of Michigan Press, 1995.

\bibitem[Nie02]{nie:thesis-habilition:fixed-param}
R.~Niedermeier.
\newblock Invitation to fixed-parameter algorithms.
\newblock Habilitation thesis, University of T\"ubingen, 2002.

\bibitem[Nie06]{nie:b:invitation-fpt}
R.~Niedermeier.
\newblock {\em Invitation to Fixed-Parameter Algorithms}.
\newblock Oxford University Press, 2006.

\bibitem[PR07]{pro-ros:j:juntas}
A.~Procaccia and J.~Rosenschein.
\newblock Junta distributions and the average-case complexity of manipulating
  elections.
\newblock {\em Journal of Artificial Intelligence Research}, 28:157--181,
  February 2007.

\bibitem[PRK07]{pro-ros-kam:c:noisy}
A.~Procaccia, J.~Rosenschein, and G.~Kaminka.
\newblock On the robustness of preference aggregation in noisy environments.
\newblock In {\em Proceedings of the 6th International Joint Conference on
  Autonomous Agents and Multiagent Systems}, pages 416--422. ACM Press, May
  2007.

\bibitem[PRZ07]{pro-ros-zoh:c:multiwinner}
A.~Procaccia, J.~Rosenschein, and A.~Zohar.
\newblock Multi-winner elections: {Complexity} of manipulation, control, and
  winner-determination.
\newblock In {\em Proceedings of the 20th International Joint Conference on
  Artificial Intelligence}, pages 1476--1481. AAAI Press, January 2007.

\bibitem[PRZ08]{pro-ros-zoh:j:proportional-representation}
A.~Procaccia, J.~Rosenschein, and A.~Zohar.
\newblock On the complexity of achieving proportional representation.
\newblock {\em Social Choice and Welfare}, 30(3):353--362, 2008.

\bibitem[Sat75]{sat:j:polsci:manipulation}
M.~Satterthwaite.
\newblock Strategy-proofness and {Arrow's} conditions: Existence and
  correspondence theorems for voting procedures and social welfare
  functions\typeout{MINOR PANIC: sat: Missing issue number}.
\newblock {\em Journal of Economic Theory}, 10(2):187--217, 1975.

\bibitem[SM96]{saa-mer:j:copeland1}
D.~Saari and V.~Merlin.
\newblock The {Copeland} method {I}: {R}elationships and the dictionary.
\newblock {\em Economic Theory}, 8(1):51--76, 1996.

\bibitem[Ste59]{ste:j:election-graph}
R.~Stearns.
\newblock The voting problem.
\newblock {\em The American Mathematical Monthly}, 66(9):761--763, 1959.

\bibitem[Wag86]{wag:j:succinct}
K.~Wagner.
\newblock The complexity of combinatorial problems with succinct input
  representations.
\newblock {\em Acta Informatica}, 23(3):325--356, 1986.

\bibitem[Zer29]{zer:j:jech}
E.~Zermelo.
\newblock Die {B}erechnung der {T}urnier-{E}rgebnisse als ein
  {M}a\-xi\-mum\-problem der {W}ahr\-schein\-lich\-keits\-rechnung.
\newblock {\em Mathematische Zeitschrift}, 29(1):436--460, 1929.

\end{thebibliography}
